*A tribute to the inventors of the various capacity formulas of quantum channels.*

# Properties of the Quantum Channel

Laszlo Gyongyosi[*1,2], Sandor Imre[1]

[1]*Department of Telecommunications, Budapest University of Technology and Economics*

2 Magyar tudosok krt, H-1111, Budapest, Hungary

[2]*Information Systems Research Group, Mathematics and Natural Sciences, Hungarian Academy of Sciences*

H-1518, Budapest, Hungary

[*]*gyongyosi@hit.bme.hu*

**Quantum information processing exploits the quantum nature of information. It offers fundamentally new solutions in the field of computer science and extends the possibilities to a level that cannot be imagined in classical communication systems. For quantum communication channels, many new capacity definitions were developed in comparison to classical counterparts. A quantum channel can be used to realize classical information transmission or to deliver quantum information, such as quantum entanglement. In this paper we overview the properties of the quantum communication channel, the various capacity measures and the fundamental differences between the classical and quantum channels.**

# Contents

# 1. Introduction

According to Moore's Law, the physical limitations of classical semiconductor-based technologies could be reached by 2020. We will then step into the Quantum Age. When first quantum computers become available on the shelf, today's encrypted information will not remain secure. Classical computational complexity will no longer guard this information. Quantum communication systems exploit the quantum nature of information offering new possibilities and limitations for engineers when designing protocols.

The capacity of a communication channel describes the capability of the channel for delivering information from the sender to the receiver, in a faithful and recoverable way. The different capacities of quantum channels have been discovered just in the '90s, and there are still many open questions about the different capacity measures. Thanks to Shannon we can calculate the capacity of classical channels within the frames of classical information theory[1] [Shannon48]. However, for quantum channels, many new capacity definitions exist in comparison to a classical communication channel. In the case of a classical channel, we can send only classical information. Quantum channels extend the possibilities, and besides the classical information we can send entanglement-assisted classical information, private classical information, and of course, quantum information. On the other hand, the elements of classical information theory cannot be applied in general for quantum information—in other words, they can be used only in some special cases. There is no general formula to describe the capacity of every quantum channel model, but one of the main results of the recent researches was the "very simplified" picture, in which the various capacities of a quantum channel (i.e., the classical, private, quantum) *are all non-additive.*

In order to navigate the reader through the emerging field of quantum mechanics based information transmission this paper is organized as follows. In the remaining part of Section 1, we overview the general model of information transmission over quantum channels. In Section 2 we introduce the reader to the representation of information stored in quantum states according to quantum information theory. In Section 3, we study the classical information transmission capability and the classical capacity of a noisy quantum channel. In Section 4 we discuss the encoding and decoding of quantum information and the properties of the quantum capacity. Section 5 provides the classical and quantum capacities of some important quantum channels. In Section 6, we study the additivity problem of quantum channel capacities. Section 7 discusses the superactivation property. Section 8 introduces some practical applications of a quantum channel. Finally, we conclude the paper in Section 9. Supplementary material is included in the Appendix.

[1] *Quantum Shannon theory has deep relevance concerning the information transmission and storage in quantum systems. It can be regarded as a natural generalization of classical Shannon theory ore more precisely classical information theory represents a special, orthogonality-restricted case of quantum information theory.*

The complete 'historical' background with the description of the most relevant works can be found in the Related Work part of each section.

## 1.1 Communication over a Quantum Channel

Communication through a quantum channel cannot be described by the results of classical information theory; it requires the generalization of classical information theory by quantum perception of the world. In the general model of communication over a quantum channel $\mathcal{N}$, the encoder encodes the message in some coded form, and the receiver decodes it, however in this case, the whole communication is realized through a quantum system.

The information sent through quantum channels is carried by quantum states, hence the encoding is fundamentally different from any classical encoder scheme. The encoding here means the preparation of a quantum system, according to the probability distribution of the classical message being encoded. Similarly, the decoding process is also different: here it means the measurement of the received quantum state. The properties of quantum communication channel, and the fundamental differences between the classical and quantum communication channel cannot be described without the elements of quantum information theory.

The model of the quantum channel represents the physically allowed transformations which can occur on the sent qubit. The result of the channel transformation is another density matrix. The physically allowed channel transformations could be very different; nevertheless they are always *Completely Positive Trace Preserving* (CPTP) transformations. The trace preserving property means that the corresponding density matrices at the input and output of the channel have the same trace.
The input of a quantum channel is a quantum state, which encodes information into a physical property. The quantum state is sent through a quantum communication channel, which in practice can be implemented e.g. by an optical-fiber channel, or by a wireless quantum communication channel. To extract any information from the quantum state, it has to be measured at the receiver's side. The outcome of the measurement of the quantum state (which might be perturbed) depends on the transformation of the quantum channel, since it can be either totally probabilistic or deterministic. In contrast to classical channels, a quantum channel transforms the information coded into quantum states, which can be e.g. the spin state of the particle, the ground and excited state of an atom, or several other physical approaches.

Section 1 is organized as follows. In the first part we discuss the process of transmission of information over a quantum channel. Then, the interaction between quantum channel output and environment will be described. Later, we introduce the reader to the description of a noisy quantum channel and its environment.

### 1.1.1 The Quantum Channel

Besides the fact that the Bloch sphere provides a very useful geometrical approach to describe the density matrices, it also can be used to analyze the capacities of the various quantum channel models. From algebraic point of view, quantum channels are linear CPTP maps, while from a geometrical viewpoint, the quantum channel $\mathcal{N}$ is an affine transformation. While, from the algebraic view the transformations are defined on density matrices, in the geometrical approach, the transformations are interpreted as Bloch vectors. Since, density matrices can be expressed in terms of Bloch vectors, hence the map of a quantum channel $\mathcal{N}$ also can be analyzed in the geometrical picture.

The image of the quantum channel's linear transform is an *ellipsoid* on the Bloch sphere (see Fig. 1.1). To preserve the condition for a density matrix $\rho$, the noise on the quantum channel $\mathcal{N}$ must be trace-preserving (TP), i.e.,

$$Tr\left(\rho\right) = Tr\left(\mathcal{N}\left(\rho\right)\right), \tag{1.1}$$

and it must be Completely Positive (CP), i.e., for any identity map $I$, the map $I \otimes \mathcal{N}$ maps a semi-positive Hermitian matrix to a semi-positive Hermitian matrix.

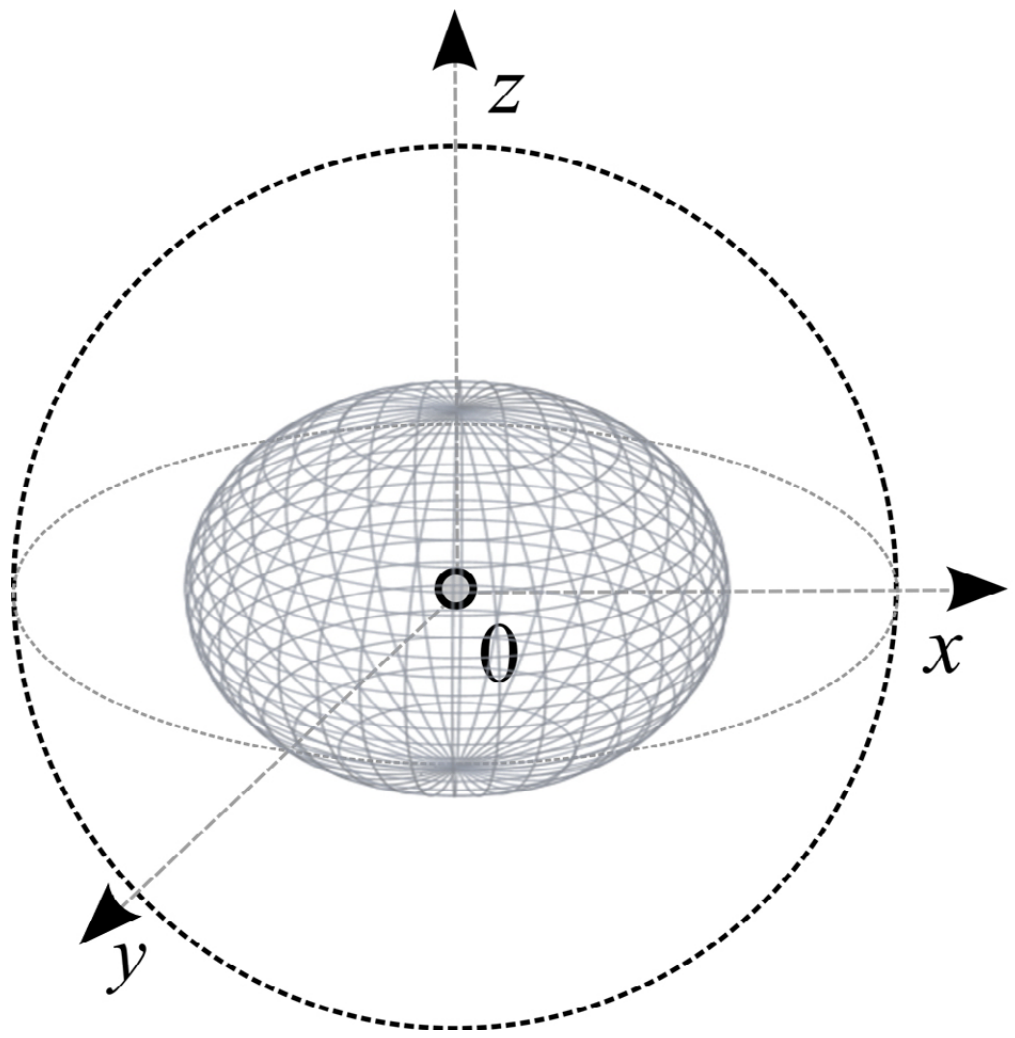

**Fig. 1.1.** Geometrically the image of the quantum channel is an ellipsoid.

In our paper, we will use the terms *unital* and *non-unital* quantum channels. This distinction means the following thing: for a unital quantum channel $\mathcal{N}$, the channel map transforms the $I$ identity transformation to the $I$ identity transformation, while this condition does not hold for a non-unital channel.
To express it, for a unital quantum channel, we have

$$\mathcal{N}\left(I\right) = I\,, \tag{1.2}$$

while for a non-unital quantum channel,

$$\mathcal{N}\left(I\right) \neq I\,. \tag{1.3}$$

As we will see, this difference can be rephrased in a geometrical interpretation, and the properties of the maps of the quantum channels can be analyzed using informational geometry.

For a unital quantum channel, the center of the geometrical interpretation of the channel ellipsoid is equal to the center of the Bloch sphere. This means that a unital quantum channel preserves the average of the system states. On the other hand, for a non-unital quantum channel, the center of the channel ellipsoid will differ from the center of the Bloch sphere. The main difference between unital and non-unital channels is that the non-unital channels do not preserve the average state in the center of the Bloch sphere. It follows from this that the numerical and algebraic analysis of non-unital quantum channels is more complicated than in the case of unital ones. While unital channels shrink the Bloch sphere in different directions with the center preserved, non-unital quantum channels shrink both the original Bloch sphere and move the center of the ball from the origin of the Bloch sphere. This fact makes our analysis more complex, however, in many cases, the physical systems cannot be described with unital quantum channel maps.

Unital channel maps can be expressed as convex combinations of the four unitary Pauli operators ($X$, $Y$, $Z$ and $I$), hence unital quantum maps are also called *Pauli channels*. Since the unital channel maps can be expressed as the convex combination of the basic unitary transformations, the unital channel maps can be represented in the Bloch sphere as different rotations with shrinking parameters. On the other hand, for a non-unital quantum map, the map cannot be decomposed into a convex combination of unitary rotations and the transformation not just shrinks the ball, but also moves its center from the origin of the Bloch sphere [Imre12].

The geometrical interpretation of a unital and a non-unital quantum channels are illustrated in Fig. 1.2.

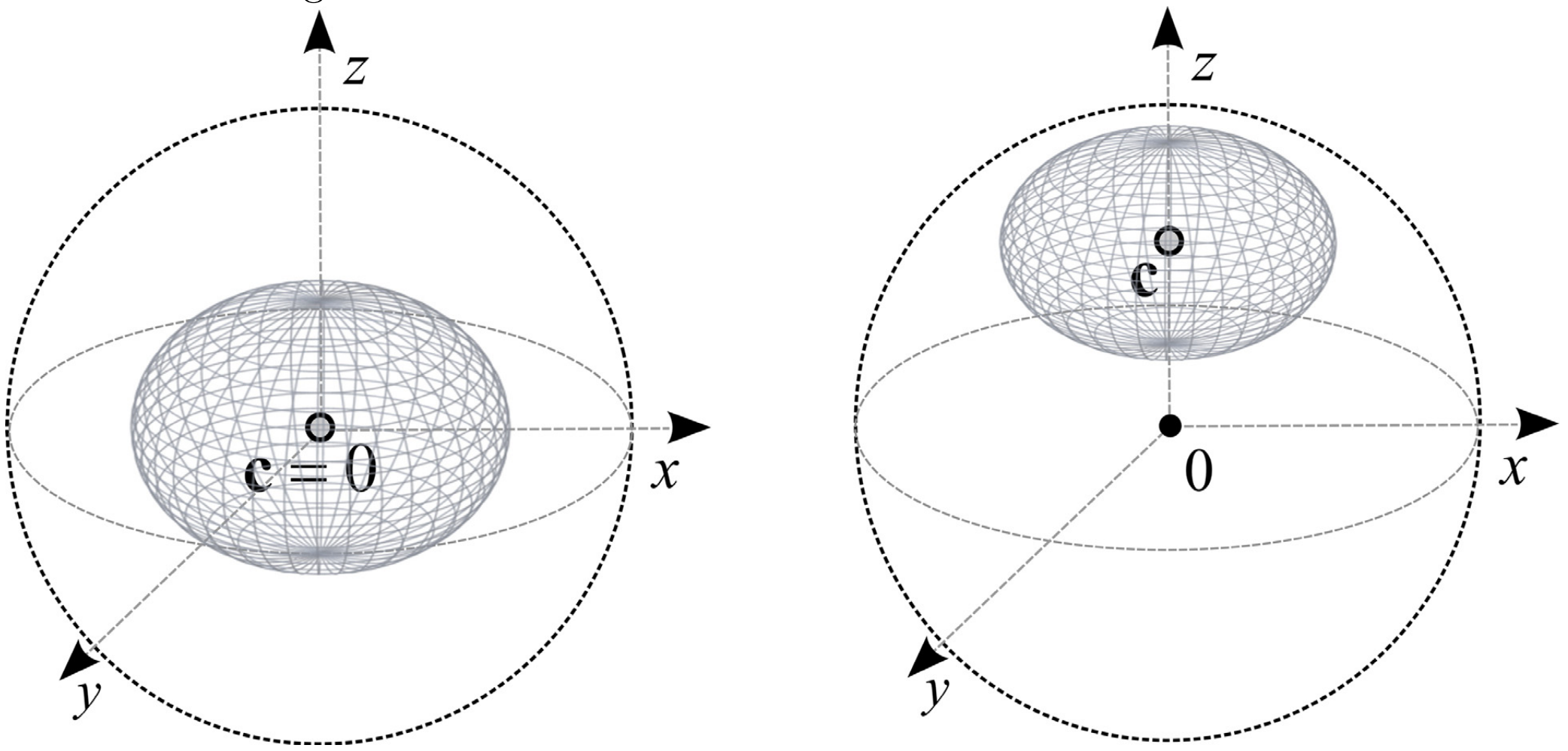

**Fig. 1.2.** The geometrical interpretation of a unital and a non-unital quantum channels.

The unital channel maps can be expressed as convex combinations of the basic unitary transformations, while non-unital quantum maps cannot be decomposed into a convex combination of unitary rotations, because of the geometrical differences between the two kinds of maps. The geometrical approaches can help to reduce the complexity of the analysis of the different quantum channel models, and as we will show, many algebraic results can be converted into geometrical problems. The connection between the channel maps and their geometrical interpretation on the Bloch sphere makes it possible to give a simpler and more elegant solution for several hard, and still unsolved problems.

### 1.1.2 Steps of the Communication

The transmission of information through classical channels and quantum channels differs in many ways. If we would like to describe the process of information transmission through a quantum communication channel, we have to introduce the three main phases of quantum communication. In the first phase, the sender, Alice, has to encode her information to compensate the noise of the channel $\mathcal{N}$ (i.e., for error correction), according to properties of the physical channel - this step is called *channel coding*. After the sender has encoded the information into the appropriate form, it has to be put on the quantum channel, which transforms it according to its channel map - this second phase is called the *channel evolution*. The quantum channel $\mathcal{N}$ conveys the quantum state to the receiver, Bob; however this state is still a superposed and probably *mixed* (according to the noise of the channel) quantum state. To extract the information which is encoded in the state, the receiver has to make a measurement - this *decoding process* (with the error correction procedure) is the third phase of the communication over a quantum channel.

In Fig. 1.3, we illustrate the channel coding phase. In case of transmission of classical information over a noisy quantum channel, Alice encodes her information into a physical attribute of a physical particle, such as the spin or polarization. For example, in the case of an electron or a half-spin particle, this can be an axis spin.

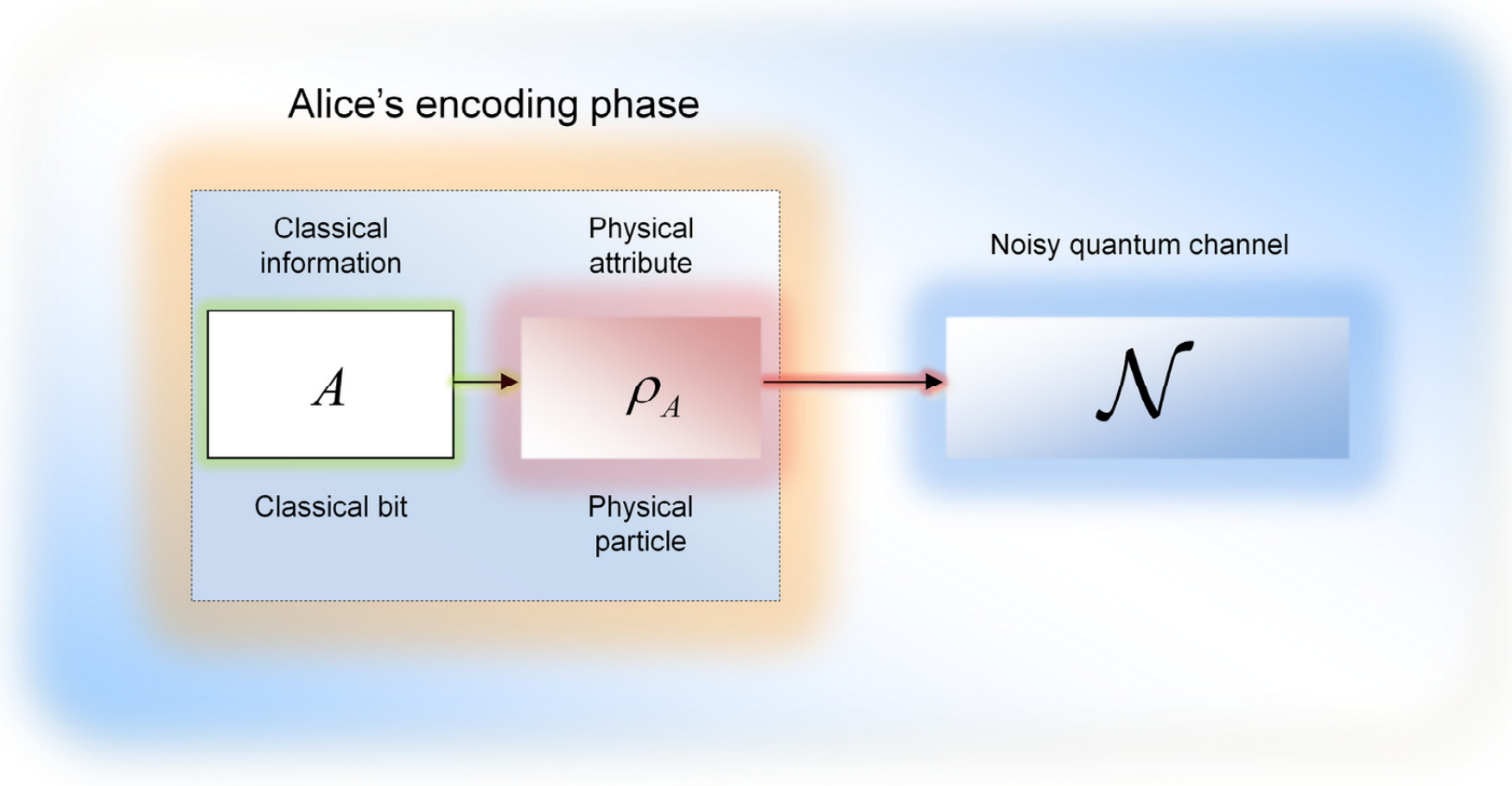

**Fig. 1.3.** The channel coding phase.

The channel transformation represents the noise of the quantum channel. Physically, the quantum channel is the medium, which moves the particle from the sender to the receiver. The noise disturbs the state of the particle, in the case of a half-spin particle, it causes spin precession. The channel evolution phase is illustrated in Fig. 1.4.

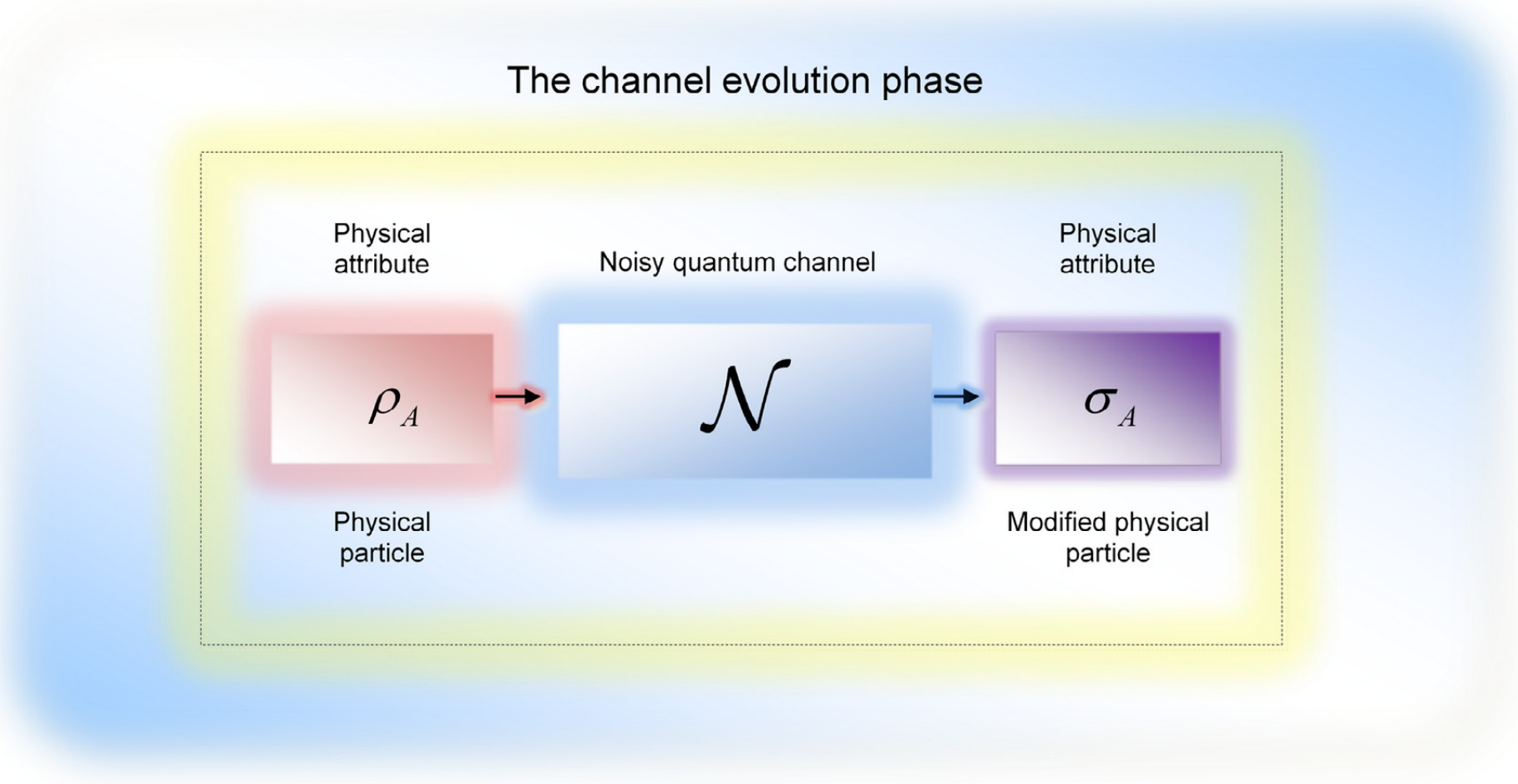

**Fig. 1.4.** The channel evolution phase.

Finally, the measurement process responsible for the decoding and the extraction of the encoded information. The previous phase determines the success probability of the recovery of the original information. If the channel $\mathcal{N}$ is completely noisy, then the receiver will get a maximally mixed quantum state. The output of the measurement of a maximally mixed state is completely undeterministic: it tells us nothing about the original information encoded by the sender. On the other hand, if the quantum channel $\mathcal{N}$ is completely noiseless, then the information which was encoded by the sender can be recovered with probability 1: the result of the measurement will be completely deterministic and completely correlated with the original message. In practice, a quantum channel realizes a map which is in between these two extreme cases.

A general quantum channel transforms the original pure quantum state into a mixed quantum state, - but not into a maximally mixed state - which makes it possible to recover the original message with a high - or low - probability, depending on the level of the noise of the quantum channel $\mathcal{N}$. The measurement phase is illustrated in Fig. 1.5.

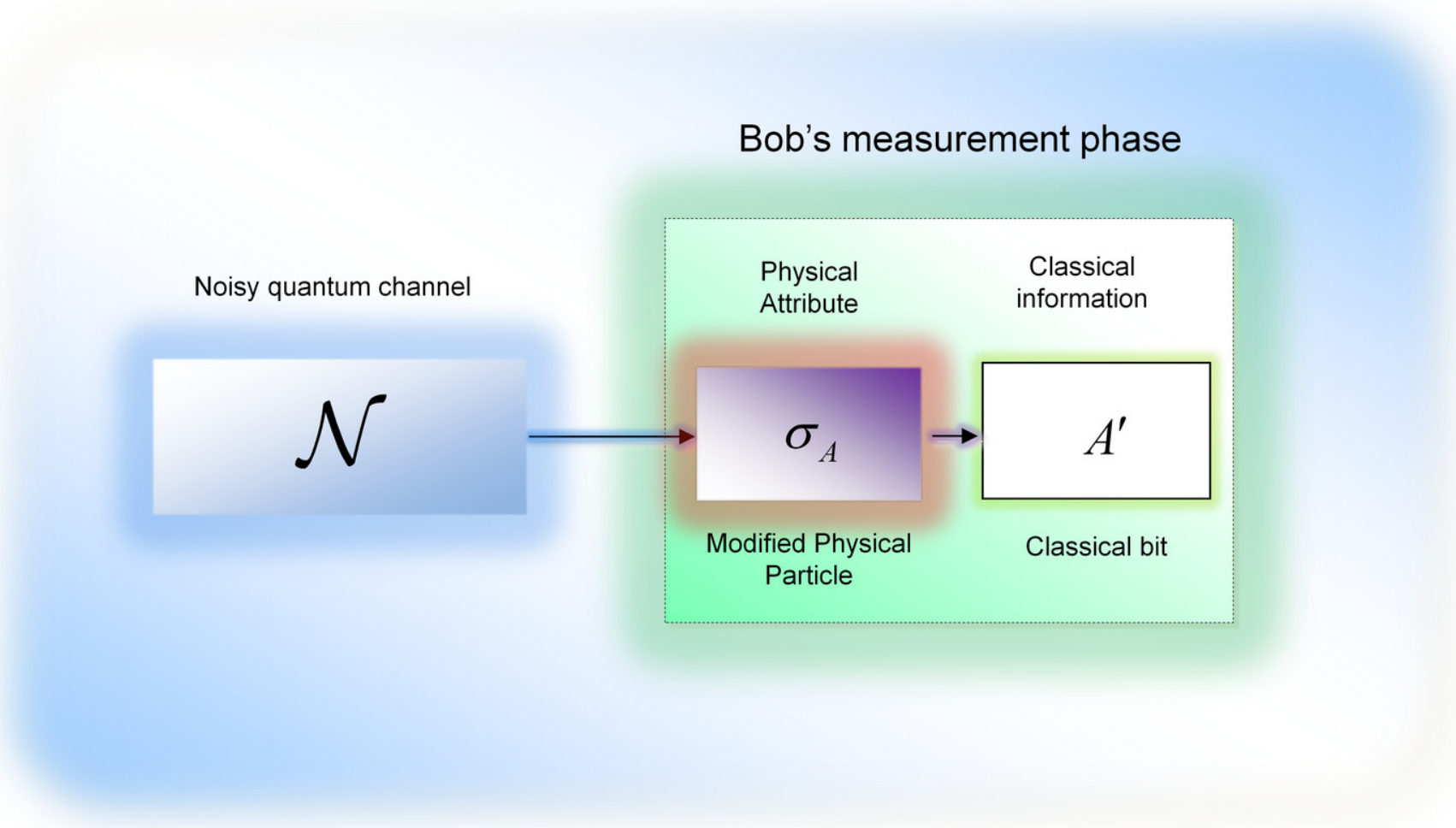

**Fig. 1.5.** The decoding process.

Quantum communication channels can be divided into many different classes. As we will present in Section 3.6, the various channel models modify the sent qubits in different ways.

## 1.2 Formal Model

As shown in Fig. 1.6, the information transmission through the quantum channel $\mathcal{N}$ is defined by the $\rho_{in}$ input quantum state and the initial state of the environment $\rho_E = |0\rangle\langle 0|$. In the initial phase, the environment is assumed to be in the pure state $|0\rangle$. The system state which consist of the input quantum state $\rho_{in}$ and the environment $\rho_E = |0\rangle\langle 0|$, is called the *composite state* $\rho_{in} \otimes \rho_E$.

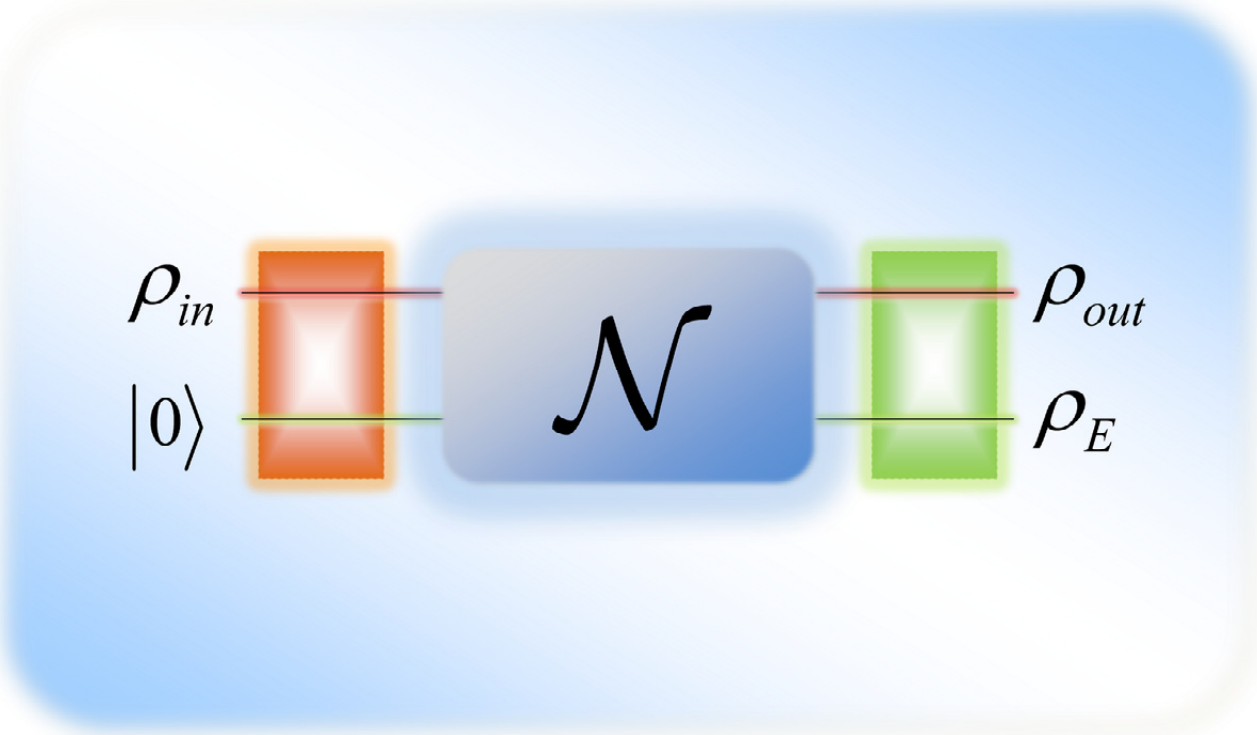

**Fig. 1.6.** The general model of transmission of information over a noisy quantum channel.

If the quantum channel $\mathcal{N}$ is used for information transmission, then the state of the composite system changes unitarily, as follows:

$$U\left(\rho_{in} \otimes \rho_E\right)U^{\dagger}, \tag{1.4}$$

where $U$ is a unitary transformation, and $U^{\dagger}U = I$. After the quantum state has been sent over the quantum channel $\mathcal{N}$, the $\rho_{out}$ output state can be expressed as:

$$\mathcal{N}\left(\rho_{in}\right) = \rho_{out} = Tr_E\left[U\left(\rho_{in} \otimes \rho_E\right)U^{\dagger}\right], \tag{1.5}$$

where $Tr_E$ traces out the environment $E$ from the joint state. Assuming the environment $E$ in the pure state $\left|0\right\rangle$, $\rho_E = \left|0\right\rangle\left\langle 0\right|$, the $\mathcal{N}\left(\rho_{in}\right)$ noisy evolution of the channel $\mathcal{N}$ can be expressed as:

$$\mathcal{N}\left(\rho_{in}\right) = \rho_{out} = Tr_E U\rho_{in} \otimes \left|0\right\rangle\left\langle 0\right|U^{\dagger}, \tag{1.6}$$

while the post-state $\rho_E$ of the environment after the transmission is

$$\rho_E = Tr_B U\rho_{in} \otimes \left|0\right\rangle\left\langle 0\right|U^{\dagger}, \tag{1.7}$$

where $Tr_B$ traces out the output system $B$. In general, the $i$-th input quantum state $\rho_i$ is prepared with probability $p_i$, which describes the ensemble $\left\{p_i, \rho_i\right\}$. The average of the *input* quantum system is

$$\sigma_{in} = \sum_i p_i \rho_i, \tag{1.8}$$

The average (or the mixture) of the *output* of the quantum channel is denoted by

$$\sigma_{out} = \mathcal{N}\left(\sigma_{in}\right) = \sum_i p_i \mathcal{N}\left(\rho_i\right). \tag{1.9}$$

## 1.3 Quantum Channel Capacity

The capacity of a communication channel describes the capability of the channel for sending information from the sender to the receiver, in a faithful and recoverable way. The perfect ideal communication channel realizes an identity map. For a quantum communication channel, it means that the channel can transmit the quantum states perfectly. Clearly speaking, the capacity of the quantum channel measures the closeness to the ideal identity transformation $I$.

To describe the information transmission capability of the quantum channel $\mathcal{N}$, we have to make a distinction between the various capacities of a quantum channel. The encoded quantum states can carry classical messages or quantum messages. In the case of classical messages, the quantum states encode the output from a *classical information source*, while in the latter the source is a *quantum information source*.

On one hand for classical communication channel $N$, only one type of capacity measure can be defined, on the other hand for a quantum communication channel $\mathcal{N}$ a number of different types of quantum channel capacities can be applied, with different characteristics. There are plenty of open questions regarding these various capacities.

In general, the *single-use* capacity of a quantum channel is not equal to the *asymptotic* capacity of the quantum channel (As we will see later, it also depends on the type of the quantum channel). The asymptotic capacity gives us the amount of information which can be transmitted in a reliable form using the quantum channel infinitely many times. This fundamental difference makes it possible to use the quantum channel for information transmission in those scenarios, which are completely unimaginable for a classical channel.

The encoding and the decoding functions mathematically can be described by the operators $\mathcal{E}$ and $\mathcal{D}$, realized on the blocks of quantum states. These superoperators describe unitary transformations on the input states together with the environment of the quantum system. The model of communication through noisy quantum channel $\mathcal{N}$ with encoding, delivery and decoding phases is illustrated in Fig. 1.7.

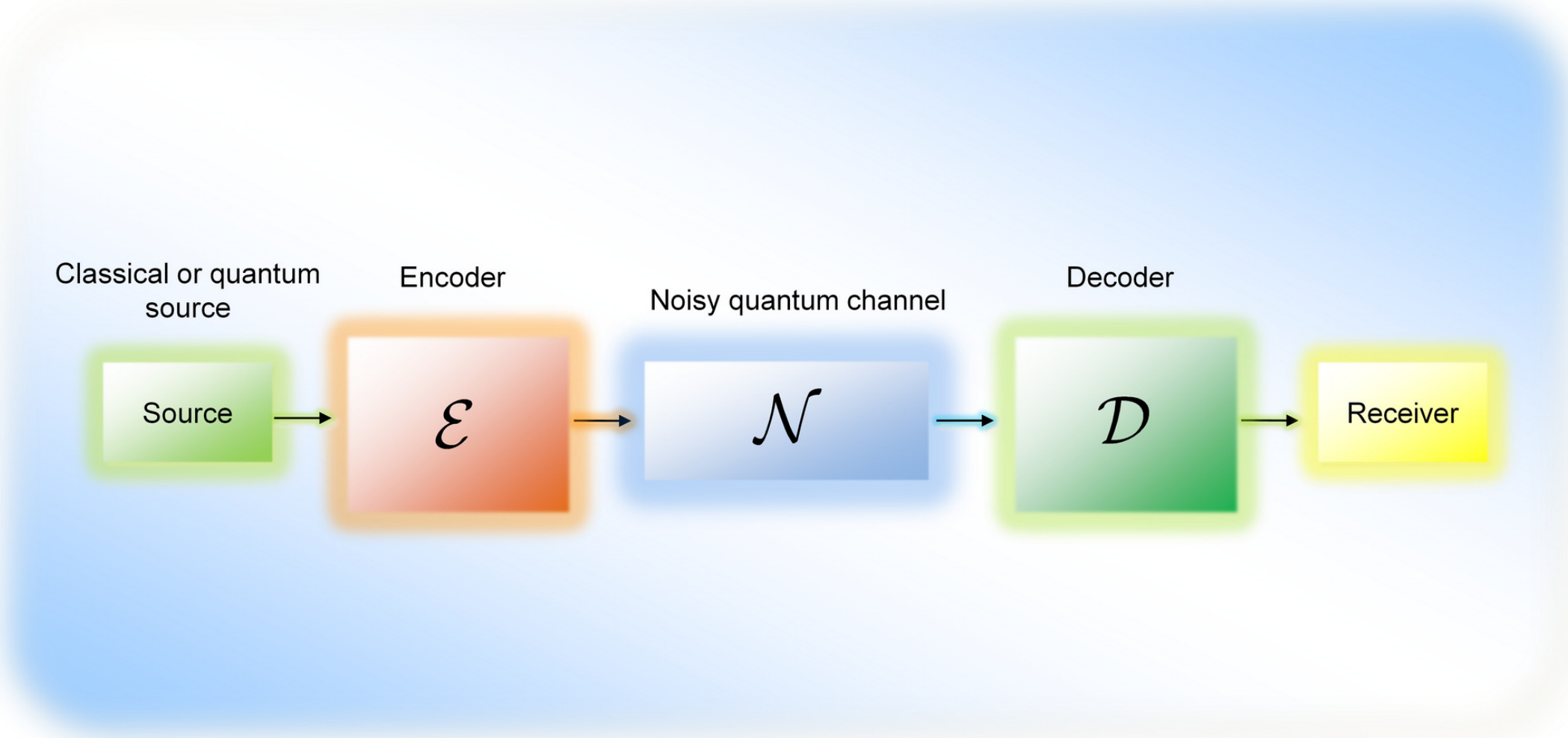

**Fig. 1.7.** Communication over noisy quantum channel. According to the noise of the quantum channel, the pure input state becomes a mixed state.

We note, in our paper we will use the terms *classical quantity* and *quantum quantity* with relation to the quantum channel $\mathcal{N}$ as follows:

- *classical quantity*: it is a measure of the *classical transmission capabilities* of a quantum channel. (For example the Holevo information, quantum mutual information, etc.)

- *quantum quantity*: it is a measure of the *quantum transmission capabilities* of a quantum channel. (For example the quantum coherent information, see Section 4.)

If we mention classical quantity we will do this with relation to the quantum channel $\mathcal{N}$, i.e., for example the Holevo information is also not a "typical" classical quantity since it is describes a quantum system not a classical one, but with relation to the quantum channel we can use the *classical* mark.

# 2. Basic Definitions

Quantum information theory also has relevance to the discussion of the capacity of quantum channels and to information transmission and storage in quantum systems. As we will see in this section, while the transmission of product states can be described similar to classical information, on the other hand, the properties of quantum entanglement cannot be handled by the elements of classical information theory. Of course, the elements of classical information theory can be viewed as a subset of the larger and more complex quantum information theory.

Section 2 is organized as follows. In the first part, we summarize the basic definitions and formulas of quantum information theory. We introduce the reader to the description of a noisy quantum channel, purification, isometric extension, Kraus representation and the von Neumann entropy. Next, we describe the encoding of quantum states and the meaning of Holevo information, the quantum mutual information and quantum conditional entropy.
The complete 'historical' background with the description of the most relevant works can be found in the Related Work subsection.

## 2.1 Preliminaries

Before starting the discussion on various capacities of quantum channels and the related consequences we summarize the basic definitions and formulas of quantum information theory intended to represent the information stored in quantum states. Those readers who are familiar with density matrices, entropies etc. may run through this section and focus only on notations (the collection of the notations can be found in the List of Notations) and can return later if it is required when processing another section.

The world of quantum information processing (QIP) is describable with the help of quantum information theory (QIT), which is the main subject of this section. We will provide an overview of the most important differences between the compressibility of classical bits and quantum bits, and between the capacities of classical and quantum communication channels. To represent classical information with quantum states, we might use pure orthogonal states. In this case there is no difference between the compressibility of classical and quantum bits. But what happens if we use non-orthogonal quantum states?

Similarly, a quantum channel can be used with pure orthogonal states to realize classical information transmission, or it can be used to transmit non-orthogonal states or even quantum entanglement. Information transmission also can be approached using the question, whether the input consists of unentangled or entangled quantum states. This leads us to say that for quantum channels many new capacity definitions exist in comparison to a classical communication channel.

Quantum information theory also has relevance to the discussion of the capacity of quantum channels and to information transmission and storage in quantum systems. While the transmission of product states can be described similar to classical information, on the other hand, the properties of quantum entanglement cannot be handled by the elements of classical information theory. Of course, the elements of classical information theory can be viewed as a subset of the larger and more complex quantum information theory.

Before we would start to our introduction to quantum information theory, we have to make a clear distinction between quantum information theory and quantum information processing. Quantum information theory is rather a generalization of the elements and functions of classical information theory to describe the properties of quantum systems, storage of information in quantum systems and the various quantum phenomena of quantum mechanics. While quantum information theory aims to provide a stable theoretical background, quantum information processing is a more general and rather experimental field: it answers what can be achieved in engineering with the help of quantum information. Quantum information processing includes the computing, error-correcting schemes, quantum communication protocols, field of communication complexity, etc.

The character of classical information and quantum information is significantly different. There are many phenomena in quantum systems which cannot be described classically, such as entanglement, which makes it possible to store quantum information in the correlation of quantum states. Entangled quantum states are named to EPR states after Einstein, Podolsky and Rosen, or Bell states, after J. Bell. Quantum entanglement was discovered in the 1930s, and it may still yield many surprises in the future. Currently it is clear that entanglement has many classically indescribable properties and many new communication approaches based on it. Quantum entanglement plays a fundamental role in advanced quantum communications, such as teleportation, quantum cryptography etc.

The elements of quantum information theory are based on the laws of quantum mechanics. The main results of quantum information processing were laid down during the end of the twentieth century, the most important results being stated by Feynman, Bennett, DiVincenzo, Devetak, Deutsch, Holevo, Lloyd, Schumacher, Shor and many more. After the basic concepts of quantum information processing had been stated, researchers started to look for efficient quantum error correction schemes and codes, and started to develop the theoretical background of fault-tolerant quantum computation. The main results from this field were presented by Bennett, Schumacher, Gottesman, Calderbank, Preskill, Knill, and Kerckhoff. On the other hand, there are still many open questions about quantum computation. The theoretical limits of quantum computers were discovered by Bennett, Bernstein, Brassard and Vazirani: quantum computers can provide at best a quadratic reduction in the complexity of search-based problems, hence if we give an NP-complete problem to quantum computer, it still cannot solve it. Recently, the complexity classes of quantum information processing have been investigated, and many new classes and lower bounds have been found.

By the end of the twentieth century, many advanced and interesting properties of quantum information theory had been discovered, and many possible applications of these results in future communication had been developed. One of the most interesting revealed connections was that between quantum information theory and the elements of geometry. The space of quantum states can be modeled as a convex set which contains points with different probability distributions, and the geometrical distance between these probability distributions can be measured by the elementary functions of quantum information theory, such as the von Neumann entropy or the quantum relative entropy function. The connection between the elements of quantum information theory and geometry leads us to the application of advanced computational geometrical algorithms to quantum space, to reveal the still undiscovered properties of quantum information processing, such as the open questions on the capacities of the quantum channels or their additivity properties. The connection between the Hilbert space of quantum states and the geometrical distance can help us to reveal the fantastic properties of quantum bits and quantum state space.

Several functions have been defined in quantum information theory to describe the statistical distances between the states in the quantum space: one of the most important is the quantum relative entropy function which plays a key role in the description of entanglement, too. This function has many different applications, and maybe this function plays the most important role in the questions regarding the capacity of quantum channels. The possible applications of the quantum relative entropy function have been studied by Schumacher and Westmoreland and by Vedral.

Quantum information theory plays fundamental role in the description of the data transmission through quantum communication channels. At the dawn of this millennium new problems have arisen, whose solutions are still not known, and which have opened the door to many new promising results such as the superactivation of zero-capacity quantum channels in 2008, and then the superactivation of the zero-error capacities of the quantum channels in 2009 and 2010. One of the earliest works on the capacities of quantum communication channels was published in the early 1970s. Along with other researchers, Holevo was showed that there are many differences between the properties of classical and quantum communication channels, and illustrated this with the benefits of using entangled input states. Later, he also stated that quantum communication channels can be used to transmit both classical and quantum information. Next, many new quantum protocols were developed, such as teleportation or superdense coding. After Alexander Holevo published his work, about thirty years later, he, with Benjamin Schumacher and Michael Westmoreland presented one of the most important result in quantum information theory, called the *Holevo-Schumacher-Westmoreland* (HSW) theorem [Holevo98], [Schumacher97]. The HSW-theorem is a generalization of the classical noisy channel coding theorem from classical information theory to a noisy quantum channel. The HSW theorem is also called the *product-state* classical channel capacity theorem of a noisy quantum channel. The understanding of the classical capacity of a quantum channel was completed by 1997 by Schumacher and Westmoreland, and by 1998 by Holevo, and it has tremendous relevance in quantum information theory, since it was the first to give a mathematical proof that a noisy quantum channel can be used to transmit classical information in a reliable form. The HSW theorem was a very important result in the history of quantum information theory, on the other hand it raised a lot of questions regarding the transmission of classical information over general quantum channels.

The quantum capacity of a quantum channel was firstly formulated by Seth Lloyd in 1996, then by Peter Shor in 2002, finally it was completed by Igor Devetak in 2003, - they result is known as the *LSD channel capacity* [Lloyd97], [Devetak03], [Shor02]. While the classical capacity of a quantum channel is described by the maximum of quantum mutual information and the Holevo information, the quantum capacity of the quantum channels is described by a completely different correlation measure: called the *quantum coherent information*. The concept of quantum coherent information plays a fundamental role in the computation of the quantum capacity which measures the asymptotic quantum capacity of the quantum capacity in general. For the complete historical background with the references see the Related Works.

### 2.1.1 Density Matrices and Trace Operator

In this section we introduce a basic concept of quantum information theory, called the *density matrix*.

Before we start to discuss the density operator, we introduce some terms. An $n \times n$ square matrix $A$ is called *positive-semidefinite* if $\langle \psi | A | \psi \rangle$ is a non-negative real number for every vector $|\psi\rangle$. If $A = A^{\dagger}$, i.e., $A$ has Hermitian matrix and the $\{\lambda_1, \lambda_2, \ldots \lambda_n\}$ eigenvalues of $A$ are all non-negative real numbers then it is positive-semidefinite. This definition has important role in quantum information theory, since *every density matrix is positive-semidefinite.* It means, for any vector $|\varphi\rangle$ the positive-semidefinite property says that

$$\langle \varphi | \rho | \varphi \rangle = \sum_{i=1}^{n} p_i \langle \varphi | \psi_i \rangle \langle \psi_i | \varphi \rangle = \sum_{i=1}^{n} p_i \left| \langle \varphi | \psi_i \rangle \right|^2 \geq 0 . \tag{2.1}$$

In (2.1) we used, the density matrix is denoted by $\rho$, and it describes the system by the classical probability weighted sum of possible states

$$\rho = \sum_i p_i |\psi_i\rangle\langle\psi_i| , \tag{2.2}$$

where $|\psi_i\rangle$ is the $i$-th system state occurring with classical probability $p_i$. As can be seen, this density matrix describes the system as a probabilistic mixture of the possible known states the so called *pure states*. For pure state $|\psi\rangle$ the density matrix is $\rho = |\psi\rangle\langle\psi|$ and the rank of the matrix is equal to one. Trivially, classical states e.g. $|0\rangle$ and $|1\rangle$ are pure, however, if we know that our system is prepared to the *superposition* $\frac{1}{\sqrt{2}}\left(\left|0\right\rangle + \left|1\right\rangle\right)$ then this state is pure, too. Clearly speaking, while superposition is a quantum linear combination of orthonormal basis states weighted by probability amplitudes, mixed states are classical linear combination of pure superpositions (quantum states) weighted by classical probabilities.

The density matrix contains all the possible information that can be extracted from the quantum system. It is possible that two quantum systems possess the same density matrices: in this case, these quantum systems are called indistinguishable, since it is not possible to construct a measurement setting, which can distinguish between the two systems.

The density matrix $\rho$ of a simple pure quantum system which can be given in the state vector representation $\left|\psi\right\rangle = \alpha\left|0\right\rangle + \beta\left|1\right\rangle$ can be expressed as the outer product of the *ket* and *bra* vectors, where bra is the transposed complex conjugate of ket, hence for $\left|\psi\right\rangle = \begin{bmatrix}\alpha \\ \beta\end{bmatrix}$, $\left\langle\psi\right| = \begin{bmatrix}\alpha^* & \beta^*\end{bmatrix}$ the density matrix is

$$\rho = \left|\psi\right\rangle\left\langle\psi\right| = \begin{bmatrix}\alpha \\ \beta\end{bmatrix}\begin{bmatrix}\alpha^* & \beta^*\end{bmatrix} = \begin{bmatrix}\alpha\alpha^* & \alpha\beta^* \\ \alpha^*\beta & \beta\beta^*\end{bmatrix} = \begin{bmatrix}\left|\alpha\right|^2 & \alpha\beta^* \\ \alpha^*\beta & \left|\beta\right|^2\end{bmatrix}. \tag{2.3}$$

The density matrix $\rho = \sum_{i=1}^{n} p_i \left|\psi_i\right\rangle\left\langle\psi_i\right|$ contains the probabilistic mixture of different pure states, which representation is based on the fact that the mixed states can be decomposed into weighted sum of pure states [Watrous06].

To reveal important properties of the density matrix, we introduce the concept of the *trace operation.* The trace of a density matrix is equal to the sum of its diagonal entries. For an $n \times n$ square matrix $A$, the $Tr$ trace operator is defined as

$$Tr\left(A\right) = a_{11} + a_{22} + \ldots + a_{nn} = \sum_{i=1}^{n} a_{ii}\,, \tag{2.4}$$

where $a_{ii}$ are the elements of the main diagonal. The trace of the matrix $A$ is also equal to the sum of the *eigenvalues* of its matrix. The eigenvalue is the factor by which the *eigenvector* changes if it is multiplied by the matrix $A$, for each eigenvectors. The *eigenvectors* of the square matrix $A$ are those non-zero vectors, whose direction remain the same to the original vector after multiplied by the matrix $A$. It means, the eigenvectors remain proportional to the original vector. For square matrix $A$, the non-zero vector $\mathbf{v}$ is called *eigenvector* of $A$, if there is a scalar $\lambda$ for which

$$A\mathbf{v} = \lambda\mathbf{v}\,, \tag{2.5}$$

where $\lambda$ is the *eigenvalue* of $A$ corresponding to the eigenvector $\mathbf{v}$.

The trace operation gives us the sum of the eigenvalues of positive-semidefinite $A$, for each eigenvectors, hence

$$Tr\left(A\right) = \sum_{i=1}^{n} \lambda_i \text{ , and } Tr\left(A^k\right) = \sum_{i=1}^{n} \lambda_i^k \,. \tag{2.6}$$

Using the eigenvalues, the *spectral decomposition* of density matrix $\rho$ can be expressed as

$$\rho = \sum_i \lambda_i \left|\varphi_i\right\rangle\left\langle\varphi_i\right| \,, \tag{2.7}$$

where $\left|\varphi_i\right\rangle$ are orthonormal vectors.

The trace is a linear map, hence for square matrices $A$ and $B$

$$Tr\left(A+B\right) = Tr\left(A\right) + Tr\left(B\right), \tag{2.8}$$

and

$$Tr\left(sA\right) = sTr\left(A\right), \tag{2.9}$$

where $s$ is a scalar. Another useful formula, that for $m \times n$ matrix $A$ and $n \times m$ matrix $B$,

$$Tr\left(AB\right) = Tr\left(BA\right), \tag{2.10}$$

which holds for any matrices $A$ and $B$ for which the product matrix $AB$ is a square matrix, since

$$Tr\left(AB\right) = \sum_{i=1}^{m}\sum_{j=1}^{n} A_{ij}B_{ji} = Tr\left(BA\right). \tag{2.11}$$

Finally, we mention that the trace of a matrix $A$ and the trace of its transpose $A^T$ are equal, hence

$$Tr\left(A\right) = Tr\left(A^T\right). \tag{2.12}$$

If we take the conjugate transpose $A^*$ of the $m \times n$ matrix $A$, then we will find that

$$Tr\left(A^* A\right) \geq 0\,, \tag{2.13}$$

which will be denoted by $\left\langle A, A \right\rangle$ and it is called the *inner product*. For matrices $A$ and $B$, the inner product $\left\langle A, B \right\rangle = Tr\left(B^* A\right)$, which can be used to define the angle between the two vectors. The inner product of two vectors will be zero if and only if the vectors are orthogonal.
As we have seen, the trace operation gives the sum of the eigenvalues of matrix $A$, this property can be extended to the density matrix, hence for each eigenvectors $\lambda_i$ of density matrix $\rho$

$$Tr\left(\rho\right) = \sum_{i=1}^{n} \lambda_i\,. \tag{2.14}$$

Now, having introduced the *trace* operation, we apply it to a density matrix. If we have an $n$-qubit system being in the state $\rho = \sum_{i=1}^{n} p_i \left|\psi_i\right\rangle\left\langle\psi_i\right|$ then

$$Tr\left(\sum_{i=1}^{n} p_i \left|\psi_i\right\rangle\left\langle\psi_i\right|\right) = \sum_{i=1}^{n} p_i Tr\left(\left|\psi_i\right\rangle\left\langle\psi_i\right|\right) = \sum_{i=1}^{n} p_i \left(\left\langle\psi_i\middle|\psi_i\right\rangle\right) = 1\,, \tag{2.15}$$

where we exploited the relation for unit-length vectors $\left|\psi_i\right\rangle$

$$\left\langle\psi_i\middle|\psi_i\right\rangle \equiv 1\,. \tag{2.16}$$

Thus the trace of any density matrix is equal to one

$$Tr\left(\rho\right) = 1\,. \tag{2.17}$$

The trace operation can help to distinguish *pure* and *mixed* states since for a given *pure* state $\rho$

$$Tr\left(\rho^2\right) = 1\,, \tag{2.18}$$

while for a *mixed* state $\sigma$,

$$Tr\left(\sigma^2\right) < 1\,. \tag{2.19}$$

where $Tr\left(\rho^2\right) = \sum_{i=1}^{n} \lambda_i^2$ and $Tr\left(\sigma^2\right) = \sum_{i=1}^{n} \omega_i^2$, where $\omega_i$ are the eigenvalues of density matrix $\sigma$.

Similarly, for a pure *entangled* system $\rho_{EPR}$

$$Tr\left(\rho_{EPR}^2\right) = 1, \tag{2.20}$$

while for any mixed subsystem $\sigma_{EPR}$ of the entangled state (i.e., for a half-pair of the entangled state), we will have

$$Tr\left(\sigma_{EPR}^2\right) < 1\,. \tag{2.21}$$

The density matrix also can be used to describe the effect of a unitary transform on the probability distribution of the system. The probability that the whole quantum system is in $\left|\psi_i\right\rangle$ can be calculated by the trace operation. If we apply unitary transform $U$ to the state $\rho = \sum_{i=1}^{n} p_i \left|\psi_i\right\rangle\left\langle\psi_i\right|$, the effect can be expressed as follows:

$$\sum_{i=1}^{n} p_i \left(U\left|\psi_i\right\rangle\right)\left(\left\langle\psi_i\right|U^\dagger\right) = U\left(\sum_{i=1}^{n} p_i \left|\psi_i\right\rangle\left\langle\psi_i\right|\right)U^\dagger = U\rho U^\dagger. \tag{2.22}$$

What will happen with the density matrix if the applied transformation is not unitary? To describe this case, we introduce a more general operator denoted by $G$, and with the help of this operator the transform can be written as

$$G(\rho) = \sum_{i=1}^{n} A_i \rho A_i^{\dagger} = \sum_{i=1}^{n} A_i \left( p_i \left| \psi_i \right\rangle \left\langle \psi_i \right| \right) A_i^{\dagger}, \tag{2.23}$$

where $\sum_{i=1}^{n} A_i A_i^{\dagger} = I$, for every matrices $A_i$. In this sense, operator $G$ describes the physically admissible or *Completely Positive Trace Preserving* (CPTP) operations. The application of a CPTP operator $G$ on density matrix $\rho$ will result in a matrix $G(\rho)$, which in this case is still a density matrix.

Now we can summarize the two most important properties of density matrices:

1. *The density matrix $\rho$ is a positive-semidefinite matrix, see (2.1).*
2. *The trace of any density matrix $\rho$ is equal to 1, see (2.15).*

### 2.1.2 Quantum Measurement

Now, let us turn to measurements and their relation to density matrices. Assuming a projective measurement device, defined by measurement operators - i.e., projectors $\{P_j\}$. The projector $P_j$ is a Hermitian matrix, for which $P_j = P_j^{\dagger}$ and $P_j^2 = P_j$. According to the *3rd Postulate of Quantum Mechanic*s the trace operator can be used to give the probability of outcome $j$ belonging to the operator $P_j$ in the following way

$$\Pr\left[\, j \middle| \rho \right] = Tr\left( P_j \rho P_j^{\dagger} \right) = Tr\left( P_j^{\dagger} P_j \rho \right) = Tr\left( P_j \rho \right). \tag{2.24}$$

After the measurement, the measurement operator $P_j$ leaves the system in a post measurement state

$$\rho_j = \frac{P_j \left[ \sum_{i=1}^{n} p_i \left| \psi_i \right\rangle \left\langle \psi_i \right| \right] P_j}{Tr\left( P_j \left[ \sum_{i=1}^{n} p_i \left| \psi_i \right\rangle \left\langle \psi_i \right| \right] P_j \right)} = \frac{P_j \rho P_j}{Tr\left( P_j \rho P_j \right)} = \frac{P_j \rho P_j}{Tr\left( P_j \rho \right)}. \tag{2.25}$$

If we have a pure quantum state $\left|\psi\right\rangle = \alpha\left|0\right\rangle + \beta\left|1\right\rangle$, where $\alpha = \left\langle 0 \middle| \psi\right\rangle$ and $\beta = \left\langle 1 \middle| \psi\right\rangle$. Using the trace operator, the measurement probabilities of $\left|0\right\rangle$ and $\left|1\right\rangle$ can be expressed as

$$\begin{aligned}\Pr\left[j = 0\middle|\psi\right] &= Tr\left(P_j\rho\right) = Tr(\left|0\right\rangle \underbrace{\left\langle 0\right|\left|\psi\right\rangle}_{\left\langle 0|\psi\right\rangle}\left\langle\psi\right|) \\ &= \left\langle 0\middle|\psi\right\rangle Tr\left(\left|0\right\rangle\left\langle\psi\right|\right) = \left\langle 0\middle|\psi\right\rangle\left\langle\psi\middle|0\right\rangle \\ &= \left\langle 0\middle|\psi\right\rangle\left(\left\langle 0\middle|\psi\right\rangle\right)^* = \alpha\cdot\alpha^* = \left|\alpha\right|^2,\end{aligned} \tag{2.26}$$

and

$$\begin{aligned}\Pr\left[j = 1\middle|\psi\right] &= Tr\left(P_j\rho\right) = Tr(\left|1\right\rangle \underbrace{\left\langle 1\right|\left|\psi\right\rangle}_{\left\langle 1|\psi\right\rangle}\left\langle\psi\right|) \\ &= \left\langle 1\middle|\psi\right\rangle Tr\left(\left|1\right\rangle\left\langle\psi\right|\right) = \left\langle 1\middle|\psi\right\rangle\left\langle\psi\middle|1\right\rangle \\ &= \left\langle 1\middle|\psi\right\rangle\left(\left\langle 1\middle|\psi\right\rangle\right)^* = \beta\cdot\beta^* = \left|\beta\right|^2,\end{aligned} \tag{2.27}$$

in accordance with our expectations. Let us assume we have an *orthonormal* basis $M = \left\{\left|x_1\right\rangle\left\langle x_1\right|,\ldots,\left|x_n\right\rangle\left\langle x_n\right|\right\}$ and an arbitrary (i.e., non-diagonal) density matrix $\rho$. The set of Hermitian operators $P_i = \left\{\left|x_i\right\rangle\left\langle x_i\right|\right\}$ satisfies the *completeness relation*, where $P_i = \left|x_i\right\rangle\left\langle x_i\right|$ is the projector over $\left|x_i\right\rangle$, i.e., quantum measurement operator $M_i = \left|x_i\right\rangle\left\langle x_i\right|$ is a valid measurement operator. The measurement operator $M_i$ projects the input quantum system $\left|\psi\right\rangle$ to the pure state $\left|x_i\right\rangle$ from the orthonormal basis $M = \left\{\left|x_1\right\rangle\left\langle x_1\right|,\ldots,\left|x_n\right\rangle\left\langle x_n\right|\right\}$. Now, the probability that the quantum state $\left|\psi\right\rangle$ is after the measurement in basis state $\left|x_i\right\rangle$ can be expressed as

$$\left\langle\psi\middle|M_i^\dagger M_i\middle|\psi\right\rangle = \left\langle\psi\middle|P_i\middle|\psi\right\rangle = \left(\sum_{j=1}^{n} x_j^*\left\langle x_j\right|\right)\left|x_i\right\rangle\left\langle x_i\right|\left(\sum_{l=1}^{n}\left|x_l\right\rangle x_l\right) = \left|x_i\right|^2. \tag{2.28}$$

In the computational basis $\left\{\left|x_1\right\rangle,\ldots,\left|x_n\right\rangle\right\}$, the state of the quantum system after the measurement can be expressed as

$$\rho' = \sum_{i=1}^{n} p_i \left|x_i\right\rangle\left\langle x_i\right|, \tag{2.29}$$

and the matrix of the quantum state $\rho'$ will be *diagonal* in the computational basis $\left\{\left|x_i\right\rangle\right\}$, and can be given by

$$\rho' = \begin{bmatrix} p_1 & 0 & \dots & 0 \\ 0 & p_2 & 0 & \vdots \\ \vdots & \vdots & \ddots & 0 \\ 0 & 0 & 0 & p_n \end{bmatrix}. \tag{2.30}$$

To illustrate it, let assume we have an initial (not diagonal) density matrix in the computational basis $\left\{\left|0\right\rangle,\left|1\right\rangle\right\}$ e.g. $\left|\psi\right\rangle = \alpha\left|0\right\rangle + \beta\left|1\right\rangle$ with $p = \left|\alpha\right|^2$ and $1 - p = \left|\beta\right|^2$ as

$$\rho = \left|\psi\right\rangle\left\langle\psi\right| = \begin{bmatrix} \left|\alpha\right|^2 & \alpha\beta^* \\ \alpha^*\beta & \left|\beta\right|^2 \end{bmatrix}, \tag{2.31}$$

and we have orthonormal basis $M = \left\{\left|0\right\rangle\left\langle 0\right|,\left|1\right\rangle\left\langle 1\right|\right\}$. In this case, the after-measurement state can be expressed as

$$\rho' = p\left|0\right\rangle\left\langle 0\right| + \left(1 - p\right)\left|1\right\rangle\left\langle 1\right| = \begin{bmatrix} \left|\alpha\right|^2 & 0 \\ 0 & \left|\beta\right|^2 \end{bmatrix} = \begin{bmatrix} p & 0 \\ 0 & 1 - p \end{bmatrix}. \tag{2.32}$$

As it can be seen, the matrix of $\rho'$ is a diagonal matrix in the computational basis $\left\{\left|0\right\rangle,\left|1\right\rangle\right\}$. Eq. (2.31) and (2.32) highlights the difference between quantum superpositions (probability amplitude weighted sum) and classical probabilistic mixtures of quantum states.

Now, let us see the result of the measurement on the input quantum system $\rho$

$$M\left(\rho\right) = \sum_{j=0}^{1} M_j \rho M_j^{\dagger} = M_0 \rho M_0^{\dagger} + M_1 \rho M_1^{\dagger}. \tag{2.33}$$

For the measurement operators $M_0 = |0\rangle\langle 0|$ and $M_1 = |1\rangle\langle 1|$ the completeness relation holds

$$\begin{aligned}\sum_{j=0}^{1} M_j M_j^\dagger &= |0\rangle\langle 0||0\rangle\langle 0| + |1\rangle\langle 1||1\rangle\langle 1| = |0\rangle\overbrace{\langle 0|0\rangle}^{=1}\langle 0| + |1\rangle\overbrace{\langle 1|1\rangle}^{=1}\langle 1| \\ &= |0\rangle\langle 0| + |1\rangle\langle 1| = \begin{bmatrix} 1 & 0 \\ 0 & 1 \end{bmatrix} = I.\end{aligned} \tag{2.34}$$

Using input system $\rho = |\psi\rangle\langle\psi|$, where $|\psi\rangle = \alpha|0\rangle + \beta|1\rangle$, the state after the measurement operation is

$$\begin{aligned}M(\rho) &= \sum_{j=0}^{1} M_j \rho M_j^\dagger \\ &= |0\rangle\langle 0|\rho|0\rangle\langle 0| + |1\rangle\langle 1|\rho|1\rangle\langle 1| = \\ &= |0\rangle\langle 0||\psi\rangle\langle\psi||0\rangle\langle 0| + |1\rangle\langle 1||\psi\rangle\langle\psi||1\rangle\langle 1| \\ &= |0\rangle\langle 0|\psi\rangle\langle 0|\psi\rangle\langle 0| + |1\rangle\langle 1|\psi\rangle\langle 1|\psi\rangle\langle 1| \\ &= |\langle 0|\psi\rangle|^2 |0\rangle\langle 0| + |\langle 1|\psi\rangle|^2 |1\rangle\langle 1| \\ &= |\alpha|^2 |0\rangle\langle 0| + |\beta|^2 |1\rangle\langle 1| = p|0\rangle\langle 0| + 1 - p|1\rangle\langle 1|.\end{aligned} \tag{2.35}$$

As we have found, after the measurement operation $M(\rho)$, the *off-diagonal* entries will have zero values, and they *have no relevance*. As follows, the initial input system $\rho = |\psi\rangle\langle\psi|$ after operation $M$ becomes

$$\rho = \begin{bmatrix} |\alpha|^2 & \alpha\beta^* \\ \alpha^*\beta & |\beta|^2 \end{bmatrix} \xrightarrow{M} \rho' = \begin{bmatrix} |\alpha|^2 & 0 \\ 0 & |\beta|^2 \end{bmatrix}. \tag{2.36}$$

#### 2.1.2.1 Orthonormal Basis Decomposition

Let assume we have orthonormal basis $\{|b_1\rangle, |b_2\rangle, \ldots, |b_n\rangle\}$, which basis can be used to rewrite the quantum system $|\psi\rangle$ in a unique decomposition

$$|\psi\rangle = b_1|b_1\rangle + b_2|b_2\rangle + \ldots + b_n|b_n\rangle = \sum_{i=1}^{n} b_i|b_i\rangle, \tag{2.37}$$

with complex $b_i$. Since $\left\langle \psi \middle| \psi \right\rangle = 1$, we can express it in the form

$$\left\langle \psi \middle| \psi \right\rangle = \sum_{i=1}^{n}\sum_{j=1}^{n} b_i^* b_j \left\langle b_i \middle| b_j \right\rangle = \sum_{i=1}^{n} \left| b_i \right|^2 = 1, \tag{2.38}$$

where $b_i^*$ is the complex conjugate of *probability amplitude* $b_i$, thus $\left| b_i \right|^2$ is the *probability* $p_i$ of measuring the quantum system $\left| \psi \right\rangle$ in the given basis state $\left| b_i \right\rangle$, i.e.,

$$p_i = \left| b_i \right|^2. \tag{2.39}$$

Using (2.2), (2.37) and (2.38) the density matrix of quantum system $\left| \psi \right\rangle$ can be expressed as

$$\begin{aligned} \rho &= \left| b_1 \right|^2 \left| b_1 \right\rangle \left\langle b_1 \right| + \left| b_2 \right|^2 \left| b_2 \right\rangle \left\langle b_2 \right| + \ldots + \left| b_n \right|^2 \left| b_n \right\rangle \left\langle b_n \right| \\ &= \sum_{i=1}^{n} \left| b_i \right|^2 \left| b_i \right\rangle \left\langle b_i \right| = \sum_{i=1}^{n} p_i \left| b_i \right\rangle \left\langle b_i \right|. \end{aligned} \tag{2.40}$$

This density matrix is a diagonal matrix with the probabilities in the diagonal entries

$$\rho = \begin{bmatrix} p_1 & \cdots & 0 & 0 \\ 0 & p_2 & 0 & \vdots \\ \vdots & 0 & \ddots & 0 \\ 0 & \cdots & 0 & p_n \end{bmatrix}. \tag{2.41}$$

The diagonal property of density matrix (2.40) in (2.41) can be checked, since the elements of the matrix can be expressed as

$$\rho_{ij} = \left\langle b_i \middle| \rho \middle| b_j \right\rangle = \left\langle b_i \right| \left( \sum_{l=1}^{n} p_i \left| b_i \right\rangle \left\langle b_i \right| \right) \left| b_j \right\rangle = \sum_{l=1}^{n} p_l \left\langle b_i \middle| b_l \right\rangle \left\langle b_l \middle| b_j \right\rangle, \tag{2.42}$$

where $\sum_{l=1}^{n} p_i = 1$.

#### 2.1.2.2 The Projective and POVM Measurement

The *projective measurement* is also known as the *von Neumann measurement* is formally can be described by the Hermitian operator $\mathcal{Z}$, which has the spectral decomposition

$$\mathcal{Z} = \sum_m \lambda_m P_m . \tag{2.43}$$

where $P_m$ is a projector to the eigenspace of $\mathcal{Z}$ with eigenvalue $\lambda_m$. For the projectors

$$\sum_m P_m = I , \tag{2.44}$$

and they are pairwise orthogonal. The measurement outcome $m$ corresponds to the eigenvalue $\lambda_m$, with measurement probability

$$\Pr\left[ m \middle| \psi \rangle \right] = \langle \psi | P_m | \psi \rangle . \tag{2.45}$$

When a quantum system is measured in an orthonormal basis $|m\rangle$, then we make a projective measurement with projector $P_m = |m\rangle\langle m|$, thus (2.43) can be rewritten as

$$\mathcal{Z} = \sum_m m P_m . \tag{2.46}$$

The $\mathcal{P}$ *POVM* (Positive Operator Valued Measurement) is intended to select among the non-orthogonal states $\left\{ |\psi_i\rangle \right\}_{i=1}^{m}$ and defined by a *set* of POVM operators $\left\{ \mathcal{M}_i \right\}_{i=1}^{m+1}$, where

$$\mathcal{M}_i = \mathcal{Q}_i^{\dagger} \mathcal{Q}_i , \tag{2.47}$$

and since we are not interested in the post-measurement state the exact knowledge about measurement operator $\mathcal{Q}_i$ is not required. For POVM operators $\mathcal{M}_i$ the completeness relation holds,

$$\sum_i \mathcal{M}_i = I\ . \tag{2.48}$$

For the POVM the probability of a given outcome $n$ for the state $\left|\psi\right\rangle$ can be expressed as

$$\Pr\left[i\middle|\psi\right\rangle\right] = \left\langle\psi\middle|\mathcal{M}_i\middle|\psi\right\rangle. \tag{2.49}$$

The POVM also can be imagined as a "black-box", which outputs a number from 1 to $m$ for the given input quantum state $\psi$, using the set of operators

$$\left\{\mathcal{M}_1,\ldots,\mathcal{M}_m,\mathcal{M}_{m+1}\right\}, \tag{2.50}$$

where $\left\{\mathcal{M}_1,\ldots,\mathcal{M}_m\right\}$ are responsible to distinguish $m$ different typically non-orthogonal states i.e., if we observe $i\in\left[1,m\right]$ on the display of the measurement device we can be sure, that the result is correct. However, because unknown non-orthogonal states can not be distinguished with probability 1, we have to introduce an extra measurement operator, $\mathcal{M}_{m+1}$, as the price of the distinguishability of the $m$ different states and if we obtain $m+1$ as measurement results we can say nothing about $\left|\psi\right\rangle$. This operator can be expressed as

$$\mathcal{M}_{m+1} = I - \sum_{i=1}^{m}\mathcal{M}_i\ . \tag{2.51}$$

Such $\mathcal{M}_{m+1}$ can be always constructed if the states in $\left\{\left|\psi_n\right\rangle\right\}_{n=1}^{m}$ are linearly independent. We note, we will omit listing operator $\mathcal{M}_{m+1}$ in further parts of the paper. The POVM measurement apparatus will be a key ingredient to distinguish quantum codewords with zero-error, and to reach the zero-error capacity of quantum channels.

The POVM can be viewed as the most general formula from among of any possible measurements in quantum mechanics. Therefore the effect of a projective measurement can be described by POVM operators, too. Or with other words, the projective measurements are the special case POVM measurement [Imre05]. The elements of the POVM are not necessarily orthogonal, and the number of the elements can be larger than the dimension of the Hilbert space which they are originally used in.

## 2.2 Geometrical Interpretation of the Density Matrices

While the *wavefunction* representation is the full physical description of a quantum system in the space-time, the tensor product of multiple copies of two dimensional Hilbert spaces is its discrete version, with discrete finite-dimensional Hilbert spaces. The geometrical representation also can be extended to analyze the geometrical structure of the transmission of information though a quantum channel, and it also provides a very useful tool to analyze the capacities of different quantum channel models.

The Bloch sphere is a geometrical conception, constructed to represent two-level quantum systems in a more expressive way than is possible with algebraic tools. The Bloch sphere has unit radius and is defined in a three-dimensional real vector space. The pure states are on the surface of the Bloch sphere, while the mixed states are in the interior of the original ball. In the Bloch sphere representation, the state of a single qubit $\left|\psi\right\rangle = \alpha\left|0\right\rangle + \beta\left|1\right\rangle$ can be expressed as

$$\left|\psi\right\rangle = e^{i\delta}\left(\cos\frac{\theta}{2}\left|0\right\rangle + e^{i\varphi}\sin\frac{\theta}{2}\left|1\right\rangle\right), \tag{2.52}$$

where $\delta$ is the global phase factor, which can be ignored from the computations, hence the state $\left|\psi\right\rangle$ in the terms of the angle $\theta$ and $\varphi$ can be expressed as

$$\left|\psi\right\rangle = \cos\frac{\theta}{2}\left|0\right\rangle + e^{i\varphi}\sin\frac{\theta}{2}\left|1\right\rangle. \tag{2.53}$$

The Bloch sphere is a very useful tool, since it makes possible to describe various, physically realized one-qubit quantum systems, such as the photon polarization, spins or the energy levels of an atom. Moreover, if we would like to compute the various channel capacities of the quantum channel, the geometrical expression of the channel capacity also can be represented by the Bloch sphere. Before we would introduce the geometrical calculation of the channel capacities, we have to start from the geometrical interpretation of density matrices. The density matrix $\rho$ can then be expressed using the Pauli matrices $\sigma_X = \begin{bmatrix} 0 & 1 \\ 1 & 0 \end{bmatrix}$, $\sigma_Y = \begin{bmatrix} 0 & -i \\ i & 0 \end{bmatrix}$ and $\sigma_Z = \begin{bmatrix} 1 & 0 \\ 0 & -1 \end{bmatrix}$ as

$$\rho = \frac{1 + r_X\sigma_X + r_Y\sigma_Y + r_Z\sigma_Z}{2}, \tag{2.54}$$

where for the Bloch vector $\mathbf{r} = \left(r_X, r_Y, r_Z\right) = \left(\sin\theta\cos\phi, \sin\theta\sin\phi, \cos\theta\right)$ we have $\left\|\left(r_X, r_Z, r_Y\right)\right\| \leq 1$, and $\sigma = \left(\sigma_X, \sigma_Y, \sigma_Z\right)^T$. In the vector representation, the previously shown formula can be expressed as

$$\rho = \frac{1 + \mathbf{r}\sigma}{2}. \tag{2.55}$$

In conclusion, every state can be expressed as linear combinations of the Pauli matrices and according to these Pauli matrices every state can be interpreted as a point in the three-dimensional real vector space. The unitary transforms can also be represented geometrically. If we apply a unitary transformation $U$ to the density matrix $\rho$, then it can be expressed as

$$\rho \to \rho' = U\rho U^{\dagger} = \frac{1 + U\mathbf{r}\sigma U^{\dagger}}{2} = \frac{1 + U\mathbf{r}U^{\dagger}\sigma}{2}, \tag{2.56}$$

where $\mathbf{r}' = U\mathbf{r}U^{\dagger}$ realizes a unitary transformation on $\mathbf{r}$ as a rotation (see Fig 2.8).

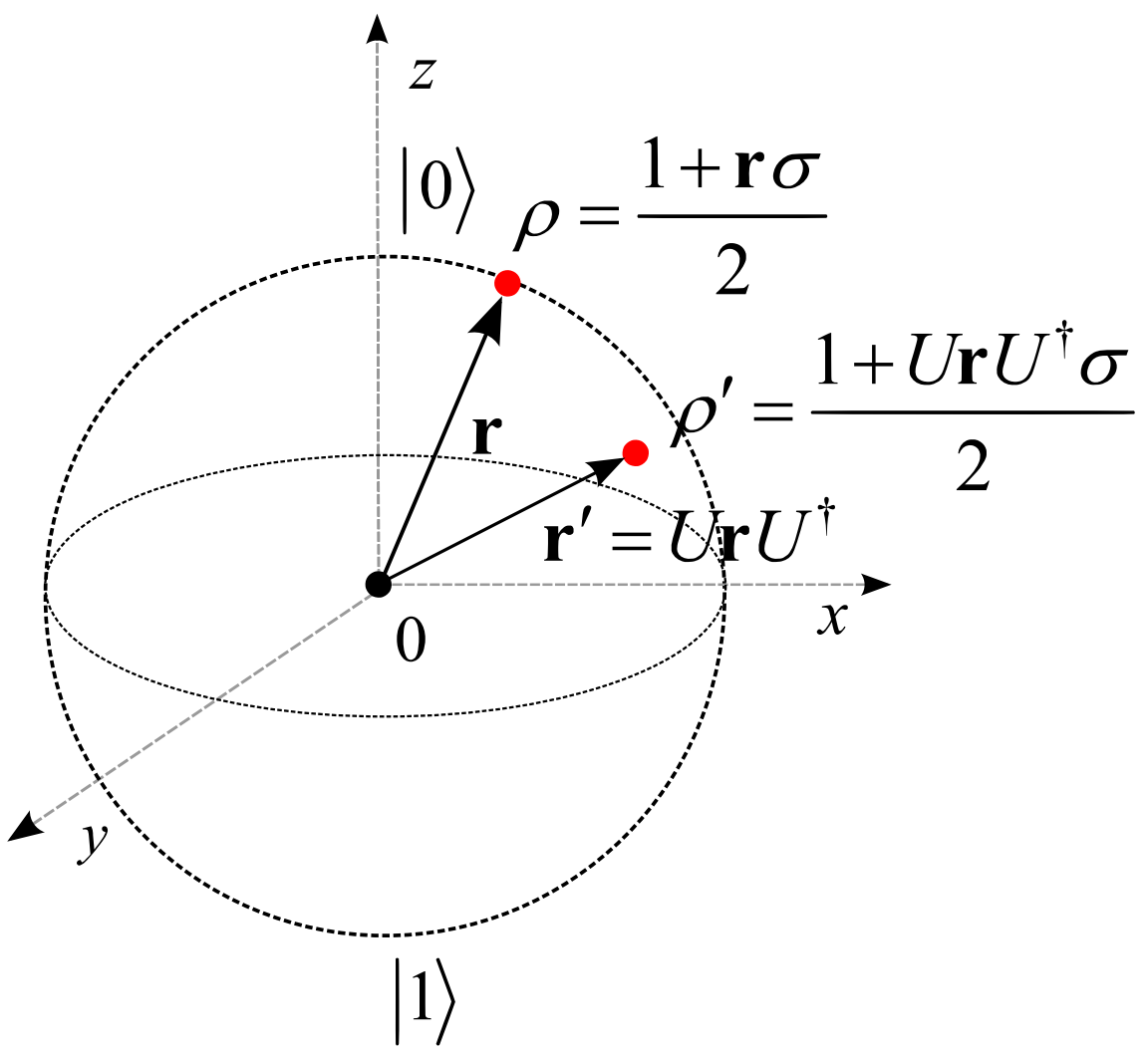

**Fig. 2.1**. Unitary transformation as rotation, using the geometrical interpretation of the density matrices. The density matrices of pure states are on the surface on the Bloch sphere.

Another important consequence of the geometrical interpretation of density matrices on the Bloch sphere is that density matrix $\rho$ can be expressed in a "weighted form" of density matrices $\rho_1$ and $\rho_2$ as follows:

$$\rho = \gamma\rho_1 + \left(1 - \gamma\right)\rho_2, \tag{2.57}$$

where $0 \leq \gamma \leq 1$, and $\rho_1$ and $\rho_2$ are pure states, and lie on a line segment connecting the density matrices in the Bloch sphere representation. Using probabilistic mixtures of the pure density matrices, any quantum state which lies between the two states can be expressed as a convex combination

$$\rho = p\rho_1 + \left(1 - p\right)\rho_2, \; 0 \leq p \leq 1. \tag{2.58}$$

This remains true for an arbitrary number of quantum states, hence this result can be expressed for arbitrary number of density matrices. The geometrical interpretation of the convex combination of quantum states is shown in Fig. 2.2.

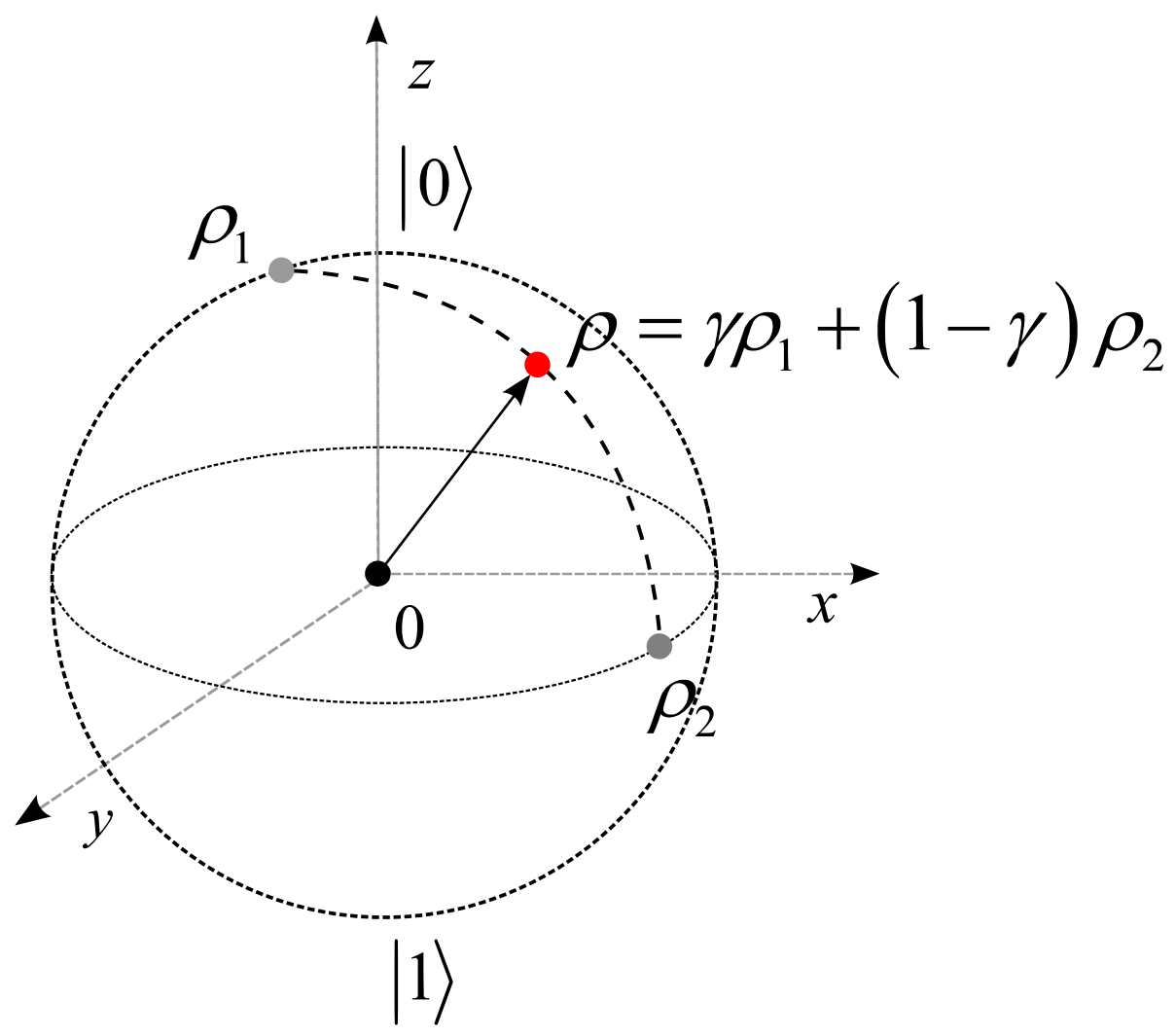

**Fig. 2.2.** Using probabilistic mixtures of the pure density matrices, any quantum state can be expressed as a convex combination.

Mixed quantum states can be represented as *statistical mixtures* of pure quantum states. The statistical representation of a pure state is unique. On the other hand we note that the decomposition of a mixed quantum state is not unique.

In Fig. 2.3 we compared a pure and a mixed quantum state. The pure state $\rho$ is on the surface of the Bloch sphere, while the mixed state $\sigma$ is inside the ball. A maximally mixed quantum state, $\sigma = \frac{1}{2}I$, can be found in the center of the Bloch sphere. As we depicted, the mixed state can be expressed as probabilistic mixture of pure states $\{\rho_1, \rho_2\}$ and $\{\rho_3, \rho_4\}$. As it has been stated by von Neumann and presented in Fig. 2.3(b), the *decomposition of a mixed state is not unique*, since it can be expressed as a mixture of $\{\rho_1, \rho_2\}$ or equivalently of $\{\rho_3, \rho_4\}$.

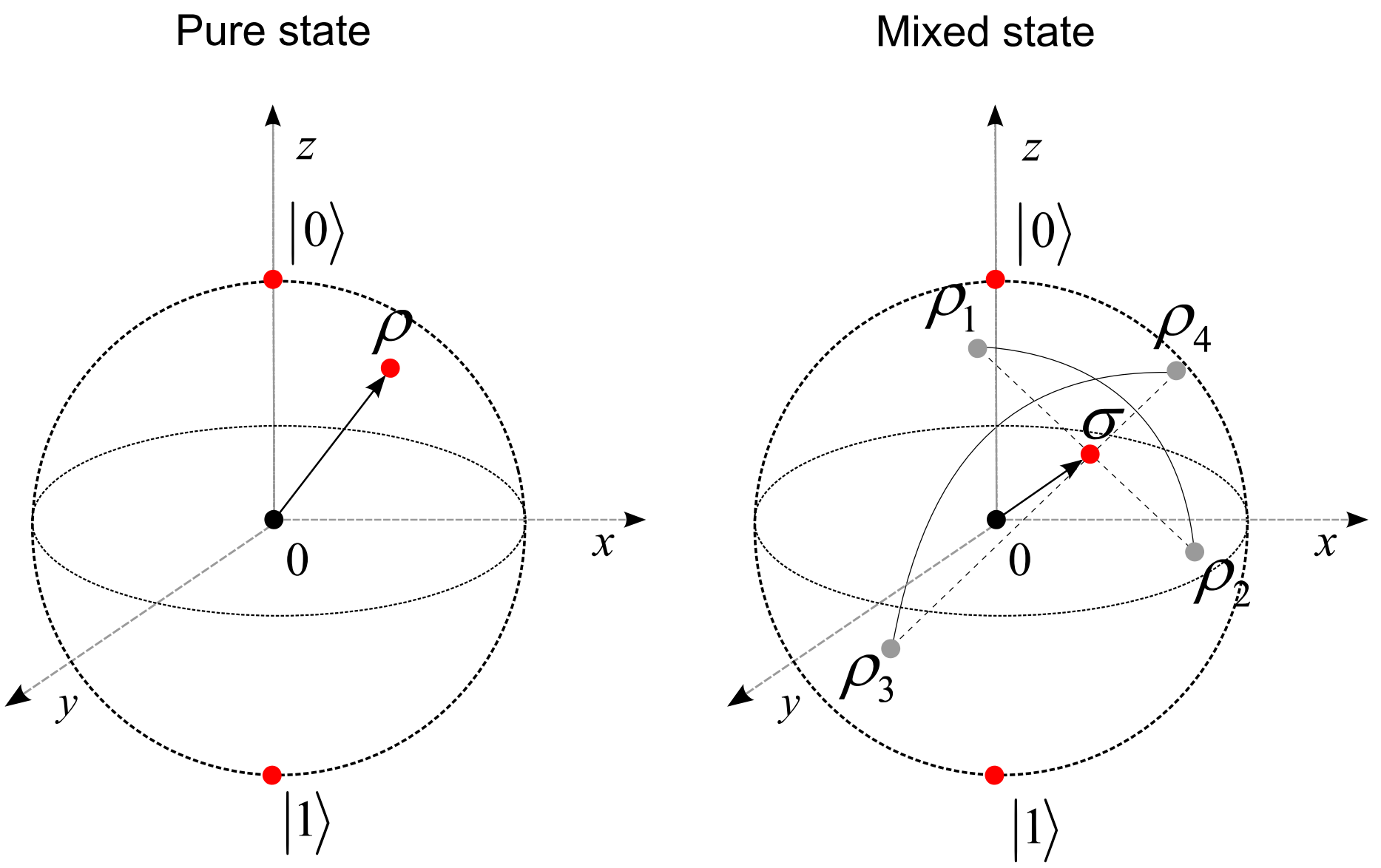

**Fig. 2.3.** A pure state and a mixed state in the Bloch sphere representation. The mixed state can be expressed as probabilistic mixture of pure quantum states, however this decomposition is not unique.

One can use a pure state $\rho$ to recover mixed state $\sigma$ from it, after the effects of environment are traced out. With the help of the partial trace operator, Bob, the receiver, can decouple the environment from his mixed state, and the original state can be recovered by discarding the effects of the environment. If Bob's state is a *probabilistic mixture* $\sigma = \sum_i p_i \left|\varphi_i\right\rangle\left\langle\varphi_i\right|$, then a global pure *purification* state $\left|\Psi\right\rangle$ exists, which from Bob's state can be expressed as

$$\sigma = Tr_{environment} \left|\Psi\right\rangle\left\langle\Psi\right|. \tag{2.59}$$

As we showed in Fig. 2.4, state $\sigma$ can be recovered from $\left|\Psi\right\rangle$ after discarding the environment. The decoupling of the environment can be achieved with the $Tr_{environment}$ operator. For any unitary transformation of the environment, the pure state $\left|\Psi\right\rangle$ is a unique state.

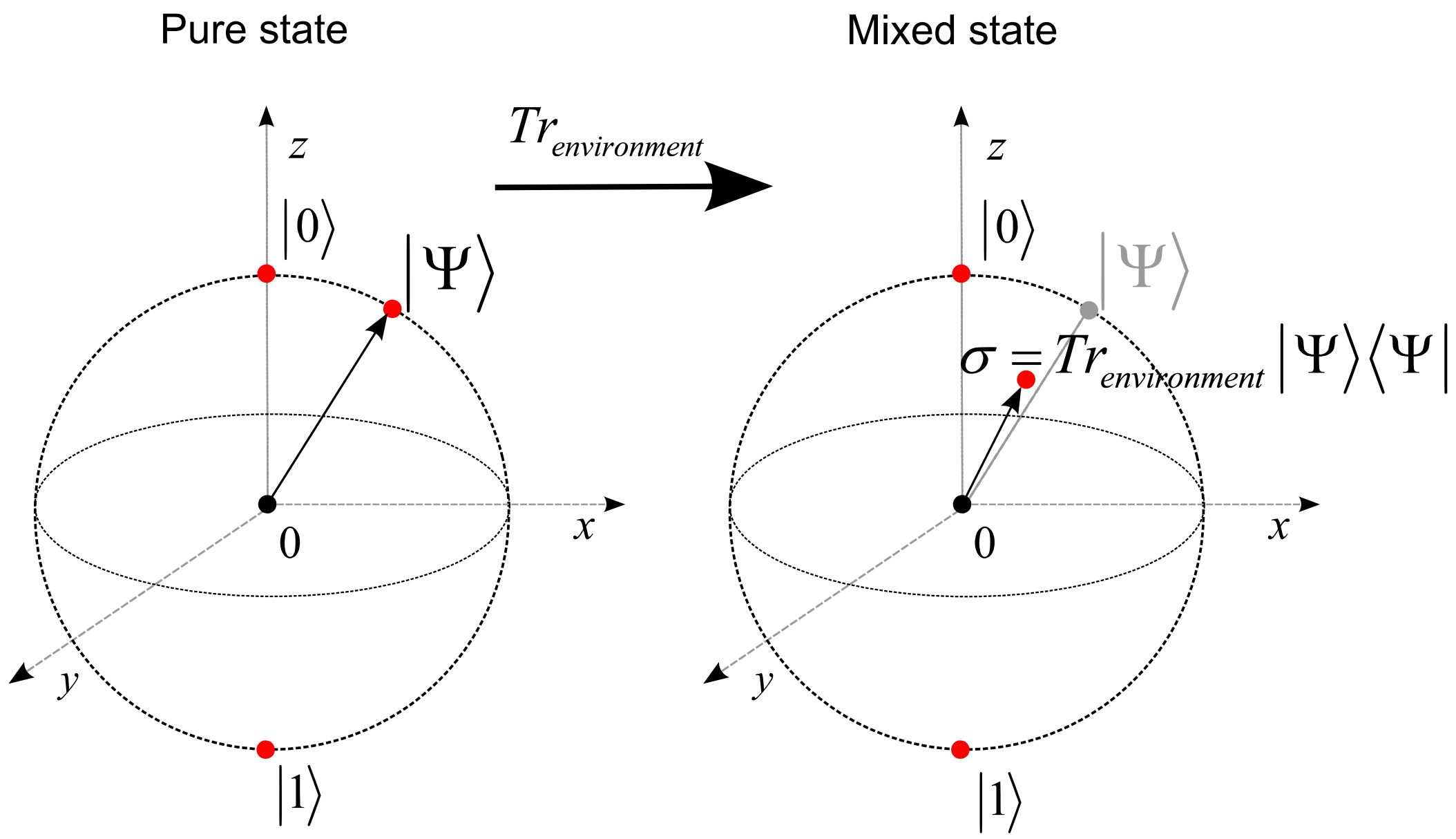

**Fig. 2.4.** The mixed state can be recovered from the pure state after the environment has been traced out.

We have seen, that the decomposition of mixed quantum states into pure quantum states is not unique, hence for example, it can be easily verified by the reader, that the decomposition of a mixed state $\sigma = \frac{1}{2}\left(|0\rangle\langle 0| + |1\rangle\langle 1|\right)$ can be made with pure states $\{|0\rangle, |1\rangle\}$, and also can be given with pure states $\left\{\frac{1}{\sqrt{2}}\left(|0\rangle + |1\rangle\right), \frac{1}{\sqrt{2}}\left(|0\rangle - |1\rangle\right)\right\}$. Here, we have just changed the basis from rectilinear to diagonal, and we have used just pure states - and it resulted in the same mixed quantum state.

## 2.3 Channel System Description

If we are interested in the origin of noise (randomness) in the quantum channel the model should be refined in the following way: Alice's register $X$, the purification state $P$, channel input $A$, channel output $B$, and the environment state $E$. The input system $A$ is described by a quantum system $\rho_x$, which occurs on the input with probability $p_X(x)$. They together form an ensemble denoted by $\{p_X(x), \rho_x\}_{x \in X}$, where $x$ is a classical variable from the register $X$. In the preparation process, Alice generates pure states $\rho_x$ according to random variable $x$, i.e., the input density operator can be expressed as $\rho_x = |x\rangle\langle x|$, where the classical states $\{|x\rangle\}_{x\in X}$ form an orthonormal basis. According to the elements of Alice's register $X$, the input system can be characterized by the quantum system

$$\rho_A = \sum_{x\in X} p_X(x)\rho_x = \sum_{x\in X} p_X(x)|x\rangle\langle x|. \tag{2.60}$$

The system description is illustrated in Fig. 2.5.

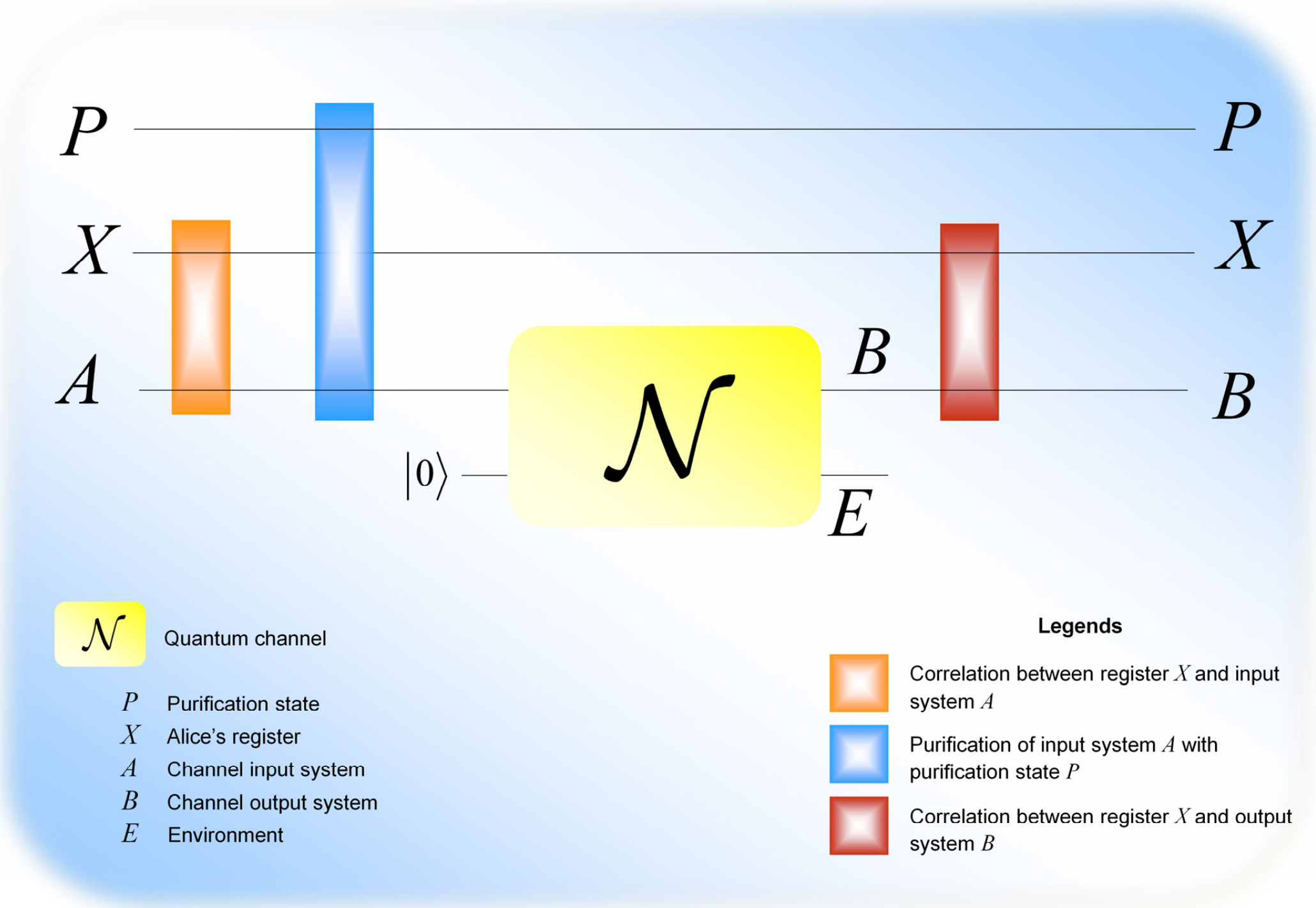

**Fig. 2.5.** Detailed model of a quantum communication channel exposing the interaction with the environment. Alice's register is denoted by $X$, the input system is $A$ while $P$ is the purification state. The environment of the channel is denoted by $E$, the output of the channel is $B$. The quantum channel has positive classical capacity if and only if the channel output system $B$ will be correlated with Alice's register $X$.

The system state $\rho_x$ with the corresponding probability distribution $p_X(x)$ can be indentified by a set of measurement operators $M = \{|x\rangle\langle x|\}_{x\in X}$. If the density operators $\rho_x$ in $\rho_A$ are mixed, the probability distribution $p_X(x)$ and the classical variable $x$ from the register $X$ cannot be indentified by the measurement operators $M = \{|x\rangle\langle x|\}_{x\in X}$, since the system state $\rho_x$ is assumed to be a mixed or in a non-orthonormal state. Alice's register $X$ and the quantum system $A$ can be viewed as a tensor product system as

$$\left\{p_X(x), |x\rangle\langle x|_X \otimes \rho_A^x\right\}_{x\in X}, \tag{2.61}$$

where the classical variable $x$ is correlated with the quantum system $\rho_x$, using orthonormal basis $\{|x\rangle\}_{x\in X}$. Alice's register $X$ represents a classical variable, the channel input system is generated corresponding to the register $X$ in the form of a quantum state, and it is described by the density operator $\rho_A^x$. The input system $A$ with respect to the register $X$, is described by the density operator

$$\rho_{XA} = \sum_{x\in X} p_X(x)|x\rangle\langle x|_X \otimes \rho_A^x, \tag{2.62}$$

where $\rho_A^x = |\psi_x\rangle\langle\psi_x|_A$ is the density matrix representation of Alice's input state $|\psi_x\rangle_A$.

### 2.3.1 Purification

The *purification* gives us a new viewpoint on the noise of the quantum channel. Assuming Alice's side $A$ and Alice's register $X$, the spectral decomposition of the density operator $\rho_A$ can be expressed as

$$\rho_A = \sum_x p_X(x)|x\rangle\langle x|_A, \tag{2.63}$$

where $p_X(x)$ is the probability of variable $x$ in Alice's register $X$. The $\{p_X(x), |x\rangle\}$ together is called an ensemble, where $|x\rangle$ is a quantum state according to classical variable $x$. Using the set of orthonormal basis vectors $\{|x\rangle_P\}_{x\in X}$ of the purification system $P$, the purification of (2.63) can be given in the following way:

$$|\varphi\rangle_{PA} = \sum_x \sqrt{p_X(x)}|x\rangle_P|x\rangle_A. \tag{2.64}$$

From the purified system state $|\varphi\rangle_{PA}$, the original system state $\rho_A$ can be expressed with the partial trace operator (see Appendix) $Tr_P(\cdot)$, which operator traces out the purification state (i.e., the environment) from the system

$$\rho_A = Tr_P(|\varphi\rangle\langle\varphi|_{PA}). \tag{2.65}$$

From joint system (2.64) and the purified state (2.65), one can introduce a new definition. The *extension* of $\rho_A$ can be given as

$$\rho_A = Tr_P\left(\omega_{PA}\right), \tag{2.66}$$

where $\omega_{PA}$ is the joint system of purification state $P$ and channel input $A$ [Wilde11], which represents a noisy state.

### 2.3.2 Isometric Extension

*Isometric extension* has utmost importance, because it helps us to understand what happens between the quantum channel and its environment whenever a quantum state is transmitted from Alice to Bob. Since the channel and the environment together form a closed physical system the isometric extension of the quantum channel $\mathcal{N}$ is the *unitary representation* of the channel

$$\mathcal{N} : U_{A\to BE}\,, \tag{2.67}$$

enabling the "one-sender and two-receiver" view: beside Alice the sender, both Bob and the environment of the channel are playing the receivers. In other words, the output of the noisy quantum channel $\mathcal{N}$ can be described only after the environment of the channel is traced out

$$\rho_B = Tr_E\left(U_{A\to BE}\left(\rho_A\right)\right) = \mathcal{N}\left(\rho_A\right). \tag{2.68}$$

### 2.3.3 Kraus Representation

The map of the quantum channel can also be expressed by means of a special tool called the *Kraus Representation.* For a given input system $\rho_A$ and quantum channel $\mathcal{N}$, this representation can be expressed as

$$\mathcal{N}\left(\rho_A\right) = \sum_i N_i \rho_A N_i^{\dagger}, \tag{2.69}$$

where $N_i$ are the Kraus operators, and $\sum_i N_i^{\dagger} N_i = I$. The isometric extension of $\mathcal{N}$ by means of the *Kraus Representation* can be expressed as

$$\rho_B = \mathcal{N}\left(\rho_A\right) = \sum_i N_i \rho_A N_i^{\dagger} \to U_{A\to BE}\left(\rho_A\right) = \sum_i N_i \otimes \left|i\right\rangle_E. \tag{2.70}$$

The action of the quantum channel $\mathcal{N}$ on an operator $|k\rangle\langle l|$, where $\{|k\rangle\}$ form an orthonormal basis also can be given in operator form using the Kraus operator $N_{kl} = \mathcal{N}(|k\rangle\langle l|)$. By exploiting the property $UU^{\dagger} = P_{BE}$, for the input quantum system $\rho_A$

$$\begin{aligned}\rho_B &= U_{A\to BE}(\rho_A) = U\rho_A U^{\dagger} \\ &= \left(\sum_i N_i \otimes |i\rangle_E\right)\rho_A\left(\sum_j N_j^{\dagger} \otimes \langle j|_E\right) = \sum_{i,j} N_i\rho_A N_j^{\dagger} \otimes |i\rangle\langle j|_E .\end{aligned} \tag{2.71}$$

If we trace out the environment, we get the equivalence of the two representations

$$\rho_B = Tr_E\left(U_{A\to BE}(\rho_A)\right) = \sum_i N_i\rho_A N_i^{\dagger} . \tag{2.72}$$

### 2.3.4 The von Neumann Entropy

Quantum information processing exploits the quantum nature of information. It offers fundamentally new solutions in the field of computer science and extends the possibilities to a level that cannot be imagined in classical communication systems. On the other hand, it requires the generalization of classical information theory through a quantum perception of the world. As Shannon entropy plays fundamental role in classical information theory, the von Neumann entropy does the same for quantum information. The von Neumann entropy $\mathrm{S}(\rho)$ of quantum state $\rho$ can be viewed as an extension of classical entropy for quantum systems. It measures the information of the quantum states in the form of the uncertainty of a quantum state. The classical Shannon entropy $H(X)$ of a variable $X$ with probability distribution $p(X)$ can be defined as

$$H(X) = -\sum_{x\in X} p(x)\log(p(x)), \tag{2.73}$$

with $1 \leq H(X) \leq \log(|X|)$, where $|X|$ is the cardinality of the set $X$.

The von Neumann entropy

$$\mathrm{S}(\rho) = -Tr(\rho\log(\rho)) \tag{2.74}$$

measures the information contained in the quantum system $\rho$. Furthermore $\mathrm{S}(\rho)$ can be expressed by means of the Shannon entropy for the eigenvalue distribution

$$\mathrm{S}(\rho) = H(\lambda) = -\sum_{i=1}^{d}\lambda_i\log(\lambda_i), \tag{2.75}$$

where $d$ is the level of the quantum system and $\lambda_i$ are the eigenvalues of density matrix $\rho$.

### 2.3.5 The Holevo Quantity

The *Holevo bound* determines the amount of information that can be extracted from a single qubit state. If Alice sends a quantum state $\rho_i$ with probability $p_i$ over an ideal quantum channel, then at Bob's receiver a mixed state

$$\rho_B = \rho_A = \sum_i p_i \rho_i \tag{2.76}$$

appears. Bob constructs a measurement $\{M_i\}$ to extract the information encoded in the quantum states. If he applies the measurement to $\rho_A$, the probability distribution of Bob's classical symbol $B$ will be $\Pr\left[b \middle| \rho_A\right] = Tr\left(M_b^\dagger M_b \rho_A\right)$. As had been shown by Holevo [Holevo73], the bound for the maximal classical mutual information between Alice and Bob is

$$I\left(A:B\right) \leq \mathrm{S}\left(\rho_A\right) - \sum_i p_i \mathrm{S}\left(\rho_i\right) \equiv \chi, \tag{2.77}$$

where $\chi$ is called the *Holevo quantity, and* (2.77) known as the *Holevo bound.*
In classical information theory and classical communication systems, the mutual information $I\left(A:B\right)$ is bounded only by the classical entropy of $H\left(A\right)$, hence $I\left(A:B\right) \leq H\left(A\right)$. The mutual information $I\left(A:B\right)$ is bounded by the classical entropy of $H\left(A\right)$, hence $I\left(A:B\right) \leq H\left(A\right)$. On the other hand, for mixed states and pure non-orthogonal states the Holevo quantity $\chi$ can be greater than the mutual information $I\left(A:B\right)$, however, it is still bounded by $H\left(A\right)$, which is the bound for the pure orthogonal states

$$I\left(A:B\right) \leq \chi \leq H\left(A\right). \tag{2.78}$$

The *Holevo bound* highlights the important fact that one qubit can contain at most one classical bit i.e., cbit of information.

### 2.3.6 Quantum Conditional Entropy

While the classical conditional entropy function is always takes a non negative value, the *quantum conditional entropy can be negative.* The quantum conditional entropy between quantum systems $A$ and $B$ is given by

$$\mathrm{S}\left(A \middle| B\right) = \mathrm{S}\left(\rho_{AB}\right) - \mathrm{S}\left(\rho_B\right). \tag{2.79}$$

If we have two uncorrelated subsystems $\rho_A$ and $\rho_B$, then the information of the quantum system $\rho_A$ does not contain any information about $\rho_B$, or reversely, thus

$$\mathrm{S}\left(\rho_{AB}\right) = \mathrm{S}\left(\rho_A\right) + \mathrm{S}\left(\rho_B\right), \tag{2.80}$$

hence we get $\mathrm{S}\left(A\middle|B\right) = \mathrm{S}\left(\rho_A\right)$, and similarly $\mathrm{S}\left(B\middle|A\right) = \mathrm{S}\left(\rho_B\right)$. The negative property of conditional entropy $\mathrm{S}\left(A\middle|B\right)$ can be demonstrated with an *entangled* state, since in this case, the joint quantum entropy of the joint state less than the sum of the von Neumann entropies of its individual components. For a pure entangled state, $\mathrm{S}\left(\rho_{AB}\right) = 0$, while $\mathrm{S}\left(\rho_A\right) = \mathrm{S}\left(\rho_B\right) = 1$ since the two qubits are in *maximally mixed* $\frac{1}{2}I$ state, which is classically totally unimaginable. Thus, in this case

$$\mathrm{S}\left(A\middle|B\right) = -\mathrm{S}\left(\rho_B\right) \leq 0, \tag{2.81}$$

and

$$\mathrm{S}\left(B\middle|A\right) = -\mathrm{S}\left(\rho_A\right) \leq 0 \text{ and } \mathrm{S}\left(\rho_A\right) = \mathrm{S}\left(\rho_B\right). \tag{2.82}$$

### 2.3.7 Quantum Mutual Information

The classical mutual information $I\left(\cdot\right)$ measures the information correlation between random variables *A* and *B*. In analogue to classical information theory, $I\left(A:B\right)$ can be described by the quantum entropies of individual states and the von Neumann entropy of the joint state as follows:

$$I\left(A:B\right) = \mathrm{S}\left(\rho_A\right) + \mathrm{S}\left(\rho_B\right) - \mathrm{S}\left(\rho_{AB}\right) \geq 0, \tag{2.83}$$

i.e., the quantum mutual information is always a non negative function. However, there is a distinction between classical and quantum systems, since the quantum mutual information can take its value above the maximum of the classical mutual information. This statement can be confirmed, if we take into account that for an pure entangled quantum system, the quantum mutual information is

$$I\left(A:B\right) = \mathrm{S}\left(\rho_A\right) + \mathrm{S}\left(\rho_B\right) - \mathrm{S}\left(\rho_{AB}\right) = 1 + 1 - 0 = 2, \tag{2.84}$$

and we can rewrite this equation as

$$I\left(A:B\right) = 2\mathrm{S}\left(\rho_A\right) = 2\mathrm{S}\left(\rho_B\right). \tag{2.85}$$

This quantum function has *non-classical* properties, such as that its value for a pure joint system $\rho_{AB}$ can be

$$I\left(A:B\right)=2\mathrm{S}\left(\rho_A\right)=2\mathrm{S}\left(\rho_B\right) \tag{2.86}$$

while

$$\mathrm{S}\left(\rho_A\right)=\mathrm{S}\left(\rho_B\right) \text{ and } \mathrm{S}\left(\rho_{AB}\right)=0\,. \tag{2.87}$$

In Sections 2.1.7 and 2.1.8 we have shown that if we use entangled states, the quantum mutual information could be *2*, while the quantum conditional entropies could be *-1*. In classical information theory, negative entropies can be obtained only in the case of mutual information of three or more systems. An important property of maximized quantum mutual information: *it is always additive for a quantum channel*.

The character of classical information and quantum information is significantly different. There are many phenomena in quantum systems which cannot be described classically, such as entanglement, which makes it possible to store quantum information in the correlation of quantum states. Similarly, a quantum channel can be used with pure orthogonal states to realize classical information transmission, or it can be used to transmit non-orthogonal states or even quantum entanglement. Information transmission also can be approached using the question, whether the input consists of unentangled or entangled quantum states. This leads us to say that for quantum channels many new capacity definitions exist in comparison to a classical communication channel. In possession of the general communication model and the quantities which are able to represent information content of quantum states we can begin to investigate the possibilities and limitations of information transmission through quantum channels.

### 2.3.8 Quantum Relative Entropy

The *quantum relative entropy* measures the informational distance between quantum states, and introduces a deeper characterization of the quantum states than the von Neumann entropy. Similarly to the classical relative entropy, this quantity measures the distinguishability of the quantum states, in practice it can be realized by POVM measurements. The relative entropy classically is a measure that quantifies how close a probability distribution $p$ is to a model or candidate probability distribution $q$. For probability distributions $p$ and $q$, the classical relative entropy is given by

$$D\left(p\|q\right)=\sum_i p_i \log\left(\frac{p_i}{q_i}\right), \tag{2.88}$$

while the quantum relative entropy between quantum states $\rho$ and $\sigma$ is

$$D(\rho\|\sigma) = Tr(\rho\log(\rho)) - Tr(\rho\log(\sigma)) = Tr[\rho(\log(\rho) - \log(\sigma))]. \quad (2.89)$$

In the definition above, the term $Tr(\rho\log(\sigma))$ is finite only if $\rho\log(\sigma) \geq 0$ for all diagonal matrix elements. If this condition is not satisfied, then $D(\rho\|\sigma)$ could be infinite, since the trace of the second term could go to infinity.

The *quantum informational distance* (i.e., quantum relative entropy) has some distance-like properties (for example, the quantum relative entropy function between a maximally mixed state and an arbitrary quantum state is symmetric, hence in this case it is not just a pseudo distance), however it is *not commutative*, thus $D(\rho\|\sigma) \neq D(\sigma\|\rho)$, and $D(\rho\|\sigma) \geq 0$ iff $\rho \neq \sigma$, and $D(\rho\|\sigma) = 0$ iff $\rho = \sigma$. Note, if $\sigma$ has zero eigenvalues, $D(\rho\|\sigma)$ may diverge, otherwise it is a finite and continuous function. Furthermore, the quantum relative entropy function has another interesting property, since if we have two density matrices $\rho$ and $\sigma$, then the following property holds for the traces used in the expression of $D(\rho\|\sigma)$

$$Tr(\rho\log(\rho)) \geq Tr(\rho\log(\sigma)). \quad (2.90)$$

The symmetric Kullback-Leibler distance is widely used in classical systems, for example in computer vision and sound processing. Quantum relative entropy reduces to the classical Kullback-Leibler relative entropy for simultaneously diagonalizable matrices.

We note, the quantum mutual information can be defined by quantum relative entropy $D(\cdot\|\cdot)$. This quantity can be regarded as the informational distance between the tensor product of the individual subsystems $\rho_A \otimes \rho_B$, and the joint state $\rho_{AB}$ as follows:

$$I(A:B) = D(\rho_{AB}\|\rho_A \otimes \rho_B) = \mathrm{S}(\rho_A) + \mathrm{S}(\rho_B) - \mathrm{S}(\rho_{AB}). \quad (2.91)$$

### 2.3.9 Quantum Rényi-entropy

As we have seen, the quantum informational entropy can be defined by the $\mathrm{S}(\rho)$ von Neumann entropy function. On the other hand, another entropy function can also be defined in the quantum domain, it is called the Rényi-entropy and denoted by $\mathrm{R}(\rho)$. This function has relevance mainly in the description of quantum entanglement. The Rényi-entropy function is defined as follows

$$\mathrm{R}(\rho) = \frac{1}{1-r} Tr(\rho^r), \tag{2.92}$$

where $r \geq 0$. $\mathrm{R}(\rho)$ is equal to the von Neumann entropy function $\mathrm{S}(\rho)$ if

$$\lim_{r \to 1} \mathrm{R}(\rho) = \mathrm{S}(\rho). \tag{2.93}$$

If parameter $r$ converges to infinity, then we have

$$\lim_{r \to \infty} \mathrm{R}(\rho) = -\log(\|\rho\|). \tag{2.94}$$

On the other hand if $r = 0$ then $\mathrm{R}(\rho)$ can be expressed from the rank of the density matrix

$$\mathrm{R}(\rho) = \log(rank(\rho)). \tag{2.95}$$

## 2.4 Related Work

The field of quantum information processing is a rapidly growing field of science, however there are still many challenging questions and problems. These most important results will be discussed in further sections, but these questions cannot be exposited without a knowledge of the fundamental results of quantum information theory.

### Early Years of quantum information theory

quantum information theory extends the possibilities of classical information theory, however for some questions, it gives extremely different answers. The advanced communications and quantum networking technologies offered by quantum information processing will revolutionize traditional communication and networking methods. Classical information theory— was founded by Claude Shannon in 1948 [Shannon48]. In Shannon's paper the mathematical framework of communication was invented, and the main definitions and theorems of classical information theory were laid down. On the other hand, classical information theory is just one part of quantum information theory. The other, missing part is the Quantum Theory, which was completely finalized in 1926.
The results of quantum information theory are mainly based on the results of von Neumann, who constructed the mathematical background of quantum mechanics [Neumann96]. An interesting—and less well known—historical fact is that quantum entropy was discovered by Neumann before the classical information theoretic concept of entropy. Quantum entropy was discovered in the 1930s, based on the older idea of entropy in classical Statistical Mechanics, while the classical information theoretic concept was discovered by Shannon only later, in 1948. It is an interesting note, since the reader might have thought that quantum entropy is an extension of the classical one, however it is not true. Classical entropy, in the context of Information Theory, is a special case of von Neumann's quantum entropy. Moreover, the name of Shannon's formula was proposed by von Neumann. Further details about the history of Quantum Theory, and the main results of physicists from the first half of the twentieth century——such as Planck, Einstein, Schrödinger, Heisenberg, or Dirac——can be found in the works of Misner et al. [Misner09], McEvoy [McEvoy04], Sakurai [Sakurai94], Griffiths [Griffiths95] or Bohm [Bohm89].

"*Is quantum mechanics useful*"— asked by Landauer in 1995 [Landauer95]. Well, having the results of this paper in our hands, we can give an affirmative answer: *definitely yes*. An interesting work about the importance of quantum mechanical processes was published by Dowling [Dowling03]. Some fundamental results from the very early days of Quantum Mechanics can be found in [Planck1901], [Thomson1901], [Einstein1905], [Gerlach1922], [de Broglie1924], [Schrödinger1926], [Heisenberg1925], [Einstein1935], [Schrödinger1935], [Dirac82]. About the early days of Information Theory see the work of Pierce [Pierce73]. A good introduction to Information Theory can be found in the work of Yeung [Yeung02]. More information about the connection of Information Theory and statistical mechanics can be found in work of Aspect from 1981 [Aspect81], in the book of Jaynes [Jaynes03] or Petz [Petz08]. The elements of classical information theory and its mathematical background were summarized in a very good book by Cover [Cover91]. On matrix analysis a great work was published by Horn and Johnson [Horn86].

A very good introduction to quantum information theory was published by Bennett and Shor [Bennett98]. The idea that the results of quantum information theory can be used to solve computational problems was first claimed by Deutsch in 1985 [Deutsch85].

Later in the 90s, the answers to the most important questions of quantum information theory were answered, and the main elements and the fundamentals of this field were discovered. Details about the simulation of quantum systems and the possibility of encoding quantum information in physical particles can be found in Feynman's work from 1982 [Feynman82]. Further information on quantum simulators and continuous-time automata can be found in the work of Vollbrecht and Cirac [Vollbrecht08].

**Quantum Coding and Quantum Compression**

The next milestone in quantum information theory is Schumacher's work from 1995 [Schumacher95a] in which he introduced the term, "*qubit.*" In [Schumacher96a-c] the main theories of quantum source coding and the quantum compression were presented. The details of quantum data compression and quantum typical subspaces can be found in [Schumacher95a]. In this paper, Schumacher extended those results which had been presented a year before, in 1994 by Schumacher and Jozsa on a new proof of quantum noiseless coding, for details see [Schumacher94]. Schumacher in 1995 also defined the quantum coding of pure quantum states; in the same year, Lo published a paper in which he extended these result to mixed quantum states, and he also defined an encoding scheme for it [Lo95]. Schumacher's results from 1995 on the compression of quantum information [Schumacher95a] were the first main results on the encoding of quantum information——*its importance and significance in quantum information theory is similar to Shannon's noiseless channel coding theorem in classical information theory.* In this work, Schumacher also gives upper and lower bounds on the rate of quantum compression. We note, that the mathematical background of Schumacher proof is very similar to Shannon's proof, as the reader can check in [Schumacher95a] and in Shannon's proof [Shannon48].

The method of sending classical bits via quantum bits was firstly completed by Schumacher et al. in their famous paper form 1995, see [Schumacher95]. In the same year, an important paper on the encoding of information into physical particles was published by Schumacher. The fundaments of noiseless quantum coding were laid down by Schumacher, one year later, in 1996 [Schumacher96]. In 1996, many important results were published by Schumacher and his colleges. These works cover the discussion of the relation of entropy exchange and coherent quantum information, which was completely unknown before 1996. The theory of processing of quantum information, the transmission of entanglement over a noisy quantum channel, the error-correction schemes with the achievable fidelity limits, or the classical information capacity of a quantum channel with the limits on the amount of accessible information in a quantum channel were all published in the same year. For further information on the fidelity limits and communication capabilities of a noisy quantum channel, see the work of Barnum et al. also from 1996 [Barnum96]. In 1997, Schumacher and Westmoreland completed their proof on the classical capacity of a quantum channel, and they published in their famous work, for details see [Schumacher97]. These results were extended in their works from 1998, see [Schumacher98a-98c]. On the experimental side of fidelity testing see the work of Radmark et al. [Radmark09].

About the limits for compression of quantum information carried by ensembles of mixed states, see the work of Horodecki [Horodecki98]. An interesting paper about the quantum coding of mixed quantum states was presented by Barnum et al. [Barnum01]. Universal quantum compression makes it possible to compress quantum information without the knowledge about the information source itself which emits the quantum states. Universal quantum information compression was also investigated by Jozsa et al. [Jozsa98], and an extended version of Jozsa and Presnell [Jozsa03]. Further information about the technique of universal quantum data compression can be found in the article of Bennett et al. [Bennett06]. The similarity of the two schemes follows from the fact that in both cases we compress quantum information, however in the case of Schumacher's method we do not compress entanglement. The two compression schemes are not equal to each other, however in some cases——if running one of the two schemes fails——they can be used to correct the errors of the other, hence they can be viewed as auxiliary protocols of each other. Further information about the mathematical background of the processes applied in the compression of quantum information can be found in Elias's work [Elias72].

A good introduction to quantum error-correction can be found in the work of Gottesman, for details see [Gottesman04a]. A paper about classical data compression with quantum side information was published by Devetak and Winter [Devetak03a]. We note that there is a connection between the compression of quantum information and the concentration of entanglement, however the working method of Schumacher's encoding and the process of entanglement concentrating are completely different. Benjamin Schumacher and Richard Jozsa published a very important paper in 1994 [Schumacher94]. Here, the authors were the first to give an explicit proof of the quantum noiseless coding theorem, which was a milestone in the history of quantum computation. Further information on Schumacher's noiseless quantum channel coding can be found in [Schumacher96].

The basic coding theorems of quantum information theory were summarized by Winter in 1999 [Winter99a]. In this work, he also analyzed the possibilities of compressing quantum information. A random coding based proof for the quantum coding theorem was shown by Klesse in 2008 [Klesse08]. A very interesting article was presented by Horodecki in 1998 [Horodecki98], about the limits for the compression of quantum information into mixed states. On the properties of indeterminate-length quantum coding see the work of Schumacher and Westmoreland [Schumacher01a].

The quantum version of the well-known Huffman coding can be found in the work of Braunstein et al. from 2000 [Braunstein2000]. Further information about the compression of quantum information and the subspaces can be found in [Fukuda10a], [Hayden08a], and [Hayden08b]. The details of quantum coding for mixed states can be found in the work of Barnum et al. [Barnum01].

**Quantum Entanglement**

Entanglement is one of the most important differences between the classical and the quantum worlds. An interesting paper on communication via one- and two-particle operators on Einstein-Podolsky-Rosen states was published in 1992, by Bennett [Bennett92c]. About the history of entanglement see the paper of Einstein, Podolsky and Rosen from 1935 [Einstein1935]. In this manuscript, we did not give a complete mathematical background of quantum entanglement—further details on this topic can be found in Nielsen's book [Nielsen2000] or by Hayashi [Hayashi06], or in an very good article published by the four Horodeckis in 2009 [Horodecki09]. We have seen that entanglement concentration can be applied to generate maximally mixed entangled states. We also gave the asymptotic rate at which entanglement concentration can be made, it is called the entropy of entanglement and we expressed it in an explicit form. A very important paper on the communication cost of entanglement transformations was published by Hayden and Winter, for details see [Hayden03a]. The method of entanglement concentration was among the first quantum protocols, for details see the work of Bennett et al. from 1996 [Bennett96b]. The method of Bennett's was improved by Nielsen in 1999, [Nielsen99]. A very important work on variable length universal entanglement concentration by local operations and its application to teleportation and dense coding was published by Hayashi and Matsumoto [Hayashi01]. The entanglement cost of antisymmetric states was studied by [Matsumoto04].

The calculation of entanglement-assisted classical capacity requires a superdense protocol-like encoding and decoding strategy,——we did not explain its working mechanism in detail, further information can be found in the work of Bennett et al. [Bennett02]. A paper about the compression of quantum-measurement operations was published by Winter and Massar in 2001 [Winter01a]. Later, in 2004, Winter extended these results [Winter04]. Here we note, these results are based on the work of Ahlswede and Winter [Ahlswede02].

The definition of a conditionally typical subspace in quantum information was given by Schumacher and Westmoreland in 1997 [Schumacher97]. Holevo also introduced it in 1998 [Holevo98].

We did not explain in detail entanglement concentrating [Bennett96b], entanglement transformations [Nielsen99], or entanglement generation, entanglement distribution and quantum broadcasting,——further information can be found in [Hayashi01], [Hayden03a], [Hsieh08], [Winter01], [Yard05a], [Yard05b]. About the classical communication cost of entanglement manipulation see the work of Lo and Popescu from 1999 [Lo99a]. The fact that noncommuting mixed states cannot be broadcast was shown by Barnum et al. in 1995, see [Barnum95].

Lo and Popescu also published a work on concentrating entanglement by local actions in 2001, for details see [Lo01]. About the purification of noisy entanglement see the article of Bennett et al. from 1996 [Bennett96c]. The entanglement purification protocol was a very important result, since it will have great importance in the quantum capacity of a quantum channel. (However, when the authors have developed the entanglement purification scheme, this connection was still not completely cleared.)

About the quantum networks for concentrating entanglement and the distortion-free entanglement concentration, further information can be found in the paper of Kaye and Mosca from 2001 [Kaye01]. In 2005, Devetak and Winter have shown, that there is a connection between the entanglement distillation and the quantum coherent information, which measure has tremendous relevance in the quantum capacity of the quantum channels, for details see [Devetak05]. An interesting paper about distortion-free entanglement concentration was published by Kohout et al. in 2009 [Kohout09]. The method presented in that paper gives an answer to streaming universal. We did not mentioned the inverse protocol of entanglement concentration which is called entanglement dilution, for further details see the works of Lo and Popescu from 1999 [Lo99a] and 2001 [Lo01], and Harrow and Lo's work from 2004 [Harrow04a]. Harrow and Lo have also given an explicit solution of the communication cost of the problem of entanglement dilution, which was an open question until 2004. Their results are based on the previous work of Hayden and Winter from 2003, for details see [Hayden03a]. The typical entanglement in stabilizer states was studied by Smith and Leung, see [Smith06]. The teleportation-based realization of a two-qubit entangling gate was shown by Gao et al. [Gao10].

**Quantum Channels**

About the statistical properties of the HSW theory and the general HSW capacity, a very interesting paper was published by Hayashi and Nagaoka in 2003 [Hayashi03]. As we have seen, some results of quantum information theory are similar to the results of classical information theory, however *many things have no classical analogue*. As we have found in this section, the Holevo theorem gives an information-theoretic meaning to the von Neumann entropy, however it does not make it possible to use it in the case of the interpretation of von Neumann entropy of physical macrosystems. Further properties of the von Neumann entropy function was studied by Audenaert in 2007 [Audenaert07].

The concept of quantum mutual information measures the classical information which can be transmitted through a noisy quantum channel (originally introduced by Adami and Cerf [Adami96]) however it cannot be used to measure the maximal transmittable quantum information. The maximized quantum mutual information is always additive, however this is not true for the Holevo information. In this case, the entanglement makes non-additive the Holevo information, but it has no effect on the quantum mutual information. Further information about the mathematical background of these “strange” phenomena can be found in the work of Adami from 1996 [Adami96] or in the book of Hayashi from 2006 [Hayashi06]. A very good book on these topics was published by Petz in 2008 [Petz08].

For the properties of Holevo information and on the capacity of quantum channels see the works of Holevo [Holevo73], [Holevo98], Schumacher and Westmoreland [Schumacher96-96f], [Schumacher97], Horodecki [Horodecki05], Datta [Datta04a], Arimoto [Arimoto72] , On the geometrical interpretation of the maps of a quantum channel see the works of Cortese [Cortese02], Petz [Petz96], [Petz07-10], Hiai [Hiai91], [Hiai2000], [Imai88].

On physical properties of quantum communication channels the work of Levitin [Levitin69], on the capacities of quantum communication channels see Bennett [Bennett97], DiVincenzo98 [DiVincenzo98], Schumacher [Schumacher97],

Fuchs [Fuchs97]. In 1997, Barnum, Smolin and Terhal also summarized the actual (in 1997) results on quantum channel, see [Barnum97b].

The mathematical background of distinguishing arbitrary multipartite basis unambiguously was shown by Duan et al. [Duan07].)

In 2010, Dupis et al. [Dupuis10] published a paper in which they described a protocol for quantum broadcast quantum channel, then Jon Yard et al. published a paper on quantum broadcast channels [Yard06]. Before these results, in 2007, an important practical result on broadcasting was shown by Guha et al. [Guha07a], [Guha07b], who demonstrated the classical capacity of practical (bosonic) quantum channels. General quantum protocols—such as super-dense coding and teleportation—are not described in this article. Further information about these basic quantum protocols can be found in the book of Hayashi from 2006 [Hayashi06], in the book of Nielsen and Chuang [Nielsen2000], or in the paper of Bennett and Wiesner [Bennett92c], and Bennett [Bennett92b], (both papers from 1992), and Bennett's paper from 1993 [Bennett93].

A very good overview of the capacity of quantum channels was presented by Smith in 2010, see [Smith10]. About the information tradeoff relations for finite-strength quantum measurements, see the works of [Fuchs2000]. On the mathematical background of quantum communication see the works of [Petz96], Ruskai et al. [Ruskai01], [Hayashi05] and [Vedral98]. The generalized Pauli channels are summarized by Ohno and Petz in [Ohno09].

The relative entropy function was introduced by Solomon Kullback and Richard Leibler in 1951 [Kullback51]. Another interpretation of the relative entropy function was introduced by Bregman, known as the class of Bregman divergences [Bregman67]. A very important paper about the role of relative entropy in quantum information theory was published by Schumacher and Westmoreland in 2000 [Schumacher2000]. The quantum relative entropy function was originally introduced by Umegaki, and later modified versions have been defined by Ohya, Petz and Watanbe [Ohya97]. Some possible applications of quantum relative entropy in quantum information processing were introduced by Vedral [Vedral2000].

About the negativity of quantum information see the works of Horodecki et al. [Horodecki05], [Horodecki07]. About the use of entanglement in quantum information theory, see the work of Li et al. from 2010, [Li10]. A method for measuring two-qubit entanglement by local operations and classical communication was shown by Bai et al. in 2005 [Bai05]. About the additivity of the capacity of quantum channels see [Fujiwara02], [King09] and [Shor02a]. A very good paper on the Holevo capacity of finite dimensional quantum channels and the role of additivity problem in quantum information theory was published by Shirokov [Shirokov06]. We note, the additivity problem also will be discussed in detail Section 6. A great summary of classical and quantum information theory can be found in the book of Desurvire from 2009 [Desurvire09]. The bounds for the quantity of information transmittable by a quantum communication channel was analyzed by Holevo in 1973, see [Holevo73]. About sending classical information via noisy quantum channels, see the works of Schumacher and Jozsa [Schumacher94], Schumacher from 1996 [Schumacher96], and Schumacher and Westmoreland from 1997 [Schumacher97] and Smith's summarize [Smith10]. The mathematical background of classical relative entropy function can be found in the works of Kullback and Leibler [Kullback51], [Kullback59], and [Kullback87]. For the details of Bregman distance see [Bregman67] and [Banerjee05]. Further information about the Kraft-McMillan inequality can be found in [Kraft49], [McMillan56] and [Cover91].

**Comprehensive Surveys**

A reader who is interested in the complete mathematical background of quantum information theory can find the details for example in Nielsen and Chuang's book [Nielsen2000]. For a general introduction to the quantum information theory and its applications see the excellent book of Hayashi [Hayashi06]. We also suggest the book of Imre from 2005, see [Imre05]. A very good introduction to quantum information theory was published by Bennett and Shor, for details see [Bennett98]. Also in 1998, Preskill summarized the actual state of quantum information theory in the form of lecture notes [Preskill98]. Preskill also summarized the conditions of reliable quantum computers, for details see [Preskill98a]. Also in 1998, a good work on the basics of quantum computations and the mathematical formalism was published by Vedral and Plenio [Vedral98] and by Nielsen [Nielsen98]. On the mathematical background of quantum information processing, see the works of Shor [Shor94], [Shor97], [Shor02], and [Shor04]. The description of classical data compression can be found in the very good book of Cover and Thomas [Cover91], or in the book of Berger [Berger71]. We also suggest the work of Stinespring [Stinespring55]. A very important result regarding the compression of classical information was published by Csiszár and Körner in 1978 [Csiszár78], and later the authors published a great book about coding theorems for discrete memoryless systems [Csiszár81]. A work on the non-additivity of Renyi entropy was published by Aubrun et al. [Aubrun09]. On the connection of quantum entanglement and classical communication through a depolarizing channel see [Bruss2000]. Regarding the results of quantum Shannon theory, we suggest the great textbook of Wilde [Wilde11]. The structure of random quantum channels, eigenvalue statistics and entanglement of random subspaces are discussed in [Collins09a], [Collins09b]. We also suggest the physics lectures of Feynman [Feynman98]. Finally, for an interesting viewpoint on “topsy turvy world of quantum computing” see [Mullins01].

# 3. Classical Capacities of a Quantum Channel

Communication over quantum channels is bounded by the corresponding capacities. Now, we lay down the fundamental theoretic results on *classical capacities of quantum channels*. These results are all required to analyze the advanced and more promising properties of quantum communications.

Section 3 is organized as follows. In the first part of this section, we introduce the reader to formal description of a noisy quantum channel. Then we start to discuss the classical capacity of a quantum channel. Next, we show the various encoder and decoder settings for transmission of classical information. We define the exact formula for the measure of maximal transmittable classical information. Finally, we discuss some important channel maps.
The most relevant works are included in the Related Work subsection.

## 3.1 Extended Formal Model

The discussed model is general enough to analyze the limitations for information transfer over quantum channels. However, later we will investigate special quantum channels which models specific physical environment. Each quantum channel can be represented as a CPTP map (*Completely Positive Trace Preserving*), hence the process of information transmission through a quantum communication channel can be described as a quantum operation.

The general model of a quantum channel describes the transmission of an input quantum bit, and its interaction with the environment (see Fig. 3.1.). Assuming Alice sends quantum state $\rho_A$ into the channel this state becomes entangled with the environment $\rho_E$, which is initially in a pure state $|0\rangle$. For a mixed input state a so called *purification state* $P$ can be defined, from which the original mixed state can be restored by a partial trace (see Appendix) of the pure system $\rho_A P$. The unitary operation $U_{AE}$ of a quantum channel $\mathcal{N}$ entangles $\rho_A P$ with the environment $\rho_E$, and outputs Bob's mixed state as $\rho_B$ (and the purification state as $P$). The purification state is a reference system, it cannot be accessed, it remains the same after the transmission.

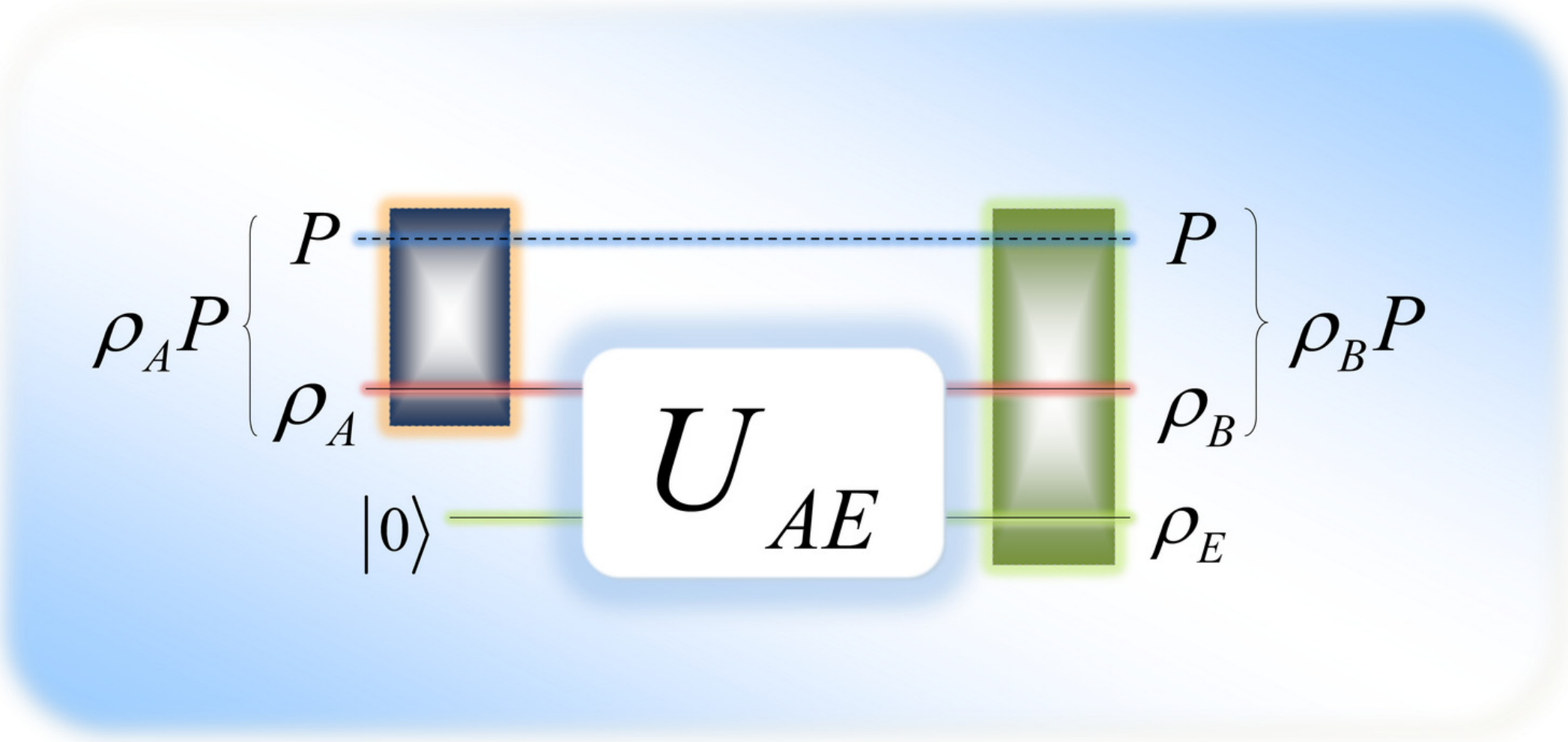

**Fig. 3.1.** The formal model of a noisy quantum communication channel. The output of the channel is a mixed state.

The output of the noisy quantum channel is denoted by $\rho_B$, the post state of the environment by $\rho_E$, while the post-purification state after the output realized on the channel output is depicted by $P$.

## 3.2 Capacity of Classical Channels

Before we start to investigate quantum channels, we survey the results of transmitting information over classical noisy channels. In order to achieve reliable (error-free) information transfer we use the so called *channel coding* which extends the payload (useful) information bits with redundancy bits so that at the receiver side Bob will be able to correct some amount of error by means of this redundancy.

The channel is given an input *A,* and maps it probabilistically (it is a *stochastic* mapping, not a unitary or deterministic transformation) to an output *B,* and the probability of this mapping is denoted by $p(B|A)$.

The *channel capacity* $C(N)$ of a *classical* memoryless communication channel $N$ gives an upper bound on the number of classical bits which can be transmitted per channel use, in reliable manner, i.e., with arbitrarily small error at the receiver.

The *capacity* of a *classical* memoryless communication channel $N$ gives an upper bound on the number of classical bits which can be transmitted per channel use, in reliable manner, i.e., with arbitrarily small error at the receiver. The simple memoryless classical channel model is shown in Fig. 3.2.

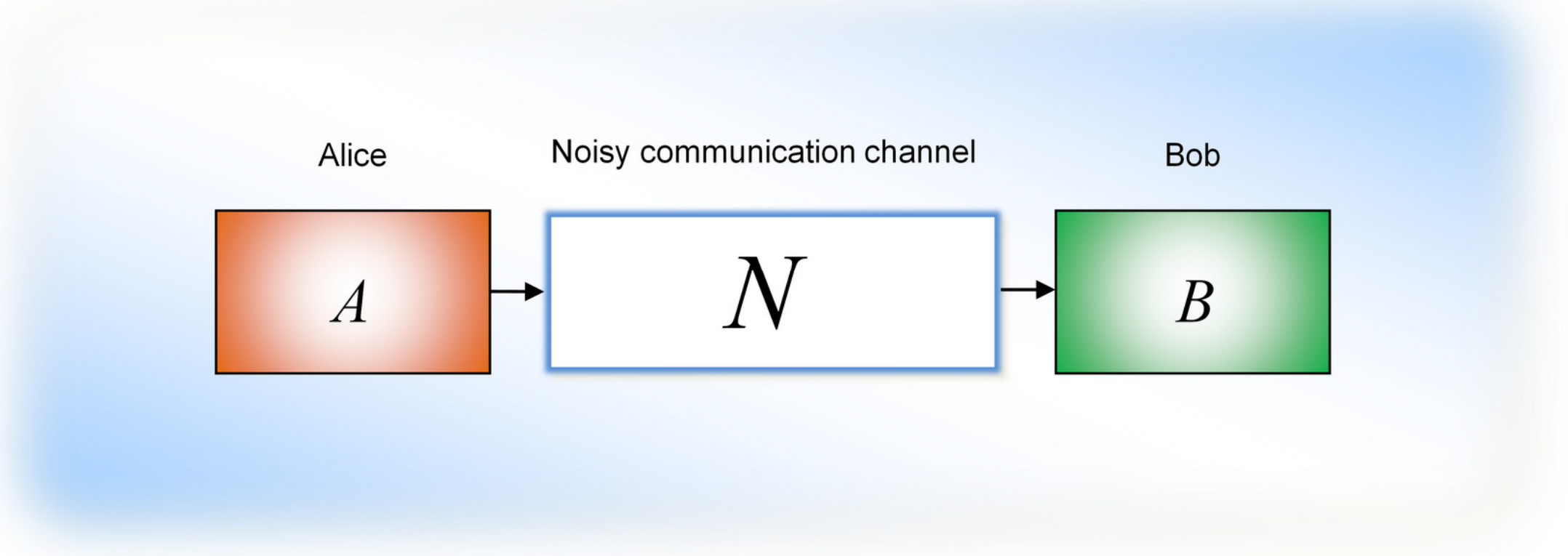

**Fig. 3.2.** Simple memoryless classical channel model for Shannon's noisy channel coding theorem.

As it has been proven by Shannon the capacity $C(N)$ of a noisy classical memoryless communication channel $N$, can be expressed by means of the maximum of the mutual information $I(A:B)$ over all possible input distributions $p(x)$ of random variable $X$

$$C(N) = \max_{p(x)} I(A:B). \tag{3.1}$$

In order to make the capacity definition more plausible let us consider Fig. 3.3. Here, the effect of environment $E$ is represented by the classical conditional entropies

$$H(A:E|B) > 0 \text{ and } H(B:E|A) > 0. \tag{3.2}$$

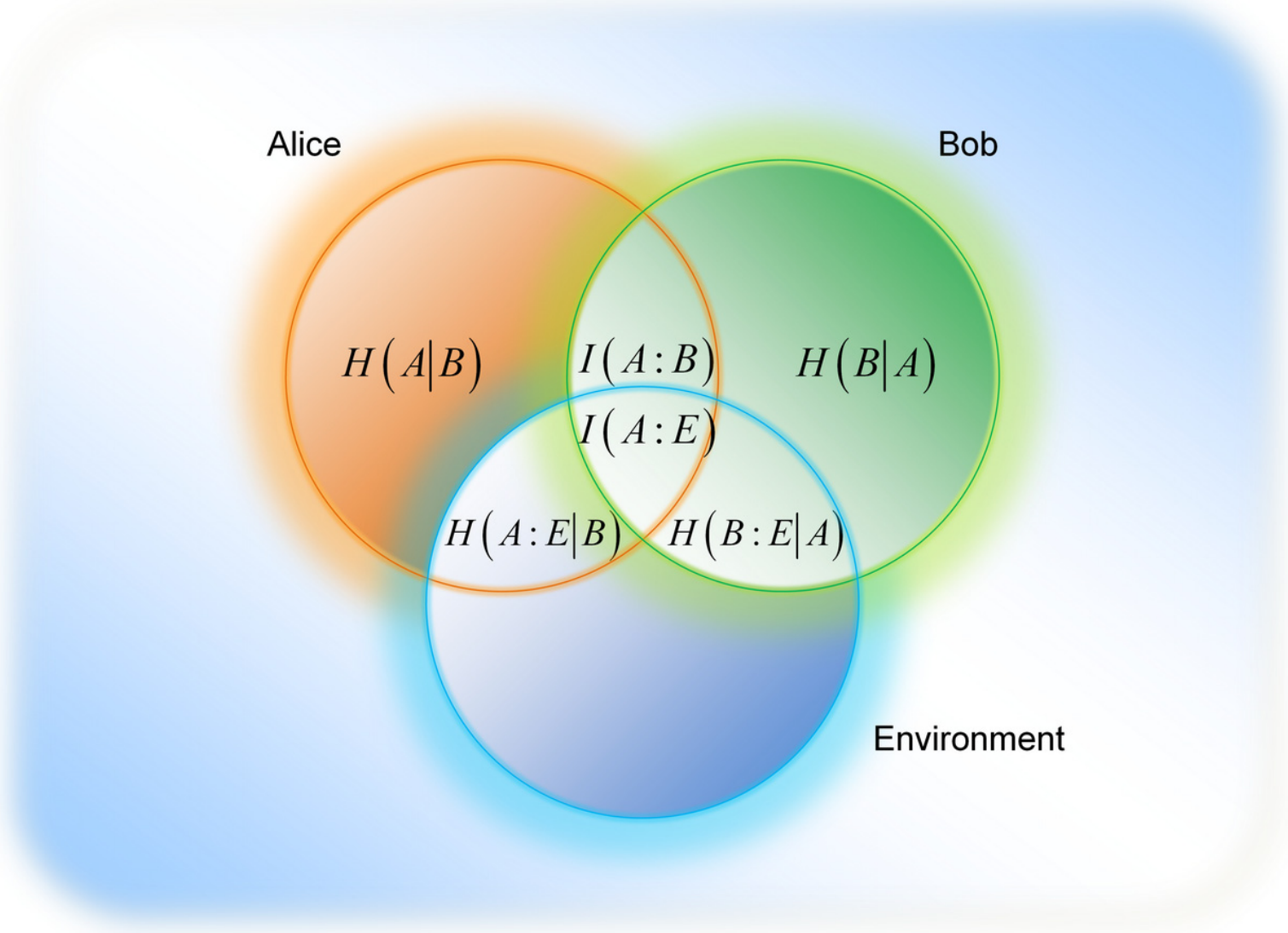

**Fig. 3.3.** The effects of the environment on the transmittable information and on the receiver's uncertainty.

Shannon's noisy coding theorem claims that forming $K$ different codewords $m = \log K$ of length from the source bits and transmitting each of them using the channel $n$ times ($m$ to $n$ coding) the rate at which information can be transmitted through the channel is

$$R = \frac{\log(K)}{n}, \tag{3.3}$$

and exponentially small probability of error at this rate can be achieved only if $R \leq C(N)$, otherwise the probability of the successful decoding exponentially tends to zero, as the number of channel uses increases. Now, having introduced the capacity of classical channel it is important to highlight the following distinction. The *asymptotic capacity* of any channel describes that rate, which can be achieved if the channel can be used $n$ times (denoted by $N^{\otimes n}$), where where $n \to \infty$. In case of $n = 1$ we speak about *single-use* (*single-letter*) capacity. Multiple channel uses can be implemented in consecutive or parallel ways, however from practical reasons we will prefer the latter one.

## 3.3 Transmission of Classical Information over Noisy Quantum Channels

As the next step during our journey towards the quantum information transfer through quantum channels (which is the most general case) we are leaving the well-known classical (macro) world and just entering into the border zone. Similar to the ancient Romans - who deployed a sophisticated wide border defense system (called *the limes* which consisted of walls, towers, rivers, etc.), instead of drawing simply a red line between themselves and the barbarians – we remain classical in terms of inputs and outputs but allow the channel operating in a quantum manner.

Quantum channels can be used in many different ways to transmit information from Alice to Bob. Alice can send classical bits to Bob, but she also has the capability of transmitting quantum bits. In the first case, we talk about the classical capacity of the quantum channel, while in the latter case, we have a different measure - the quantum capacity. The map of the channel is denoted by $\mathcal{N}$ , which is trace preserving if

$$Tr\left(\mathcal{N}\left(\rho\right)\right) = Tr\left(\rho\right) \tag{3.4}$$

for all density matrices $\rho$, and positive if the eigenvalues of $\mathcal{N}\left(\rho\right)$ are non-negative whenever the eigenvalues of $\rho$ are non-negative.

Compared to classical channels – which have only one definition for capacity – the transmittable classical information and thus the corresponding capacity definition can be different when one considers quantum channels. This fact splits the classical capacity of quantum channels into three categories, namely the (*unentangled*) *classical (*also known as the *product-state* classical capacity, or the HSW (Holevo-Schumacher-Westmoreland) capacity*) capacity* $C\left(\mathcal{N}\right)$, *private classical capacity* $P\left(\mathcal{N}\right)$ and *entanglement-assisted classical capacity* $C_E\left(\mathcal{N}\right)$ (see Fig. 3.4.).

The (*unentangled*) *classical capacity* $C\left(\mathcal{N}\right)$ is a natural extension of the capacity definition from classical channels to the quantum world. For the sake of simplicity the term *classical capacity* will refer to the *unentangled* version in the forthcoming pages of this paper. (The entangled version will be referred as the entanglement-assisted classical capacity. As we will see, the HSW capacity is defined for product state inputs; however it is possible to extend it for entangled input states)

The *private classical capacity* $P(\mathcal{N})$ has deep relevance in secret quantum communications and quantum cryptography. It describes the rate at which Alice is able to send classical information through the channel in secure manner. Security here means that an eavesdropper will not be able to access the encoded information without revealing her/himself.

The *entanglement-assisted classical capacity* $C_E(\mathcal{N})$ measures the classical information which can be transmitted through the channel, if Alice and Bob have already shared entanglement before the transmission. A well-known example of such protocols is "*superdense coding*" [Imre05]. Next, we discuss the above listed various classical capacities of quantum channels in detail.

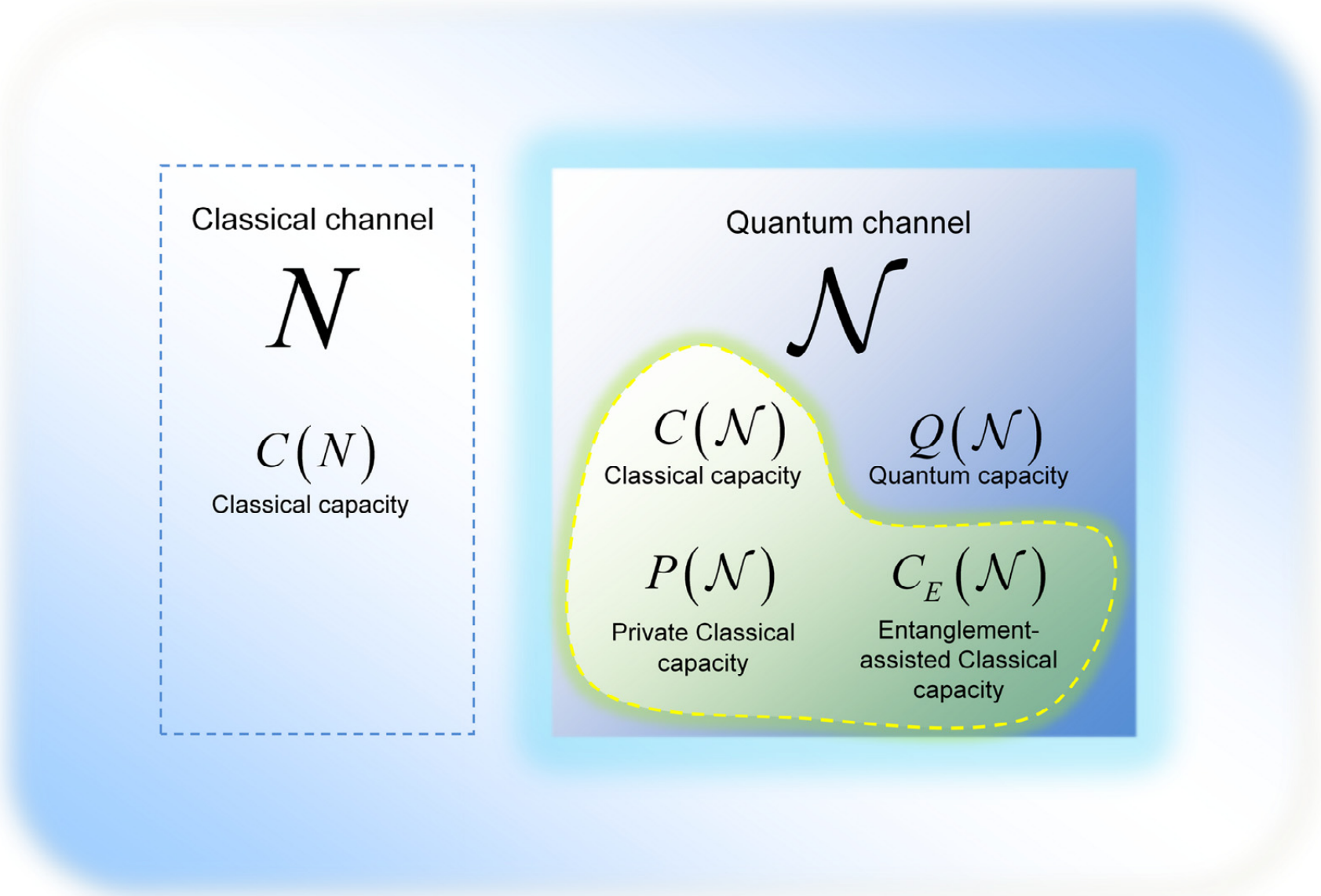

**Fig. 3.4.** The "zoo" (taxonomy) of different capacities of classical and the quantum communication channels. The classical capacities of the quantum channel are separated by a dashed line from the quantum capacity.

As the first obvious generalization of classical channel capacity definition is if we maximize the quantum mutual information over all possible input ensembles

$$C(\mathcal{N}) = \max_{\text{all } p_i, \rho_i} I(A:B). \tag{3.5}$$

Next, we start to discuss the classical information transmission capability of a noisy quantum channel.

### 3.3.1 The Holevo-Schumacher-Westmoreland Capacity

The HSW (Holevo-Schumacher-Westmoreland) theorem defines the maximum of classical information which can be transmitted through a noisy quantum channel $\mathcal{N}$ if the input contains product states (i.e., entanglement is not allowed, also known as the product-state classical capacity) and the output is measured by joint measurement setting (see the *second* measurement setting in subsection 3.3.2.1.). In this setting, for the quantum noisy communication channel $\mathcal{N}$, the classical capacity can be expressed as follows

$$\begin{aligned} C(\mathcal{N}) &= \max_{all\ p_i,\rho_i} \chi = \max_{all\ p_i,\rho_i} \left[ \mathrm{S}(\sigma_{out}) - \sum_i p_i \mathrm{S}(\sigma_i) \right] \\ &= \max_{all\ p_i,\rho_i} \left[ \mathrm{S}\left( \mathcal{N}\left( \sum_i p_i \rho_i \right) \right) - \sum_i p_i \mathrm{S}(\mathcal{N}(\rho_i)) \right] \\ &= \chi(\mathcal{N}), \end{aligned} \tag{3.6}$$

where the maximum is taken over all ensembles $\{p_i, \rho_i\}$ of input quantum states, while for $\sigma_{out}$ see (1.9), while $\chi(\mathcal{N})$ is the Holevo capacity of $\mathcal{N}$. Trivially follows, that the $\chi(\mathcal{N})$ capacity reaches its maximum for a perfect noiseless quantum channel $\mathcal{N} = I$.

If Alice chooses among a set of quantum codewords, then is it possible to transmit these codewords through the noisy quantum channel $\mathcal{N}$ to Bob with arbitrary small error, if

$$R < C(\mathcal{N}) = \max_{all\ p_i,\rho_i} \left[ \mathrm{S}\left( \mathcal{N}\left( \sum_i p_i \rho_i \right) \right) - \sum_i p_i \mathrm{S}(\mathcal{N}(\rho_i)) \right]; \tag{3.7}$$

if Alice adjusts $R$ to be under $\max_{all\ p_i,\rho_i} \chi$, then she can transmit her codewords with arbitrarily small error. If Alice chooses $R > C(\mathcal{N})$, then she cannot select a quantum code of arbitrary size, which was needed for her to realize an error-free communication. The HSW channel capacity guarantees an error-free quantum communication only if $R < C(\mathcal{N}) = \max_{all\ p_i,\rho_i} \chi$ is satisfied for her code rate $R$.

### 3.3.2 Various Classical Capacities of a Quantum Channel

The asymptotic channel capacity is the "true measure" of the various channel capacities, instead of the single-use capacity, which characterizes the capacity only in a very special case. The three classical capacities of the quantum channel (see Fig. 3.4) of quantum channels will be discussed next.

In the regularization step, the channel capacity is computed as a limit. In possession of this limit, we will use the following lower bounds for the single-use capacities. In Section 3.3.1 we have also seen, the *Holevo-Schumacher-Westmoreland* theorem gives an explicit answer to the maximal transmittable classical information over the quantum channel. Next, we show the connection between these results. As we will see in subsection 3.3.2.1, four different measurement settings can be defined for the measurement of the *classical* capacity of the quantum channel. Here we call the attention of the reader that Holevo bound (2.77) limits the classical information stored in a quantum bit. HSW theorem can be regarded a similar scenario but a quantum channel deployed between Alice and Bob introduces further uncertainty before extracting the classical information. Obviously if we assume an ideal channel the two scenarios become the same.

Now, we present an example allowing the comparison of classical capacity of a simple channel model in classical and quantum context. The binary symmetric channel inverts the input cbits with probability $p$ and leaves it unchanged with (1-$p$). The equivalent quantum bit flip channel (see Section 3.6) applies the Pauli $X$ and the identity transforms $I$.

Considering the worst case $p = 0.5$ all the sent information vanishes in the classical channel $C(N) = 1 - H\left(p\right) = 0$. However, the HSW theorem enables the optimization not only over the input probabilities but over input ensembles $\left\{p_i, \rho_i\right\}$. If we set $\rho_i$ to the eigenvectors of Pauli $X$ deriving them from its spectral decomposition

$$X = 1\left|+\right\rangle\left\langle+\right| + (-1)\left|-\right\rangle\left\langle-\right|, \tag{3.8}$$

where $\left|\pm\right\rangle = \frac{\left|0\right\rangle \pm \left|1\right\rangle}{\sqrt{2}}$, $C\left(\mathcal{N}\right) = 1$ can be achieved. This results is more than surprising, encoding into quantum states in certain cases may improve the transfer of classical information between distant points i.e., the increased degree of freedom enables reducing the uncertainty introduced by the channel.

#### 3.3.2.1 Measurement Settings

Similar to classical channel encoding, the quantum states can be transmitted in codewords $n$ qubit of length using the quantum channel consecutively $n$-times or equivalently we can send codewords over $n$ copies of quantum channel $\mathcal{N}$ denoted by $\mathcal{N}^{\otimes n}$. For the sake of simplicity we use $n=2$ in the figures belonging to the following explanation. In order to make the transient smoother between the single-shot and the asymptotic approaches we depicted the scenario using *product input states* and *single* (or independent) measurement devices at the output of the channel in Fig. 3.5. In that case the $C(\mathcal{N})$ classical capacity of quantum channel $\mathcal{N}$ with input $A$ and output $B$ can be expressed by the maximization of the $I(A:B)$ quantum mutual information as follows:

$$C(\mathcal{N}) = \max_{all\ p_i,\rho_i} I(A:B). \tag{3.9}$$

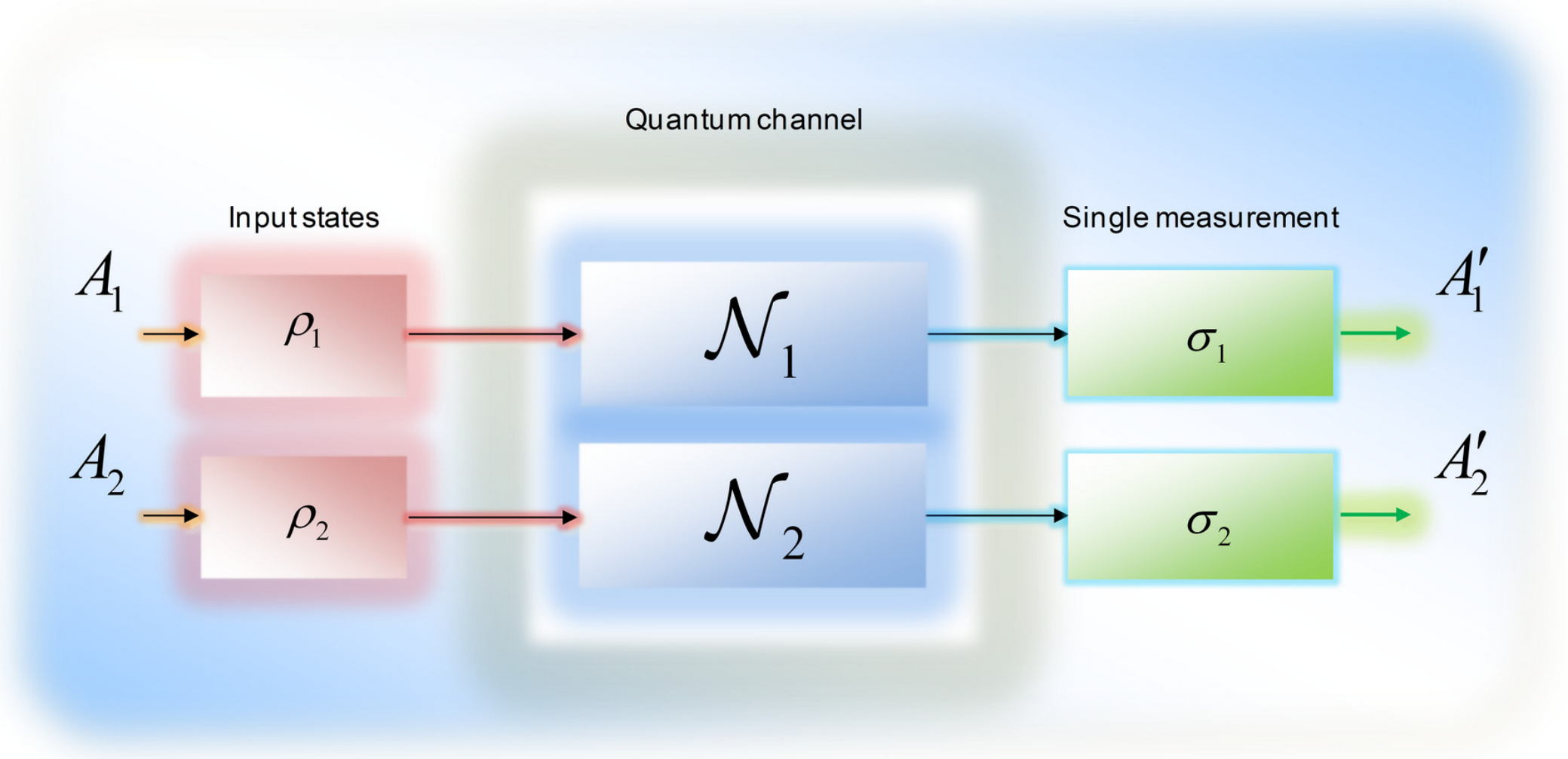

**Fig. 3.5.** Transmission of classical information over quantum channel with product state inputs and single measurements. Environment is not depicted.

From (3.9) also follows that for this setting the single-letter $C^{(1)}(N)$ and the asymptotic $C(\mathcal{N})$ classical capacities are equal:

$$C^{(1)}(N) = C(\mathcal{N}) = \max_{all\ p_i,\rho_i} I(A:B). \tag{3.10}$$

On the other hand, if we have *product state inputs* but we change the measurement setting from the single measurement setting to *joint measurement* setting, then the classical channel capacity cannot be given by (3.9), hence

$$C\left(\mathcal{N}\right) \neq \max_{all\ p_i,\rho_i} I\left(A:B\right). \tag{3.11}$$

If we would like to step forward, we have to accept the fact, that the quantum mutual information cannot be used to express the asymptotic version: the *maximized* quantum mutual information is *always additive* (see Section 2.3.) - but not the Holevo information. As follows, if we would take the regularized form of quantum mutual information to express the capacity, we will find that the asymptotic version is equal to the single-use version, since:

$$\lim_{n\to\infty} \frac{1}{n} \max_{all\ p_i,\rho_i} I\left(A:B\right) = \max_{all\ p_i,\rho_i} I\left(A:B\right). \tag{3.12}$$

From (3.12) follows, that if we have *product inputs* and *joint measurement* at the outputs, we cannot use the $\max_{all\ p_i,\rho_i} I\left(A:B\right)$ maximized quantum mutual information function to express $C\left(\mathcal{N}\right)$. If we would like to compute the classical capacity $C\left(\mathcal{N}\right)$ for that case, we have to leave the quantum mutual information function, and instead of it we have to use the maximized Holevo information $\max_{all\ p_i,\rho_i} \chi$.

This new $C\left(\mathcal{N}\right)$ capacity (according to the *Holevo-Schumacher-Westmoreland* theorem, see Section 3.3.1) can be expressed by the Holevo capacity $\chi\left(\mathcal{N}\right)$, which will be equal to the maximization of Holevo information of channel $\mathcal{N}$:

$$C\left(\mathcal{N}\right) = \chi\left(\mathcal{N}\right) = \max_{all\ p_i,\rho_i} \chi. \tag{3.13}$$

The Holevo capacity and the asymptotic channel capacity will be equal in this case.

The HSW theorem gives an explicit answer for the classical capacity of the *product state input* with *joint measurement* setting, and expresses $C\left(\mathcal{N}\right)$ as follows:

$$C\left(\mathcal{N}\right) = \chi\left(\mathcal{N}\right) = \max_{all\ p_i,\rho_i} \left[ \mathrm{S}\left( \mathcal{N}\left( \sum_i p_i\rho_i \right) \right) - \sum_i p_i \mathrm{S}\left( \mathcal{N}\left(\rho_i\right) \right) \right]. \tag{3.14}$$

The relation discussed above holds for the restricted channel setting illustrated in Fig. 3.6, where the input consists of product states, and the output is measured by a joint measurement setting.

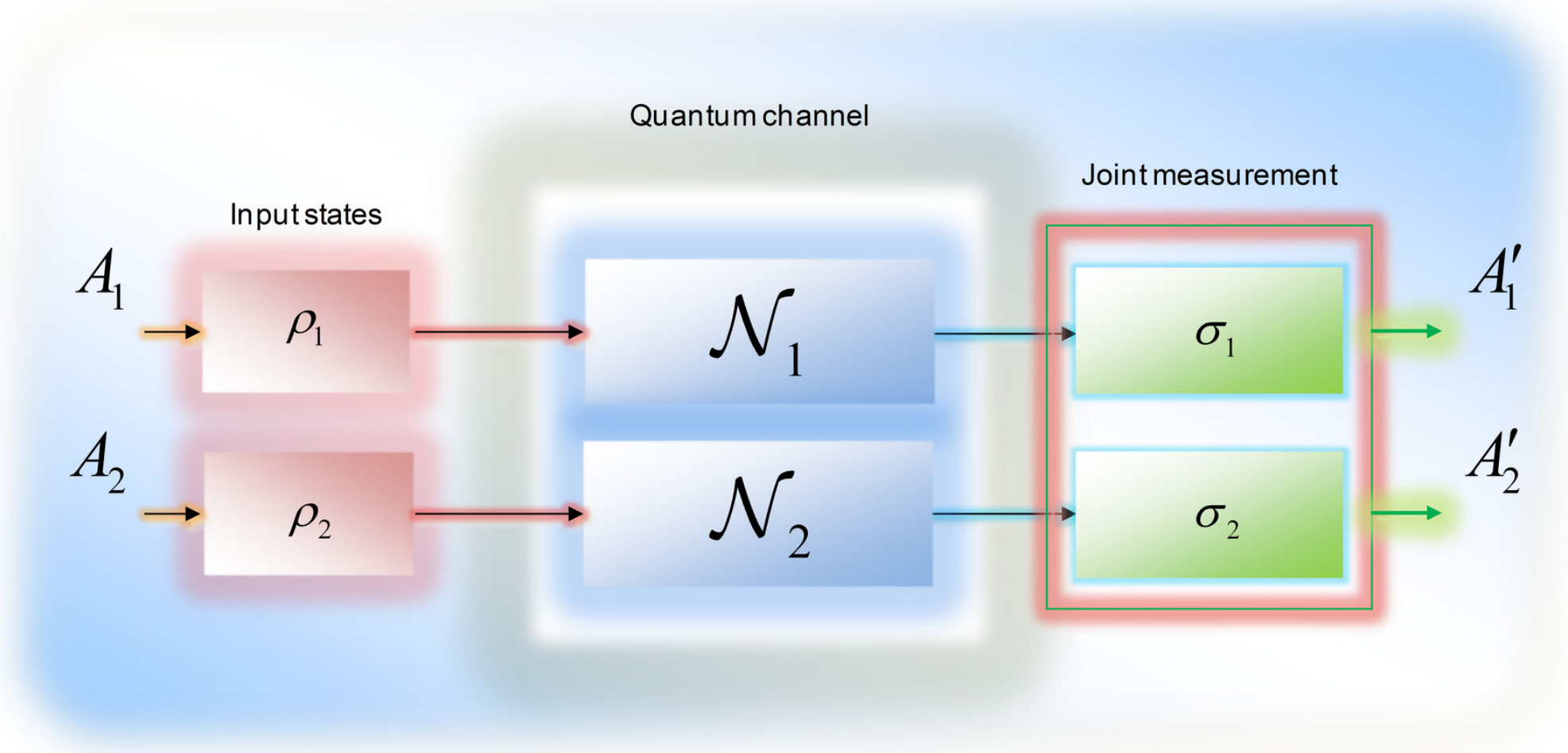

**Fig. 3.6.** Transmission of classical information over quantum channel with product state inputs and joint measurements. Environment is not depicted.

However, if *entangled inputs* are allowed with the *joint measurement setting* - then this equality does not hold anymore. As a conclusion, the relation between the maximized Holevo information $\chi(\mathcal{N})$ of the channel of the channel and the asymptotic classical channel capacity $C(\mathcal{N})$:

$$\chi(\mathcal{N}) \leq C(\mathcal{N}). \tag{3.15}$$

This means that we have to redefine the asymptotic formula of $C(\mathcal{N})$ for entangled inputs and joint measurement setting, to measure the maximum transmittable classical information through a quantum channel.

In the 1990s, it was conjectured that the formula of (3.14) can be applied to describe the channel capacity for entangled inputs with the *single measurement* setting; however it was an open question for a long time. Single measurement *destroys* the possible benefits arising from the entangled inputs, and joint measurement is required to achieve the benefits of entangled inputs [King2000].

In 2009 Hastings have used *entangled input states* and showed that the entangled inputs (with the *joint measurement*) can increase the amount of classical information which can be transmitted over a noisy quantum channel. In this case, $C(\mathcal{N}) \neq \chi(\mathcal{N})$ and the $C(\mathcal{N})$ can be expressed with the help of Holevo capacity as follows, using the asymptotic formula of $\chi(\mathcal{N})$:

$$C(\mathcal{N}) = \lim_{n\to\infty} \frac{1}{n} \chi\left(\mathcal{N}^{\otimes n}\right). \tag{3.16}$$

The channel construction for this relation is illustrated in Fig. 3.7. The entangled input is formally denoted by $\Psi_{12}$.

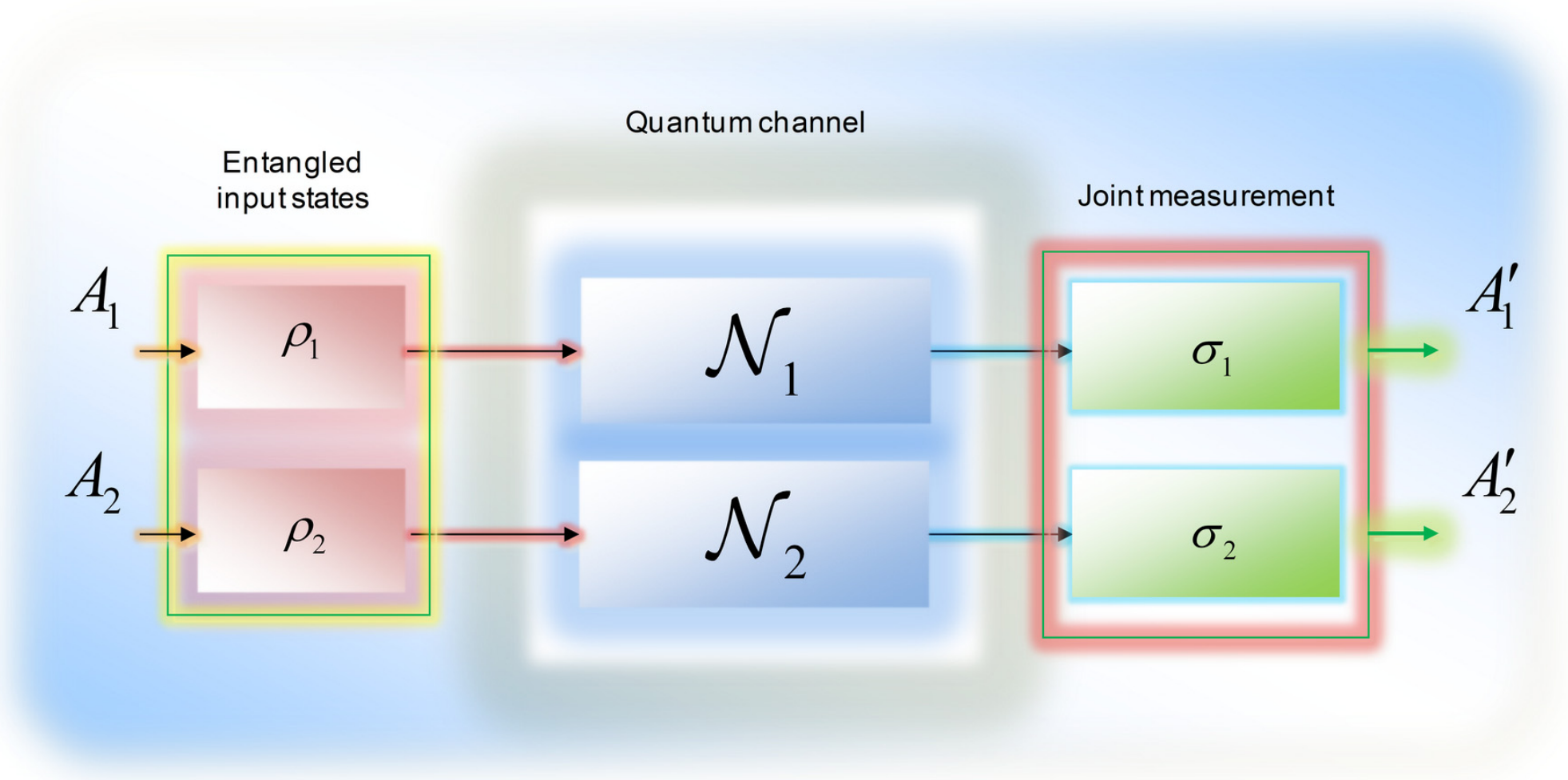

**Fig. 3.7.** Transmission of classical information over quantum channel with entangled inputs $\Psi_{12}$ and joint measurements. Environment is not depicted.

We also show the channel construction of the fourth possible construction to measure the classical capacity of a quantum channel. In this case, we have entangled input states, however we use a single measurement setting instead of a joint measurement setting.

To our knowledge, currently there is no quantum channel model where the channel capacity can be increased with this setting, since in this case the benefits of entanglement vanish because of the joint measurement setting has been changed into the single measurement setting. We illustrated this setting in Fig. 3.8.

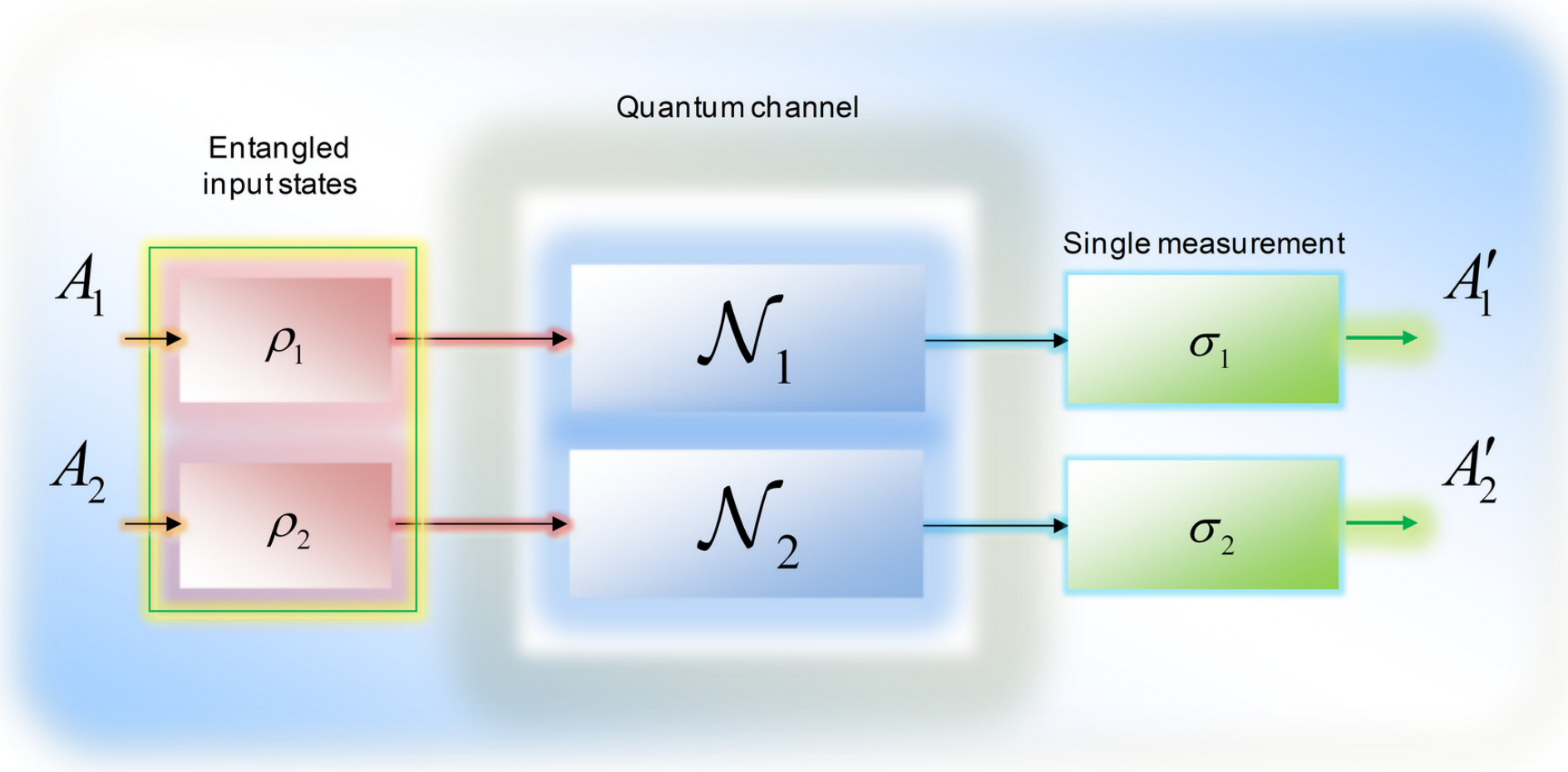

**Fig. 3.8.** Transmission of classical information over quantum channel with entangled inputs and single measurements. Environment is not depicted.

We have seen in (3.14), that if we have *product input states* and we change from a single to a *joint measurement* setting, then the classical capacity of $\mathcal{N}$ cannot be expressed by the maximized quantum mutual information function, because it is always additive (see Section 2.3.), hence

$$C\left(\mathcal{N}\right) \neq \lim_{n\to\infty} \frac{1}{n} \max_{all\ p_i,\rho_i} I\left(A:B\right). \tag{3.17}$$

If we allow *entangled input states* and *joint measurement* (see (3.16)), then we have to use the $C\left(\mathcal{N}\right)$ asymptotic formula of the previously derived Holevo capacity, $\chi\left(\mathcal{N}\right)$ which yields

$$C\left(\mathcal{N}\right) = \lim_{n\to\infty} \frac{1}{n} \chi\left(\mathcal{N}^{\otimes n}\right) \neq \chi\left(\mathcal{N}\right). \tag{3.18}$$

The general sketch of the asymptotic $C\left(\mathcal{N}\right)$ classical capacity is illustrated in Fig. 3.9. The $n$ independent uses of the quantum channel are denoted by $\mathcal{N}^{\otimes n}$.

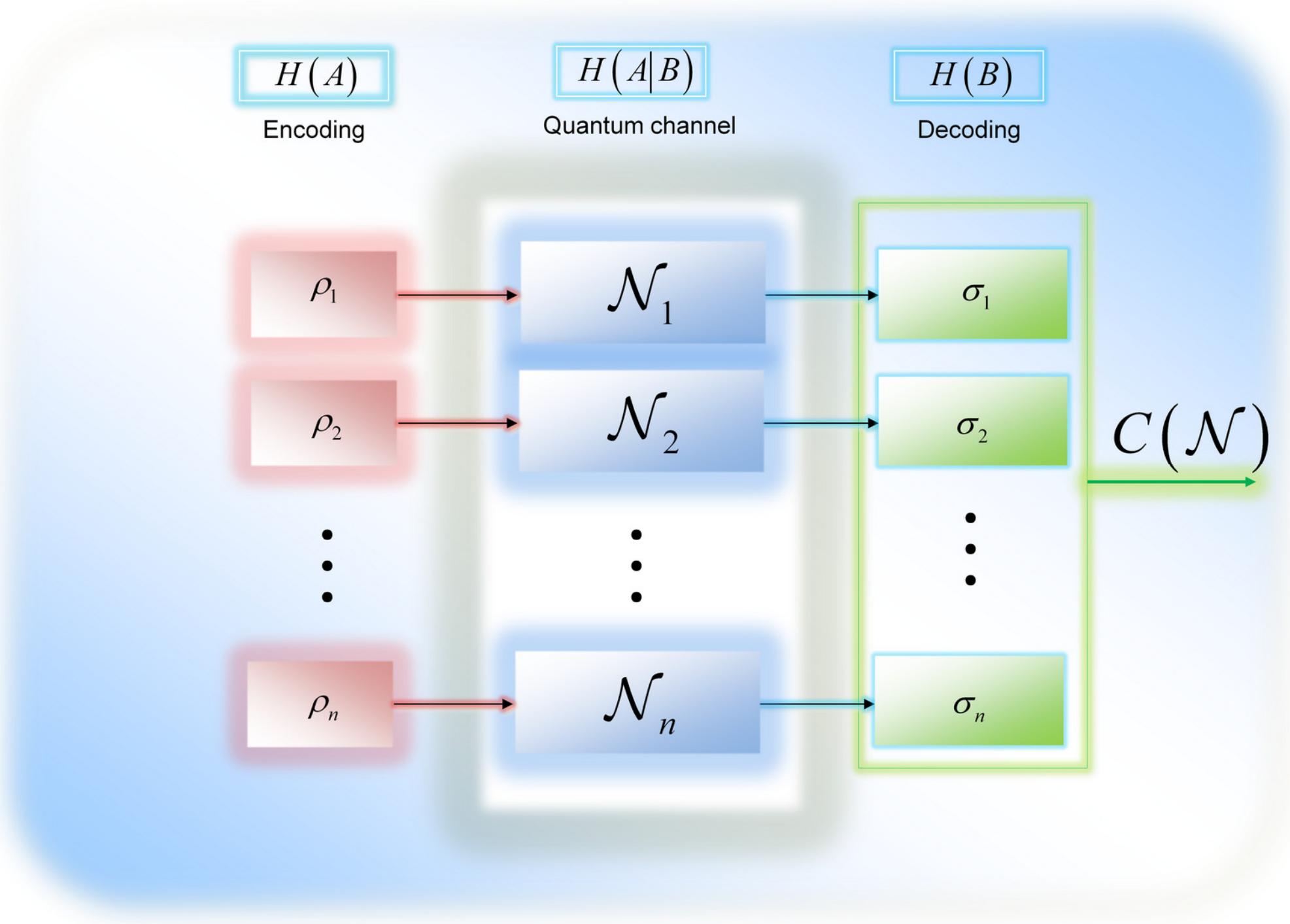

**Fig. 3.9.** The asymptotic classical capacity of a quantum channel. The classical capacity measures the maximum achievable classical information by Bob using noisy quantum channel.

The strange nature of the asymptotic formula will results in some very interesting, and classically unimaginable phenomena. We will discover these fields in Sections 5 and 6.

### 3.3.3 Brief Summary

The Holevo quantity measures the classical information, which remains in the encoded quantum states after they have transmitted through a noisy quantum channel. During the transmission, some information passes to the environment from the quantum state, which results in the increased entropy of the sent quantum state. The HSW theorem states very similar to Holevo's previous result. As in the case of the Holevo quantity, the HSW capacity measures the classical capacity of a noisy quantum channel - however, as we will see in Section 4, the Holevo quantity also can be used to express the quantum capacity of the quantum channel, which is a not trivial fact. The HSW capacity maximizes the Holevo quantity over a set of possible input states, and expresses the classical information, which can be sent through *reliably* in the form of *product input states* over the noisy quantum channel, hence HSW capacity is also known as *product state channel capacity*. In this case, the input states are not entangled; hence there is no entanglement between the multiple uses of the quantum channel. As we have seen in subsection 3.3.2.1, if the input of the channel consists of product states and we use *single measurement* setting, then the classical capacity can be expressed as the maximized of the quantum mutual information. On the other hand, if the single measurement has been changed to *joint measurement*, this statement is not true anymore; - this capacity will be equal to HSW capacity, see (3.14). Moreover, if we step forward, and we allow *entanglement* among the input states, then we cannot use anymore the HSW capacity, which was defined in (3.6) . In this case we have to take its asymptotic formula, which was shown in (3.16). The bound of the HSW theorem is shown in Fig. 3.10.

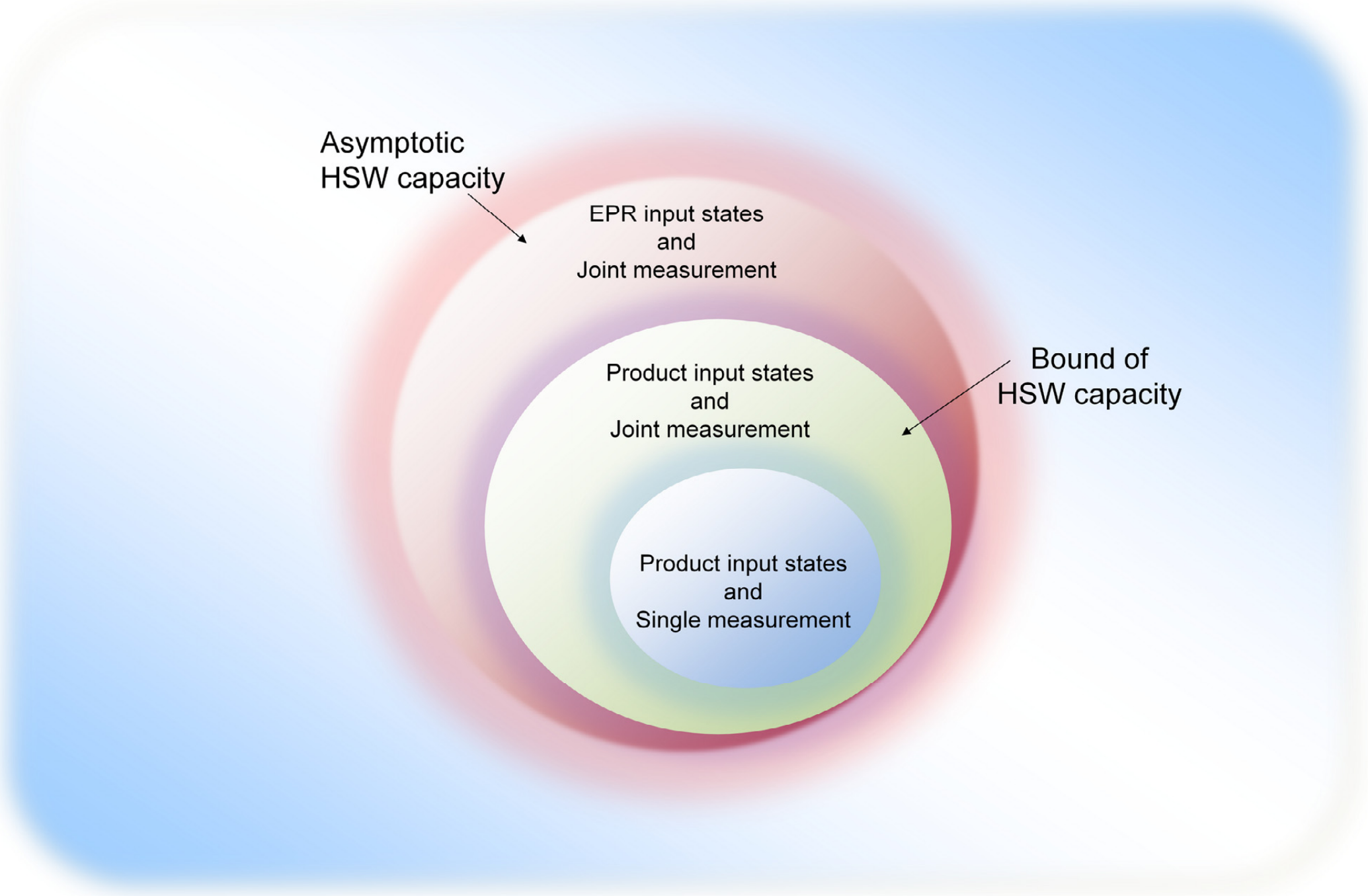

**Fig. 3.10.** The bound of HSW capacity. The HSW theorem was defined for the measure of product state capacity. The capacity formula for entangled input states and the joint measurement can be described by the asymptotic version of the HSW capacity.

Next we discuss the private classical capacity of quantum channels.

### 3.3.4 The Private Classical Capacity

The private classical capacity $P(\mathcal{N})$ of a quantum channel $\mathcal{N}$ describes the maximum rate at which the channel is able to send *classical information* through the channel reliably and *privately* (i.e., without any information leaked about the original message to an eavesdropper). Privately here means that an eavesdropper will not be able to access the encoded information without revealing her/himself i.e., the private classical capacity describes the maximal secure information that can be obtained by Bob on an eavesdropped quantum communication channel.

The generalized model of the private communication over quantum channels is illustrated in Fig. 3.11. The first output of the channel is denoted by $\sigma_B = \mathcal{N}(\rho_A)$, the second "receiver" is the eavesdropper $E$, with state $\sigma_E$. The single-use private classical capacity from these quantities can be expressed as the maximum of the difference between two mutual information quantities. The eavesdropper, Eve, attacks the quantum channel, and she steals $I(A:E)$ from the information $I(A:B)$ sent by Alice to Bob, therefore the *single-use* private classical capacity (or *private information*) of $\mathcal{N}$ can be determined as

$$P^{(1)}(\mathcal{N}) = \max_{all\ p_i,\rho_i} \left( I(A:B) - I(A:E) \right). \tag{3.19}$$

while the *asymptotic* private classical capacity is

$$P(\mathcal{N}) = \lim_{n\to\infty} \frac{1}{n} P^{(1)}\left(\mathcal{N}^{\otimes n}\right) = \lim_{n\to\infty} \frac{1}{n} \max_{all\ p_i,\rho_i} \left( I(A:B) - I(A:E) \right). \tag{3.20}$$

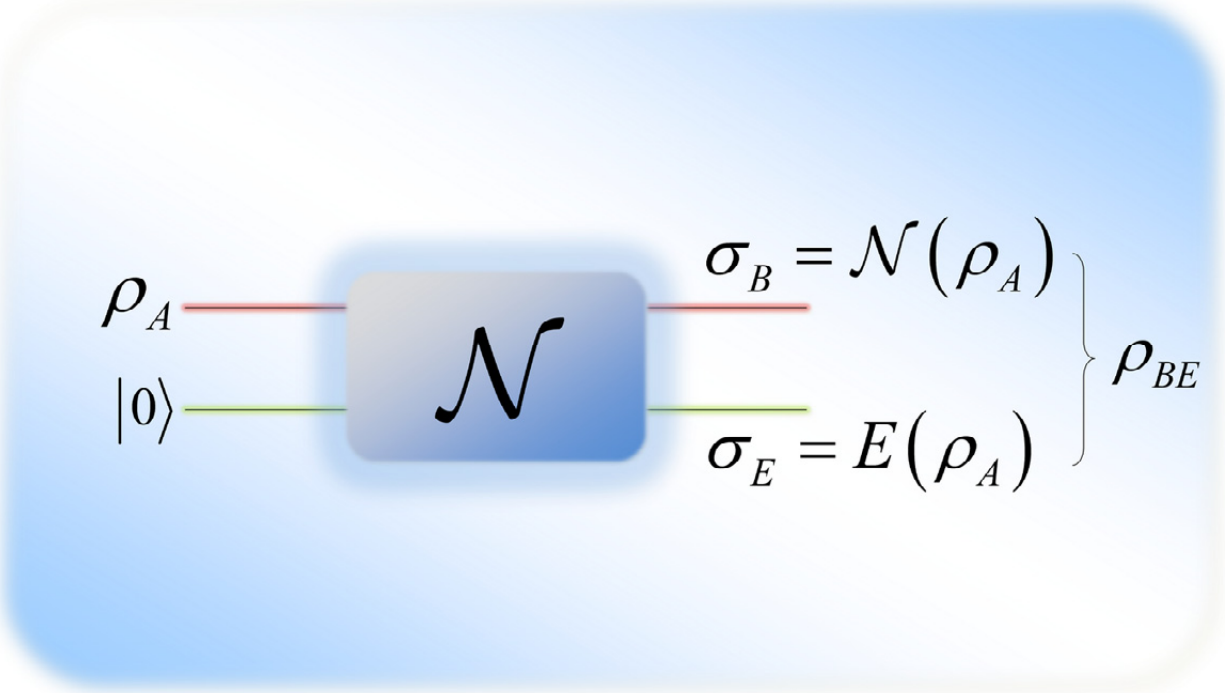

**Fig. 3.11.** The model of private classical communication of a quantum channel.

The $P(\mathcal{N})$ asymptotic and the $P^{(1)}(\mathcal{N})$ single-use private classical capacity can be expressed as the difference of two quantum mutual information functions (Section 2.3.), see (3.19) and (3.20). Here, we give an equivalent definition for private classical capacity $P(\mathcal{N})$ and show, that it also can be rewritten using the Holevo quantity $\mathcal{X}$, as follows:

$$P(\mathcal{N}) = \lim_{n\to\infty} \frac{1}{n} \max_{all\ p_i,\rho_i} \left(\mathcal{X}_{AB} - \mathcal{X}_{AE}\right), \quad (3.21)$$

where

$$\mathcal{X}_{AB} = \mathrm{S}\left(\mathcal{N}_{AB}\left(\rho_{AB}\right)\right) - \sum_i p_i \mathrm{S}\left(\mathcal{N}_{AB}\left(\rho_i\right)\right) \quad (3.22)$$

and

$$\mathcal{X}_{AE} = \mathrm{S}\left(\mathcal{N}_{AE}\left(\rho_{AE}\right)\right) - \sum_i p_i \mathrm{S}\left(\mathcal{N}_{AE}\left(\rho_i\right)\right) \quad (3.23)$$

measure the Holevo quantities between Alice and Bob, and Alice and the eavesdropper Eve, respectively, while $\rho_{AB} = \sum_i p_i\rho_i$ and $\rho_{AE} = \sum_i p_i\rho_i$.

An important corollary from (3.20), while the quantum mutual information itself is additive (see the properties of quantum mutual information function in Section 2.3.), the difference of two quantum mutual information functions is not (i.e., we need the asymptotic version to compute the "true" private classical capacity of a quantum channel.)

### 3.3.5 The Entanglement-assisted Classical Capacity

The last capacity regarding classical communication over quantum channels is called *entanglement-assisted classical capacity* $C_E(\mathcal{N})$, which measures the classical information which can be transmitted through the channel, if Alice and Bob have shared entanglement before the transmission i.e., entanglement is applied not between the input states like in case of the HSW (i.e., the product-state capacity) theorem. This capacity measures classical information, and it can be expressed with the help of the *quantum mutual information function* (see Section 2.3.) as

$$C_E(\mathcal{N}) = \max_{all\ p_i,\rho_i} I(A:B). \tag{3.24}$$

The main difference between the classical capacity $C(\mathcal{N})$ and the entanglement-assisted classical capacity $C_E(\mathcal{N})$, is that in the latter case the maximum of the transmittable classical information is equal to the maximized quantum mutual information, - hence the entanglement-assisted classical capacity $C_E(\mathcal{N})$ can be derived from the *single-use* version $C_E^{(1)}(\mathcal{N})$. From (3.24) the reader can conclude, there is no need for the asymptotic version to express the entanglement-assisted classical capacity, i.e.:

$$C_E(\mathcal{N}) = C_E^{(1)}(\mathcal{N}) = \max_{all\ p_i,\rho_i} I(A:B). \tag{3.25}$$

It also can be concluded, that shared entanglement does not change the additivity of maximized quantum mutual information - or with other words, it remains true if the parties use shared entanglement for the transmission of classical information over $\mathcal{N}$. In Fig. 3.12 we illustrate the general model of entanglement-assisted classical capacity $C_E(\mathcal{N})$.

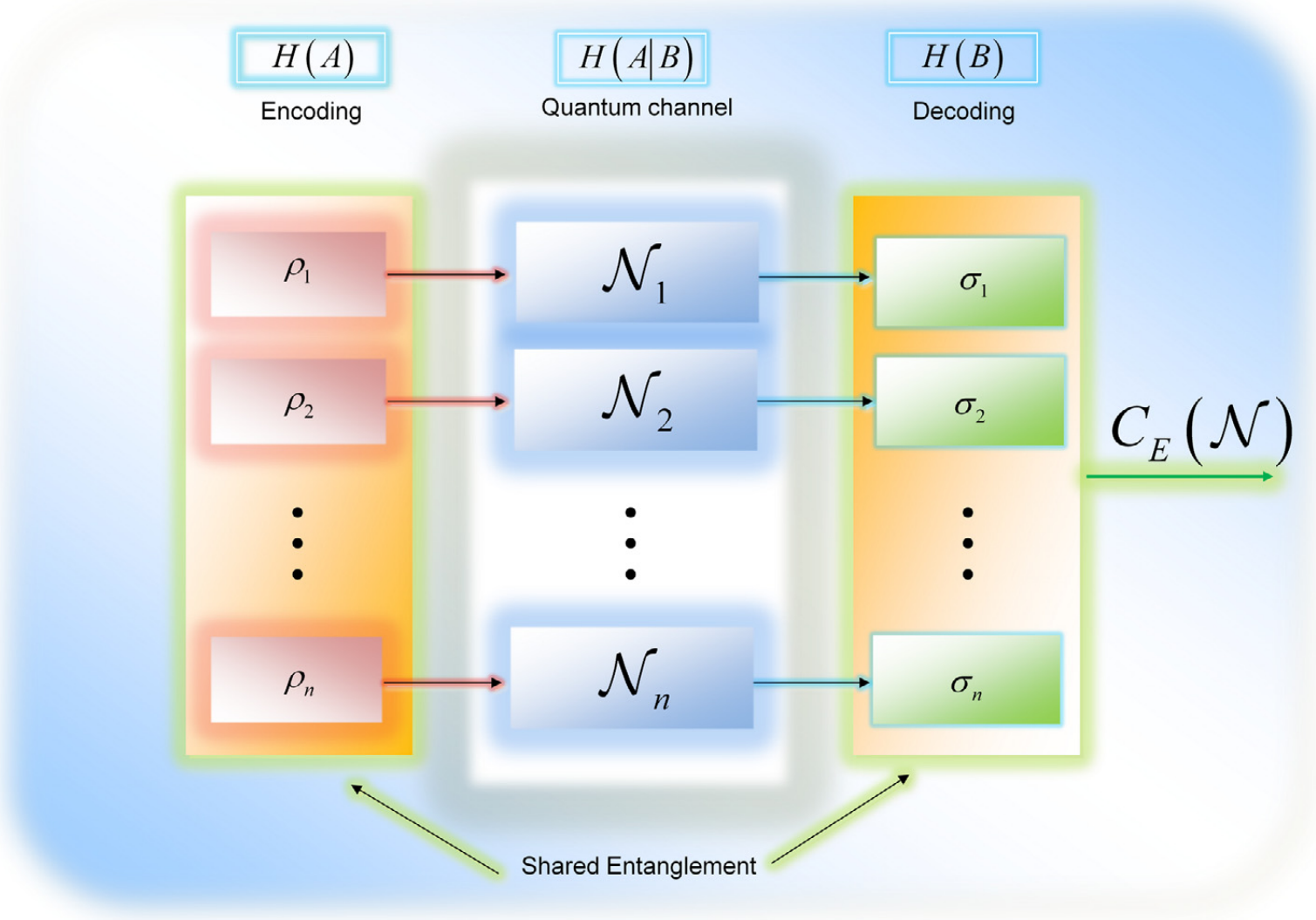

**Fig. 3.12.** The entanglement-assisted capacity of a quantum channel. This capacity measures the maximum of transmittable classical information through a quantum channel, if shared a priori entanglement between the parties is allowed.

We note an important property of shared entanglement: while it does not provide any benefits in the improving of the classical capacity of the quantum channel, (see (3.24)), it can be used to increase the single-use classical capacity. It was shown, that with the help of shared entanglement the transmission of a single quantum bit can be realized with higher success probability, - this strategy is known as the CHSH (*Clauser-Horne-Shimony-Holt*) game, for details see [Imre05].

**3.3.5.1 Brief Summary of Classical Capacities**

Here, we give a brief summarization on the classical capacities defined in Section 3-3. For the *asymptotic* capacity of a quantum channel, we have

$$C\left(\mathcal{N}\right) \geq \chi\left(\mathcal{N}\right). \tag{3.26}$$

According to the results of Holevo-Schumacher-Westmoreland, the asymptotic classical capacity is not equal to the single-use classical capacity. The *asymptotic* formula of the classical capacity $C\left(\mathcal{N}\right)$ can be expressed by the help of the Holevo capacity $\chi\left(\mathcal{N}\right)$ as

$$C\left(\mathcal{N}\right) = \lim_{n\to\infty} \frac{1}{n} \chi\left(\mathcal{N}^{\otimes n}\right). \tag{3.27}$$

The difference between the single-use formula and the asymptotic formula holds for the private capacity $P\left(\mathcal{N}\right)$. Unlike these capacities, in the case of entanglement-assisted classical capacity $C_E\left(\mathcal{N}\right)$, we will find something else in the expression. In this case, we have

$$C_E\left(\mathcal{N}\right) = C_E^{(1)}\left(\mathcal{N}\right) = \max_{all\ p_i,\rho_i} I\left(A:B\right), \tag{3.28}$$

and so we can conclude, *there is no regularization.* Since there is no regularization needed, it also means that the entanglement-assisted classical capacity $C_E\left(\mathcal{N}\right)$ will always be additive. This makes it easier to compute the entanglement-assisted capacity than the other formulas, in which regularization is needed.

Originally, it was conjectured that in the general case, the Holevo information $\chi$ is additive too, for the same channels. Later, a counterexample was found by Hastings. As has been shown, in this case the additivity of the Holevo information fails.

Similarly, for the $P(\mathcal{N})$ private classical capacity, - which also measures classical information we have

$$P(\mathcal{N}) \geq \max_{all\ p_i,\rho_i} \left( I(A:B) - I(A:E) \right), \tag{3.29}$$

and finally, for the classical capacity $C(\mathcal{N})$ of $\mathcal{N}$

$$\max_{all\ p_i,\rho_i} I(A:B) \leq C(\mathcal{N}) \leq \lim_{n\to\infty} \frac{1}{n}\chi\left(\mathcal{N}^{\otimes n}\right). \tag{3.30}$$

As can be seen, in case of the classical and private classical capacities the regularization is needed, since the asymptotic and the single-use formulas are not equal.

## 3.4 The Classical Zero-Error Capacity

Shannon's results on capacity [Shannon48] guarantees transmission rate only in average when using multiple times of the channel. The zero-error capacity of the quantum channel describes the amount of (classical or quantum) information which can be transmitted *perfectly* (*zero probability of error*) through a noisy quantum channel. The zero-error capacity of the quantum channel could have an overriding importance in future quantum communication networks.

The zero-error capacity stands a very strong requirement in comparison to the standard capacity where the information transmission can be realized with asymptotically small *but non-vanishing* error probability, since in the case of zero-error communication the *error probability of the communication has to be zero*, hence the transmission of information has to be perfect and no errors are allowed. While in the case of classical non zero-error capacity for an $n$-length code the error probabilities after the decoding process are $\Pr[error] \to 0$ as $n \to \infty$, in case of an $n$-length zero-error code, $\Pr[error] = 0$.

In this subsection we give the exact definitions which required for the characterization of a quantum zero-error communication system. We will discuss the classical and quantum zero-error capacities and give the connection between zero-error quantum codes and the elements of graph theory.

### 3.4.1 Classical Zero-Error Capacities of Quantum Channels

In this section we review the background of zero-error capacity $C_0(\mathcal{N})$ of a quantum channel $\mathcal{N}$. Let us assume that Alice has information source $\{X_i\}$ encoded into quantum states $\{\rho_i\}$ which will be transmitted through a quantum channel $\mathcal{N}$ (see Fig. 3.13.). The quantum states will be measured by a set of POVM operators $\mathcal{P}=\{\mathcal{M}_1,\ldots,\mathcal{M}_k\}$ at the receiver (see Section 2.1.2.2). The classical zero-error quantum capacity $C_0(\mathcal{N})$ for product input states can be reached if and only if the input states are *pure* states, similarly to the HSW capacity $C(\mathcal{N})$.

The zero-error transmission of quantum states requires perfect distinguishability. To achieve this perfect distinguishability of the zero-error quantum codewords, they have to be *pairwise orthogonal. Non-adjacent codewords can be distinguished perfectly.* Two inputs are called *adjacent* if they can result in the same output. The number of possible non-adjacent codewords determines the rate of maximal transmittable classical information through $\mathcal{N}$.

In the $d$ dimensional Hilbert space (e.g. $d$=2 for qubits) at most $d$ pairwise distinguishable quantum states exist, thus for a quantum system which consist of $n$ pieces of $d$ dimensional quantum states at most $d^n$ pairwise distinguishable $n$-length quantum codewords are available. Obviously if two quantum codewords are not orthogonal, then they cannot be distinguished perfectly. We note, if we would like to distinguish between $K$ *pairwise orthogonal* quantum codewords (the length of each codewords is $n$) in the $d^n$ dimensional Hilbert space, then we have to define the POVM set (see Fig. 3.15)

$$\mathcal{P}=\left\{\mathcal{M}^{(1)},\ldots,\mathcal{M}^{(K)}\right\}, \tag{3.31}$$

where $\mathcal{M}^{(i)}$ are set of $d$-dimensional projectors on the individual quantum systems (e.g. qubits) which distinguish the $n$-length codewords

$$\mathcal{M}^{(i)}=\{\mathcal{M}_1,\ldots,\mathcal{M}_m\} \tag{3.32}$$

where $m=d^n$. The probability that Bob gives measurement outcome $j$ from quantum state $\rho_i$ is

$$\Pr\left[\ j \middle| \rho_i \right] = Tr\left( \mathcal{M}_j \mathcal{N}\left( \rho_i \right) \right). \tag{3.33}$$

The $i$-th *codeword* $\left| \psi_{X_i} \right\rangle$ encodes the $n$-length classical codeword $X_i = \left\{ x_{i,1}, x_{i,2}, \ldots, x_{i,n} \right\}$ consisting of $n$ product input quantum states:

$$\left| \psi_{X_i} \right\rangle = \left[ \left| \psi_{i,1} \right\rangle \otimes \left| \psi_{i,2} \right\rangle \otimes \left| \psi_{i,3} \right\rangle \cdots \otimes \left| \psi_{i,n} \right\rangle \right],\ i = 1..K\,. \tag{3.34}$$

where $\rho_i = \left| \psi_{X_i} \right\rangle \left\langle \psi_{X_i} \right|$.

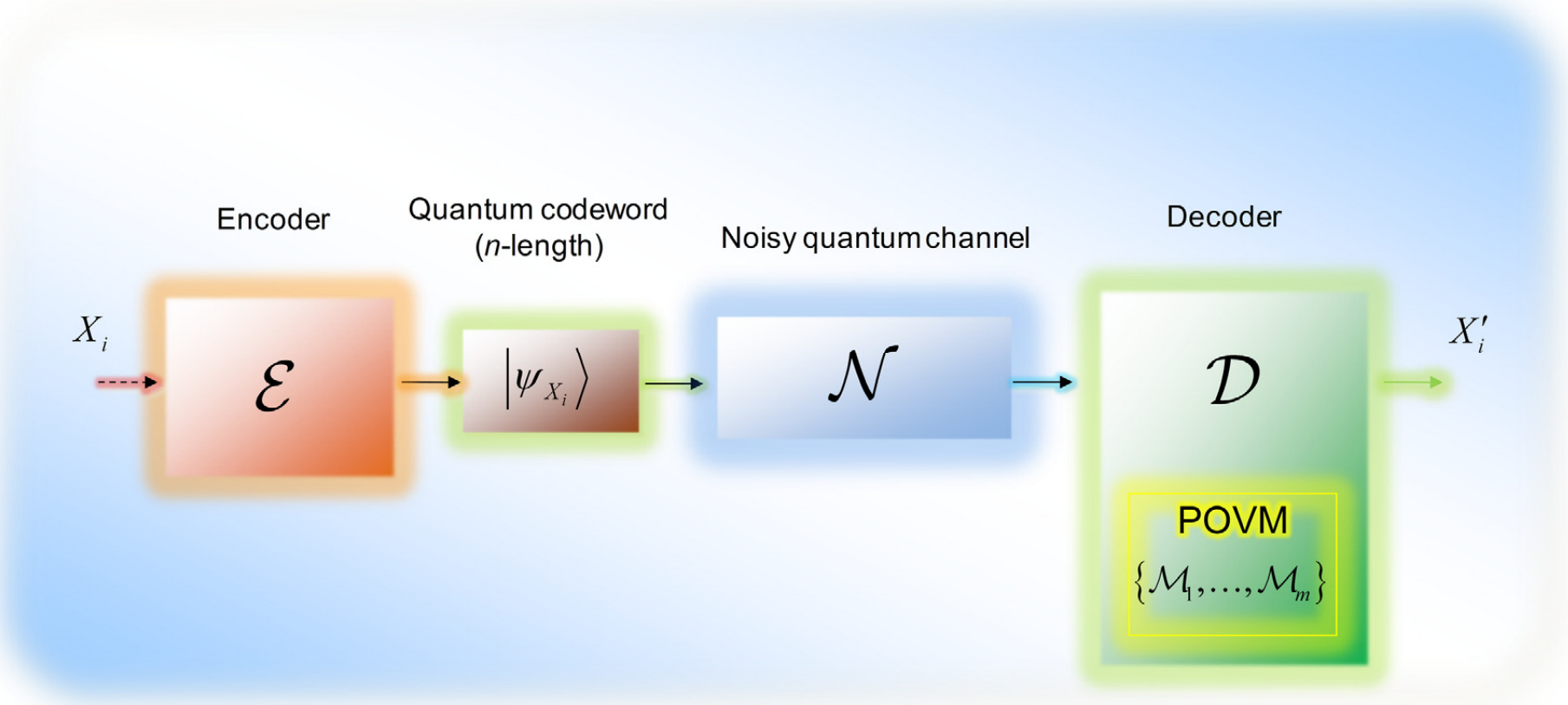

**Fig. 3.13.** A quantum zero-error communication system.

The quantum block code consist of codewords

$$\begin{gathered} \left| \psi_{X_1} \right\rangle = \left[ \left| \psi_{1,1} \right\rangle \otimes \left| \psi_{1,2} \right\rangle \otimes \left| \psi_{1,3} \right\rangle \cdots \otimes \left| \psi_{1,n} \right\rangle \right] \\ \vdots \qquad\qquad\qquad \vdots \\ \left| \psi_{X_K} \right\rangle = \left[ \left| \psi_{K,1} \right\rangle \otimes \left| \psi_{K,2} \right\rangle \otimes \left| \psi_{K,3} \right\rangle \cdots \otimes \left| \psi_{K,n} \right\rangle \right], \end{gathered} \tag{3.35}$$

where $K$ is the number of classical ($n$ length) messages.

The decoder will produce the output codeword $X_i' = \left\{ x_{i,1}', x_{i,2}', \ldots, x_{i,n}' \right\}$ generated by the POVM measurement operators, where the POVM $\mathcal{M}^{(i)}$ can distinguish $m$ messages $\left\{ X_1', X_2', \ldots X_m' \right\}$ ($n$-length) at the output. Bob would like to determine each message $i \in [1, K]$ with unit probability. The zero probability of error means that for the input code $\left| \psi_{X_i} \right\rangle$ the decoder has to identify the classical output codeword $X_i'$ with classical input codeword $X_i$ perfectly for each

possible $i$, otherwise the quantum channel has no zero-error capacity; that is, for the zero-error quantum communication system

$$\Pr\left[ X_i' \middle| X_i \right] = 1 \,. \tag{3.36}$$

### 3.4.2 Formal Definitions of Quantum Zero-Error Communication

In this subsection we review the most important definitions of quantum zero-error communication systems.

The *non-adjacent* elements are important for zero-error capacity, since *only non-adjacent codewords can be distinguished perfectly.* Two inputs are called *adjacent* if they can result in the same output, while for *non-adjacent* inputs, the output of the encoder is unique. The number of possible non-adjacent codewords determines the rate of maximal transmittable classical information through quantum channels.

Formally, the *non-adjacent* property of two quantum states $\rho_1$ and $\rho_2$ can be given as

$$Set_1 \cap Set_2 = \varnothing \,, \tag{3.37}$$

where

$$Set_i = \left\{ \Pr\left[ X_j' \middle| X_i \right] = Tr\left( \mathcal{M}_j \mathcal{N}\left( \left| \psi_{X_i} \right\rangle \left\langle \psi_{X_i} \right| \right) \right) > 0,\ j \in \left\{ 1, \ldots, m \right\} \right\},\ i = 1,2 \,, \tag{3.38}$$

using POVM decoder $\mathcal{P} = \left\{ \mathcal{M}_1, \ldots, \mathcal{M}_m \right\}$. In a relation of a noisy quantum channel $\mathcal{N}$, the non-adjacent property can be rephrased as follows. Two input quantum states $\rho_1$ and $\rho_2$ are non-adjacent with relation to $\mathcal{N}$, if $\mathcal{N}\left( \rho_1 \right)$ and $\mathcal{N}\left( \rho_2 \right)$ are *perfectly distinguishable.* The notation $\rho_1 \underset{\mathcal{N}}{\perp} \rho_2$ also can be used to denote the non-adjacent inputs of quantum channel $\mathcal{N}$.

A quantum channel $\mathcal{N}$ has greater than zero zero-error capacity if and only if a subset of quantum states $\Omega = \left\{ \rho_i \right\}_{i=1}^{l}$ and POVM $\mathcal{P} = \left\{ \mathcal{M}_1, \ldots, \mathcal{M}_m \right\}$ exists where for at *least two states* $\rho_1$ and $\rho_2$ from subset $\Omega$, the relation (3.37) holds; that is, the non-adjacent property with relation to the POVM measurement is satisfied. For the quantum channel $\mathcal{N}$, the two inputs $\rho_1$ and $\rho_2$ are non-adjacent if and only if the quantum channel takes the input states $\rho_1$ and $\rho_2$ into orthogonal subspaces

$$\mathcal{N}(\rho_1) \underset{\mathcal{N}}{\perp} \mathcal{N}(\rho_2); \tag{3.39}$$

that is, the quantum channel has positive classical zero-error capacity $C_0(\mathcal{N})$ if and only if this property holds for the output of the channel for a given POVM $\mathcal{P} = \{\mathcal{M}_1, \ldots, \mathcal{M}_m\}$. The previous result can be rephrased as follows. Using the trace preserving property of the quantum channel (see Section 1.1), the two quantum states $\rho_1$ and $\rho_2$ are non-adjacent if and only if for the channel output states $\mathcal{N}(\rho_1), \mathcal{N}(\rho_2)$,

$$Tr(\mathcal{N}(\rho_1)\mathcal{N}(\rho_2)) = 0, \tag{3.40}$$

and if $\rho_1$ and $\rho_2$ are non-adjacent input states then

$$Tr(\rho_1\rho_2) = 0. \tag{3.41}$$

Non-adjacent inputs produce distinguishable outputs, as depicted in Fig. 3.14. For these inputs, the outputs of $\mathcal{N}$ will be orthogonal.

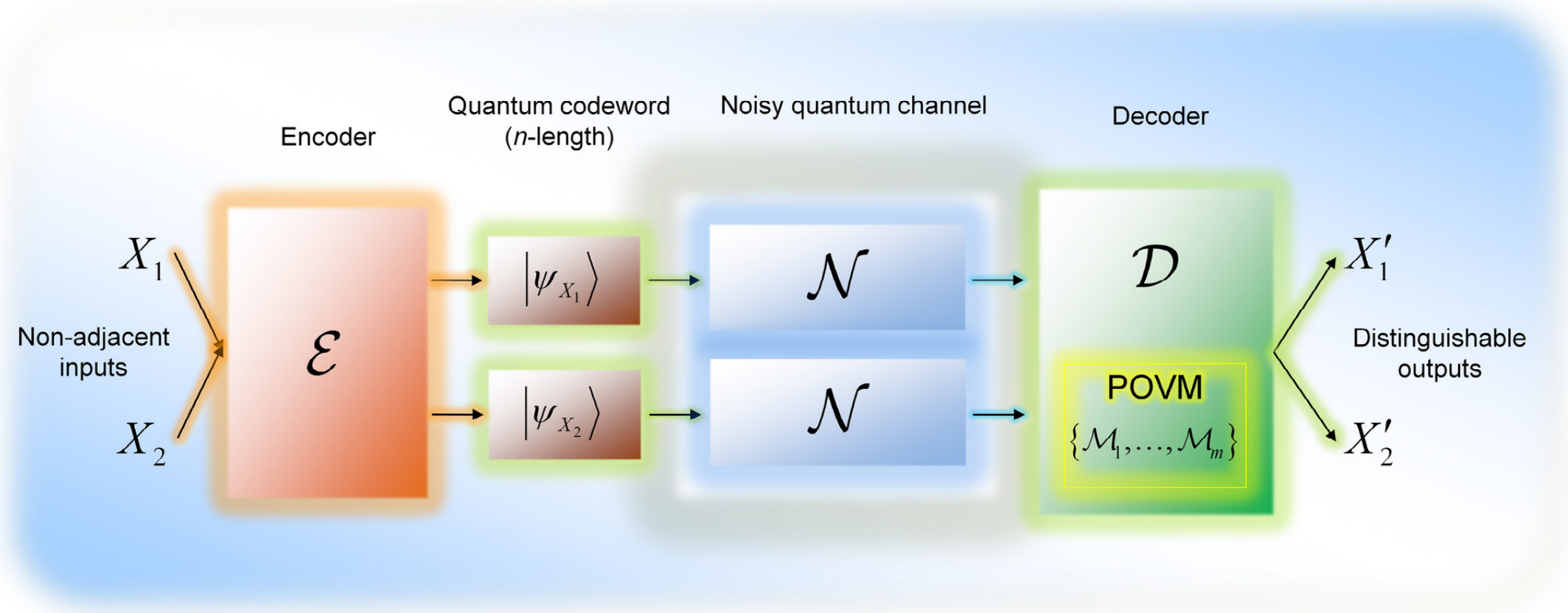

**Fig. 3.14.** The non-adjacent inputs can be distinguished at the output. The quantum zero-error communication requires non-adjacent quantum codewords.

Let the two *non-adjacent* input codewords of the $\mathcal{N}$ be denoted by $|\psi_{X_1}\rangle$ and $|\psi_{X_2}\rangle$. These quantum codewords encode messages $X_1 = \{x_{1,1}, x_{1,2}, \ldots, x_{1,n}\}$ and $X_2 = \{x_{2,1}, x_{2,2}, \ldots, x_{2,n}\}$. For this setting, we construct the following POVM

operators for the given complete set of POVM $\mathcal{P} = \left\{ \mathcal{M}_1, \ldots, \mathcal{M}_m \right\}$ and the two input codewords $\left| \psi_{X_1} \right\rangle$ and $\left| \psi_{X_2} \right\rangle$ as follows (see Fig. 3.15.)

$$\mathcal{M}^{(1)} = \left\{ \mathcal{M}_1, \ldots, \mathcal{M}_k \right\} \tag{3.42}$$

and

$$\mathcal{M}^{(2)} = \left\{ \mathcal{M}_{k+1}, \ldots, \mathcal{M}_m \right\}. \tag{3.43}$$

The groups of operators, $\mathcal{M}^{(1)}$ and $\mathcal{M}^{(2)}$, will identify and distinguish the input codewords $\left| \psi_{X_1} \right\rangle$ and $\left| \psi_{X_2} \right\rangle$. Using this setting the two non-adjacent codewords $\left| \psi_{X_1} \right\rangle$ and $\left| \psi_{X_2} \right\rangle$ can be distinguished with probability one at the output since

$$\begin{aligned} \Pr\left[ X_i' \middle| X_1 \right] &= 1,\ i = 1, \ldots, k, \\ \Pr\left[ X_i' \middle| X_2 \right] &= 1,\ i = k+1, \ldots, m, \end{aligned} \tag{3.44}$$

where $X_i'$ is a number between 1 and $m$, (according to the possible number of POVM operators) which identifies the measured unknown quantum codeword and consequently

$$\begin{aligned} \Pr\left[ X_i' \middle| X_1 \right] &= 0,\ i = k+1, \ldots, m, \\ \Pr\left[ X_i' \middle| X_2 \right] &= 0,\ i = 1, \ldots, k. \end{aligned} \tag{3.45}$$

For input message $\left| \psi_{X_1} \right\rangle$ and $\left| \psi_{X_2} \right\rangle$ with the help of set $\mathcal{M}^{(1)}$ and $\mathcal{M}^{(2)}$ these probabilities are

$$\begin{aligned} \Pr\left[ X_1' \middle| X_1 \right] &= Tr\left( \mathcal{M}^{(1)} \mathcal{N}\left( \left| \psi_{X_1} \right\rangle \left\langle \psi_{X_1} \right| \right) \right) = 1, \\ \Pr\left[ X_2' \middle| X_2 \right] &= Tr\left( \mathcal{M}^{(2)} \mathcal{N}\left( \left| \psi_{X_2} \right\rangle \left\langle \psi_{X_2} \right| \right) \right) = 1, \end{aligned} \tag{3.46}$$

where $\mathcal{M}^{(1)}$ and $\mathcal{M}^{(2)}$ are orthogonal projectors, $\mathcal{M}^{(1)}$ and $\mathcal{M}^{(2)}$ are defined in (3.42) and (3.43)), and $\mathcal{M}^{(1)} + \mathcal{M}^{(2)} + \mathcal{M}^{(2+1)} = I$ (see Section 2.1.2.2), to make it possible for the quantum channel to take the input states into orthogonal subspaces; that is, $\mathcal{N}\left(\left|\psi_{X_1}\right\rangle\left\langle\psi_{X_1}\right|\right) \perp \mathcal{N}\left(\left|\psi_{X_2}\right\rangle\left\langle\psi_{X_2}\right|\right)$ has to be satisfied. The POVM measurement has to be restricted to projective measurement. As follows, the $\mathcal{P} = \left\{\mathcal{M}^{(1)}, \mathcal{M}^{(2)}\right\}$ POVM measurement can be replaced with the set of *von Neumann* operators, $\mathcal{Z} = \left\{\mathcal{P}^{(1)}, \mathcal{P}^{(2)}\right\}$ , where $\mathcal{P}^{(1)} + \mathcal{P}^{(2)} = I$. This result also can be extended for arbitrarily number of operators, depending on the actual system. The non-adjacent property also can be interpreted for arbitrary length of quantum codewords. For a given quantum channel $\mathcal{N}$, the two $n$-length input quantum codewords $\left|\psi_{X_1}\right\rangle$ and $\left|\psi_{X_2}\right\rangle$, which are tensor products of $n$ quantum states, then *input* codewords $\left|\psi_{X_1}\right\rangle$ and $\left|\psi_{X_2}\right\rangle$ are non-adjacent in relation with $\mathcal{N}$ if and only if *at least one* pair of quantum states $\left\{\left|\psi_{1,i}\right\rangle, \left|\psi_{2,i}\right\rangle\right\}$ from the two $n$-length sequences is perfectly distinguishable. Formally, at least one *input* quantum state pair $\left\{\left|\psi_{1,i}\right\rangle, \left|\psi_{2,i}\right\rangle\right\}$ with $i$, $1 \leq i \leq n$, exists in $\left|\psi_{X_1}\right\rangle$ and $\left|\psi_{X_2}\right\rangle$, for which $\mathcal{N}\left(\left|\psi_{1,i}\right\rangle\left\langle\psi_{1,i}\right|\right)$ is non-adjacent to $\mathcal{N}\left(\left|\psi_{2,i}\right\rangle\left\langle\psi_{2,i}\right|\right)$. These statements with input codewords $\left|\psi_{X_1}\right\rangle$ and $\left|\psi_{X_2}\right\rangle$ are summarized in (3.47). The two quantum codewords are distinguishable, because there is a state-pair in the sequences which are completely distinguishable (depicted by the frames).

$$\begin{aligned}
\mathcal{N}\left(\left|\psi_{X_1}\right\rangle\left\langle\psi_{X_1}\right|\right) &= \left[\mathcal{N}\left(\left|\psi_{1,1}\right\rangle\left\langle\psi_{1,1}\right|\right) \otimes \cdots \otimes \boxed{\mathcal{N}\left(\left|\psi_{1,i}\right\rangle\left\langle\psi_{1,i}\right|\right)} \otimes \cdots \otimes \mathcal{N}\left(\left|\psi_{1,n}\right\rangle\left\langle\psi_{1,n}\right|\right)\right], \\
\mathcal{N}\left(\left|\psi_{X_2}\right\rangle\left\langle\psi_{X_2}\right|\right) &= \left[\mathcal{N}\left(\left|\psi_{2,1}\right\rangle\left\langle\psi_{2,1}\right|\right) \otimes \cdots \otimes \boxed{\mathcal{N}\left(\left|\psi_{2,i}\right\rangle\left\langle\psi_{2,i}\right|\right)} \otimes \cdots \otimes \mathcal{N}\left(\left|\psi_{2,n}\right\rangle\left\langle\psi_{2,n}\right|\right)\right].
\end{aligned} \tag{3.47}$$

In (3.47) the two $n$-length codewords $\left|\psi_{X_1}\right\rangle$ and $\left|\psi_{X_2}\right\rangle$ are distinguishable at the output of $\mathcal{N}$, since $\mathcal{N}\left(\left|\psi_{1,i}\right\rangle\left\langle\psi_{1,i}\right|\right)$ is non-adjacent to $\mathcal{N}\left(\left|\psi_{2,i}\right\rangle\left\langle\psi_{2,i}\right|\right)$, i.e., the distinguishability of output codewords $\mathcal{N}\left(\left|\psi_{X_1}\right\rangle\left\langle\psi_{X_1}\right|\right)$ and $\mathcal{N}\left(\left|\psi_{X_2}\right\rangle\left\langle\psi_{X_2}\right|\right)$ depends on the distinguishability of the quantum states of the codewords.

Because we have stated that the two codewords can be distinguished at the channel output, the following relation has to be hold for their trace, according to (3.40), and their non-adjacency can be verified as follows:

$$
\begin{aligned}
& Tr\left(\mathcal{N}\left(\left|\psi_{X_1}\right\rangle\left\langle\psi_{X_1}\right|\right)\mathcal{N}\left(\left|\psi_{X_2}\right\rangle\left\langle\psi_{X_2}\right|\right)\right) \\
& = Tr\left(\left(\bigotimes_{i=1}^{n}\mathcal{N}\left(\left|\psi_{1,i}\right\rangle\left\langle\psi_{1,i}\right|\right)\right)\left(\bigotimes_{i=1}^{n}\mathcal{N}\left(\left|\psi_{2,i}\right\rangle\left\langle\psi_{2,i}\right|\right)\right)\right) \\
& = \prod_{i=1}^{n} Tr\left(\mathcal{N}\left(\left|\psi_{1,i}\right\rangle\left\langle\psi_{1,i}\right|\right)\mathcal{N}\left(\left|\psi_{2,i}\right\rangle\left\langle\psi_{2,i}\right|\right)\right) = 0.
\end{aligned}
\tag{3.48}
$$

As follows from (3.48), a quantum channel $\mathcal{N}$ has non-zero zero-error capacity if and only if there exists at least two non-adjacent input quantum states $\rho_1$ and $\rho_2$. These two non-adjacent quantum states make distinguishable the two, $n$-length quantum codewords at the output of quantum channel $\mathcal{N}$, and these input codewords will be called as *non-adjacent quantum codewords*. The decoding of non-adjacent codewords to achieve zero-error communication over a quantum channel is depicted in Fig. 3.15.

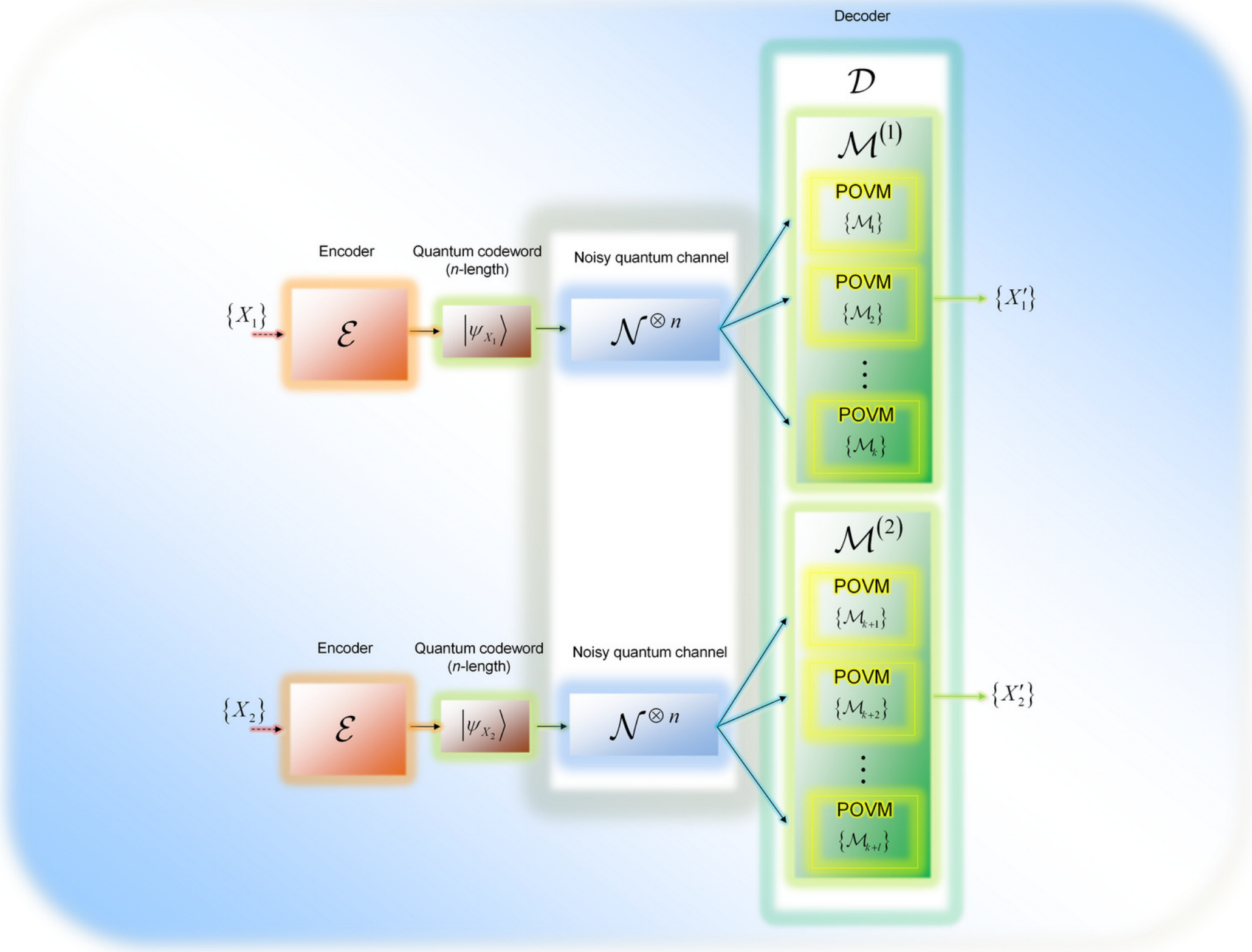

**Fig. 3.15.** Each of the non-adjacent input codewords is distinguished by a set of measurement operators to achieve the zero-error quantum communication.

The joint measurement of the quantum states of an output codeword is *necessary* and *sufficient* to distinguish the input codewords with zero-error. *Necessary*, because the joint measurement is required to distinguish orthogonal general (i.e., non zero-error code) tensor product states [Bennett99a]. Sufficient, because the non-adjacent quantum states have orthogonal *supports* at the output of the noisy quantum channel, i.e., $Tr\left(\rho_i\rho_j\right) = 0$ [Medeiros05]. (The *support* of a matrix $A$ is the orthogonal complement of the kernel of the matrix. The *kernel* of $A$ is the set of all vectors $v$, for which $Av = 0$.) In the joint measurement, the $\left\{\mathcal{M}_i\right\}$, $i = 1,\ldots,m$ projectors are $d^n \times d^n$ matrices, while if we were to use a single measurement then the size of these matrices would be $d \times d$.

In Fig. 3.16 we compared the difference between single and joint measurement settings for a given $n$-length quantum codeword $\left|\psi_X\right\rangle = \left[\left|\psi_1\right\rangle \otimes \left|\psi_2\right\rangle \otimes \left|\psi_3\right\rangle \cdots \otimes \left|\psi_n\right\rangle\right]$. In the case of single measurement Bob measures each of the $n$ quantum states of the $i$-th codeword states individually. In case of the joint measurement Bob waits until he receives the $n$ quantum states, then measures them together.

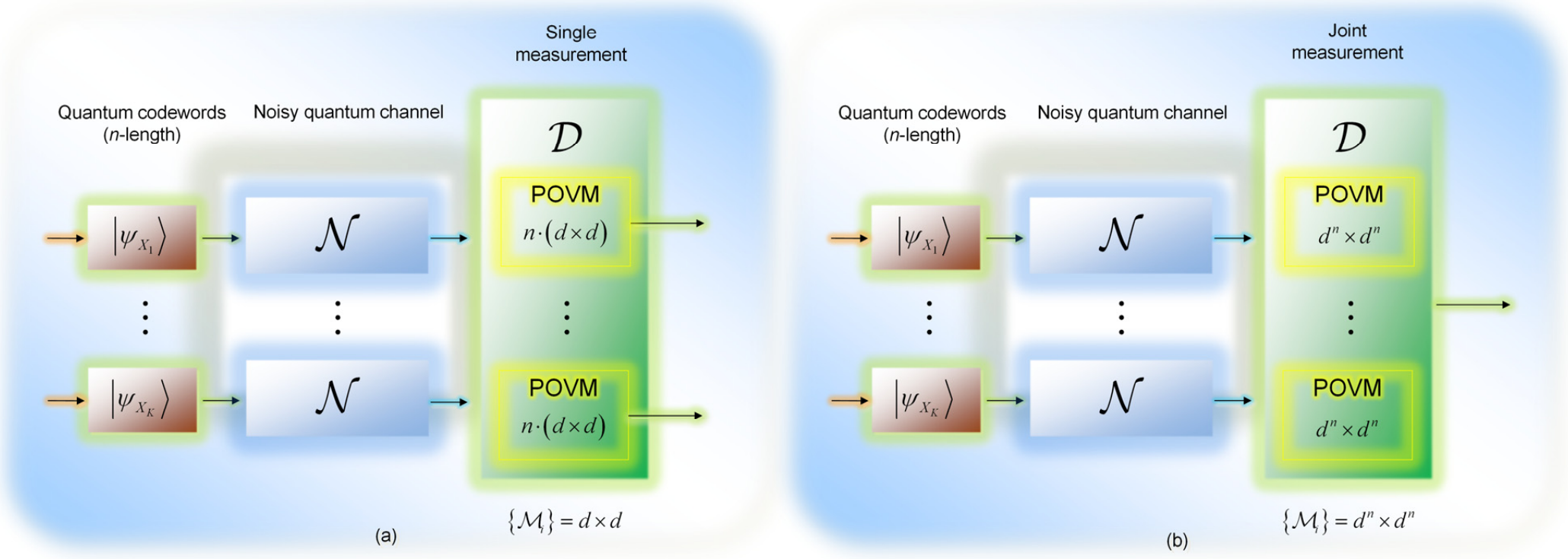

**Fig. 3.16.** Comparison of single (a) and joint (b) measurement settings. The joint measurement is necessary to attain the quantum zero-error communication.

Next we study the achievable rates for zero error classical communication over quantum channels.

### 3.4.3 Achievable Zero-Error Rates in Quantum Systems

Theoretically (without making any assumptions about the physical attributes of the transmission), the *classical single-use zero-error capacity* $C_0^{(1)}\left(\mathcal{N}\right)$ of the noisy quantum channel can be expressed as

$$C_0^{(1)}\left(\mathcal{N}\right) = \log\left(K\left(\mathcal{N}\right)\right), \tag{3.49}$$

where $K\left(\mathcal{N}\right)$ is the maximum number of different messages which can be sent over the channel with a *single use* of $\mathcal{N}$ (or in other words the maximum size of the set of *mutually non-adjacent* inputs).

The asymptotic *zero-error capacity* of the noisy quantum channel $\mathcal{N}$ can be expressed as

$$C_0\left(\mathcal{N}\right) = \lim_{n\to\infty} \frac{1}{n} \log\left(K\left(\mathcal{N}^{\otimes n}\right)\right), \tag{3.50}$$

where $K\left(\mathcal{N}^{\otimes n}\right)$ is the maximum number of $n$-length classical messages that the quantum channel can transmit with zero error and $\mathcal{N}^{\otimes n}$ denotes the $n$-uses of the channel.
The $C_0\left(\mathcal{N}\right)$ asymptotic classical zero-error capacity of a quantum channel is *upper bounded* by the HSW capacity, that is,

$$C_0^{(1)}\left(\mathcal{N}\right) \le C_0\left(\mathcal{N}\right) \le C\left(\mathcal{N}\right). \tag{3.51}$$

Next, we study the connection of zero-error quantum codes and graph theory.

### 3.4.4 Connection with Graph Theory

The problem of finding *non-adjacent* codewords for the zero-error information transmission can be rephrased in terms of graph theory. The adjacent codewords are also called *confusable*, since these codewords can generate the same output with a given non-zero probability. Since we know that two input codewords $\left|\psi_{X_1}\right\rangle$ and $\left|\psi_{X_2}\right\rangle$ are *adjacent* if there is a channel output codeword $\left|\psi_{X'}\right\rangle$ which can be resulted by either of these two, that is $\Pr\left[X'\middle|X_1\right] > 0$ and $\Pr\left[X'\middle|X_2\right] > 0$.

The non-adjacent property of two quantum codewords can be analyzed by the *confusability graph* $\mathcal{G}_n$, where $n$ denotes the *length of the block code*.

Let us take as many vertices as the number of input messages $K$, and connect two vertices if these input messages are adjacent. For example, using the quantum version of the famous *pentagon graph* we show how the classical zero-error capacity $C_0\left(\mathcal{N}\right)$ of the quantum channel $\mathcal{N}$ changes if we use block codes of length $n$=1 and $n$=2. In the pentagon graph an input codeword from the set of non-orthogonal qubits $\{|0\rangle,|1\rangle,|2\rangle,|3\rangle,|4\rangle\}$ is connected with two other adjacent input codewords, and the number of total codewords is 5 [Lovász79].

The $\mathcal{G}_1$ *confusability* graph of the pentagon structure for block codes of length $n$=1 is shown in Fig. 3.17. The vertices of the graph are the possible input messages, where $K = 5$. The *adjacent* input messages are connected by a line. The non-adjacent inputs $|2\rangle$ and $|4\rangle$ are denoted by gray circles, and there is no connection between these two input codewords.

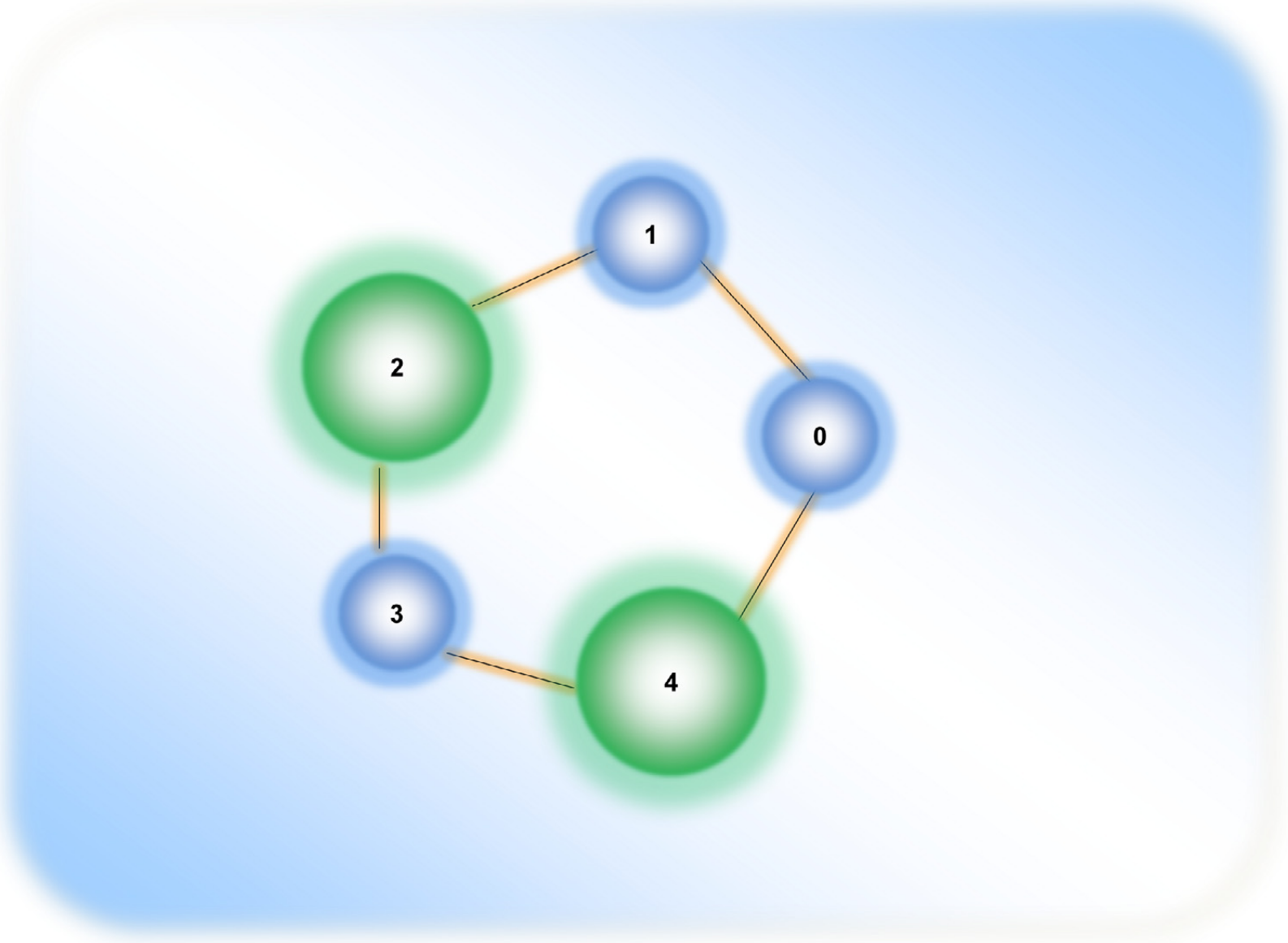

**Fig. 3.17.** The confusability graph of a zero-error code for one channel use. The two possible non-adjacent codewords are denoted by the large shaded circles.

For the block codes of length $n$=1, the maximal transmittable classical information with zero error is

$$C_0(\mathcal{N}) = \log(2) = 1, \tag{3.52}$$

since only two non-adjacent vertices can be found in the graph. We note, other possible codeword combinations also can be used to realize the zero-error transmission, in comparison with the confusability graph in Fig. 3.17, for example $|1\rangle$ and $|3\rangle$ also non-adjacent, etc. On the other hand, the maximum number of non-adjacent vertices (two, in this case) cannot be exceeded, thus $C_0\left(\mathcal{N}\right)=1$ remains in all other possible cases, too.

Let assume that we use $n=2$ length of block codes. First, let us see how the graph changes. The non-adjacent inputs are denoted by the large gray shaded circles. The connections between the possible codewords (which can be used as a block code) are denoted by the thick line and the dashed circle. The confusability graph $\mathcal{G}_2$ for $n=2$ length of block codes is shown in Fig. 3.18. The two half-circles together on the left and right sides represent one circle and the two half circles at the top and bottom of the figure also represent one circle; thus there are five dashed circles in the figure.

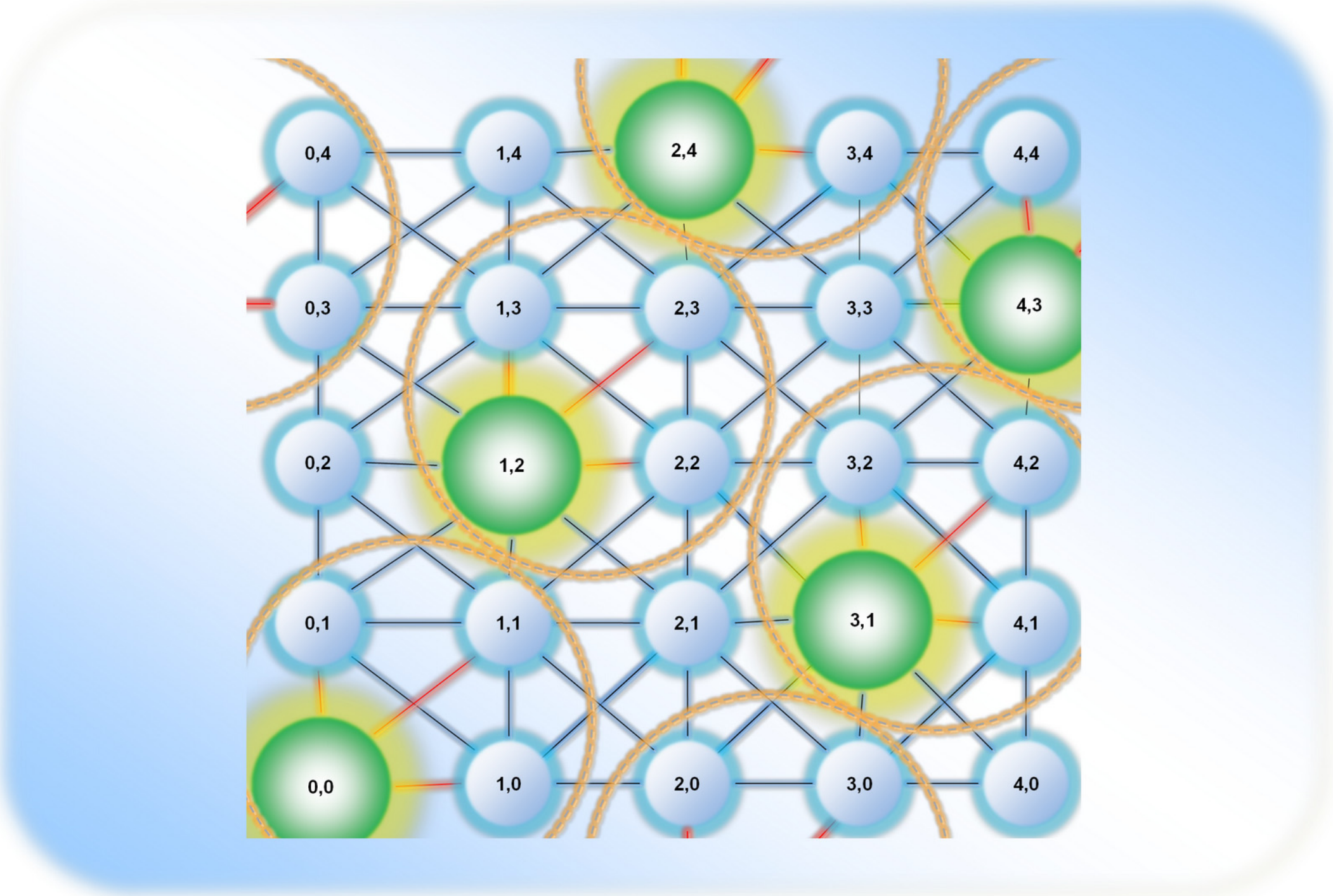

**Fig. 3.18.** The graph of zero-error code for two channel uses of a quantum channel. The possible zero-error codewords are depicted by the thick lines and dashed circles.

It can be seen that the complexity of the structure of the graph has changed dramatically, although we have made only a small modification: we increased the lengths of the block codes from $n=1$ to $n=2$. The five two-length codewords and zero-error quantum block codes which can realize the zero-error transmission can be defined as follows using the computational basis $\left\{|0\rangle,|1\rangle,|2\rangle,|3\rangle,|4\rangle\right\}$.

The arrows indicate those codewords which are connected in the graph with the given codeword; that is, these subsets can be used as quantum block codes as depicted in Fig. 3.18.

$$\begin{aligned}
&\left|\psi_{X_1}\right\rangle = \left[\left|0\right\rangle \otimes \left|0\right\rangle\right] \rightarrow \\
&\quad \left\{\left|\psi_{X_1}\right\rangle = \left[\left|0\right\rangle \otimes \left|0\right\rangle\right], \left|\psi_{X_2}\right\rangle = \left[\left|0\right\rangle \otimes \left|1\right\rangle\right], \left|\psi_{X_3}\right\rangle = \left[\left|1\right\rangle \otimes \left|0\right\rangle\right], \left|\psi_{X_4}\right\rangle = \left[\left|1\right\rangle \otimes \left|1\right\rangle\right]\right\}, \\
&\left|\psi_{X_2}\right\rangle = \left[\left|1\right\rangle \otimes \left|2\right\rangle\right] \rightarrow \\
&\quad \left\{\left|\psi_{X_1}\right\rangle = \left[\left|1\right\rangle \otimes \left|2\right\rangle\right], \left|\psi_{X_2}\right\rangle = \left[\left|2\right\rangle \otimes \left|2\right\rangle\right], \left|\psi_{X_3}\right\rangle = \left[\left|1\right\rangle \otimes \left|3\right\rangle\right], \left|\psi_{X_4}\right\rangle = \left[\left|2\right\rangle \otimes \left|3\right\rangle\right]\right\}, \\
&\left|\psi_{X_3}\right\rangle = \left[\left|2\right\rangle \otimes \left|4\right\rangle\right] \rightarrow \\
&\quad \left\{\left|\psi_{X_1}\right\rangle = \left[\left|2\right\rangle \otimes \left|4\right\rangle\right], \left|\psi_{X_2}\right\rangle = \left[\left|3\right\rangle \otimes \left|4\right\rangle\right], \left|\psi_{X_3}\right\rangle = \left[\left|2\right\rangle \otimes \left|0\right\rangle\right], \left|\psi_{X_4}\right\rangle = \left[\left|3\right\rangle \otimes \left|0\right\rangle\right]\right\}, \\
&\left|\psi_{X_4}\right\rangle = \left[\left|3\right\rangle \otimes \left|1\right\rangle\right] \rightarrow \\
&\quad \left\{\left|\psi_{X_1}\right\rangle = \left[\left|3\right\rangle \otimes \left|1\right\rangle\right], \left|\psi_{X_2}\right\rangle = \left[\left|4\right\rangle \otimes \left|1\right\rangle\right], \left|\psi_{X_3}\right\rangle = \left[\left|3\right\rangle \otimes \left|2\right\rangle\right], \left|\psi_{X_4}\right\rangle = \left[\left|4\right\rangle \otimes \left|2\right\rangle\right]\right\}, \\
&\left|\psi_{X_5}\right\rangle = \left[\left|4\right\rangle \otimes \left|3\right\rangle\right] \rightarrow \\
&\quad \left\{\left|\psi_{X_1}\right\rangle = \left[\left|4\right\rangle \otimes \left|3\right\rangle\right], \left|\psi_{X_2}\right\rangle = \left[\left|0\right\rangle \otimes \left|3\right\rangle\right], \left|\psi_{X_3}\right\rangle = \left[\left|4\right\rangle \otimes \left|4\right\rangle\right], \left|\psi_{X_4}\right\rangle = \left[\left|0\right\rangle \otimes \left|4\right\rangle\right]\right\}.
\end{aligned} \tag{3.53}$$

The classical zero-error capacity which can be achieved by $n = 2$ length block codes is

$$C_0\left(\mathcal{N}^{\otimes 2}\right) = \frac{1}{2}\log\left(5\right) = 1.1609\,. \tag{3.54}$$

From an engineering point of view this result means, that for the pentagon graph, the maximum rate at which classical information can be transmitted over a noisy quantum channel $\mathcal{N}$ with a zero error probability, can be achieved with quantum block code length of two.

For the classical zero-error capacities of some typical quantum channels see Section 5.

## 3.5 Entanglement-assisted Classical Zero-Error Capacity

In the previous subsection we discussed the main properties of zero-error capacity using product input states. Now, we add the entanglement to the picture. Here we discuss how the encoding and the decoding setting will change if we bring entanglement to the system and how it affects the classical zero-error capacity of a quantum channel.

If entanglement allowed between the communicating parties then the single-use and asymptotic *entanglement-assisted* classical zero-error capacities are defined as

$$C_0^{E(1)}\left(\mathcal{N}\right) = \log\left(K^E\left(\mathcal{N}\right)\right) \tag{3.55}$$

and

$$C_0^{E}\left(\mathcal{N}\right) = \lim_{n\to\infty}\frac{1}{n}\log\left(K^E\left(\mathcal{N}^{\otimes n}\right)\right). \tag{3.56}$$

where $K^E\left(\mathcal{N}^{\otimes n}\right)$ is the maximum number of $n$-length mutually non-adjacent classical messages that the quantum channel can transmit with zero error using *shared entanglement*.

Before we start to discuss the properties of the entanglement-assisted zero-error quantum communication, we introduce a new type of graph, called the *hypergraph* $\mathcal{G}_H$. The hypergraph is very similar to our previously shown *confusability* graph $\mathcal{G}_n$. The hypergraph contains a set of vertices and hyperedges. The vertices represent the *inputs* of the quantum channel $\mathcal{N}$, while the hyperedges contain all the channel inputs which could cause the same channel output with non-zero probability.

We will use some new terms from graph theory in this subsection; hence we briefly summarize these definitions:

- *maximum independent set of* $\mathcal{G}_n$: the maximum number of non-adjacent inputs ($K$),

- *clique of* $\mathcal{G}_n$: $\kappa_i$, the set of possible inputs of a given output in a confusability graph (which inputs could result in the same output with non-zero probability),

- *complete graph*: if all the vertices are connected with one another in the graph; in this case there are no non-adjacent inputs; i.e., the channel has no zero-error capacity.

In Fig. 3.19(a) we show a hypergraph $\mathcal{G}_H$, where the inputs of the channel are the vertices and the hyperedges represent the channel outputs. Two inputs are non-adjacent if they are in a different loop. The two non-adjacent inputs are depicted by the greater grey shaded vertices. In Fig. 3.19(b) we give the confusability graph $\mathcal{G}_n$ for a single channel use ($n = 1$), for the same input set. The cliques in the $\mathcal{G}_n$ confusability graph are depicted by $\kappa_i$.

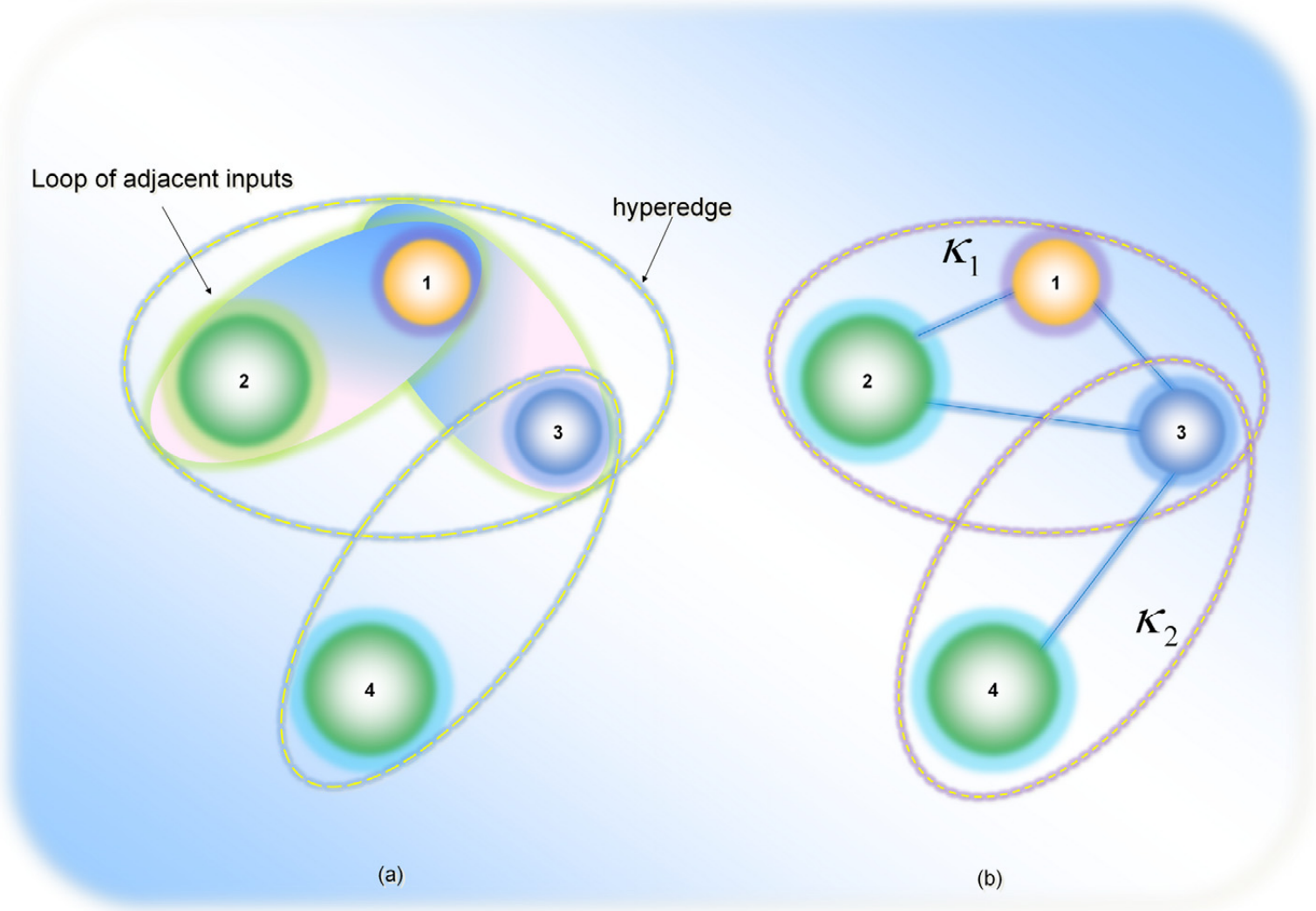

**Fig. 3.19.** The hypergraph and the confusability graph of a given input system with four inputs. The hyperedges of the hypergraph are labeled by the output. The number of non-adjacent inputs is two.

As follows from Fig. 3.19, both the hypergraph and the confusability graph can be used to determine the non-adjacent inputs. However, if the number of inputs starts to increase, the number of hyperedges in the hypergraph will be significantly lower than the number of edges in the confusability graph of the same system (see Fig. 3.21). In short, the entanglement-assisted zero-error quantum communication protocol works as follows according to Fig. 3.20 [Cubitt10]. Before the communication, Alice and Bob share entanglement between themselves. The $d$-dimensional shared system between Alice and Bob will be denoted by $\rho_{AB} = \left|\Phi_{AB}\right\rangle\left\langle\Phi_{AB}\right|$, where

$$\left|\Phi_{AB}\right\rangle = \frac{1}{\sqrt{d}}\sum_{i=0}^{d-1}\left|i\right\rangle_A\left|i\right\rangle_B \tag{3.57}$$

is a rank-$d$ maximally entangled qudit state (also called as *edit*). If Alice would like to send a message $q \in \{1,\ldots,K\}$, where $K$ is the number of messages, to Bob, she has to measure her half of the entangled system using a complete orthogonal basis $B_q = \left\{\left|\psi_{x'}\right\rangle\right\}$, $x' \in \kappa_q$, where $x'$ is a vertice in the hypergraph $\mathcal{G}_H$ from clique $\kappa_q$. The *orthonormal representation of a graph is a map:* the vertice $x'$ represents the unit vector $\left|\psi_{x'}\right\rangle$ such that if $x$ and $x'$ are *adjacent* then $\left\langle\psi_x\middle|\psi_{x'}\right\rangle = 0$ (*i.e., they are orthogonal in the orthonormal representation*) and $\kappa_q$ is the clique corresponding to message $q$ in the hypergraph $\mathcal{G}_H$. The hypergraph has $K$ cliques of size $d$, $\left\{\kappa_1,\ldots,\kappa_K\right\}$ (i.e., each message $q \in \left\{1,\ldots,K\right\}$ is represented by a $d$-size clique in the hypergraph $\mathcal{G}_H$.) After the measurement, Bob's state will collapse to $\left|\psi_x\right\rangle^*$. Bob will measure his state in $B_q = \left\{\left|\psi_x\right\rangle\right\}$ to get the final state $\left|\psi_{x'}\right\rangle^*$. Bob's output is denoted by $y$. Bob's possible states are determined by those vertices $x'$, for which $p\left(y\middle|x'\right) > 0$, and these *adjacent* states are *mutually orthogonal*; i.e., for any two $x_1'$ and $x_2'$, $\left\langle\psi_{x_1'}\middle|\psi_{x_2'}\right\rangle = 0$. Finally, Alice makes her measurement using $B_q = \left\{\left|\psi_{x'}\right\rangle\right\}$, then Bob measures his state $\left|\psi_x\right\rangle^*$ in $B_q = \left\{\left|\psi_{x'}\right\rangle\right\}$ to produce $\left|\psi_{x'}\right\rangle^*$.

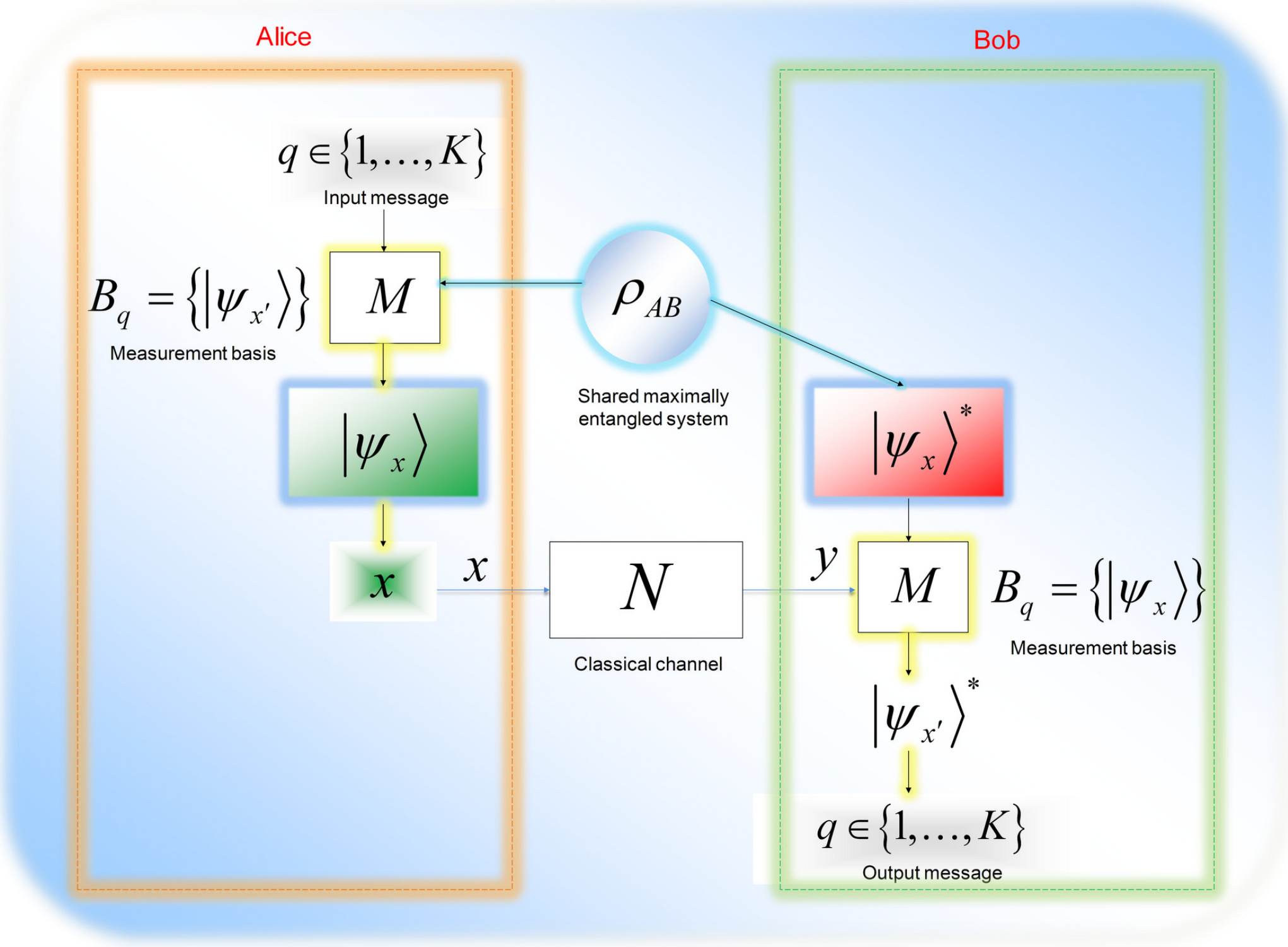

**Fig. 3.20.** The steps of the entanglement-assisted zero-error quantum communication protocol.

In order to make the above explanations more plausible, let us provide an example. Supposed Alice's set contains $K = 6$ codewords and she shares a rank-four (*i.e., d*=4) maximally entangled qudit state with Bob

$$\Phi_{AB} = \frac{1}{\sqrt{4}} \sum_{i=0}^{3} \left| i \right\rangle_A \left| i \right\rangle_B , \tag{3.58}$$

however, in the general case $d$ can be chosen as large as Alice and Bob would like to use. Alice measures her system from the maximally entangled state, and she chooses a basis among the $K$ possible states, according to which message $q$ she wants to send Bob. Alice's measurement outcome is depicted by $x$, which is a random value. Alice sends $q$ and $x$ to the classical channel $N$. In the next phase, Bob performs a projective measurement to decide which $x$ value was made to the classical channel by Alice. After Bob has determined it, he can answer which one of the possible $K$ messages had been sent by Alice with the help of the maximally entangled system. Alice makes her measurement on her side using one of the six possible bases $B_q = \left\{ \left| \psi_{x'} \right\rangle \right\}$ on her half of the state $\rho_{AB}$. Her system collapses to $\left| \psi_x \right\rangle \in B_q$, while Bob's system collapses to $\left| \psi_x \right\rangle^*$, conditioned on $x$. Alice makes $x$ to the classical channel $N$; Bob will receive classical message $y$. From the channel output $y = N\left(x\right)$, where $N$ is the classical channel between Alice and Bob, Bob can determine the mutually adjacent inputs (i.e., those inputs which could produce the given output). If Bob makes a measurement in basis $B_q = \left\{ \left| \psi_x \right\rangle \right\}$, then he will get $\left| \psi_{x'} \right\rangle^*$, where these states for a given set of $x'$ corresponding to possible $x$ are *orthogonal states*, so he can determine $x$ and the original message $q$. The channel output gives Bob the information that some set of mutually adjacent inputs were used on Alice's side. On his half of the entangled system, the states will be mutually orthogonal. A measurement on these mutually orthogonal states will determine Bob's state and he can tell Alice's input with certainty.

Using this protocol, the number of mutually non-adjacent input messages is

$$K^E \geq 6 , \tag{3.59}$$

while if Alice and Bob would like to communicate with zero-error but without shared entanglement, then $K = 5$. As follows, for the single-use classical zero-error capacities we get

$$C_0^{(1)} = \log\left(5\right) \tag{3.60}$$

and

$$C_0^{E(1)} = \log\left(K^E\right) = \log\left(6\right), \tag{3.61}$$

while for the asymptotic entanglement-assisted classical zero-error capacity,

$$C_0^E \geq \log\left(K^E\right) = \log\left(6\right). \tag{3.62}$$

According to Alice's $K^E = 6$ messages, the hypergraph can be partitioned into six cliques of size $d = 4$. The adjacent vertices are denoted by a common loop. The overall system contains $6 \times 4 = 24$ basis vectors. These vectors are grouped into $K^E = 6$ orthogonal bases. Two input vectors are connected in the graph if they are adjacent vectors; i.e., they can produce the same output. The hypergraph $\mathcal{G}_H$ of this system is shown in Fig. 3.21. The mutually non-adjacent inputs are denoted by the great shaded circles. An important property of the entanglement-assisted classical zero-error capacity is that the number of maximally transmittable messages is not equal to the number of non-adjacent inputs. While the hypergraph has five independent vertices, the maximally transmittable messages are greater than or equal to six. The confusability graph of this system for a single use of quantum channel $\mathcal{N}$ would consist of $6 \times 4 \times 9 = 216$ connections, while the hypergraph has a significantly lower number ($6 \times 6 = 36$) of hyperedges. The adjacent vertices are depicted by the loops connected by the thick lines. The six possible messages are denoted by the six, four dimensional (i.e., each contains four vertices) cliques $\left\{\kappa_1, \ldots, \kappa_K\right\}$. The cliques (dashed circles) show the set of those input messages which could result in the same output with a given probability $p > 0$.

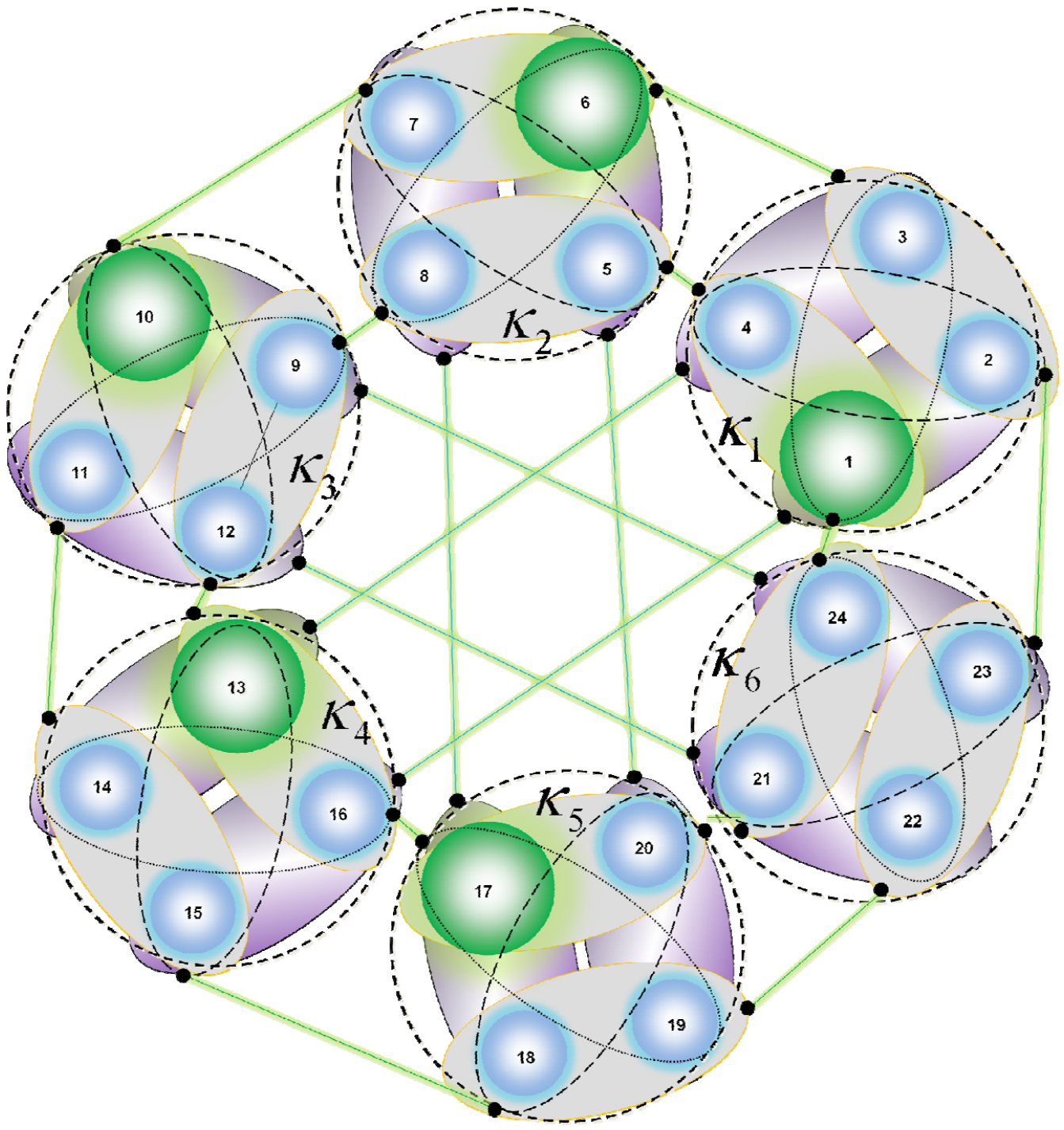

**Fig. 3.21.** The hypergraph of an entanglement-assisted zero-error quantum code. The non-adjacent inputs are depicted by the great shaded circles. The adjacent vertices are depicted by loops connected by the thick lines.

We note, the cliques are defined in the $\mathcal{G}_n$ confusability graph representation, but we also included them on the hypergraph $\mathcal{G}_H$. The adjacent vertices which share a loop represent mutually orthogonal input states. For these mutually orthogonal inputs the output will be the same.

The complete theoretical background of this example, i.e., the proof of the fact, that entanglement can increase the asymptotic classical zero-error capacity $C_0\left(\mathcal{N}\right)$ of a quantum channel was described in [Cubitt10].

We have seen in this subsection that shared entanglement between Alice and Bob can help to increase the maximally transmittable classical messages using noisy quantum channels with zero error probability. According to the *Cubitt-Leung-Matthews-Winter* theorem (CLMW theorem) [Cubitt10] there exist entanglement-assisted quantum communication protocol which can send one of $K$ messages with *zero error*; hence for the entanglement-assisted asymptotic classical zero-error capacity

$$\log\left(K\right) \leq C_0 = \lim_{n\to\infty}\frac{1}{n}\log\left(K\left(\mathcal{N}^{\otimes n}\right)\right) < C_0^E = \lim_{n\to\infty}\frac{1}{n}\log K^E\left(\mathcal{N}^{\otimes n}\right) \geq \log\left(K^E\right)$$

. (3.63)

Entanglement is very useful in zero-error quantum communication, since with the help of entanglement the maximum amount of perfectly transmittable information can be achieved.

As was show by Leung et al. [Leung10], using special input codewords (based on a special Pauli graph), entanglement can help to increase the classical zero-error capacity to the maximum achievable HSW capacity; that is, there exists a special combination for which the entanglement-assisted classical zero-error capacity $C_0^E\left(\mathcal{N}\right)$ is

$$C_0^E\left(\mathcal{N}\right) = \log\left(9\right), \tag{3.64}$$

while the classical zero-error capacity is

$$C_0\left(\mathcal{N}\right) = \log\left(7\right), \tag{3.65}$$

i.e., with the help of entanglement-assistance the number of possible input messages ($K$) can be increased.
Another important discovery is that for this special input system the entanglement-assisted classical zero-error capacity, $C_0^E\left(\mathcal{N}\right)$, is equal to the maximal transmittable classical information over $\mathcal{N}$; that is

$$C_0^E\left(\mathcal{N}\right) = C\left(\mathcal{N}\right) = \log\left(9\right). \tag{3.66}$$

In the asymptotic setting the maximum achievable capacities as functions of block code length are summarized in Fig. 3.22.

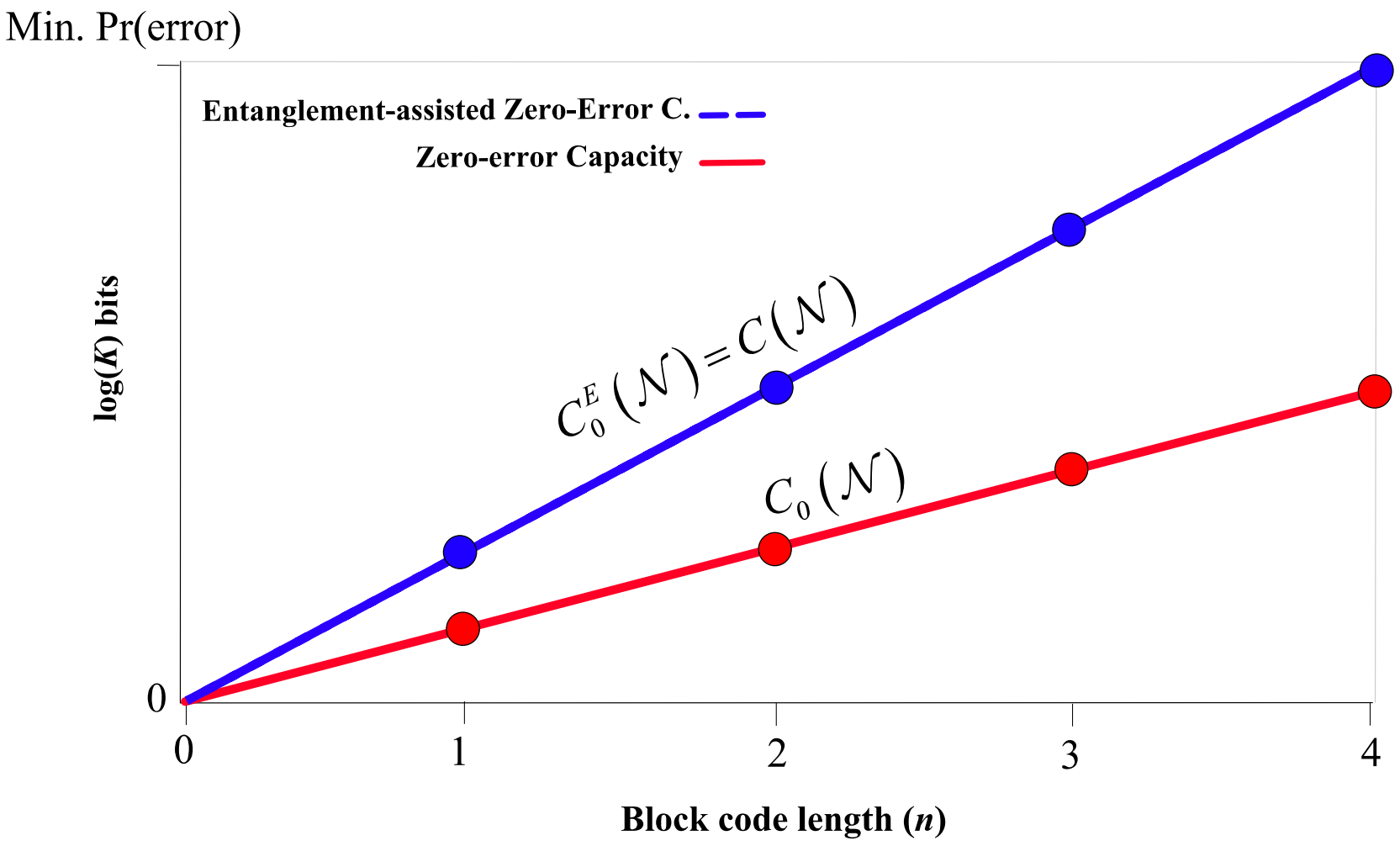

**Fig. 3.22.** The asymptotic classical zero-error capacities without entanglement and with entanglement assistance using a special Pauli graph.

The maximal amount of transmittable classical information which can be sent through a noisy quantum channel $\mathcal{N}$ without error increases with the length of the input block code, and with the help of EPR input states (for this special Pauli graph-based code) the classical HSW capacity can be reached, which is also the upper bound of the classical zero-error capacity.

## 3.6 Some Important Channel Maps

Here, we give a brief survey of some important quantum channel maps. We discuss the density matrix representation of these channel models and their geometric illustration on the Bloch sphere. For the corresponding definitions related to the state-vector description we advise to the reader to [Imre05].

### 3.6.1 The Flipping Channel Models

The *bit flip* channel can be defined by means of $\sigma_X$, the Pauli $X$ transformation. The bit flip channel changes the probability amplitudes of the input qubit. The map of the bit flip channel can be expressed as

$$\mathcal{N}\left(\rho\right) = p\left(\sigma_X \rho \sigma_X\right) + \left(1-p\right)\rho . \tag{3.67}$$

In the geometric representation (see Fig. 3.23), this channel map shrinks the original Bloch sphere along the $y$ and $z$ axes, by the factor $1-2p$.

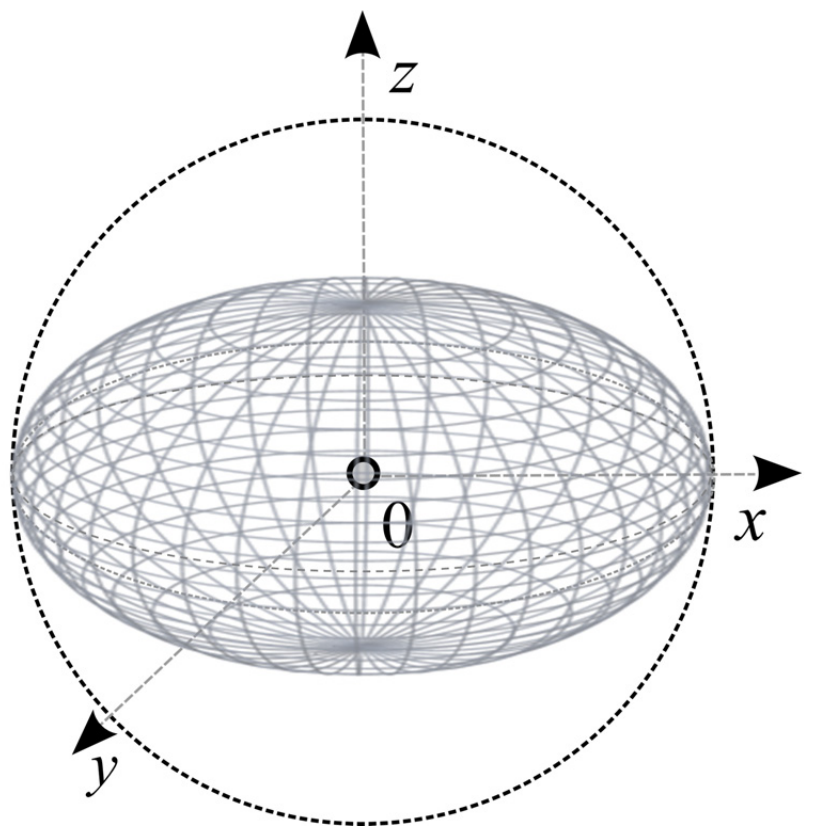

**Fig. 3.23.** The bit flip channel shrinks the Bloch sphere along the $y$ and $z$ axes.

Similarly to the bit flip channel, the *phase flip* quantum channel applies the Pauli $Z$ transformation $\sigma_Z$. The phase flip channel changes the sign of the relative phase of the input qubit. The map $\mathcal{N}$ of this channel can be expressed as

$$\mathcal{N}(\rho) = p(\sigma_Z \rho \sigma_Z) + (1-p)\rho , \tag{3.68}$$

where $p$ describes the probability that the channel does a phase-flip error on the input qubit. In the Bloch sphere representation means that the width of the original Bloch sphere will be reduced by a factor of $1-2p$ in its equatorial plane. The surface of channel ellipsoid consists of the set of channel output $\rho$ vectors.
In the geometric representation (see Fig. 3.24), the phase flip channel map shrinks the original Bloch sphere along the $x$ and $y$ axes, by the factor $1-2p$.

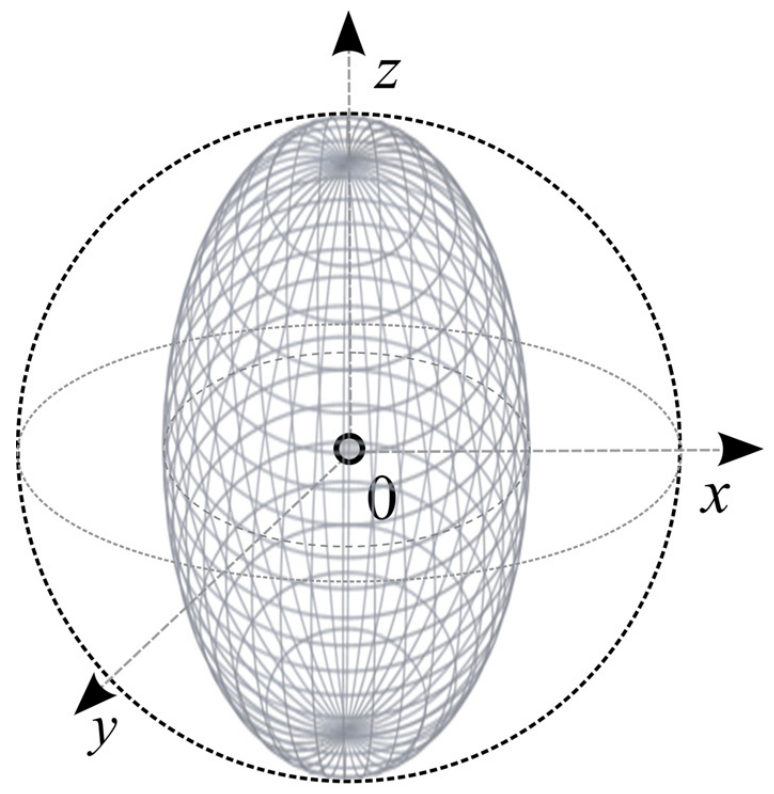

**Fig. 3.24.** The geometric interpretation of the phase flip channel.

In the physical realizations, this channel map is also referred as the *phase-damping* channel model.
If the channel simultaneously realizes bit flip and phase flip transformations on the input quantum state then the channel is called a *bit-phase flip* channel. The effect of the bit-phase flip channel can be described by the $\sigma_Y$, the Pauli $Y$ transformation

$$\mathcal{N}\left(\rho\right) = p\left(\sigma_Y \rho \sigma_Y\right) + \left(1-p\right)\rho. \tag{3.69}$$

This channel also shrinks the Bloch sphere by the factor $1-2p$ along the direction of the $x$ and $z$ axes. The geometric interpretation of the *bit-phase flip* channel is depicted in Fig. 3.25.

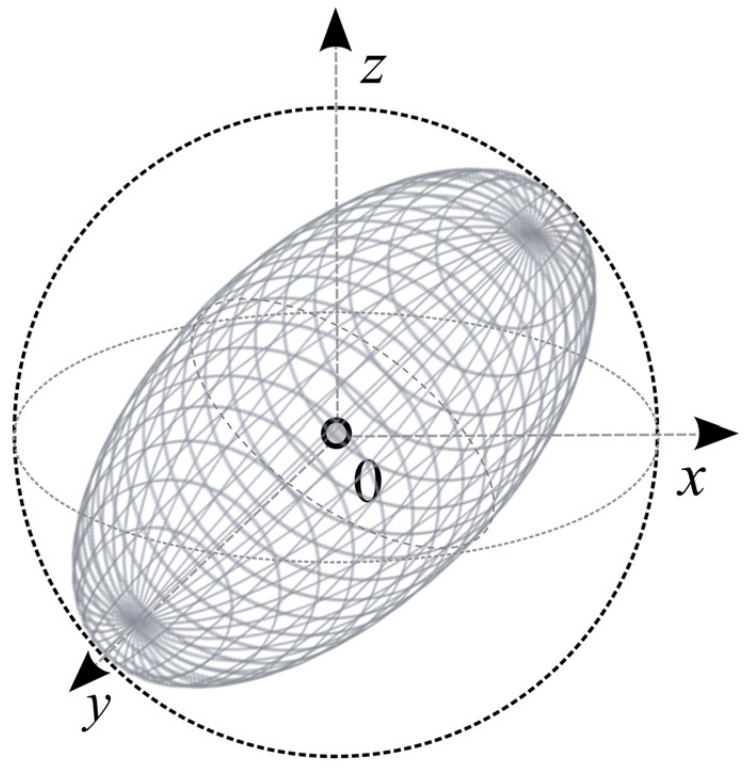

**Fig. 3.25.** The image of the bit-phase flip channel.

Next, we discuss the depolarizing quantum channel model.

### 3.6.2 The Depolarizing Channel Model

The last discussed unital channel model is the *depolarizing* channel which performs the following transformation

$$\mathcal{N}\left(\rho_i\right) = p\frac{I}{2} + \left(1-p\right)\rho_i, \tag{3.70}$$

where $p$ is the *depolarizing parameter* of the channel, and if Alice uses two orthogonal states $\rho_0$ and $\rho_1$ for the encoding then the mixed input state is

$$\rho = \left( \sum_i p_i \rho_i \right) = p_0 \rho_0 + \left(1 - p_0\right) \rho_1 . \tag{3.71}$$

After the unital channel has realized the transformation $\mathcal{N}$ on state $\rho$, we will get the following result

$$\begin{aligned} \mathcal{N}(\rho) &= \mathcal{N}\left( \sum_i p_i \rho_i \right) = \mathcal{N}\left( p_0 \rho_0 + \left(1 - p_0\right) \rho_1 \right) \\ &= p \frac{1}{2} I + \left(1 - p\right)\left( p_0 \rho_0 + \left(1 - p_0\right) \rho_1 \right) \\ &= \begin{pmatrix} p \frac{1}{2} + \left(1 - p\right) p_0 & 0 \\ 0 & p \frac{1}{2} + \left(1 - p\right)\left(1 - p_0\right) \end{pmatrix}. \end{aligned} \tag{3.72}$$

Geometrically, the map of the depolarizing quantum channel shrinks the original Bloch sphere in every direction by $1 - p$. The effect of the depolarizing quantum channel is shown in Fig. 3.26.

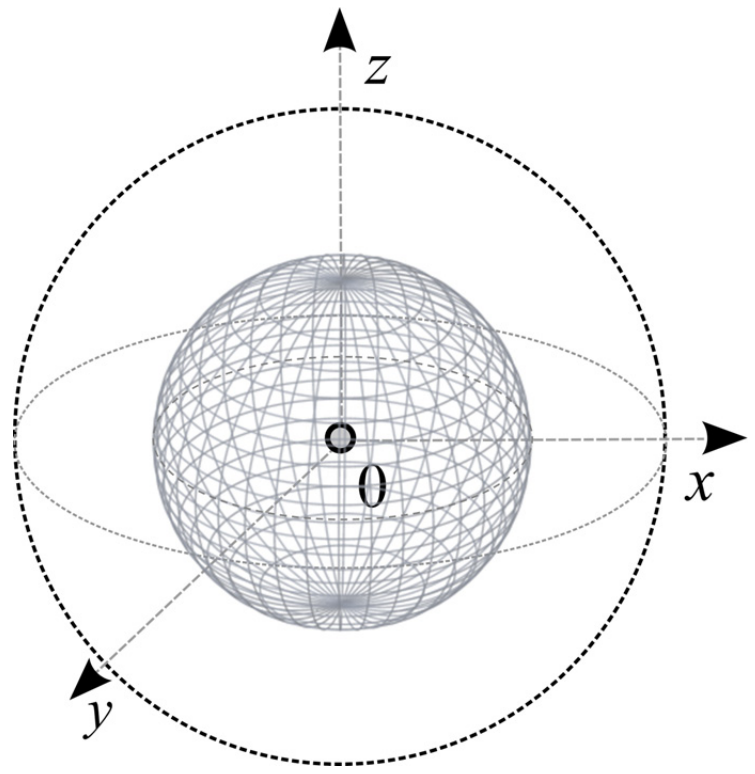

**Fig. 3.26.** The depolarizing channel shrinks the original Bloch sphere in every direction.

One of the most important type of channel map to describe decoherence is called the amplitude damping channel (or decay) map. We will discuss it in the next subsection.

### 3.6.3 The Amplitude Damping Channel

The result of decoherence is also a mixed quantum state, such as in the case of the previously discussed channel maps, however in this case, the density matrices of these mixed states will differ. In the case of decoherence, the non-diagonal values of the density matrix completely vanish.

The *amplitude damping* channel map shrinks the Bloch sphere in the two directions of the equatorial plane – similarly to the phase flip channel, but it also moves the center of the ellipsoid from the center of the Bloch sphere. Therefore, this channel map is not unital. The height of the scaled ellipsoid will be given by the scaling factor $1-2p$. The direction of the shift can be upward or downward. On the other hand, the width of the ellipsoid will differ from the previous cases, namely by to the factor $\sqrt{1-2p}$. The geometric interpretation of the amplitude damping quantum channel is illustrated in Fig. 3.27. (We emphasize that the ball can be shrunk to the opposite direction along the $z$ axis, too.)

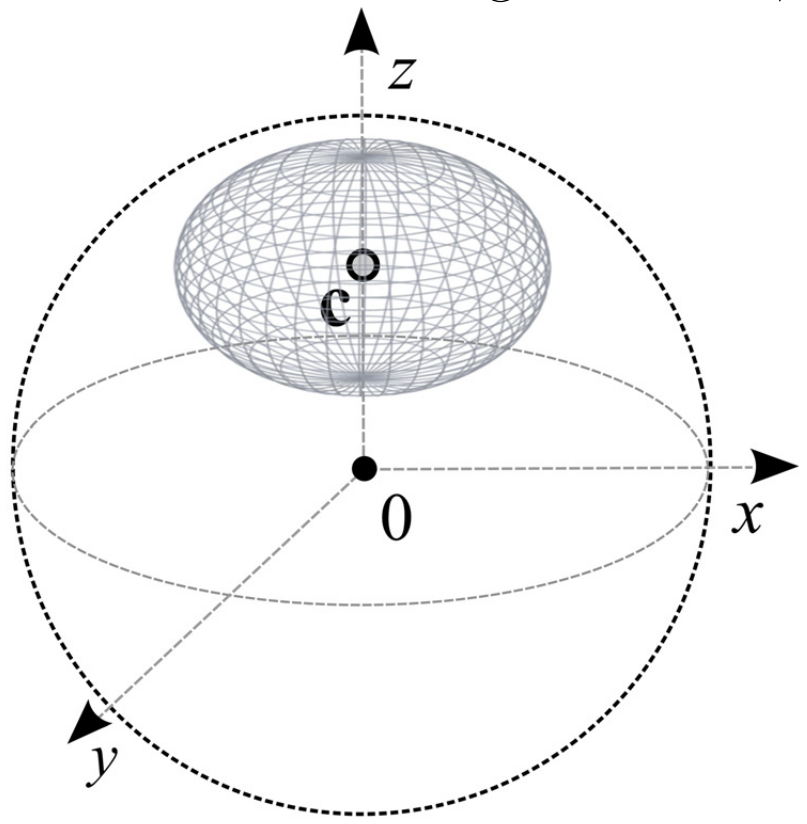

**Fig. 3.27.** The picture of the amplitude damping quantum channel.

As we have seen in Section 2, any quantum channel $\mathcal{N}$ can be described in the Kraus-representation [King09], using a set of Kraus matrices $\mathcal{A}=\{A_i\}$ in the following form

$$\mathcal{N}(\rho)=\sum_i A_i \rho A_i^{\dagger}, \tag{3.73}$$

where $\sum_i A_i^{\dagger} A_i = I$. For an amplitude damping quantum channel

$$A_1=\begin{bmatrix}\sqrt{p} & 0\\ 0 & 1\end{bmatrix}, \text{ and } A_2=\begin{bmatrix}0 & 0\\ \sqrt{1-p} & 0\end{bmatrix}, \tag{3.74}$$

where $p$ represents the probability that the channel leaves input state $\left|0\right\rangle$ unchanged. Obviously the channel flips the input state from $\left|0\right\rangle$ to $\left|1\right\rangle$ with probability $1-p$. [Cortese02], [Nielsen2000]. For $p=0$, the channel output is $\left|1\right\rangle$ with probability 1. However, the channel leaves untouched the input state $\left|1\right\rangle$, hence the output of the channel will be $\left|1\right\rangle$.

For a non-unital quantum channel, the set of Kraus operators $\mathcal{A}=\left\{A_i\right\}$ can be transformed to the *King-Ruskai-Szarek-Werner* (KRSW) ellipsoid channel model with parameters $\left\{t_k,\lambda_k\right\}$, $k=1,2,3$. The effect of $\left\{t_k\neq 0\right\}$ is that the average output $\sigma=\sum_i p_i\rho_i$ of the channel moves away from the origin of the Bloch sphere, meaning that the center of the smallest enclosing quantum informational ball is not equal to the origin of the Bloch sphere. The affine map of the amplitude damping channel can be expressed using Bloch vectors $\mathbf{r}_{in}$ and $\mathbf{r}_{out}$ in the following way

$$\mathbf{r}_{out}=\begin{pmatrix}\mathbf{r}_{out}^{(x)}\\ \mathbf{r}_{out}^{(y)}\\ \mathbf{r}_{out}^{(z)}\end{pmatrix}=\begin{pmatrix}\sqrt{1-p} & 0 & 0\\ 0 & \sqrt{1-p} & 0\\ 0 & 0 & 1-\frac{p}{2}\end{pmatrix}\begin{pmatrix}\mathbf{r}_{in}^{(x)}\\ \mathbf{r}_{in}^{(y)}\\ \mathbf{r}_{in}^{(z)}\end{pmatrix}+\begin{pmatrix}0\\ 0\\ \frac{p}{2}\end{pmatrix}. \tag{3.75}$$

The amplitude damping channel can be visualized in the KRSW ellipsoid channel model

$$\begin{aligned}&t_x=0,\ t_y=0,\ t_z=1-p,\\ &\lambda_x=\sqrt{p},\ \lambda_y=\sqrt{p},\ \lambda_z=p,\end{aligned} \tag{3.76}$$

where $p\in\left[0,1\right]$ is the channel parameter.

### 3.6.4 The Dephasing Channel Model

The second type of decoherence map discussed is unitary and results in relative phase differences between the computational basis states: the channel map which realizes it is called the *dephasing* map. In contrast to the amplitude damping map, it realizes a unitary transformation. The unitary representation of the dephasing quantum channel for a given input $\rho = \sum_{i,j} \rho_{ij} |i\rangle\langle j|$ can be expressed as

$$\mathcal{N}(\rho) = \sum_{i} \rho_{ii} |E_i\rangle\langle E_i|, \tag{3.77}$$

where $|E_i\rangle$ are the environment states. The dephasing quantum channel acts on the density operator $\rho$ as follows

$$\mathcal{N}(\rho_i) = p\sigma_Z \rho \sigma_Z + (1-p)\rho_i, \tag{3.78}$$

where $\sigma_Z$ is the Pauli $Z$-operator. The dephasing quantum channel is also degradable, for definition of degradable quantum channel see Section 6.3.1. The image of the dephasing channel map is similar to that of the phase flip channel map, however, the shrinkage of the original Bloch sphere is greater (see Fig. 3.28). The dephasing channel transforms an arbitrary superposed pure quantum state $\alpha|0\rangle + \beta|1\rangle$ into a mixture

$$\mathcal{N}(\rho) \to \rho' = \begin{bmatrix} |\alpha|^2 & \alpha\beta^* e^{-\gamma(t)} \\ \alpha^* \beta e^{-\gamma(t)} & |\beta|^2 \end{bmatrix}, \tag{3.79}$$

where $\gamma(t)$ is a positive real parameter, which characterizes the coupling to the environment, using the time parameter $t$.

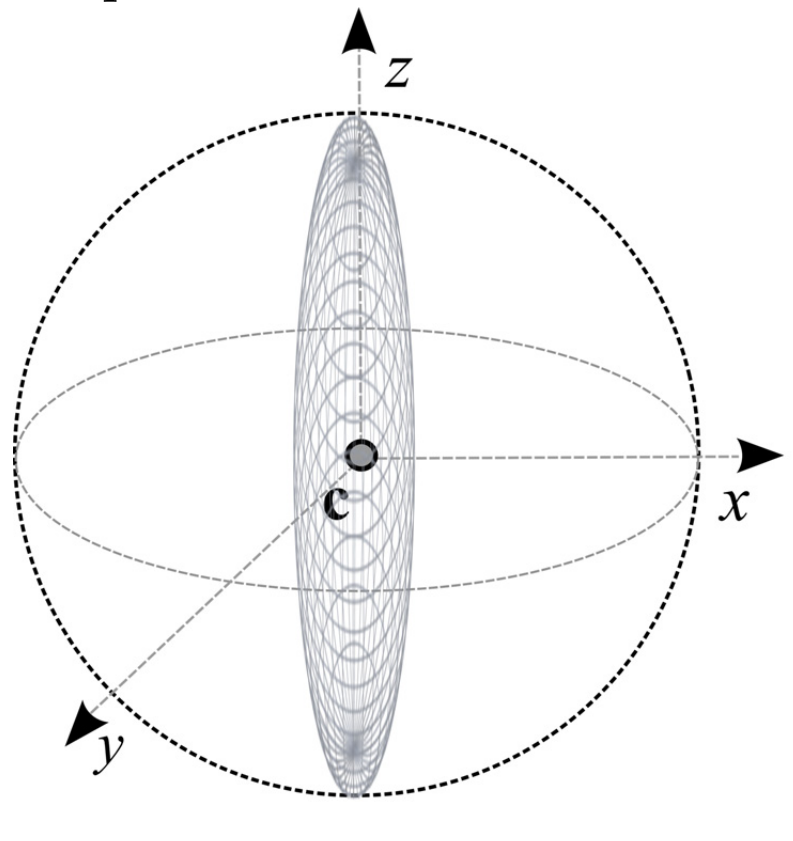

**Fig. 3.28.** The image of the dephasing channel map.

Finally, next we study the pancake map, which also represents decoherence.

### 3.6.5 The Pancake Map

To give an example for physically not allowed (nonphysical, non-CP) transformations, we introduce the *pancake map* in Fig. 3.29. The non-CP property means, that there exists no Completely Positive Trace Preserving map, which preserves some information along the equatorial spanned by the $x$ and $y$ axes of the Bloch sphere, while it completely demolishes any information along the $z$ axis. This map is called the pancake map, and it realizes a physically not allowed (non-CP) transformation. The effect of the pancake map is similar to the bit-phase flip channel, however, this channel defines a non-CP transform: it "smears" the original Bloch sphere along the equatorial spanned by the $x$ and $y$ axes.

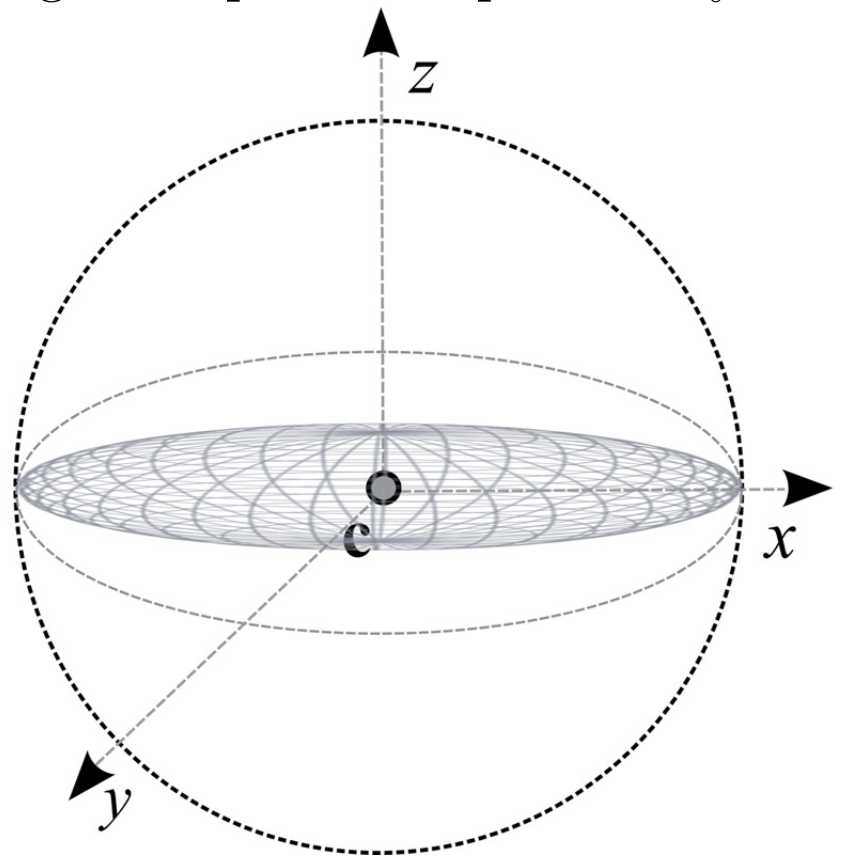

**Fig. 3.29.** The pancake map is a physically not allowed map.

On the other hand, the pancake map—besides the fact that is a non-physical map—can be used theoretically to transfer some information, and some information can be transmitted through these kinds of channel maps. The reason behind decoherence is *Nature*. She cannot be perfectly eliminated from quantum systems in practice. The reduction of decoherence is also a very complex task, hence it brings us on the engineering side of the problem: the quantum systems have to be designed in such a way that the unwanted interaction between the quantum states and the environment has to be minimal [Shor95], [Shor96]. Currently - despite the efficiency of these schemes - the most important tools to reduce decoherence are quantum error-correcting codes and decoupling methods.

## 3.7 Related Work

The classical world with the classical communication channel can be viewed as a special case of a quantum channel, since classical information can be encoded into the qubits—just as into classical bits. Classical information can also be encoded in quantum states. In this section we summarize the most important works related to the classical capacity of the quantum channels.

**The Early Days**

At the end of the twentieth century, the capacities of a quantum channel were still an open problem in quantum information theory. Before the several, and rather different, capacities of the quantum channel were recognized, the "academic" opinion was that quantum channels could be used only for the transmission of classical information encoded in the form of quantum states [Holevo73], [Holevo73a]. As has been found later, the classical capacity of the quantum channel can be measured in several different settings. It was shown that the classical capacity depends on whether the input states are entangled or not, or whether the output is measured by single or by joint measurement setting [Bennett97], [Fuchs2000], [King09]. In a specified manner, the classical capacity has been defined for measuring the maximal asymptotic rate at which classical information can be transmitted through the quantum channel, with an arbitrarily high reliability [Barnum97a], [Schumacher97].

The first proposed capacity measure was the *classical capacity* of a quantum channel—denoted by $C(\mathcal{N})$—measures the maximum transmittable classical information—in the form of product or entangled quantum states. The idea of transmitting classical information through a quantum channel was formulated in the 1970s. The Holevo bound was introduced by Holevo in 1973, however the theorem which describes the classical capacity of the quantum channel in an explicit way appeared just about *three decades later*, in the mid 1990s.

The maximal accessible classical information from a quantum source firstly has been characterized by Levitin [Levitin69] and Holevo [Holevo73], [Holevo73a] in the early days, which were some of the first and most important results in quantum information theory regarding the classical capacity of quantum channels. More information about the connection between the Holevo bound and the accessible information (which quantifies the information of the receiver after the measurement) can be found in [Holevo73], [Holevo73a]. Later this result was developed and generalized by Holevo, Schumacher, and Westmoreland, and became known in quantum information theory as the *HSW channel capacity* [Schumacher97], [Holevo98]. The HSW theorem uses the Holevo information to describe the amount of classical information which can be transmitted through a noisy quantum channel, and it makes possible to apply different measurement constructions on the sender and on the receiver's side. The proofs of the HSW theorem, such as the direct coding theorem and the converse theorem, with the complete mathematical background can be found in the work of Holevo [Holevo98] and of Schumacher and Westmoreland [Schumacher97]. About the efficiency problems of implementation and construction of joint POVM (Positive Operator Valued Measure) measurement setting, as it was shown in the same works of the authors.

One of the most important result on the mechanism of the encoding of quantum information into physical particles was discovered by Glauber in the very early years of quantum information processing [Glauber1963] and a great summarize from more than four-decades later [Glauber05]. Also from this era and field, important results on the encoding and decoding processes of quantum information were shown in the works of Gordon [Gordon1964] and Helstrom [Helstrom76]. About detection of quantum information and the process of measurement see [Fannes73], or the work of Helstrom from 1976 [Helstrom76], or Herbert's work from 1982 [Herbert82]. Before their results, Levitin published a paper about the quantum measure of the amount of information in 1969 [Levitin69], which was a very important basis for further work.

**Classical Capacity of a Quantum Channel**

The amount of classical information which can be transmitted through a noisy quantum channel in a reliable form with product input states, using the quantum channel many times, was determined by the HSW theorem [Holevo98], [Schumacher97]. This coding theorem is an analogue to Shannon's classical channel coding theorem, however it extends its possibilities. The inventors of the HSW theorem—Holevo, Schumacher and Westmoreland—proved and concluded independently the same result. Holevo's result from 1998 can be found in [Holevo98], Schumacher and Westmoreland's results can be found in [Schumacher97]. They, with Hausladen et al. in 1995 [Hausladen95], and in 1996 [Hausladen96], have also confirmed that the maximal classical information which can be transmitted via pure quantum states is bounded by the Holevo information.

A different approach to the proof of the HSW theorem was presented by Nielsen and Chuang in 2000 [Nielsen2000]. An interesting connection between the mathematical background of the compressibility of quantum states and the HSW theorem was shown by Devetak in 2003 [Devetak03], who proved that a part of the mathematical background constructed for the compression of quantum information can be used to prove the HSW theorem. Another interesting approach for proving the HSW theorem and bounds on the error probability was presented by Hayashi and Nagaoka in 2003 [Hayashi03]. The additivity property of qubit channels which require four inputs to achieve capacity was analyzed by Hayashi et al. in [Hayashi05].

Very important connections regarding the transmission of classical information over noisy quantum channels was derived in the work of Schumacher and Westmoreland in 1997 [Schumacher97], and two years later, a very important work was published on the relevance of optimal signal ensembles in the classical capacity of a noisy quantum channels [Schumacher99]. (We also suggest their work on the characterizations of classical and quantum communication processes [Schumacher99a].) The classical information capacity of a class of most important practical quantum channels (Gaussian quantum channels) was shown by Wolf and Eisert [Wolf05] or the work of Lupo et al. [Lupo11]. The generalized minimal output entropy conjecture for Gaussian channels was studied by Giovannetti et al. [Giovannetti10].

About the role of feedback in quantum communication, we suggest the works of Bowen [Bowen04] and 2005 [Bowen05], the article of Bowen et al. [Bowen05a], and the work of Harrow [Harrow04a]. The works of Bowen provide a great introduction to the role of quantum feedback on the classical capacity of the quantum channel, it was still an open question before. As he concluded, the classical capacity of a quantum channel using quantum feedback is equal to the entanglement-assisted classical capacity, the proof was given in Bowen and Nagarajan's paper [Bowen05a].

We have also seen that the noise of a quantum channel can be viewed as a result of the entanglement between the output and the reference system called the purification state (see purification in (2.64)). Some information leaks to the environment, and to the purification state, which purification state cannot be accessed. As is implicitly woven into this section, a noisy quantum channel can be viewed as a special case of an ideal quantum communication channel. The properties of the general quantum channel model and the quantum mutual information function can be found in the work of Adami and Cerf [Adami96].

A great analysis of completely-positive trace preserving (CPTP) maps was published by Ruskai et al. [Ruskai01]. Further information on the classical capacity of a quantum channel can be found in [Bennett98], [Holevo98], [King09], [Nielsen2000].

**Entanglement-assisted Classical Capacity**

In the early 1970s, it was also established that the classical capacity of a quantum channel can be higher with *shared entanglement*—this capacity is known as the *entanglement-assisted classical capacity* of a quantum channel, which was completely defined by Bennett et al. just in 1999 [Bennett99], and is denoted by $C_E\left(\mathcal{N}\right)$. The preliminaries of the definition of this quantity were laid down by Bennett and Wiesner in 1992 [Bennett92c]. Later, in 2002 Holevo published a review paper about the entanglement-assisted classical capacity of a quantum channel [Holevo02a].

Entanglement-assisted classical communication requires a super-dense protocol-like encoding and decoding strategy [Bennett02]. About the classical capacity of a noiseless quantum channel assisted by noisy entanglement, an interesting paper was published by Horodecki et al. in 2001 [Horodecki01]. In the same work the authors have defined the "noisy version" of the well-known superdense coding protocol, which originally was defined by Bennett in 1992 [Bennett92c] for ideal (hence noiseless) quantum channels. As can be found in the works of Bennett et al. from 1999 [Bennett99] and from 2002 [Bennett02], the *entanglement-assisted classical capacity* opened the possibility to transmit more classical information using shared entanglement (in case of single-use capacity). As can be checked by the reader, the treatment of entanglement-assisted classical capacity is based on the working mechanism of the well-known superdense coding protocol—however, classical entanglement-assisted classical capacity used a noisy quantum channel instead of an ideal one.

Bennett, in two papers from 1999 [Bennett99] and 2002 [Bennett02] showed that the *quantum mutual information* function (see Adami and Cerf's work [Adami96]) can be used to describe the classical entanglement-assisted capacity of the quantum channel i.e., the *maximized quantum mutual information of a quantum channel and the entanglement-assisted classical capacity are equal.* The connection between the quantum mutual information and the entanglement-assisted capacity can be found in the works of Bennett et al. [Bennett99] and [Bennett02]. In the latter work, the formula of the quantum-version of the well-known classical Shannon formula was generalized for the classical capacity of the quantum channel. In these two papers the authors also proved that the entanglement-assisted classical capacity is an upper bound of the HSW channel capacity. Holevo gave an explicit upper bound on the classical information which can be transmitted through a noisy quantum channel, it is known as the Holevo-bound. The Holevo-bound states that the most classical information which can be transmitted in a qubit (i.e., two level quantum system) through a noiseless quantum channel in a reliable form, is one bit. However, as was shown later by Bennett et al. in 1999 [Bennett99], the picture changes, if the parties use shared entanglement (known as the *Bennett-Shor-Smolin-Thapliyal, or the BSST-*theorem). As follows, the BSST-theorem gives a closer approximation to the maximal transmittable classical information (i.e., to the "single-use" capacity) over quantum channels, hence it can be viewed as the *true "quantum version" of the well known classical Shannon capacity formula* (since it is a maximization formula), instead of the "non entanglement-assisted" classical capacity. Moreover, the inventors of the BSST-theorem have also found a very important property of the entanglement-assisted classical capacity: *its single-use version is equal to the asymptotic version*, which implies the fact that no regularization is needed. (As we have seen in this section, we are not so lucky in the case of general classical and private classical capacities. As we will show in Section 4, we are "unlucky" in the case of quantum capacity, too.) They have also found that no classical feedback channel can increase the entanglement-assisted classical capacity of a quantum channel, and this is also true for the classical (i.e., the not entanglement-assisted one) capacity of a quantum channel. These results were also confirmed by Holevo in 2002 [Holevo02a]. It was a very important discovery in the history of the classical capacity of the quantum channel, and due to the BSST-theorem, the analogue with classical Shannon's formula *has been finally completed*. Later, it was discovered that in special cases the entanglement-assisted capacity of a quantum channel can be improved [Harrow04], [Patrón09]. The Holevo information can be attained even with pure input states, and the concavity of the Holevo information also shown. The concavity can be used to compute the classical HSW capacity of quantum channels, since the maximum of the transmittable information can be computed by a local maximum among the input states. Moreover, as was shown by Bennett et al. in 2002, the concavity holds for the entanglement-assisted classical capacity, too [Bennett02], [Bennett09]— the concavity, along with the non-necessity of any computation of an asymptotic formula, and the use of classical feedback channels to improve the capacity, *makes* *the entanglement-assisted classical capacity the most generalized classical capacity*—and it has the same role as Shannon's formula in classical information theory [Bennett09]. The fact that the classical feedback channel does not increase the classical capacity and the entanglement-assisted classical capacity of the quantum channel, follows from the work of Bennett et al., and the proof of the BSST-theorem [Bennett02]. Wang and Renner's work [Wang10] introduces the reader to the connection between the single-use classical capacity and hypothesis testing.

**The Private Classical Capacity**

The third classical capacity of the quantum channel is the *private classical capacity*, denoted by $P\left(\mathcal{N}\right)$. The concept of private classical capacity was introduced by Devetak in 2003 [Devetak03], and one year later by Cai et al. in 2004 [Cai04]. Private classical capacity measures classical information, and it is always at least as large as the single-use quantum capacity (or the quantum coherent information) of any quantum channel. As shown in [Devetak05a], for a degradable quantum channel (see Section 6.3.1) the coherent information (see Section 4) is additive [Devetak05a],—however for a general quantum channel these statements do not hold. The additivity of private information would also imply the fact that shared entanglement cannot help to enhance the private classical capacity for degradable quantum channels (see Section 6.3.1). The complete proof of the private classical capacity of the quantum channel was made by Devetak [Devetak03], who also cleared up the connection between private classical capacity and the quantum capacity. As was shown by Smith et al. [Smith08d], the private classical capacity of a quantum channel is additive for degradable quantum channels, and closely related to the quantum capacity of a quantum channel (moreover, Smith has shown that the private classical capacity is equal to the quantum coherent information for degradable channels), since in both cases we have to "protect" the quantum states: in the case of private classical capacity the enemy is called Eve (the eavesdropper), while in the latter case the name of the enemy is "environment." As was shown in [Devetak03], the eavesdropper in private coding acts as the environment in quantum coding of the quantum state, and vice-versa. This "gateway" or "dictionary" between the classical capacity and the quantum capacity of the quantum channel was also used by Devetak [Devetak03], by Devetak and Shor [Devetak05a] and by Smith and Smolin [Smith08d], using a different interpretation.

About the coherent communication with continuous quantum variables over the quantum channels a work was published Wilde et al. in [Wilde07] and [Wilde10]. On the noisy processing of private quantum states, see the work of Renes et al. [Renes07]. A further application of private classical information in communicating over adversarial quantum channels was shown by Leung et al. [Leung08]. Further information about the private classical capacity can be found in [Devetak03], [Devetak05], [Bradler09], [Li09], [Smith08d], [Smith09a], [Smith09b]. Another important work on non-additive quantum codes was shown by Smolin et al. [Smolin07]. A great summary on the main results of Quantum Shannon Theory was published by Wilde [Wilde11]. For further information on quantum channel capacities and advanced quantum communications see the book of Imre and Gyongyosi [Imre12] and [Gyongyosi12]. We also suggest the great work of Bennett et al. on the quantum reverse Shannon theorem [Bennett09]. A work on the connection of secure communication and Gaussian-state quantum Illumination was published by Shapiro [Shapiro09].

**The Zero-Error Classical Capacity**

The properties of *zero-error* communication systems are discussed in Shannon's famous paper on the zero-error capacity of a noisy channel [Shannon56], in the work of Körner and Orlitsky on zero-error information theory [Körner98], and in the work of Bollobás on modern graph theory [Bollobas98]. We also suggest the famous proof of Lovász on the Shannon capacity of a graph [Lovász79]. The proof of the classical zero-error capacity of quantum channel can be found in Medeiros's work [Medeiros05]. Here, he has shown, that the classical zero-error capacity of the quantum channel is also bounded above by the classical HSW capacity. The important definitions of quantum zero-error communication and the characterization of quantum states for the zero-error capacity were given by Medeiros et al., in [Medeiros06]. On the complexity of computation of zero-error capacity of quantum channels see the work of Beigi and Shor [Beigi07]. The fact, that the zero-error classical capacity of the quantum channel can be increased with entanglement, was shown by Cubitt et al. in 2010 [Cubitt10]. The role of entanglement in the asymptotic rate of zero-error classical communication over quantum channels was shown by Leung et al. in 2010 [Leung10]. For further information about the theoretical background of entanglement-assisted zero-error quantum communication see [Cubitt10] and for the properties of entanglement, the proof of the Bell-Kochen-Specker theorem in [Bell1966], [Kochen1967].

# 4. The Quantum Capacity of a Quantum Channel

Having discussed the general model of quantum channels and introduced various classical capacities in this section we focus on the *quantum information* transfer over quantum channels. Two new quantities will be explained. By means of *fidelity F* one can describe the differences between two quantum states e.g. between the input and output states of a quantum channel. On the other hand *quantum coherent information* represents the quantum information loss to the environment during quantum communication similarly as mutual information did for a classical channel *N*. Exploiting this latter quantity we can define the maximal quantum information transmission rate through quantum channels – the quantum capacity $Q(\mathcal{N})$ analogously to Shannon's noisy channel theorem. As we have seen in Section 3, the classical capacity of a quantum channel is described by the maximum of quantum mutual information and the Holevo information. The quantum capacity of the quantum channels is described by the maximum of *quantum coherent information.* The concept of quantum coherent information plays a fundamental role in the computation of the *LSD (Lloyd-Shor-Devetak)* channel capacity [Lloyd97], [Devetak03], [Shor02] which measures the asymptotic quantum capacity of the quantum capacity in general.

Section 4 is organized as follows. First, we discuss the transmission of quantum information over a nosy quantum channel. Next, we define the quantum coherent information and overview its main properties. Finally the formula for the measure of maximal transmittable quantum information over a quantum channel will be introduced. The description of the most relevant works can be found in the Related Work subsection.

## 4.1 Preserving Quantum Information

The encoding and decoding quantum information have many similarities to the classical case, however, there exist some fundamental differences, as we will reveal in this section. In the case of quantum communication, the source is a quantum information source and the *quantum information* is encoded into quantum states. When transmitting quantum information, the information is encoded into non-orthogonal superposed or entangled quantum states chosen from the ensemble $\{\rho_k\}$ according to a given probability $\{p_k\}$. If the states $\{\rho_k\}$ are pure and mutually orthogonal, we talk about classical information; that is, in this case the quantum information reduces to classical.

Formulating the process more precisely (see Fig. 4.1) the encoding and the decoding mathematically can be described by the operators $\mathcal{E}$ and $\mathcal{D}$ realized on the blocks of quantum states. The input of the encoder consists of $m$ pure quantum states, and the encoder maps the $m$ quantum states into the joint state of $n$ intermediate systems. Each of them is sent through an independent instance of the quantum channel $\mathcal{N}$ and decoded by the decoder $\mathcal{D}$, which results in $m$ quantum states again. The output of the decoder $\mathcal{D}$ is typically mixed, according to the noise of the quantum channel. The rate of the code is equal to $m/n$.

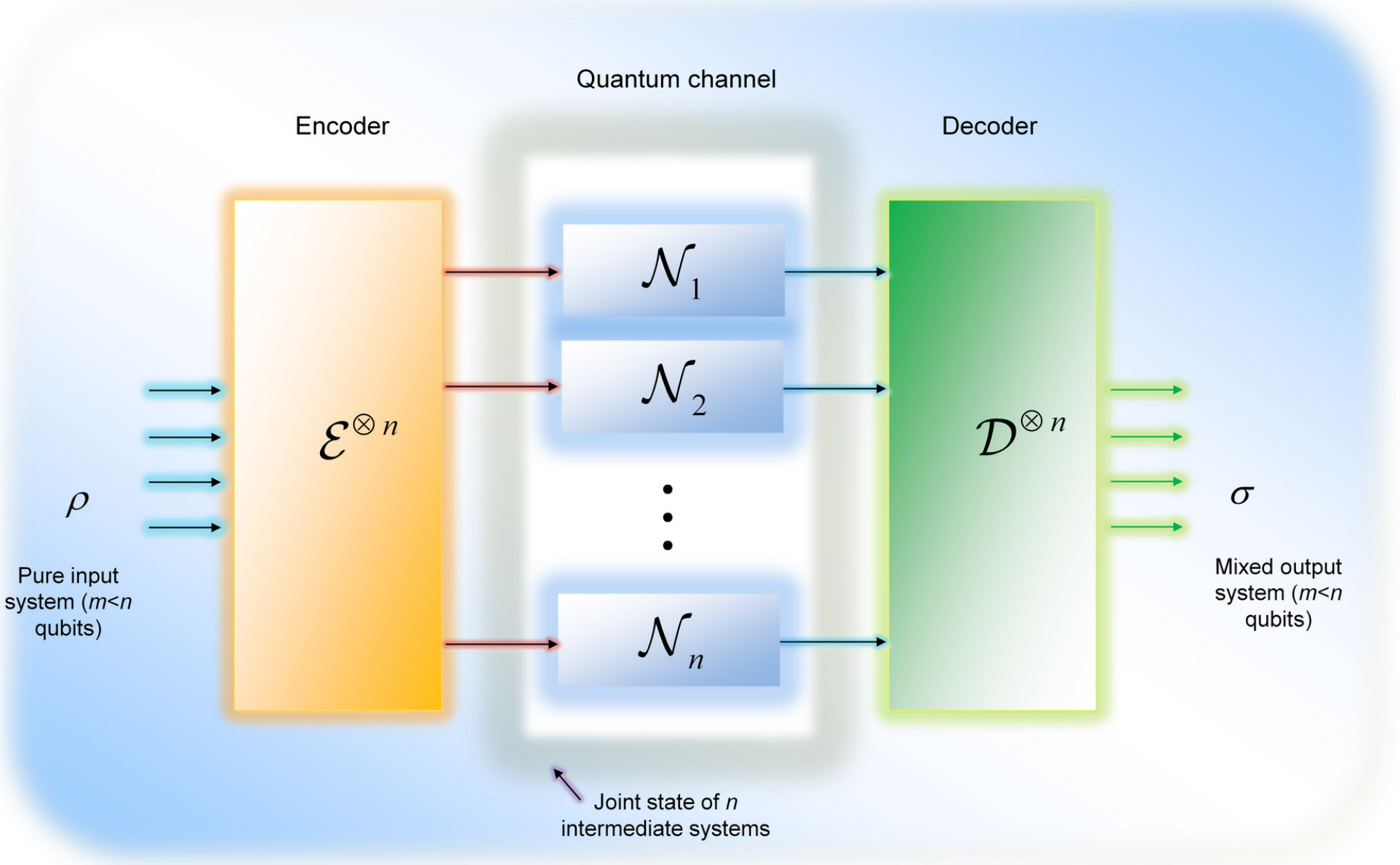

**Fig. 4.1.** Transmission of quantum information through the quantum channel. The encoder produces a joint state of $n$ intermediate systems. The encoded qubits are passed through the independent instances of the quantum channel.

Theoretically quantum states have to preserve their original superposition during the whole transmission, without the disturbance of their actual properties. Practically, quantum channels are entangled with the environment which results in mixed states at the output. Mixed states are classical probability weighted sum of pure states where these probabilities appear due to the interaction with the environment (i.e., noise). Therefore, we introduce a new quantity, which is able to describe the quality of the transmission of the superposed states through the quantum channel. The fidelity (see Appendix) for two pure quantum states is defined as

$$F\left(\left|\varphi\right\rangle,\left|\psi\right\rangle\right)=\left|\left\langle\varphi\middle|\psi\right\rangle\right|^2. \tag{4.1}$$

The fidelity of quantum states can describe the relation of Alice pure channel input state $\left|\psi\right\rangle$ and the received mixed quantum system $\sigma=\sum_{i=0}^{n-1}p_i\rho_i=\sum_{i=0}^{n-1}p_i\left|\psi_i\right\rangle\left\langle\psi_i\right|$ at the channel output as

$$F\left(\left|\psi\right\rangle,\sigma\right)=\left\langle\psi\middle|\sigma\middle|\psi\right\rangle=\sum_{i=0}^{n-1}p_i\left|\left\langle\psi\middle|\psi_i\right\rangle\right|^2. \tag{4.2}$$

Fidelity can also be defined for *mixed* states $\sigma$ and $\rho$

$$F\left(\rho,\sigma\right)=\left[Tr\left(\sqrt{\sqrt{\sigma}\rho\sqrt{\sigma}}\right)\right]^2=\sum_i p_i\left[Tr\left(\sqrt{\sqrt{\sigma_i}\rho_i\sqrt{\sigma_i}}\right)\right]^2. \tag{4.3}$$

Let us assume that we have a quantum system denoted by $A$ and a reference system $P$. Initially, the quantum system $A$ and the reference system $P$ are in a *pure entangled* state, denoted by $\left|\psi^{PA}\right\rangle$. The density matrix $\rho_A$ of system $A$ can be expressed by a partial trace over $P$, as follows

$$\rho_A=Tr_P\left(\left|\psi^{PA}\right\rangle\left\langle\psi^{PA}\right|\right). \tag{4.4}$$

The entanglement between the initial quantum system and the reference state is illustrated in Fig. 4.2.

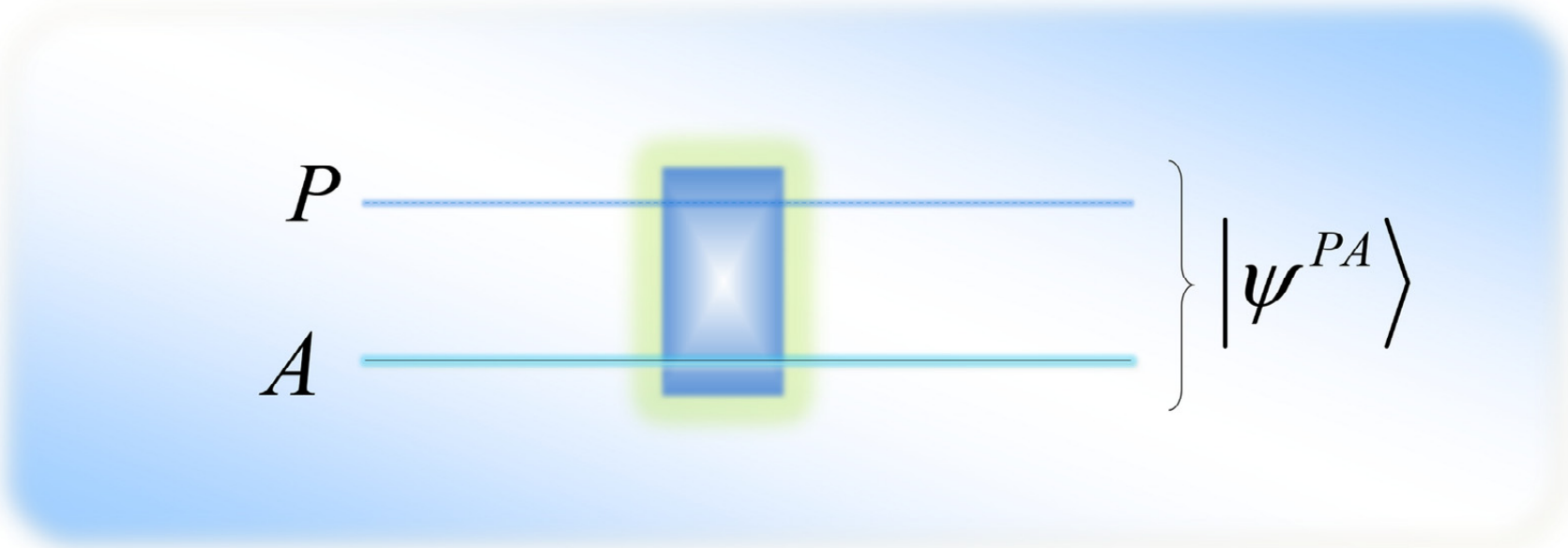

**Fig. 4.2.** Initially, the quantum system and the reference system are in a pure entangled state.

In the next step, $\rho_A$ will be transmitted through the quantum channel $\mathcal{N}$, while the reference state $P$ is *isolated from the environment* (see Section 2.), hence it has not been not modified during the transmission. After the quantum system $\rho_A$ is transmitted through the quantum channel, the final state will be

$$\rho^{PB} = \left(\mathcal{I}^P \otimes \mathcal{N}^A\right)\left(\left|\psi^{PA}\right\rangle\left\langle\psi^{PA}\right|\right), \tag{4.5}$$

where $\mathcal{I}^P$ is the identity transformation realized on the reference system $P$. After the system $A$ is sent through the quantum channel, both the quantum system $A$ and the entanglement between $A$ and $P$ are affected, as we illustrated in Fig. 4.3. The resultant output system is denoted by $B$.

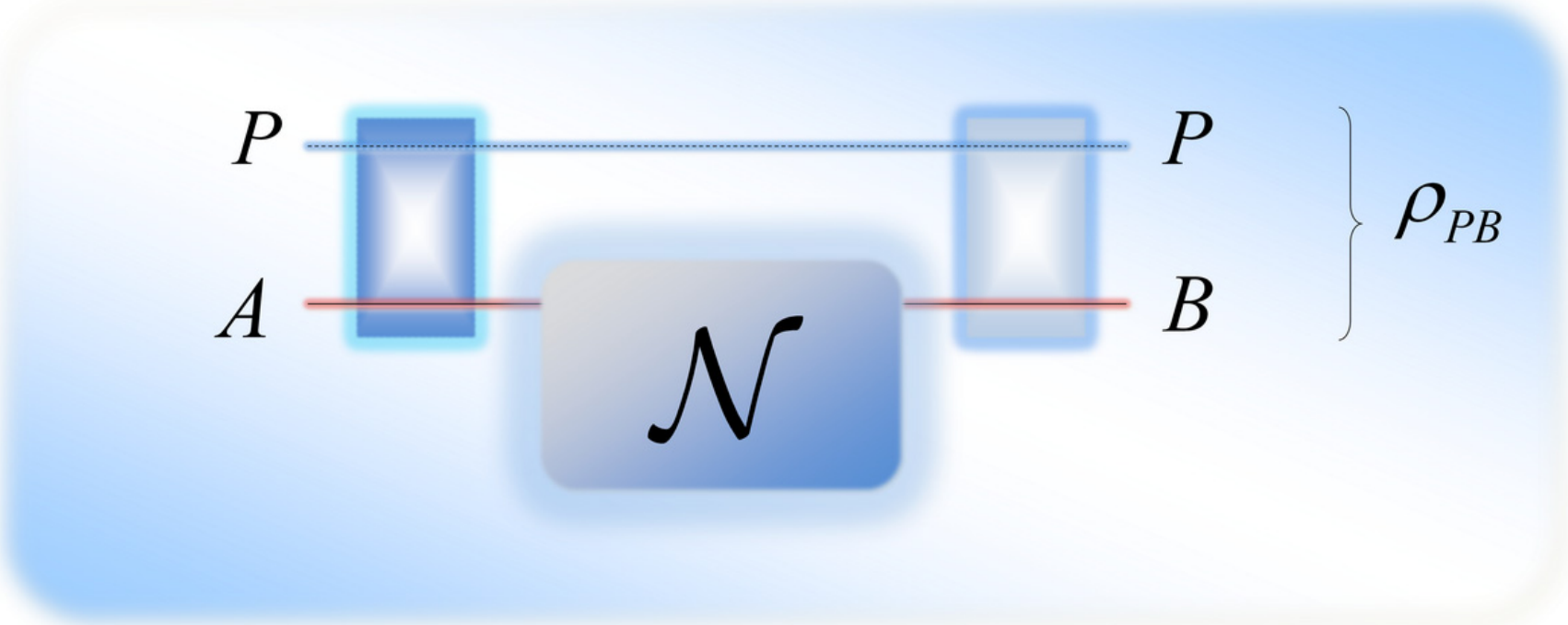

**Fig. 4.3.** After the system $A$ is sent through the quantum channel $\mathcal{N}$, both the quantum system $A$ and the entanglement between $A$ and $P$ are affected.

Now, we can study the preserved entanglement between the two systems $A$ and $P$. Entanglement fidelity $F_E$ measures the fidelity between the initial pure system $\left|\psi^{PA}\right\rangle$ and the mixed output quantum system $\rho_{PB}$ as follows

$$F_E = F_E\left(\rho_A, \mathcal{N}\right) = F\left(\left|\psi^{PA}\right\rangle, \rho_{PB}\right) = \left\langle\psi^{PA}\right|\left(\mathcal{I}^P \otimes \mathcal{N}^A\right)\left(\left|\psi^{PA}\right\rangle\left\langle\psi^{PA}\right|\right)\left|\psi^{PA}\right\rangle. \tag{4.6}$$

It is important to highlight the fact that $F_E$ depends on $\left|\psi^{PA}\right\rangle$ i.e., on the reference system. The whole process is shown in Fig. 4.4. The sender's goal is to transmit quantum information, i.e., to preserve entanglement between $A$ and the inaccessible reference system $P$. Alice can apply many independent channel uses of the same noisy quantum channel $\mathcal{N}$ to transmit the quantum information.

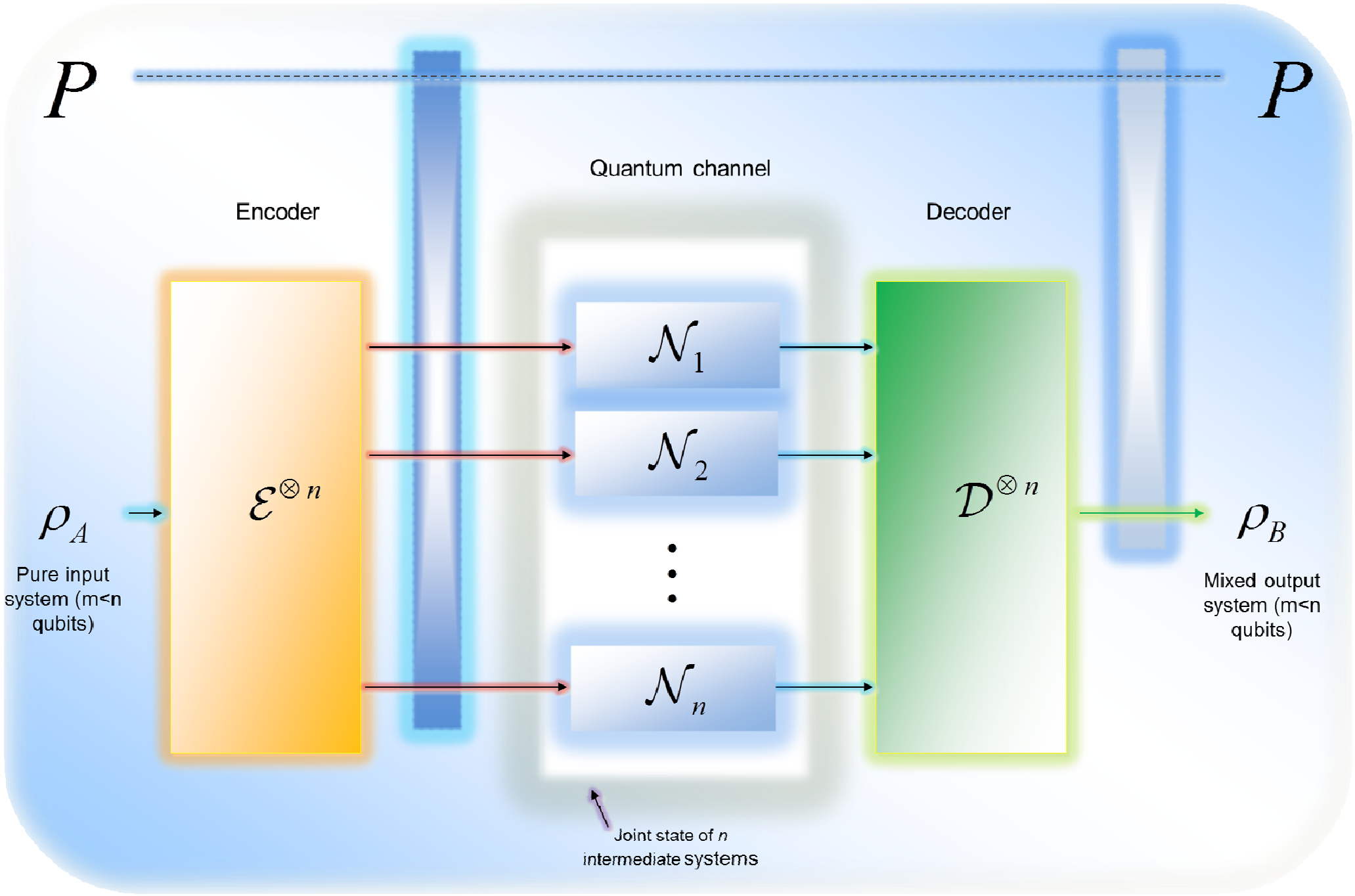

**Fig. 4.4.** Transmission of quantum information with multiple uses of the quantum channel.

Similar to encoding classical information into the quantum states, the quantum messages can be transmitted over copies of a quantum channel. In this case, we have $n$ copies of a quantum channel $\mathcal{N}$.

## 4.2 Quantum Coherent Information

In case of the classical capacity $C(\mathcal{N})$, the correlation between the input and the output is measured by the Holevo information and the quantum mutual information function. In case of the quantum capacity $Q(\mathcal{N})$, we have a completely different correlation measure with completely different behaviors: it is called the *quantum coherent information.* There is a *very important distinction* between the maximized quantum mutual information and maximized quantum coherent information: *the maximized quantum mutual information of a quantum channel* $\mathcal{N}$ *is always additive* (see Section 2.3.)*, but the quantum coherent information is not.*

The $\mathrm{S}_E$ *entropy exchange* between the initial system $PA$ and the output system $PB$ is defined as follows. The entropy that is acquired by $PA$ when input system $A$ is transmitted through the quantum channel $\mathcal{N}$ can be expressed with the help of the von Neumann entropy function as follows

$$\mathrm{S}_E = \mathrm{S}_E\left(\rho_A : \mathcal{N}\left(\rho_A\right)\right) = \mathrm{S}\left(\rho_{PB}\right), \tag{4.7}$$

or in other words the von Neumann entropy of the output system $\rho_{PB}$. As can be observed, the value of entropy exchange depends on $\rho_A$ and $\mathcal{N}$ and is independent from the purification system $P$. Now, we introduce the environment state $E$, and we will describe the map of the quantum channel as a unitary transformation. The environment is initially in a pure state $\left|0\right\rangle$. After the unitary transformation $U_{A \to BE}$ has been applied to the initial system $A\left|0\right\rangle$, it becomes

$$U_{A \to BE}\left(A\left|0\right\rangle\right) = BE\,. \tag{4.8}$$

The map of the quantum channel as a unitary transformation on the input system and the environment is shown in Fig. 4.5.

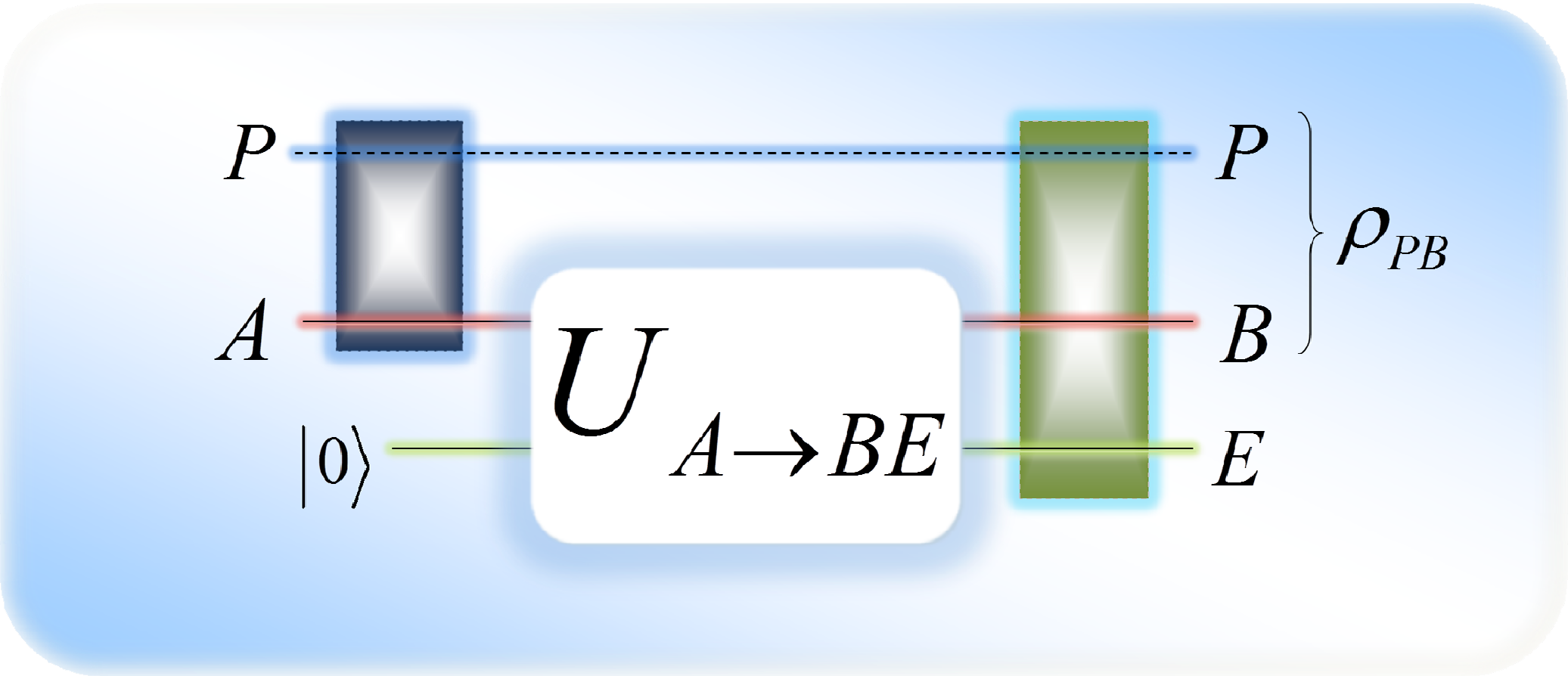

**Fig. 4.5.** The map of the quantum channel as a unitary transformation on the input system and the environment. The unitary transformation entangles $AP$ with the environment $E$, which is initially in a pure state.

Now, from the entropy of the *final state* of the environment $\rho_E$, the *entropy exchange* $S_E$ can be expressed as

$$S(\rho_{PB}) = S(\rho_E) = S_E. \tag{4.9}$$

$S_E$ measures the increase of entropy of the environment $E$, or with other words, the entanglement between $PA$ and $E$, after the unitary transformation $U_{A\to BE}$ had been applied to the system. This entropy exchange $S_E$ is analogous to the classical conditional entropy; however in this case we talk about quantum instead of classical information.

Using the notations of Fig. 4.3, the quantum coherent information can be expressed as

$$\begin{aligned} I_{coh}(\rho_A : \mathcal{N}(\rho_A)) &= S(\mathcal{N}(\rho_A)) - S_E(\rho_A : \mathcal{N}(\rho_A)) \\ &= S(\rho_B) - S(\rho_{PB}) \\ &= S(\rho_B) - S(\rho_E), \end{aligned} \tag{4.10}$$

where $S_E(\rho_A : \mathcal{N}(\rho_A))$ is the entropy exchange as defined in (4.7).

Using the definition of quantum coherent information (4.10), it can be verified that quantum coherent information takes its maximum if systems $A$ and $P$ are *maximally entangled* and the quantum channel $\mathcal{N}$ is *completely noiseless.* This can be presented easily

$$\mathrm{S}\left(\rho_B\right) = \mathrm{S}\left(\rho_A\right), \tag{4.11}$$

since the input state $\rho_A$ is maximally mixed, and

$$\mathrm{S}\left(\rho_{PB}\right) = 0, \tag{4.12}$$

because $\left|\psi^{PA}\right\rangle\left\langle\psi^{PA}\right|$ will remain pure after the state has been transmitted through the ideal quantum channel. If the input system $\left|\psi^{PA}\right\rangle\left\langle\psi^{PA}\right|$ is not a maximally entangled state, or the quantum channel $\mathcal{N}$ is not ideal, then the value of quantum coherent information will decrease.

Considering another expressive picture, quantum coherent information measures the quantum capacity as the difference between the von Neumann entropies of two channel output states. The first state is received by Bob, while the second one is received by a “second receiver” - called the environment. If we express the transformation of a quantum channel as the partial trace of the overall system, then

$$\mathcal{N}\left(\rho_A\right) = Tr_E\left(U\rho_A U^{\dagger}\right), \tag{4.13}$$

and similarly, for the “effect” of the environment $E$, we will get

$$E\left(\rho_A\right) = \rho_E = Tr_B\left(U\rho_A U^{\dagger}\right). \tag{4.14}$$

The results of (4.13) and (4.14) are summarized in Fig. 4.6.

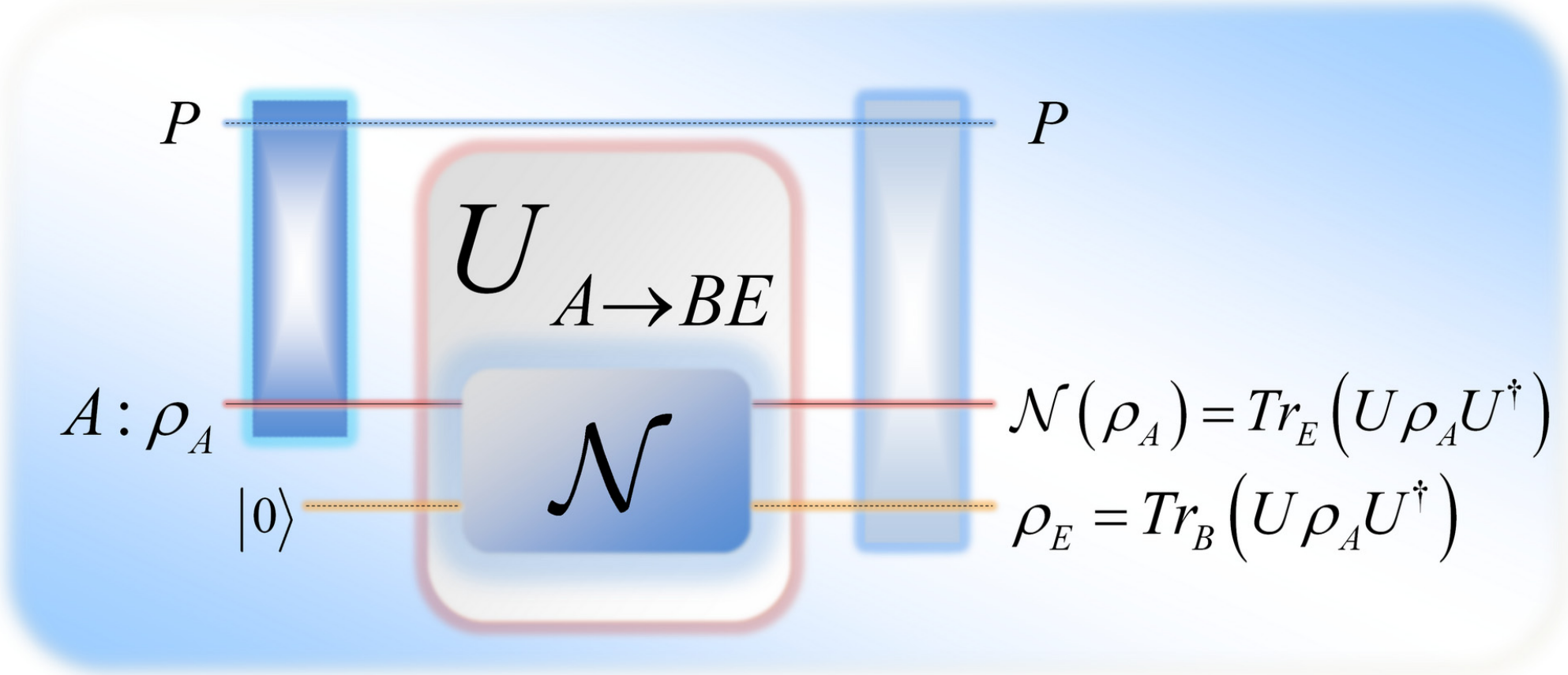

**Fig. 4.6.** The conceptional meaning of quantum coherent information. The unitary transformation represents the channel and the environment. The first receiver is Bob, the second is the environment. The state of the environment belonging to the unitary transformation is represented by dashed line. The outputs can be computed as the partial traces of the joint system.

It can be concluded that the quantum coherent information measures the capability of transmission of entanglement over a quantum channel. For the exact value of quantum coherent information of some important quantum channels see Section 5.

## 4.3 Connection between Classical and Quantum Information

As it has been shown by Schumacher and Westmoreland [Schumacher2000], the $I_{coh}$ quantum coherent information also can be expressed with the help of Holevo information, as follows

$$I_{coh}\left(\rho_A:\mathcal{N}\left(\rho_A\right)\right)=\left(\mathcal{X}_{AB}-\mathcal{X}_{AE}\right), \tag{4.15}$$

where

$$\mathcal{X}_{AB}=\mathrm{S}\left(\mathcal{N}_{AB}\left(\rho_{AB}\right)\right)-\sum_i p_i\mathrm{S}\left(\mathcal{N}_{AB}\left(\rho_i\right)\right) \tag{4.16}$$

and

$$\mathcal{X}_{AE} = \mathrm{S}\left(\mathcal{N}_{AE}\left(\rho_{AE}\right)\right) - \sum_i p_i \mathrm{S}\left(\mathcal{N}_{AE}\left(\rho_i\right)\right) \tag{4.17}$$

measure the Holevo quantities between Alice and Bob, and between Alice and environment $E$, where $\rho_{AB} = \sum_i p_i \rho_i$ and $\rho_{AE} = \sum_i p_i \rho_i$ are the average states. The definition of (4.15) also draws a very important connection: *the amount of transmittable quantum information can be derived by the Holevo information*, which measures classical information.

As follows, the *single-use* quantum capacity $Q^{(1)}\left(\mathcal{N}\right)$ can be expressed as

$$\begin{aligned} Q^{(1)}\left(\mathcal{N}\right) &= \max_{all\ p_i,\rho_i}\left(\mathcal{X}_{AB} - \mathcal{X}_{AE}\right) = \\ &= \max_{all\ p_i,\rho_i} \mathrm{S}\left(\mathcal{N}_{AB}\left(\sum_{i=1}^{n} p_i\left(\rho_i\right)\right)\right) - \sum_{i=1}^{n} p_i \mathrm{S}\left(\mathcal{N}_{AB}\left(\rho_i\right)\right) \\ &\quad - \mathrm{S}\left(\mathcal{N}_{AE}\left(\sum_{i=1}^{n} p_i\left(\rho_i\right)\right)\right) + \sum_{i=1}^{n} p_i \mathrm{S}\left(\mathcal{N}_{AE}\left(\rho_i\right)\right), \end{aligned} \tag{4.18}$$

where $\mathcal{N}\left(\rho_i\right)$ represents the $i$-th output density matrix obtained from the quantum channel input density matrix $\rho_i$.

The *asymptotic* quantum capacity $Q\left(\mathcal{N}\right)$ can be expressed by

$$\begin{aligned} Q\left(\mathcal{N}\right) &= \lim_{n\to\infty} \frac{1}{n} Q^{(1)}\left(\mathcal{N}^{\otimes n}\right) \\ &= \lim_{n\to\infty} \frac{1}{n} \max_{all\ p_i,\rho_i} I_{coh}\left(\rho_A : \mathcal{N}^{\otimes n}\left(\rho_A\right)\right) \\ &= \lim_{n\to\infty} \frac{1}{n} \max_{all\ p_i,\rho_i}\left(\mathcal{X}_{AB} - \mathcal{X}_{AE}\right). \end{aligned} \tag{4.19}$$

As shown in Fig. 4.7, the quantum coherent information can be computed as the difference between the Holevo information between Alice and Bob, and the Holevo information, which is passed through the environment.

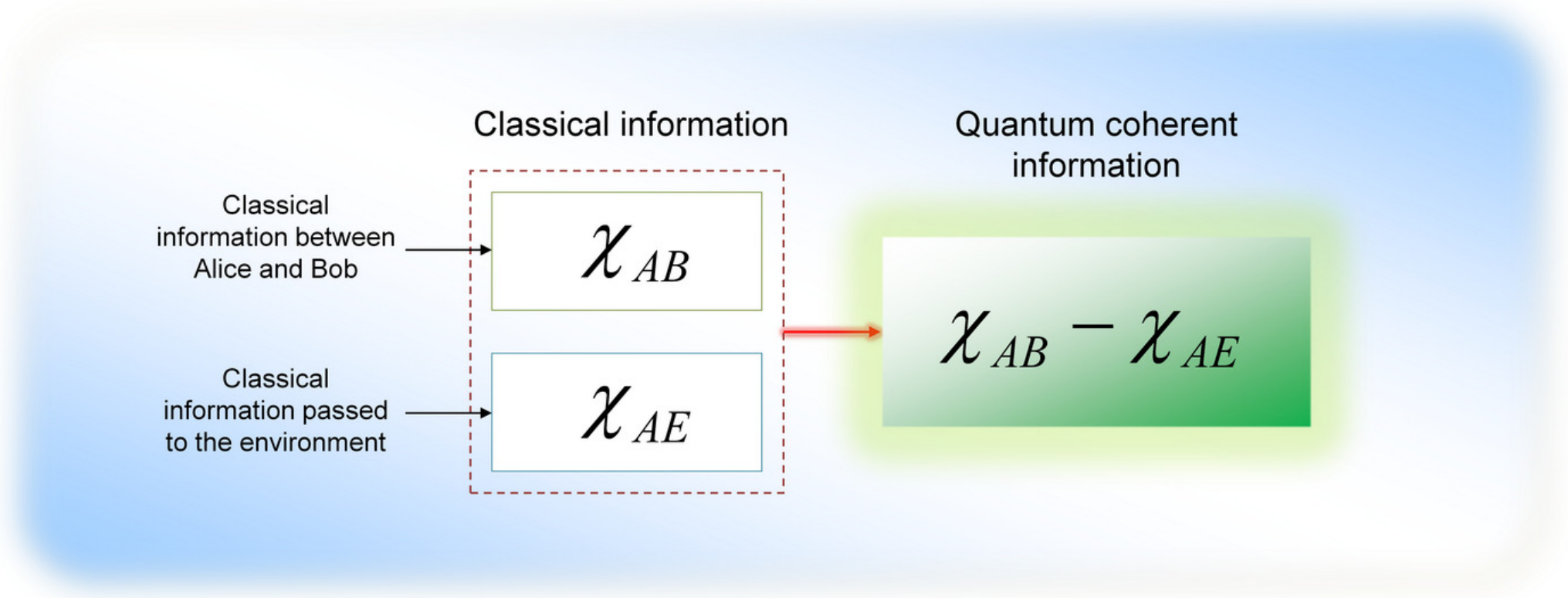

**Fig. 4.7.** Computation of the quantum coherent information from classical Holevo quantity. Using the results of Schumacher and Westmoreland, the first classical quantity measures the transmitted classical information form Alice to Bob, the second quantity measures the classical information which passed from Alice to the environment during the transmission.

As a summary, the quantum capacity $Q(\mathcal{N})$ of a quantum channel $\mathcal{N}$ can also be expressed by $\mathcal{X}_{AB}$, the *Holevo quantity* of Bob's output and by $\mathcal{X}_{AE}$, the information leaked to the environment during the transmission.

### 4.3.1 Quantum Coherent Information and Quantum Mutual Information

Finally let us make an interesting comparison between quantum coherent information and quantum mutual information. For classical information transmission, the *quantum mutual information* can be expressed according to Section 2

$$I(A:B) = \mathrm{S}(\rho_A) + \mathrm{S}(\rho_B) - \mathrm{S}(\rho_{AB}). \tag{4.20}$$

However, in case of *quantum coherent information* (4.10) the term $\mathrm{S}(\rho_A)$ vanishes. The channel transformation $\mathcal{N}$ modifies Alice's original state $\rho_A$, hence Alice's original density matrix cannot be used to express $\mathrm{S}(\rho_A)$, *after Alice's qubit has been sent through* the quantum channel $\mathcal{N}$. After the channel has modified Alice's quantum state, the initially sent qubit vanishes from the system, and we will have a different density matrix, denoted by $\rho_B = \mathcal{N}(\rho_A)$. The coherent information can expressed as $\mathrm{S}(\rho_B)-\mathrm{S}(\rho_{AB})$, where $\rho_B$ is the transformed state of Bob, and $\mathrm{S}(\rho_{AB})$ is the joint von Neumann entropy.

As follows, we will have $\mathrm{S}(\rho_B)-\mathrm{S}(\rho_{AB})$, which is equal to the *negative conditional entropy* $\mathrm{S}(A|B)$, (see Section 2) thus

$$I_{coh}(\rho_A : \mathcal{N}(\rho_A)) = \mathrm{S}(\rho_B) - \mathrm{S}(\rho_{AB}) = -\mathrm{S}(A|B). \tag{4.21}$$

This very interesting result is summarized in Fig. 4.8.

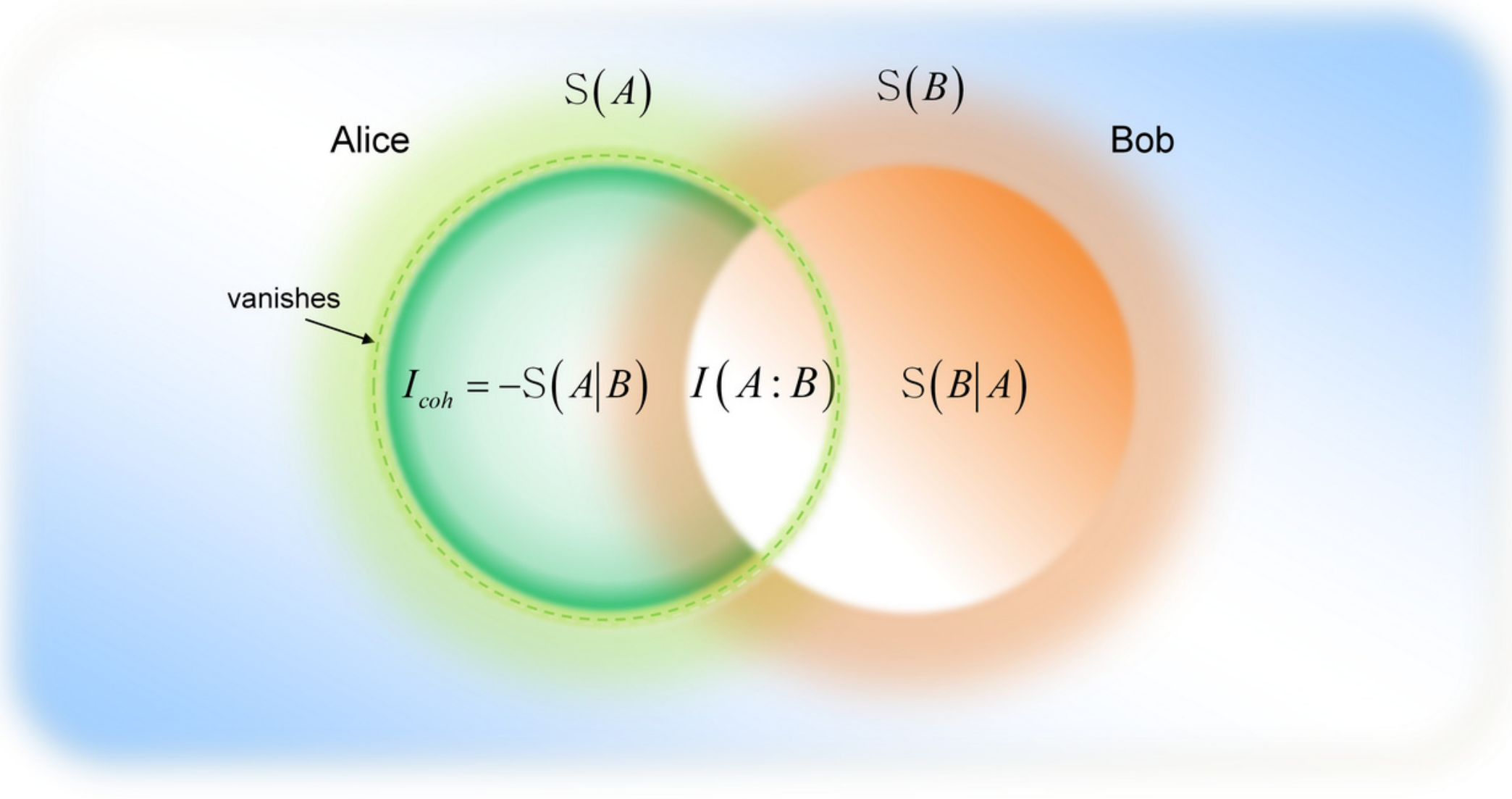

**Fig. 4.8.** The expression of quantum coherent information. The source entropy of Alice's state vanishes after the state is passed to Bob.

As we have seen in this section, there is a *very important difference* between the maximized quantum *mutual information* and the maximized *quantum coherent information* of a quantum channel. While the former is always additive, it does not remain true for the latter. *The quantum coherent information* is defined as follows

$$I_{coh}(\mathcal{N}) = \mathrm{S}(\rho_B) - \mathrm{S}(\rho_E), \tag{4.22}$$

where $\rho_B$ refers to the output of the quantum channel $\mathcal{N}$, while $\rho_E$ is the state of the environment. The term $\mathrm{S}(\rho_B)$ measures how much information Bob has, while $S(\rho_E)$ measures how much information environment has. As follows, the quantum coherent information $I_{coh}(\mathcal{N})$ measures that "*how much more information Bob has than the environment*" about the original input quantum state.

### 4.3.2 Quantum Coherent Information of an Ideal Channel

Now, we have arrived at the question of whether the $Q(\mathcal{N})$ quantum capacity of $\mathcal{N}$, as defined previously by the $I_{coh}$ quantum coherent information, is an appropriate measure to describe the whole quantum capacity of a quantum channel. The answer is yes for an ideal channel.

If we have a completely noiseless channel, then channel $\mathcal{N}_{AB} = I$ leads us to coherent information

$$Q(I) = I_{coh}(I) = S(\mathcal{N}_{AB}(\rho)) - S(\mathcal{N}_{Env.}(|0\rangle\langle 0|)) = S(\rho). \tag{4.23}$$

This simpler equation can be used to calculate the $Q(\mathcal{N}_{AB})$ quantum capacity of a quantum channel (i.e., without maximization) only when we have a completely noiseless idealistic channel $\mathcal{N}_{AB} = I$. It also implies the following: to achieve the maximal coherent information for an idealistic quantum channel $\mathcal{N}_{AB} = I$, the input quantum states have to be maximally mixed states or one half of an EPR state, since in these cases, the von Neumann entropies will be maximal (see Fig. 4.9). As we illustrated in Fig. 4.10, if the input of this channel is a pure quantum system, then the quantum coherent information completely vanishes.

On the other hand, if the environment of the communication system interacts with the quantum state, the quantum capacity could vanish, but not the classical capacity of the channel. In this case, the quantum channel $\mathcal{N}_{AB} = I$ can transmit pure orthogonal states faithfully, but it cannot transmit the superposed or entangled states. Furthermore, if the interaction is more significant, it could result in an extremely noisy quantum channel for which the $C(\mathcal{N}_{AB})$ classical capacity of $\mathcal{N}_{AB}$ could also vanish.

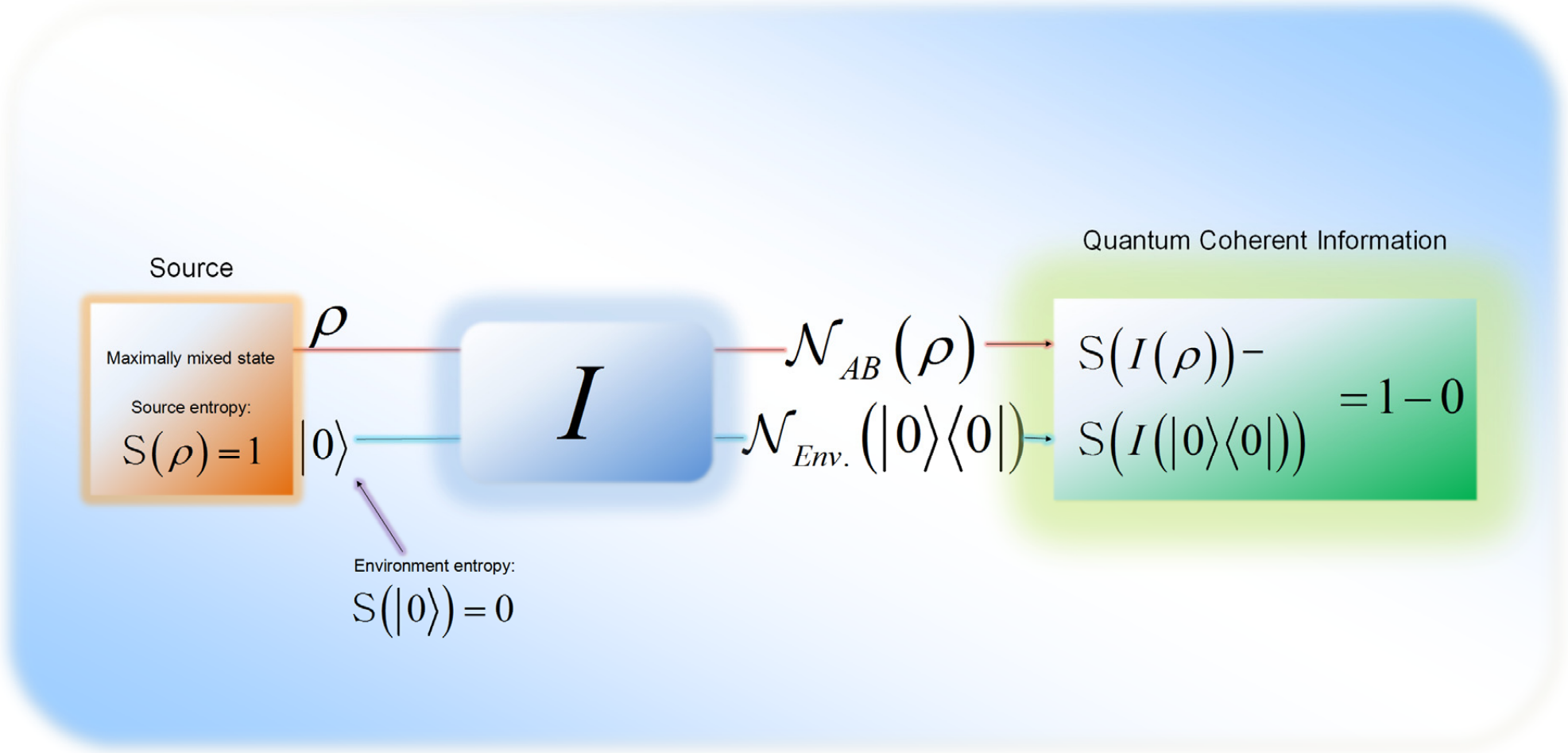

**Fig. 4.9.** The maximization of quantum coherent information with a noiseless quantum channel. The environment is initially in a pure state, so the maximization process requires a maximally mixed quantum state or one half of an EPR state.

In Fig. 4.10, we illustrated the case in which we have a pure input state and an ideal quantum channel. In that case, interestingly, $\mathcal{N}_{AB} = I$ cannot transmit quantum information.

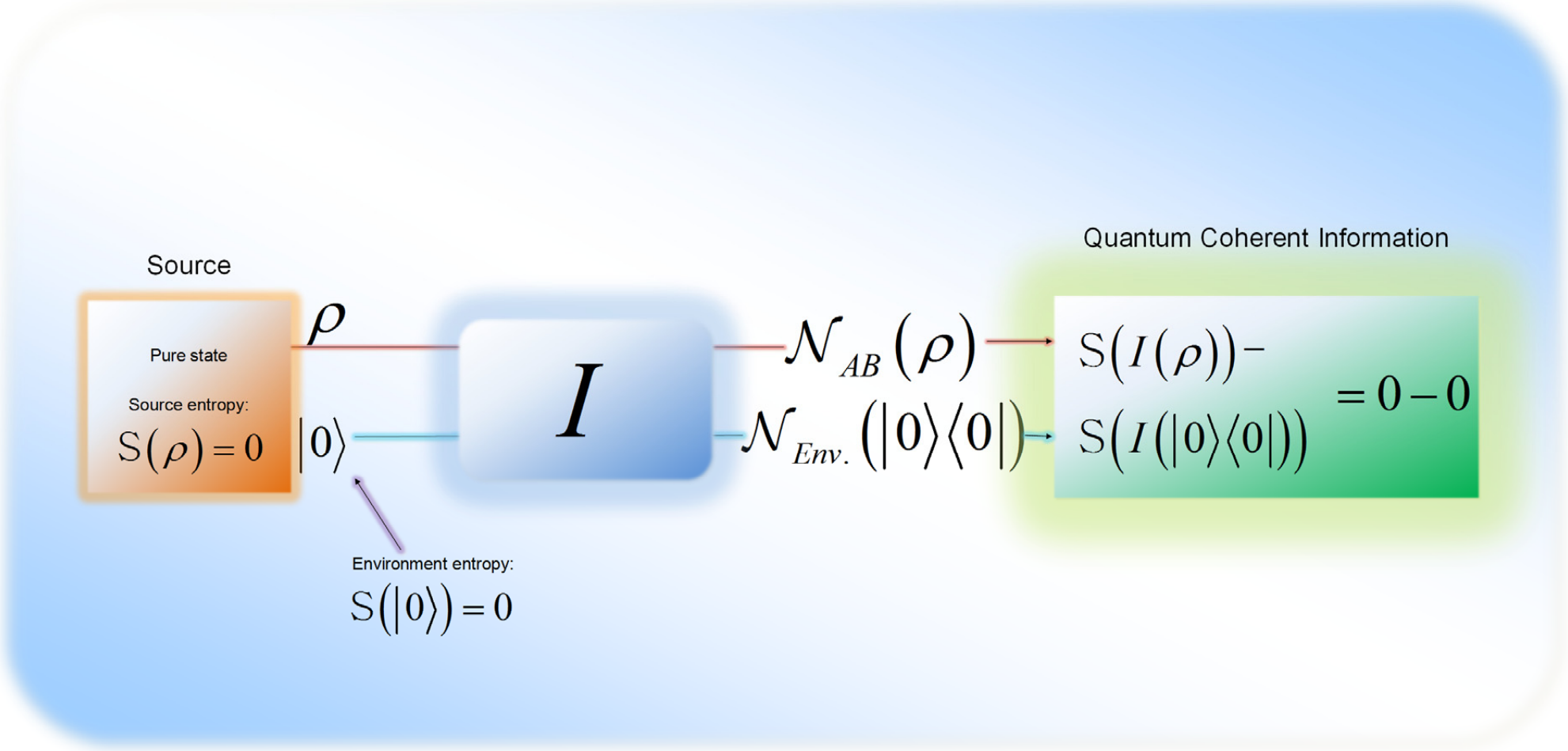

**Fig. 4.10.** For a pure input quantum state the quantum coherent information of the channel completely vanishes.

In the next section, we reveal the connection between quantum coherent information and the quantum capacity of the quantum channel.

## 4.4 The Lloyd-Shor-Devetak Formula

The concept of quantum coherent information can be used to express the *asymptotic* quantum capacity $Q(\mathcal{N})$ of quantum channel $\mathcal{N}$ called the *Lloyd-Shor-Devetak (LSD)* capacity as follows

$$\begin{aligned} Q(\mathcal{N}) &= \lim_{n\to\infty} \frac{1}{n} Q^{(1)}\left(\mathcal{N}^{\otimes n}\right) \\ &= \lim_{n\to\infty} \frac{1}{n} \max_{all\ p_i,\rho_i} I_{coh}\left(\rho_A : \mathcal{N}^{\otimes n}\left(\rho_A\right)\right) \\ &= \lim_{n\to\infty} \frac{1}{n} \max_{all\ p_i,\rho_i} \left(\mathrm{S}(\rho_B) - \mathrm{S}(\rho_E)\right), \end{aligned} \tag{4.24}$$

where $Q^{(1)}(\mathcal{N})$ represents the *single-use* quantum capacity.

The asymptotic quantum capacity can also be expressed using the Holevo information, since as we have seen previously, the quantum coherent information can be derived from the Holevo information

$$Q(\mathcal{N}) = \lim_{n\to\infty} \frac{1}{n} \max_{all\ p_i,\rho_i} \left(\mathcal{X}_{AB} - \mathcal{X}_{AE}\right), \tag{4.25}$$

where $\mathcal{X}_{AB}$ denotes the classical information sent from Alice to Bob, and $\mathcal{X}_{AE}$ describes the classical information passed from Alice to the environment during the transmission.

Quantum coherent information plays a fundamental role in describing the maximal amount of transmittable quantum information through a quantum channel $\mathcal{N}$, and - as the Holevo quantity has deep relevance in the classical HSW capacity of a quantum channel - the quantum coherent information will play a crucial role in the LSD capacity of $\mathcal{N}$.

## 4.5 The Assisted Quantum Capacity

There is another important quantum capacity called *assisted capacity* which measures the quantum capacity for a channel pair that contains different channel models – and it will have relevance in the *superactivation* of quantum channels (see Section 7). If we have a quantum channel $\mathcal{N}$, then we can find a symmetric channel $\mathcal{A}$, that results in the following assisted quantum capacity

$$Q_{\mathcal{A}}\left(\mathcal{N}\right) = Q\left(\mathcal{N} \otimes \mathcal{A}\right). \tag{4.26}$$

We note, that the symmetric channel has unbounded dimension in the strongest case, and this quantity cannot be evaluated in general. $Q_{\mathcal{A}}\left(\mathcal{N}\right)$ makes it possible to realize the superactivation of zero-capacity (in terms of LSD capacity) quantum channels. For example if we have a zero-capacity *Horodecki channel* and a zero-capacity symmetric channel, then their combination can result in positive joint capacity, as we will show in Section 7.

## 4.6 The Zero-Error Quantum Capacity

Finally, let us shortly summarize the quantum counterpart of classical zero-error capacity. In the case of quantum zero-error capacities $Q_0^{(1)}\left(\mathcal{N}\right)$ and $Q_0\left(\mathcal{N}\right)$, the encoding and decoding process differs from the classical zero-error capacity: the encoding and decoding are carried out by the *coherent* encoder and *coherent* POVM decoder, whose special techniques make it possible to preserve the quantum information during the transmission [Harrow04], [Hsieh08].

The *single-use* and *asymptotic* quantum zero-error capacity is defined in a similar way

$$Q_0^{(1)}\left(\mathcal{N}\right) = \log\left(K\left(\mathcal{N}\right)\right), \tag{4.27}$$

and

$$Q_0\left(\mathcal{N}\right) = \lim_{n\to\infty} \frac{1}{n} \log\left(K\left(\mathcal{N}^{\otimes n}\right)\right), \tag{4.28}$$

where $K\left(\mathcal{N}^{\otimes n}\right)$ is the maximum number of $n$-length mutually non-adjacent quantum messages that the quantum channel can transmit with zero error. The quantum zero-error capacity is upper bounded by LSD channel capacity $Q\left(\mathcal{N}\right)$; that is, the following relation holds between the quantum zero-error capacities:

$$Q_0\left(\mathcal{N}\right) \leq Q\left(\mathcal{N}\right). \tag{4.29}$$

## 4.7 Relation between Classical and Quantum Capacities of Quantum Channels

Before introducing some typical quantum channel maps let us summarize the main properties of various capacities in conjunction with a quantum channels. First of all, the quantum capacity of $\mathcal{N}$ cannot exceed the maximal classical capacity that can be measured with entangled inputs and joint measurement; at least, it is not possible in general. On the other hand, for some quantum channels, it is conjectured that the maximal *single-use* classical capacity - hence the capacity that can be reached with *product* inputs and a *single* measurement setting - is lower than the *quantum capacity* for the same quantum channel.
For all quantum channels

$$C(\mathcal{N}) \geq Q(\mathcal{N}), \tag{4.30}$$

where $C(\mathcal{N})$ is the classical capacity of the quantum channel that can be achieved with entangled input states and a joint measurement setting.
On the other hand, it is conjectured that for some quantum channels,

$$C(\mathcal{N}) < Q(\mathcal{N}) \tag{4.31}$$

holds as long as the classical capacity $C(\mathcal{N})$ of the quantum channel is measured by a classical encoder and a single measurement setting. (As we have seen in Section 3, the classical capacities of a quantum channel can be measured in different settings, and the strongest version can be achieved with the combination of entangled inputs and joint measurement decoding.)
The relation between the various classical capacities of a quantum channel and this relation's quantum capacity are shown in Fig. 4.11.

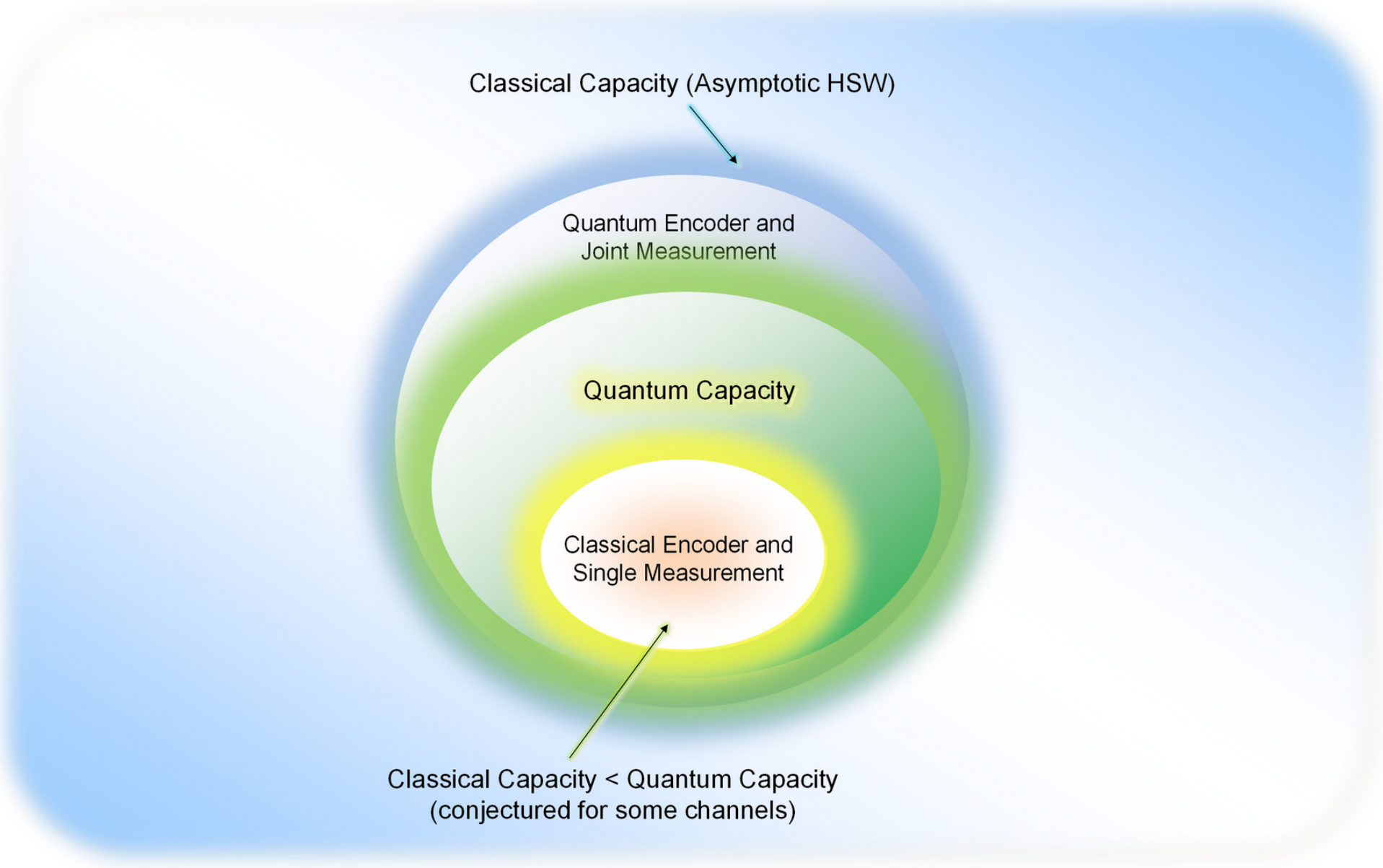

**Fig. 4.11.** The relation between classical and the quantum capacity of the same quantum channel.

We summarized the fundamental differences between classical and quantum capacities in Table 4.1.

| *Capacity* | *Type of information* | *Correlation measure between input and output* | *Measure of the Asymptotic channel capacity* |
|---|---|---|---|
| Classical | Classical information | Holevo information (Maximum of Quantum Mutual Information) | Holevo-Schumacher-Westmoreland formula |
| Private Classical | Private information (Classical) | Private information (Difference of Quantum Mutual Information functions) | Li-Winter-Zou-Guo, Smith-Smolin formula |
| Entanglement Assisted Classical | Classical information | Quantum mutual information | Bennett-Shor-Smolin-Thapliyal formula (Equal to single-use quantum mutual information.) |
| Quantum | Quantum information | Quantum Coherent Information | Lloyd-Shor-Devetak formula |

**Table 4.1.** The measure of classical and quantum capacities of the quantum channels are different, both in the case of single-use and in the asymptotic formulas.

It can be concluded from the table that in case of a quantum communication channel we have to count with so many capacities. Each of these capacities is based on different correlation measures: the *classical correlation* between the input and the output is measured by the quantum mutual information and the Holevo information. The private classical capacity is measured by the private information, which is the *maximization of the difference of two quantum mutual information functions.* For entanglement assisted capacity the correlation between input and output is also measured by the *maximized quantum mutual information*, however in this case we do not have to compute the asymptotic version to get the true capacity. Finally, the *quantum correlation* between the input and output is measured by the *quantum coherent information.*

# 4.8 Related Work

In this section we summarize the most important works regarding on the quantum capacity of the quantum channels.

The quantum capacity is one of the most important result of quantum information theory. The classical capacity of quantum channels was discovered in early years, in the beginning of the 1970s, and the researchers from this era — such as Holevo and Levitin—suggested that physical particles can encode only classical information [Levitin69], [Holevo73], [Holevo73a]. The first step in the encoding of quantum information into a physical particle was made by Feynman, in his famous work from 1982 [Feynman82]. However, the researchers did not see clearly and did not understand completely the importance of quantum capacity until the late 1990s. As we have shown in Section 3, a quantum channel can be used to transmit classical information and the amount of maximal transmittable information depends on the properties of the encoder and decoder setting, or whether the input quantum states are mixed or pure. Up to this point, we have mentioned just the transmission of classical information through the quantum channel—here we had broken this picture. The HSW theorem was a very useful tool to describe the amount of maximal transmittable classical information over a noisy quantum channel, however we cannot use it to describe the amount of maximal transmittable *quantum information.*

### Quantum Coherent Information

The computation of quantum capacity is based on the concept of *quantum coherent information*, which measures the ability of a quantum channel to preserve a quantum state. The definition of quantum coherent information (in an exact form) was originally introduced by Schumacher and Nielsen in 1996 [Schumacher96c]. This paper is a very important milestone in the history of the quantum capacity, since here the authors were firstly shown that the concept of quantum coherent information can be used to measure the quantum information (hence not the classical information) which can be transmitted through a quantum channel. The first,—but yet not complete—definitions of the quantum capacity of the quantum channel can be found in Shor's work from 1995 [Shor95], in which Shor has introduced a scheme for reducing decoherence in quantum computer memory, and in Schumacher's articles from one year later [Schumacher96b], [Schumacher96c]. Shor's paper from 1995 mainly discusses the problem of implementation of quantum error correcting schemes - the main focus was not on the exact definition of quantum capacity. Later, Shor published an extended version with a completed proof in 2002 [Shor02]. To transmit quantum information the parties have to encode and decode coherently. An interesting engineering problem is how the receiver could decode quantum states in superposition without the destruction of the original superposition [Wilde07].

The quantum capacity of a quantum channel finally was formulated completely by the *LSD-theorem*, named after Lloyd, Shor and Devetak [Lloyd97], [Shor02], [Devetak03], and they have shown that the rate of quantum communication can be expressed by the quantum coherent information. The LSD-channel capacity states that the asymptotic quantum capacity of the quantum channel is greater than (or equal to in some special cases) the single-use capacity; hence it is not equal to the quantum coherent information.

More information about the properties of fidelity and about the connection with other distance measures can be found in Fuch's works [Fuchs96], [Fuchs98]. An important article regarding the fidelity of mixed quantum states was published by Jozsa in 1994 [Jozsa94]. Fidelity also can be measured between entangled quantum states—a method to compute the fidelity of entanglement was published by Schumacher in 1996 [Schumacher96b]. Here, the upper bound of the quantum capacity was also mentioned, in the terms of quantum coherent information. Nielsen in 2002 [Nielsen02] defined a connection between the different fidelity measures.

**Proofs on Quantum Capacity**

The exact measure of quantum capacity was an open question for a long time. The fact that the quantum capacity cannot be increased by classical communication was formally proven by Bennett et al. [Bennett96a], who discussed the mixed state entanglement and quantum error correction. Barnum, in 2000 [Barnum2000], defined the connection between the fidelity and the capacity of a quantum channel, and here he also showed the same result as Bennett et al. did in 1996, namely that the quantum capacity cannot increased by classical communication [Bennett96a]. The works of Barnum et al. [Barnum2000] and Schumacher et al. [Schumacher98a] from the late 1990s gave very important results to the field of quantum information theory, since these works helped to clarify exactly the maximum amount of transmittable quantum information over very noisy quantum channels [Wilde11].

Seth Lloyd gave the first proof in 1997 on the quantum capacity of a noisy quantum channel. The details of Lloyd's proof can be found in [Lloyd97], while Shor's results in detail can be found in [Shor02]. On the basis of Shor's results, a proof on the quantum capacity was given by Hayden et al. in 2008 [Hayden08b].

The next step in the history of the quantum capacity of the quantum channel was made by Devetak [Devetak03]. Devetak also gave a proof for the quantum capacity using the private classical capacity of the quantum channel, and he gave a clear connection between the quantum capacity and the private classical capacity of the quantum channel.

As in the case of the discoverers of the HSW-theorem, the discoverers gave different proofs. The quantum capacity of a quantum channel is generally lower than the classical one, since in this case the quantum states encode quantum information. The quantum capacity requires the transmission of arbitrary quantum states, hence not just "special" orthogonal states—which is just a subset of a more generalized case, in which the states can be arbitrary quantum states.

On the several different encoder, decoder and measurement settings for quantum capacity see the work of Devetak and Winter [Devetak05], Devetak and Shor's work [Devetak05a], and the paper of Hsieh et al. [Hsieh08].

In this paper we have not mentioned the definition of unit resource capacity region and private unit resource capacity region, which can be found in detail in the works of Hsieh and Wilde [Hsieh10], and Wilde and Hsieh [Wilde10]. In 2005, Devetak and Shor published a work which analyzes the simultaneous transmission of classical and quantum information [Devetak05a].

On the quantum capacities of bosonic channels a work was published by Wolf, Garcia and Giedke, see [Wolf06]. In 2007, Wolf and Pérez-García published a paper on the quantum capacities of channels with small environment, the details can be found in [Wolf07]. They have also determined the quantum capacity of an amplitude damping quantum channel (for the description of amplitude damping channel, see Section 3.6.3), for details see the same paper from 2007 [Wolf07]. The properties of quantum coherent information and reverse coherent information were studied by Patrón in 2009 [Patrón09]. The proofs of the LSD channel capacity can be found in [Lloyd97], [Shor02], [Devetak03]. The quantum communication protocols based on the transmission of quantum information were intensively studied by Devetak [Devetak04a], and the work of the same authors on the generalized framework for quantum Shannon theory, from 2008 [Devetak08].

# 5. Classical and Quantum Capacities of some Channels

In this short section, we discuss the classical and quantum capacities of some important quantum channels.

Section 5 is organized as follows. First we discuss the classical and quantum capacities of some important channels. In the second part we introduce the reader to the geometric interpretation of quantum channel capacity. Next, we give the zero-error capacities of important quantum channels. Finally we summarize the related works.

## 5.1. Capacities of Various Quantum Channels

In order to present some examples for the above discussed results we study in this subsection the classical and quantum capacities of the following quantum channels:

1. *erasure quantum channel,*
2. *mixed erasure/phase-erasure quantum channel,*
3. *amplitude damping channel (see Section 3.6.3),*
4. *classical ideal quantum channel.*

First we derive the classical capacities of these channels in closed forms. Then we give the quantum capacities and compare them. The *erasure* quantum channel $\mathcal{N}_p$ erases the input state $\rho$ with probability $p$ or transmits the state unchanged with probability $\left(1-p\right)$

$$\mathcal{N}_p\left(\rho\right) \to \left(1-p\right)\rho + \left(p\left|e\right\rangle\left\langle e\right|\right), \tag{5.1}$$

where $\left|e\right\rangle$ is the erasure state. The classical capacity of the erasure quantum channel $\mathcal{N}_p$ can be expressed as

$$C\left(\mathcal{N}_p\right) = \left(1-p\right)\log\left(d\right), \tag{5.2}$$

where $d$ is the dimension of the input system $\rho$. As follows from (5.2), the classical capacity of $\mathcal{N}_p$ vanishes at $p = 1$, while if $0 \leq p < 1$ then the channel $\mathcal{N}_p$ can transmit some classical information.

The quantum capacity of the erasure quantum channel $\mathcal{N}_p$ is

$$Q\left(\mathcal{N}_p\right) = \left(1 - 2p\right)\log\left(d\right). \tag{5.3}$$

$Q\left(\mathcal{N}_p\right)$ vanishes at $p = 1/2$, but it can transmit some quantum information if $0 \leq p < 1/2$.

In Fig. 5.1 the classical (dashed line) and quantum capacity (solid line) of the erasure quantum channel as a function of erasure probability are shown.

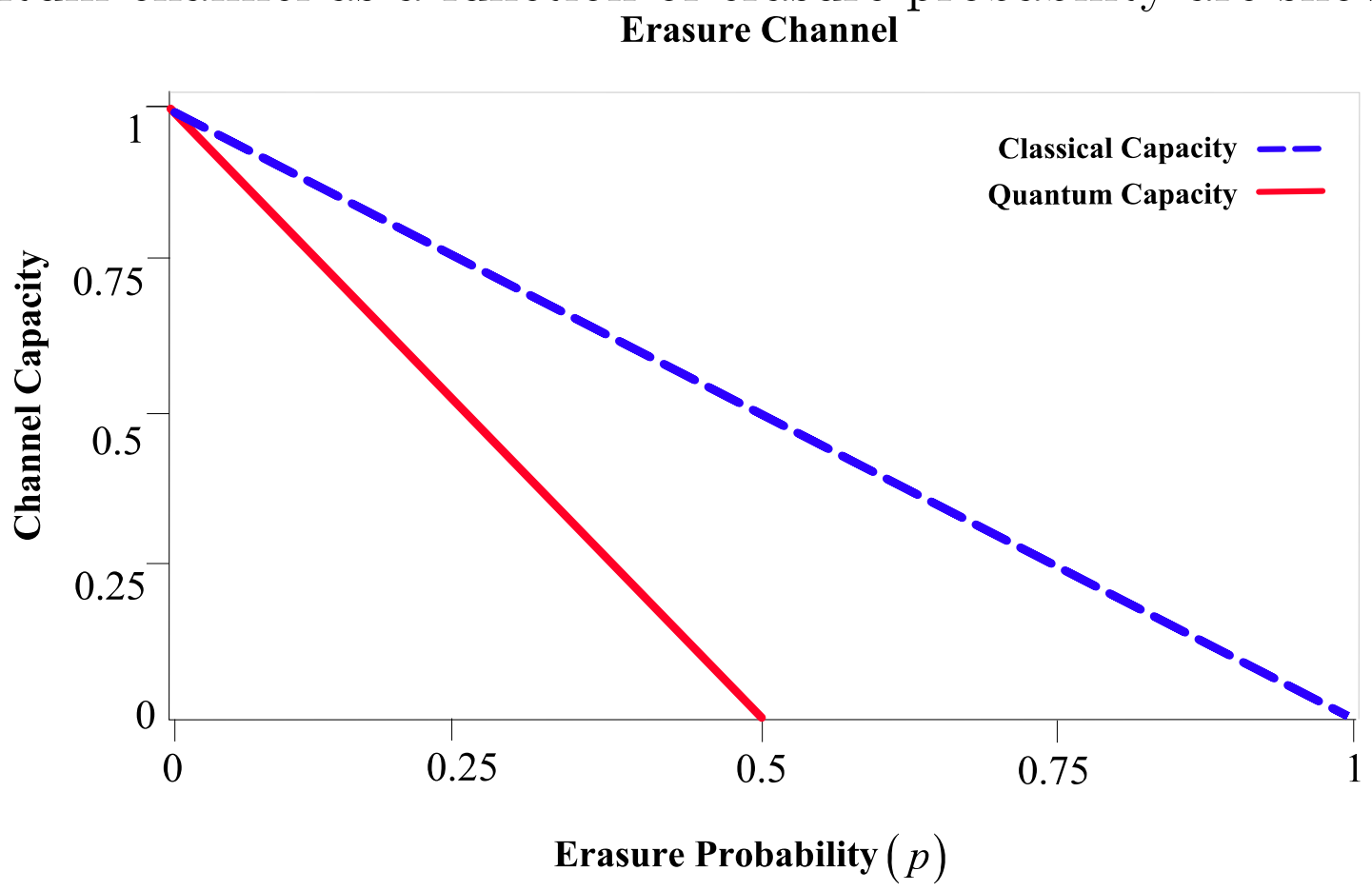

**Fig. 5.1.** The classical and quantum capacities of the erasure quantum channel as a function of erasure probability.

The *phase-erasure* quantum channel $\mathcal{N}_\delta$ erases the phase of the input quantum state with probability $p$ without causing any disturbance in the amplitude. Using input density matrix $\rho$, the map of the phase-erasure quantum channel can be expressed as

$$\mathcal{N}\left(\rho\right) \rightarrow \left(1 - p\right)\rho \otimes \left|0\right\rangle\left\langle 0\right| + p\frac{\rho + Z\rho Z^{\dagger}}{2} \otimes \left|1\right\rangle\left\langle 1\right|, \tag{5.4}$$

where $Z$ realizes the phase transformation on the input quantum system $\rho$, while the second qubit is used as a flag qubit.

The classical capacity of the $\mathcal{N}_\delta$ phase-erasure quantum channel using phase erasing probability $q$ is

$$C\left(\mathcal{N}_\delta\right) = 1, \tag{5.5}$$

since the phase error has no effect on the distinguishability of orthogonal input quantum states $\left|0\right\rangle$ and $\left|1\right\rangle$. On the other hand, if we talk about quantum capacity $Q\left(\mathcal{N}_\delta\right)$ of $\mathcal{N}_\delta$ the picture changes:

$$Q\left(\mathcal{N}_\delta\right) = \left(1-q\right)\log\left(d\right). \tag{5.6}$$

From the erasure quantum channel and the phase-erasure quantum channel a third type of quantum channel can be constructed – the *mixed erasure/phase-erasure quantum channel*. This channel erases the input quantum system with probability $p$, erases the phase with probability $q$, and leaves the input unchanged with probability $1-p-q \geq 0$. Using (5.2) and (5.5), the classical capacity of the mixed erasure/phase-erasure quantum channel, $\mathcal{N}_{p+q}$, can be expressed as

$$C\left(\mathcal{N}_{p+q}\right) = \left(1-p\right)\log\left(d\right) = C\left(\mathcal{N}_p\right). \tag{5.7}$$

Furthermore, combining (5.3) and (5.6), the quantum capacity of the mixed erasure/phase-erasure quantum channel, $\mathcal{N}_{p+q}$, we get

$$Q\left(\mathcal{N}_{p+q}\right) = \left(1-q-2p\right)\log\left(d\right). \tag{5.8}$$

The classical (dashed line) and quantum capacities (solid line) of the mixed erasure/phase-erasure quantum channel as a function of total erasure probability $p+q$ are illustrated in Fig. 5.2.

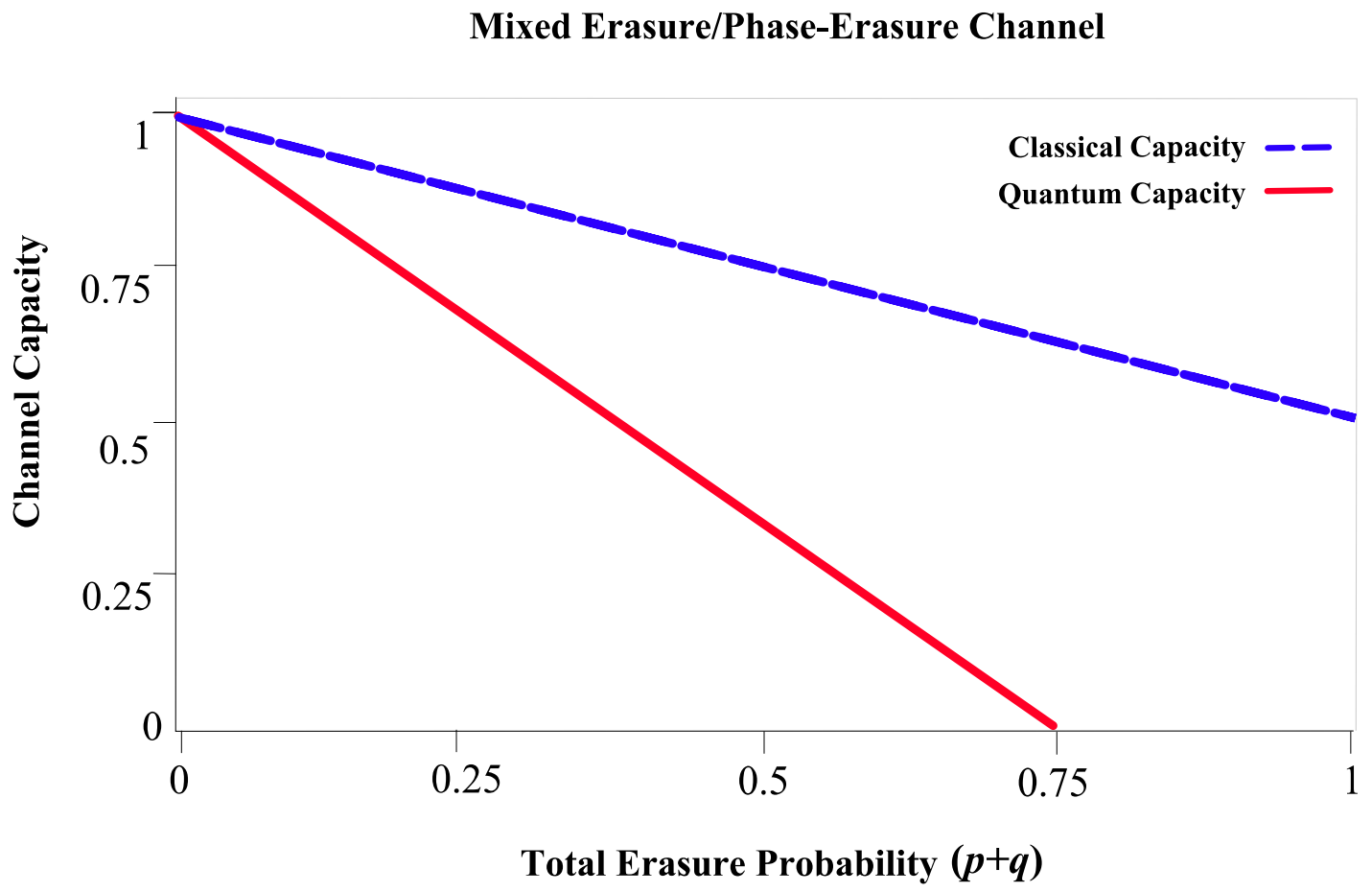

**Fig. 5.2.** The classical and quantum capacities of the mixed erasure/phase-erasure quantum channel as a function of total erasure probability.

Finally, we give the quantum capacity of another important quantum channel, called the amplitude damping channel (see Section 3.6.3). The classical capacity of the amplitude damping quantum channel can be expressed as

$$C\left(A_{\gamma}\right) = \max_{\tau} H\left(\tau\right) + \left[-H\left(\tau\left(\gamma\right)\right) + H\left(\tau\left(1-\gamma\right)\right)\right], \tag{5.9}$$

where $\tau \in \left[0,1\right]$ is a special parameter called the *population* parameter, and $H$ is the Shannon entropy function, and $H\left(\tau\right) = -\tau \log\left(\tau\right) - \left(1-\tau\right)\log\left(1-\tau\right)$. As follows from (5.9) the classical capacity $C\left(A_{\gamma}\right)$ of the amplitude damping channel completely vanishes if $\gamma = 1$, otherwise (if $0 \leq \gamma < 1$) the channel can transmit classical information. On the other hand for the quantum capacity $Q\left(A_{\gamma}\right)$ the capacity behaves differ.

The quantum capacity of this channel can be expressed as a maximization:

$$Q\left(A_{\gamma}\right) = \max_{\tau}\left[H\left(\tau\left(\gamma\right)\right) - H\left(\tau\left(1-\gamma\right)\right)\right]. \tag{5.10}$$

The classical (dashed line) and the quantum capacity (solid line) of the amplitude damping quantum channel as a function of the damping parameter $\gamma$ are shown in Fig. 5.3.

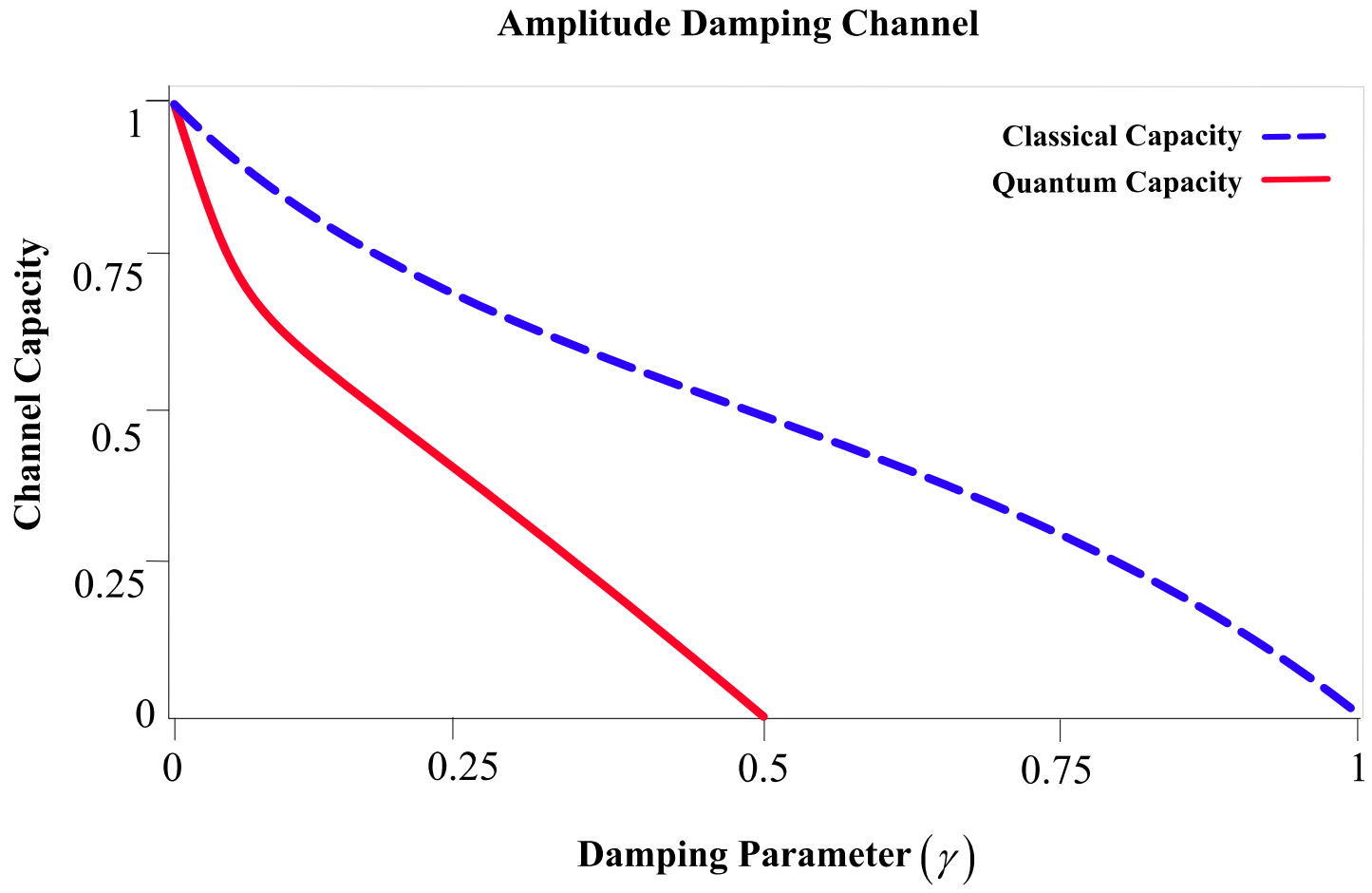

**Fig. 5.3.** The classical and quantum capacities of the amplitude damping quantum channel as a function of the damping parameter.

It can be concluded that the working mechanism of the amplitude damping channel is similar to the erasure channel (see (5.2) and (5.3)), since if the damping parameter value is equal to or greater than $0.5$, the quantum capacity of the channel completely vanishes. We obtained the same result for the erasure channel; however in that case the erasure probability $p$ was the channel parameter.

## 5.2 Quantum Capacity of the Classical Ideal Quantum Channel

In this section, we provide a non-trivial example that demonstrates the correctness of (4.15). Let us assume that there is a *classical ideal quantum channel* between Alice and Bob and that this channel works as follows. It first measures the qubit input of Alice in a computational basis and then sends the classical information to Bob. Bob then receives the classical bit. For the classical capacity of this noiseless quantum channel, the following trivial relation holds:

$$C^{(1)} = \max \mathcal{X}_{AB} = 1\,. \tag{5.11}$$

On the other hand, we have yet to take into account the second term of (4.15), $\mathcal{X}_{AE}$. Well, if $\mathcal{X}_{AE}$ is zero, then it would lead to a counterexample in (4.15), since it would entail a quantum channel that transmits one classical bit from Alice to Bob and zero bits to its environment. This would lead to the quantum capacity

$$Q^{(1)}(\mathcal{N}) = \max_{all\ p_i,\rho_i}\left(\mathcal{X}_{AB} - \mathcal{X}_{AE}\right) = 1 - 0 = 1. \tag{5.12}$$

However, this channel transmits classical information (see (5.11)); i.e., it cannot possibly have any quantum capacity. For a noiseless quantum channel which can transmit classical information perfectly (i.e., a *classical ideal quantum channel*),

$$\mathcal{X}_{AE} = 1, \tag{5.13}$$

and the quantum capacity of the noiseless classical ideal quantum channel completely vanishes, since

$$Q^{(1)}(\mathcal{N}) = \max_{all\ p_i,\rho_i}\left(\mathcal{X}_{AB} - \mathcal{X}_{AE}\right) = 1 - 1 = 0. \tag{5.14}$$

as follows, (5.14) indicates that for the noiseless classical ideal quantum channel, the environment receives the same information that Bob receives. To demonstrate this exchange, let us assume that Alice has an input density matrix $\rho_A$, and we enact a noiseless classical ideal quantum channel on this matrix. It will lead to

$$\mathcal{N}(\rho_A) = I(\rho_A) = |0\rangle\langle 0|\rho_A|0\rangle\langle 0| + |1\rangle\langle 1|\rho_A|1\rangle\langle 1|. \tag{5.15}$$

*Isometric Extension of the Classical Ideal Quantum Channel*

Now, we can express the isometric extension of the $\mathcal{N}_{AB}(\rho)$ classical noiseless qubit channel between Alice and Bob, which was already discussed in (5.15) as follows:

$$\begin{aligned} U_{A\to BE} &= \left(|0\rangle_B\langle 0|_A + |1\rangle_B\langle 1|_A\right)\otimes|E\rangle_E = \left(|E\rangle_E|0\rangle_B\langle 0|_A + |E\rangle_E|1\rangle_B\langle 1|_A\right) = \\ &= \left(|0\rangle_E|0\rangle_B\langle 0|_A + |1\rangle_E|1\rangle_B\langle 1|_A\right) = \left(|0\rangle_E\langle 0|_A + |1\rangle_E\langle 1|_A\right)\otimes|B\rangle_B, \end{aligned} \tag{5.16}$$

where $|E\rangle_E$ is the environment state and $|B\rangle_B$ is the channel output state. As follows, (5.16) clearly demonstrates that the channel between Alice and the environment is equal to what was given in (5.15), that is, the quantum channel between Alice and Bob is the same, because the isometric extension of $\mathcal{N}_{AB}(\rho)$ is just the same as $\mathcal{N}_{AE}(\rho)$,

$$\mathcal{N}_{AB}\left(\rho\right)=\mathcal{N}_{AE}\left(\rho\right), \tag{5.17}$$

which was proven in (5.16).
At this point, we return to our initial problem, stated previously in (5.12). After obtaining these results, we can write the following channels between Alice and Bob and Alice and the environment for the noiseless classical channel as

$$\mathcal{N}_{AB}\left(\rho_A\right)=\mathcal{N}_{AE}\left(\rho_A\right)=\left|0\right\rangle\left\langle 0\right|\rho_A\left|0\right\rangle\left\langle 0\right|+\left|1\right\rangle\left\langle 1\right|\rho_A\left|1\right\rangle\left\langle 1\right|. \tag{5.18}$$

We can also demonstrate that if we trace out Bob's system from $U_{A\to BE}$ using (2.72) as

$$\mathcal{N}_{AE}=Tr_B\left(U_{A\to BE}\left(\rho_A\right)\right), \tag{5.19}$$

i.e., if we apply $\mathcal{N}_{AE}$ on the input system $\rho$, the resulting output quantum system $\rho_E$ will be

$$\rho_E=\mathcal{N}_{AE}\left(\rho_A\right)=Tr_B\left(U_{A\to BE}\left(\rho_A\right)\right)=\sum_{i,j}Tr\left(N_i\rho_A N_j^\dagger\right)\left|i\right\rangle\left\langle j\right|. \tag{5.20}$$

Moreover, we get the same result if we trace out the environment from $U_{A\to BE}$, since

$$\mathcal{N}_{AB}=Tr_E\left(U_{A\to BE}\left(\rho_A\right)\right), \tag{5.21}$$

and

$$\rho_B=\mathcal{N}_{AB}\left(\rho_A\right)=Tr_E\left(U_{A\to BE}\left(\rho_A\right)\right)=\sum_{i,j}Tr\left(N_i\rho_A N_j^\dagger\right)\left|i\right\rangle\left\langle j\right|. \tag{5.22}$$

The correspondence between (5.20) and (5.22) also confirms our statement: the channel between Alice and Bob, and between Alice and the environment are completely the same

$$\mathcal{N}_{AB}\left(\rho_A\right)=\sum_i N_i\rho_A N_i^\dagger=\mathcal{N}_{AE}\left(\rho_A\right)=\sum_{i,j}N_i\rho_A N_j^\dagger\otimes\left|i\right\rangle\left\langle j\right|_E . \tag{5.23}$$

This condition leads us to zero quantum capacity, because we proved that (5.14) holds, and we proved that the quantum capacity of the *classical* ideal qubit channel, which measures its qubit input on a computational basis and then transmits the classical outcome noiselessly to the receiver, is trivially equal to zero. As we have shown in (5.18), the classical information transmitted from Alice to Bob is identical to the information transmitted from Alice to the environment.

## 5.3 Geometric Interpretation of Quantum Channels Maps

The map of a quantum channel compresses the Bloch sphere, by an affine map. This affine map must be Completely Positive Trace Preserving (CPTP), which shrinks the Bloch sphere along the *x, y* and *z* axes. Now, we introduce a new geometric representation - called the *tetrahedron* - which also can be used to illustrate the various channel maps. We also show that there is a connection with the Bloch sphere representation.

### 5.3.1 The Tetrahedron Representation

Assuming the single-qubit case the quantum channel's output is represented by a $2\times 2$ density matrix and the operation is a trace preserving Completely Positive map. The map of the quantum channel on a single-qubit in the Bloch sphere representation, can be given by the affine map

$$\mathcal{N}\left(\mathbf{r}\right) = \mathbf{r}_{\mathcal{N}} = A\mathbf{r} + \vec{b}, \tag{5.24}$$

where $A$ is a $3\times 3$ diagonal matrix with entries $\vec{\eta} = \left(\eta_x, \eta_y, \eta_z\right)$ which characterizes the tetrahedron $\mathcal{T}$ , $\vec{b}$ is a three-dimensional vector representing the shift of the center of the Bloch sphere, $\mathbf{r}$ is the initial Bloch vector of the sent pure quantum state, and $\mathbf{r}_{\mathcal{N}}$ is the Bloch vector of the channel output state.

The entries of $A$ specify the tetrahedron $\mathcal{T}$ in the parameter space of $\left\{\eta_x, \eta_y, \eta_z\right\}$, where $\eta_i \in \mathcal{T}$ if

$$\left|\eta_x \pm \eta_y\right| \leq \left|1 \pm \eta_z\right|. \tag{5.25}$$

The tetrahedron $\mathcal{T}$ is the *convex hull* of the points representing $\mathcal{I}$ , and the three Pauli rotations, thus every transformation corresponding to a point in the tetrahedron $\mathcal{T}$ can be described as a statistical mixture of the Pauli-transformations $\mathcal{I}, \sigma_x, \sigma_y$ and $\sigma_z$ where $\mathcal{I}$ is the identity transformation, and $\sigma_x, \sigma_y, \sigma_z$ are rotations by $\pi$ around the *x, y* and *z* axes. Fig. 5.4 illustrates the tetrahedron $\mathcal{T}$ for of the physically allowed transformations of the quantum channel. The vertices of $\mathcal{T}$ are the Pauli-transformations $\mathcal{I}, \sigma_x, \sigma_y$ and $\sigma_z$.

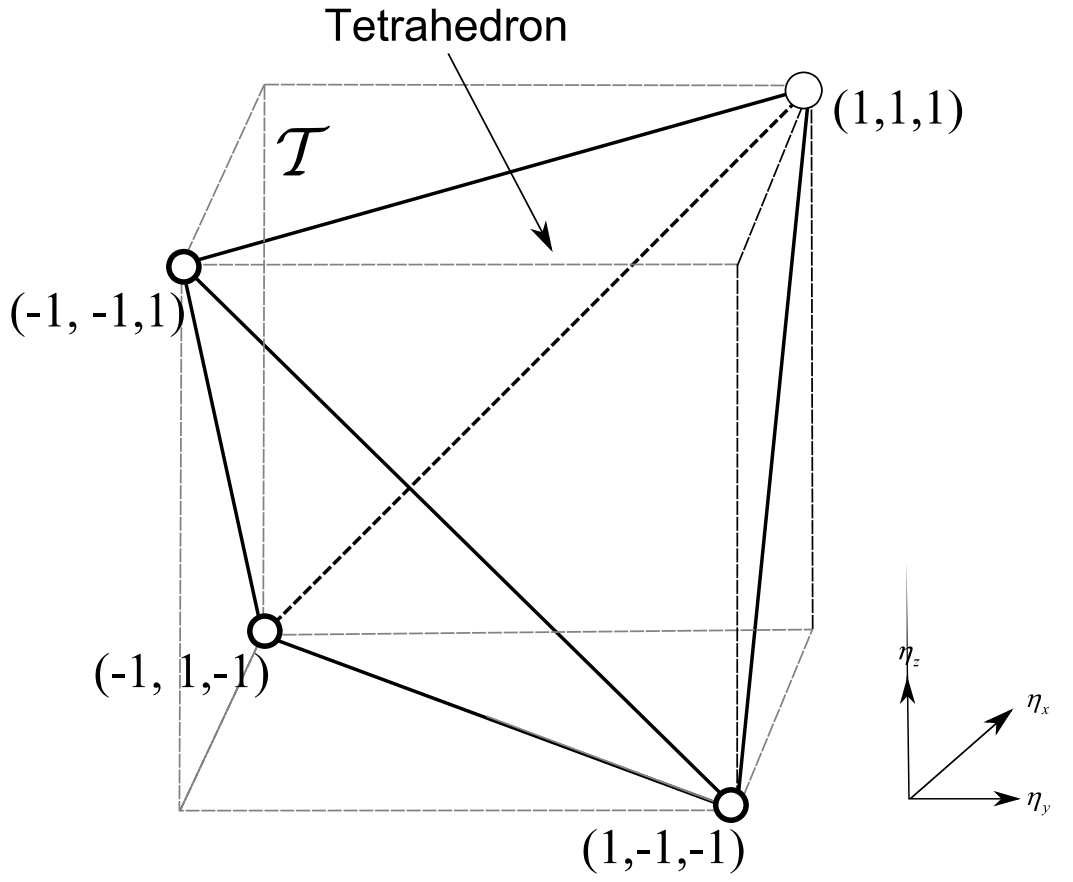

**Fig. 5.4.** The Pauli transformations can be represented by the tetrahedron. The various physically allowed channel maps of the quantum channel also can be described by this representation.

The vertices of $\mathcal{T}$ correspond with the four maps which can be described as

$$\rho \to \rho' = \sum_{j=0}^{3} \varepsilon_j \sigma_j \rho \sigma_j^\dagger, \tag{5.26}$$

where $\sigma_0$ is the identity matrix $\mathcal{I}$, while for $j = 1,2,3$, $\sigma_j$ denotes the *j*-th Pauli operator (*X*, *Z*, *Y*), and $\varepsilon_0 + \varepsilon_1 + \varepsilon_2 + \varepsilon_3 = 1$, where $\varepsilon_1, \varepsilon_2$ and $\varepsilon_3$ are non-negative parameters. The general transformation $\mathcal{N}$ of the quantum channel can be described as a convex sum of these maps

$$\rho' = \mathcal{N}\left(\rho\left(x,y,z\right)\right) = \varepsilon_1\sigma_1\rho\sigma_1 + \varepsilon_2\sigma_2\rho\sigma_2 + \varepsilon_3\sigma_3\rho\sigma_3 + \left(1 - \varepsilon_1 - \varepsilon_2 - \varepsilon_3\right)\rho. \tag{5.27}$$

The points forming the vertices of $\mathcal{T}$ represent unitary maps for which only one operator is required in the operator sum representation, see (5.27), while the edges of $\mathcal{T}$ depict the two-operator maps, and the faces of $\mathcal{T}$ assign the maps described by three operators. The points inside $\mathcal{T}$ require all four operators.

Since the quantum channel performs a *CP* map, the map $\mathcal{N}$ has to be *physically allowed* on all other quantum states. A unital quantum channel does not change the center of the Bloch sphere, thus $\vec{b} = 0$, and in this case the matrix $A$ is diagonal filled in with the elements of the distortion vector $\vec{\eta} = \left(\eta_x, \eta_y, \eta_z\right)$. Assuming the maximally entangled two qubit system $\left|\beta_{00}\right\rangle\left\langle\beta_{00}\right|$, where $\left|\beta_{00}\right\rangle = \sum_i \left|i\right\rangle\left|i\right\rangle$, for a CP-map $\mathcal{N}$, the condition $I \otimes \mathcal{N}\left(\left|\beta_{00}\right\rangle\left\langle\beta_{00}\right|\right) \geq 0$ has to be satisfied which leads to the channel output matrix

$$\rho' = \frac{1}{2}\begin{pmatrix} 1+\eta_z & 0 & 0 & \eta_x+\eta_y \\ 0 & 1-\eta_z & \eta_x-\eta_y & 0 \\ 0 & \eta_x-\eta_y & 1-\eta_z & 0 \\ \eta_x+\eta_y & 0 & 0 & 1+\eta_z \end{pmatrix}, \tag{5.28}$$

which is positive if and only if

$$\left(1+\eta_z\right)^2 - \left(\eta_x+\eta_y\right)^2 \geq 0, \text{ and } \left(1-\eta_z\right)^2 - \left(\eta_x-\eta_y\right)^2 \geq 0. \tag{5.29}$$

As we have seen in this section, the tetrahedron representation focuses specially on the physically allowed transformations compared to the Bloch sphere. On the other hand, we have also highlighted the fact that there is a connection (see (5.24) ) between Bloch sphere and tetrahedron representations.

#### 5.3.1.1 Description of Channel Maps in the Tetrahedron

Now, let us investigate these quantum channel maps in the previously defined tetrahedron representation. On the edges of the tetrahedron, we can find the unital *bit flip*, *phase flip* and the *coarse graining* transformations. The bit flip and phase flip channel maps transform the original Bloch sphere into a distorted ellipsoid, which touches the original Bloch sphere at the points $\left\{\frac{1}{\sqrt{2}}\left(\left|0\right\rangle + \left|1\right\rangle\right), \frac{1}{\sqrt{2}}\left(\left|0\right\rangle - \left|1\right\rangle\right)\right\}$ and points $\left\{\left|0\right\rangle, \left|1\right\rangle\right\}$, respectively. Another important channel is the coarse graining channel; it transforms the whole Bloch sphere into a unit length vertical line segment centered at the origin of the Bloch sphere. The locations of bit flip, phase flip, bit-phase flip and the coarse graining channels are shown in Fig. 5.5.

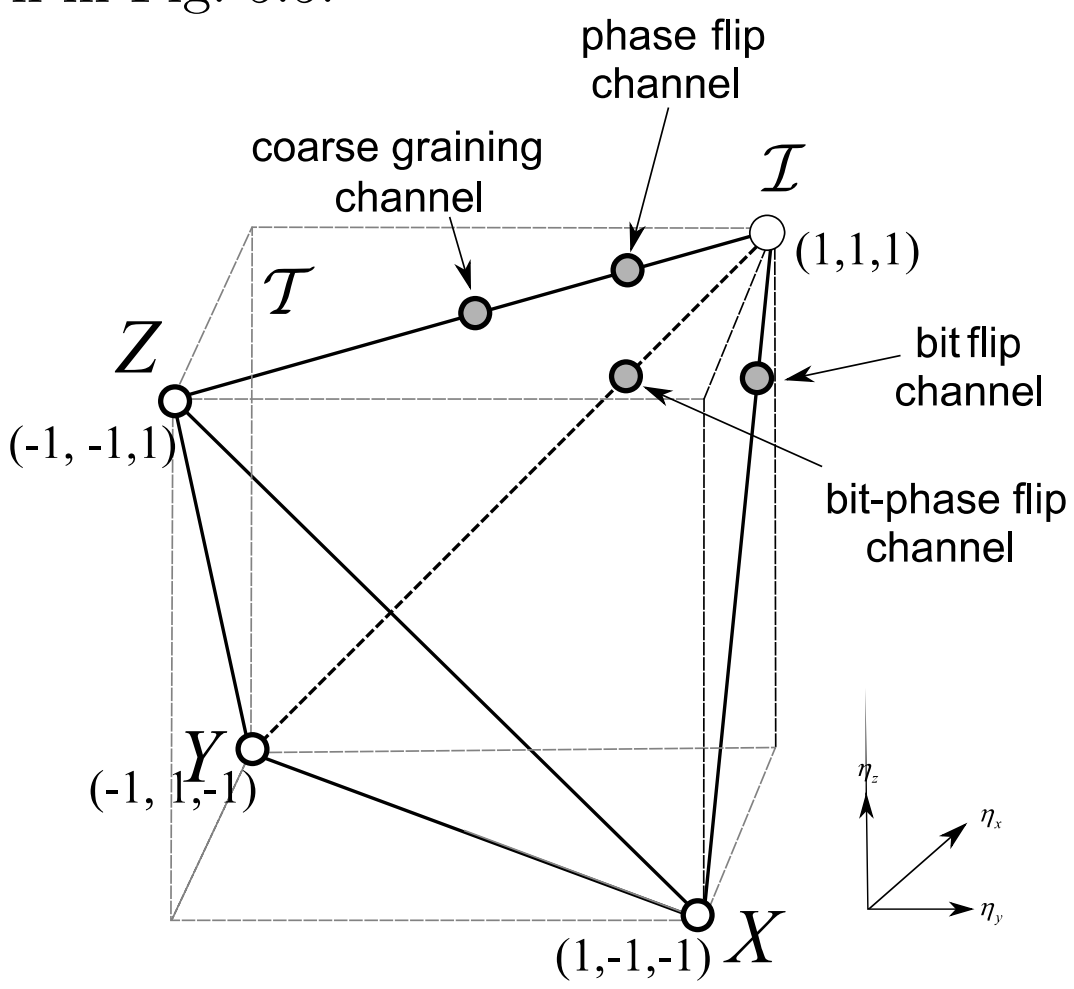

**Fig. 5.5.** The bit flip, phase flip and coarse graining channels in the tetrahedron representation.

Using the distortion parameters $\left\{\eta_x,\eta_y,\eta_z\right\}$ of the tetrahedron $\mathcal{T}$, the $I$ identity transformation can be expressed as

$$\vec{\eta}_{identity} = \left(1,1,1\right). \tag{5.30}$$

The distortion vector $\vec{\eta}$ of the bit flip, phase flip, bit-phase flip and the coarse graining quantum channels in function of channel parameter $p$ can be expressed as

$$\vec{\eta}_{bit\ flip} = \left(1,1-2p,1-2p\right), \tag{5.31}$$

$$\vec{\eta}_{phase\ flip} = \left(1-2p,1-2p,1\right), \tag{5.32}$$

$$\vec{\eta}_{bit-phase\ flip} = \left(1-2p,1,1-2p\right) \tag{5.33}$$

and

$$\vec{\eta}_{coarse\ gr.} = \left(0,0,1\right). \tag{5.34}$$

In the case of bit flip, phase flip and bit-phase flip channels the "worst case scenario" occurs at $p = \frac{1}{2}$. In these cases, these channels maps are degenerated, which results in a line with unit length. Furthermore, as can be observed in Fig. 5.5, the coarse graining channel can be viewed as the "worst case scenario" of a phase flip channel.

Inside the tetrahedron, we can find the *linear channel* map model and the *depolarizing* channel model, which are both unital. The linear channel transform maps the original Bloch sphere to a vertical line segment,

$$\vec{\eta}_{linear} = \left(0, 0, q\right), \tag{5.35}$$

while the completely depolarizing channel maps the whole Bloch sphere to one point, namely to the center of the Bloch sphere. On the other hand, while in the case of coarse graining channel the length of the line is unit, in case of the linear quantum channel the length of the line depends on the channel parameter $q$.

The output of a completely depolarizing channel is a maximally mixed state, the channel shrinks both coordinates equally. In case of linear channels, the channel map results in a line, $2q$ of lengths, while the degenerated maps occurs at $q = 0$ and $q = 1$. If $q = 0$, the channel will be represented by a point in the center of the tetrahedron, or in other words in this case, the linear quantum channel is equivalent to a *completely depolarizing* quantum channel.

Using the tetrahedron representation, the *depolarizing* quantum channel can be expressed as

$$\vec{\eta}_{depol.} = \left[1 - x\right]\left(1, 1, 1\right), \tag{5.36}$$

where $x$ determines the level of the shrinking of the original Bloch sphere. In case of $x = 1$ we have a completely depolarizing channel

$$\vec{\eta}_{comp.\ depol.} = \left(0, 0, 0\right), \tag{5.37}$$

which map can be found in the center of the tetrahedron, as illustrated in Fig. 5.6. In this figure, we also show the linear channel and the completely depolarizing channel maps in the tetrahedron representation.

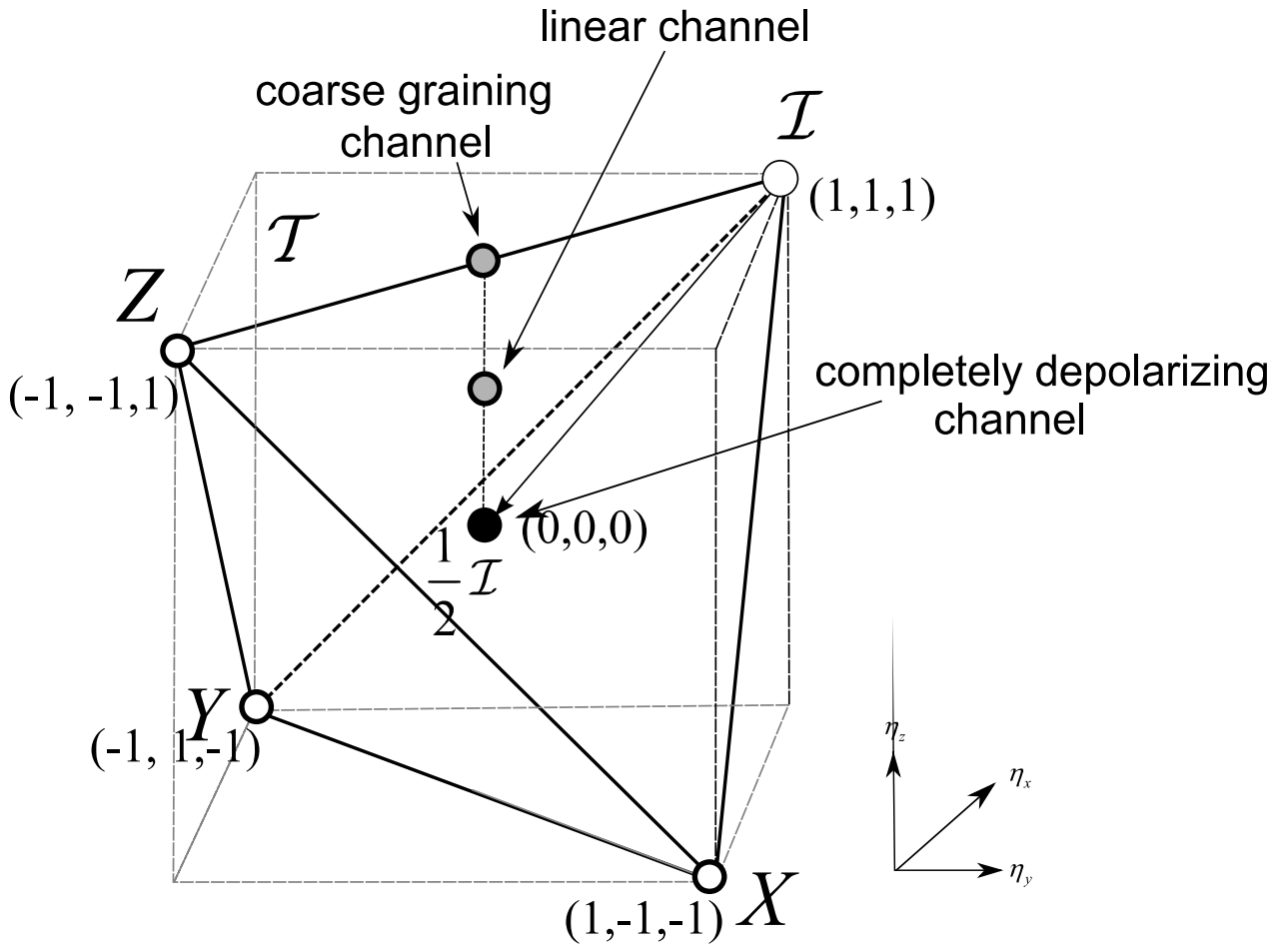

**Fig. 5.6.** The linear channel and the completely depolarizing channel maps are inside the tetrahedron.

As we have seen, the unital quantum channels can be represented in the tetrahedron view. The most important geometric property of the unital channel maps, in comparison to the non-unital maps, is that they do not change the center of the Bloch sphere. On the other hand, the non-unital quantum channel models - such as the amplitude damping quantum channel - cannot be represented in the tetrahedron.

#### 5.3.1.2 Non-Unital Quantum Channel Maps

The description of the non-unital quantum channel maps requires a more complex mathematical background. But, where does the problem arise from? The problem here is that for non-unital transformations, the center of the transformed Bloch sphere will differ from the center of the original Bloch sphere. This fact might seem to be negligible at first, but it also has an important corollary: namely, in case of non-unital channels we have to define a more complex geometric structure.

The cube which contains the tetrahedron $\mathcal{T}$, defines the Positive (i.e., not Completely Positive) maps. The tetrahedron with constraints (5.29), defines the convex polytope of CP unital maps. If the map of the quantum channel is Positive, then the channel ellipsoid lies in the original Bloch sphere. The unital quantum channels are a subset of the Positive maps, since they hold the center of the channel ellipsoid and they lie inside the original Bloch sphere. The amplitude damping channel model has great relevance to practical optical communications, since this channel model describes the energy dissipation due to losing a particle. Moreover, this channel can also take a mixed input state to a pure output state.

Beside the fact we cannot represent the amplitude damping channel on the tetrahedron, we can give the distortion vector $\vec{\eta}$ of the amplitude damping channel as follows

$$\vec{\eta}_{ampl.\ damp.} = \left(\sqrt{1-2p}, \sqrt{1-2p}, 1-2p\right). \tag{5.38}$$

Before discussing the most relevant channel models in detail we introduce the reader the geometric interpretation of quantum informational distance.

## 5.4 Geometric Interpretation of Capacity of Quantum Channels

As we already know, the depolarizing channel (see Section 3.6.2) is a unital. Geometrically, this means that the channel maps an identity transformation to an identity transformation, hence $\mathcal{N}\left(I\right) = I$. This property also implies some symmetries in the geometric picture of the capacity of unital quantum channels.

We show the channel ellipsoid of a *unital* quantum channel and the smallest quantum informational ball (colored in grey) containing the channel ellipsoid in Fig. 5.7 [Gyongyosi11]. The information theoretic radius of this quantum informational ball describes the capacity of the analyzed quantum channel. The distorted structure of the quantum informational ball is the consequence of the distance calculations which are based on the quantum relative entropy function.

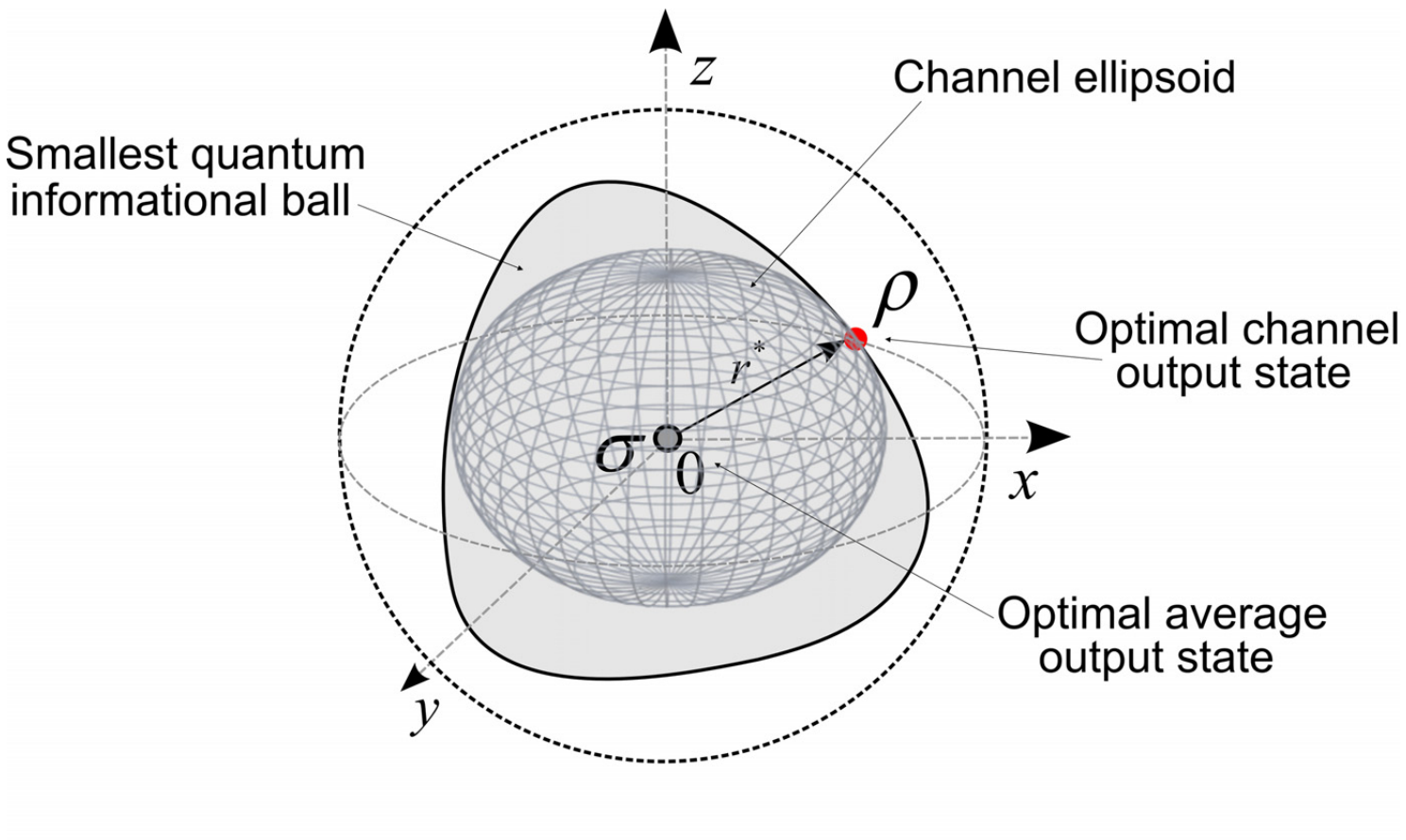

**Fig. 5.7.** The channel ellipsoid of the unital quantum channel model. The center of the channel ellipsoid and the smallest quantum informational ball is equal to the center of the of the Bloch sphere.

Unital quantum channels have another important geometric property, since the average state $\sigma = \sum_k p_k \rho_k$ of the optimal *output* ensembles $\{p_k, \rho_k\}$ is equal to the center of the Bloch sphere. To make the picture clear we emphasize that here we are interested in the output of the channel i.e., our focus is on the *output* of $\mathcal{N}$, which can be analyzed by the *optimal average* output state $\sigma$, and the *optimal channel output* state $\rho$. The HSW capacity of *unital* channels is equal to the quantum informational distance between the optimal output state $\rho$ (which maximizes the channel capacity) and the origin of the Bloch sphere, thus we have

$$D\left(\rho \middle\| \frac{1}{2} I\right) = 1 - \mathrm{S}(\rho), \tag{5.39}$$

where $D(\cdot\|\cdot)$ is the quantum relative entropy function, which for the quantum states $\rho$ and $\sigma$ can be expressed as

$$D(\rho\|\sigma) = Tr(\rho \log(\rho)) - Tr(\rho \log(\sigma)) = Tr[\rho(\log(\rho) - \log(\sigma))]. \tag{5.40}$$

In the definition above, the term $Tr(\rho \log(\sigma))$ is finite only if $\rho \log(\sigma) \geq 0$ for all diagonal matrix elements. If this condition is not satisfied, then $D(\rho\|\sigma)$ could be infinite, since the trace of the second term could go to infinity.
For an ideal quantum channel with an identity $\mathcal{N} = I$ map and an optimal pure density matrix $\rho$, the HSW channel capacity is

$$C(\mathcal{N}) = \min_{\sigma} \max_{\rho} D(\rho\|\sigma) = \max_{\rho} D\left(\rho \middle\| \frac{1}{2} I\right) = 1 - \overbrace{\mathrm{S}(\rho)}^{0} = 1 - 0 = 1, \tag{5.41}$$

since the von Neumann entropy of the pure output state is $\mathrm{S}(\rho) = 0$, while the unital channel model preserves the origin of the Bloch sphere. If we take the average of all optimal pure density matrices $\rho_i$, then the HSW capacity of $\mathcal{N}$ is

$$\begin{aligned} C(\mathcal{N}) &= \max_{all\ p_i,\rho_i} \chi = \max_{all\ p_i,\rho_i} \left[ \mathrm{S}\left( I\left( \sum_i p_i \rho_i \right) \right) - \sum_i p_i \mathrm{S}\left( I(\rho_i) \right) \right] \\ &= \max_{all\ p_i,\rho_i} \left[ \mathrm{S}\left( I\left( \sum_i p_i \rho_i \right) \right) \right] = \sum_{all\ p_i} p_i D\left( \rho_i \middle\| \frac{1}{2} I \right) \\ &= 1 - \sum_i p_i \mathrm{S}\left( I(\rho_i) \right) = 1 - 0. \end{aligned} \tag{5.42}$$

For a general (i.e., not ideal) unital quantum channel $\mathcal{N}$ the following relation holds

$$C(\mathcal{N}) = \max_{all\ p_i,\rho_i} \chi = 1 - \sum_i p_i \mathrm{S}\left( \mathcal{N}(\rho_i) \right). \tag{5.43}$$

### 5.4.1 Geometric Interpretation of the Quantum Informational Distance

The aim of this section is to discuss the geometric interpretation of HSW channel capacity in general, using quantum relative entropy as a distance measure function. We will demonstrate that the HSW channel capacity can be defined by using the quantum relative entropy function as a distance measure.

Based on the results of Nielsen et al. [Nielsen07-07a, 08-08a] and Nock and Nielsen [Nock05] the quantum relative entropy function $D(\rho \| \sigma)$ (see Section 2) can be described by means of a strictly convex and differentiable *generator function* $\mathbf{F}$ as

$$\mathbf{F}(\rho) = -\mathrm{S}(\rho) = Tr(\rho \log \rho), \tag{5.44}$$

where $-\mathrm{S}$ is the negative von Neumann entropy. The extended generator function $\mathbf{F}_E(\cdot)$ can be defined as

$$\mathbf{F}_E(\rho) = Tr(\rho \log \rho - \rho). \tag{5.45}$$

The *quantum relative entropy* $D(\rho \| \sigma)$ which measures the *informational* distance between quantum states with density matrices in the Bloch sphere $\rho = \rho(x, y, z)$ and $\sigma = \sigma(\tilde{x}, \tilde{y}, \tilde{z})$ can be calculated using generator function $\mathbf{F}$ in the following way

$$\begin{aligned} D(\rho \| \sigma) &= Tr\left( \rho \left( \log \rho - \log \sigma \right) \right) \\ &= \mathbf{F}(\rho) - \mathbf{F}(\sigma) - \left\langle \rho - \sigma, \nabla \mathbf{F}(\sigma) \right\rangle, \end{aligned} \tag{5.46}$$

where $\mathbf{F} : S\left(\mathbb{C}^d\right) \to \mathbb{R}$, and $S\left(\mathbb{C}^d\right)$ denotes the open convex domain*,* while $\left\langle \rho, \sigma \right\rangle = Tr\left(\rho\sigma^*\right) = x\tilde{x} + y\tilde{y} + z\tilde{z}$ is the *inner product* of the quantum states, and $\nabla\mathbf{F}\left(\cdot\right)$ is the gradient (i.e., the derivate of the generator function) for the quantum informational distance defined as

$$\nabla\mathbf{F}\left(x\right) = \log\left(x\right), \tag{5.47}$$

and the inverse gradient $\nabla^{-1}\mathbf{F}\left(\cdot\right)$ is

$$\nabla^{-1}\mathbf{F}\left(x\right) = e^x . \tag{5.48}$$

Similarly, the extended quantum informational function can be defined as

$$\begin{aligned} D\left(\rho\|\sigma\right) &= Tr\left(\rho\left(\log\left(\rho\right) - \log\left(\sigma\right)\right) - \rho + \sigma\right) \\ &= \mathbf{F}\left(\rho\right) - \mathbf{F}\left(\sigma\right) - \left\langle \rho - \sigma, \nabla\mathbf{F}\left(\sigma\right)\right\rangle. \end{aligned} \tag{5.49}$$

In general, function $D$ is defined by strictly convex and differentiable generator function $\mathbf{F} : S \to \mathbb{R}$ over an open convex domain $S\left(\mathbb{C}^d\right)$, however, it is not a metric, hence symmetry and triangle inequality may fail. In geometric interpretation, quantum relative entropy $D\left(\rho\|\sigma\right)$ between quantum states $\rho$ and $\sigma$ can be measured as the vertical distance between $\rho$ and the hyperplane $H_\sigma$ tangent to relative entropy function at quantum state $\sigma$ i.e., quantum relative entropy function can be expressed in geometric interpretation

$$D\left(\rho\|\sigma\right) = \mathbf{F}\left(\rho\right) - H_\sigma\left(\rho\right). \tag{5.50}$$

In Fig. 5.8, we have illustrated the geometric interpretation of quantum informational distance between quantum states $\rho$ and $\sigma$. Since, we have depicted the quantum informational distance $D\left(\rho\|\sigma\right)$, as the vertical distance between the generator function $\mathbf{F}$ and $H\left(\sigma\right)$, the hyperplane tangent to $\mathbf{F}$ at $\sigma$ [Nielsen07]. The point of intersection of quantum state $\rho$ on $H\left(\sigma\right)$ is denoted by $H_\sigma\left(\rho\right)$. The tangent hyperplane to hypersurface $\mathbf{F}\left(\rho\right)$ at quantum state $\sigma$ is

$$H_\sigma(\rho) = \mathbf{F}(\sigma) + \langle \rho - \sigma, \nabla \mathbf{F}(\sigma) \rangle. \tag{5.51}$$

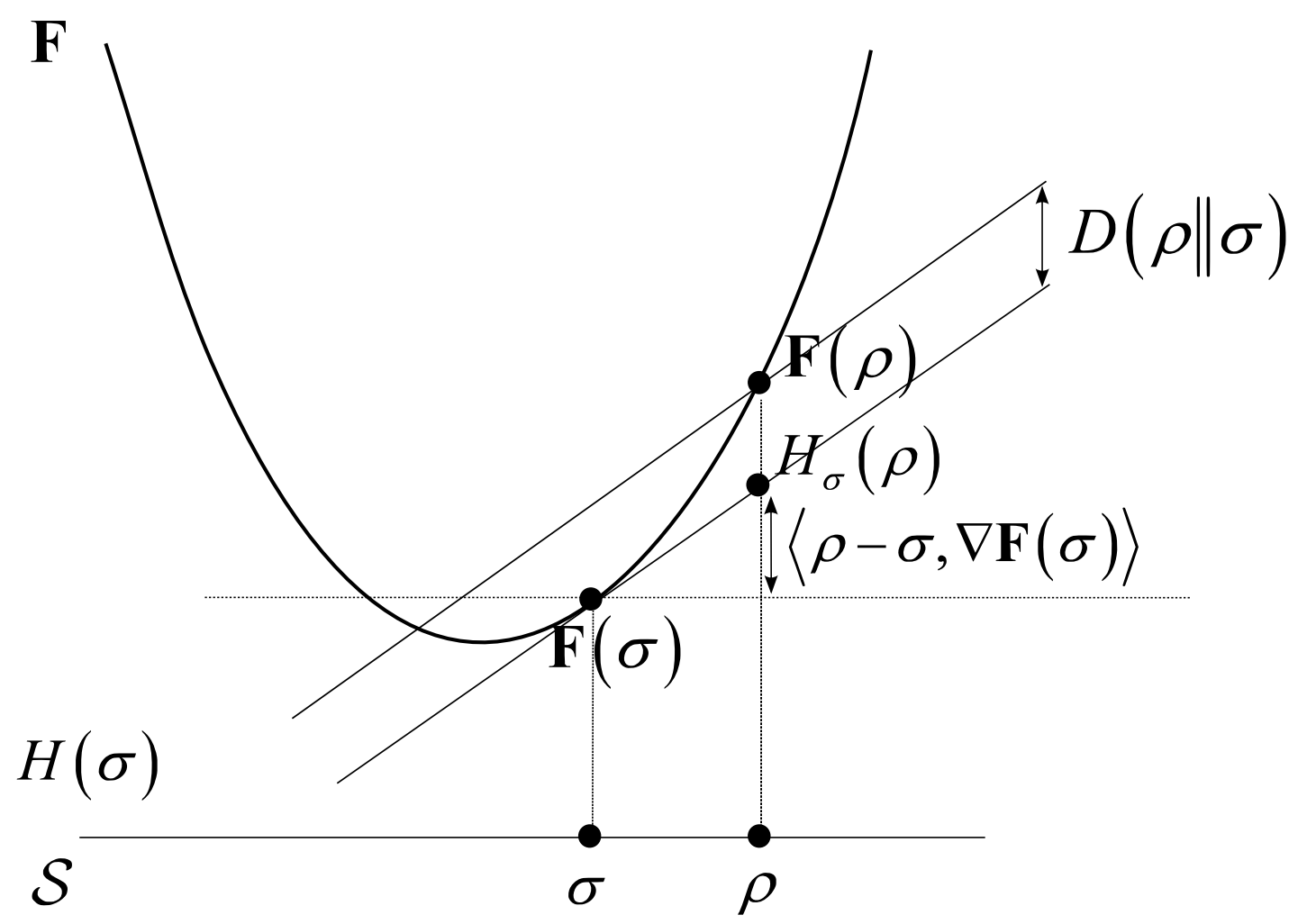

**Fig. 5.8.** Geometric interpretation of quantum informational distance between quantum states.

For mixed quantum states, the associated quantum informational distance is not symmetric, i.e., $D(\rho\|\sigma) \neq D(\sigma\|\rho)$. The strict convexity of generator function $\mathbf{F}$ implies that, for any quantum state $\rho$ and $\sigma$, $D(\rho\|\sigma) \geq 0$, with $D(\rho\|\sigma) = 0$ if and only if $\rho = \sigma$. The quantum informational distance function $D(\rho\|\sigma)$ is convex in its first argument $\rho$, but not necessarily in its second argument $\sigma$. It is worth highlighting the fact that the quantum generator function has a classical analogy, because for classical probability distributions $p$, the generator function $\mathbf{F}$ is the negative Shannon entropy

$$\mathbf{F}(x) = x \log x = -x \log \frac{1}{x} = \int p(x) \log p(x)\, dx, \tag{5.52}$$

and

$$\nabla \mathbf{F}(x) = 1 + \log x. \tag{5.53}$$

Similarly, for classical probability distributions $p$ and $q$, the informational distance can be expressed as [Nielsen07], [Nock05]

$$D\left(p(x)\middle\| q(x)\right) = \int \left(\mathbf{F}(p) - \mathbf{F}(q) - \left\langle p - q, \nabla \mathbf{F}(q) \right\rangle\right) dx = \int p(x) \log \frac{p(x)}{q(x)} dx. \tag{5.54}$$

Finally we show the connection between the Euclidean distance and an Euclidean $\mathbf{F}$ generator function. The proof can be extended to quantum informational distances, using the quantum generator function $\mathbf{F}$. If the generator function $\mathbf{F}$ is the *squared* Euclidean distance, then the strictly convex and differentiable generator function over $\mathbb{R}^d$ can be expressed as

$$\mathbf{F}(x) = x^2 = \sum_{i=0}^{d-1} x_i^2 = x^T x, \text{ with } \nabla \mathbf{F}(x) = 2x\,. \tag{5.55}$$

In this case, $D(\rho\|\sigma)$ can be formulated as

$$\begin{aligned} D(\rho\|\sigma) &= \mathbf{F}(\rho) - \mathbf{F}(\sigma) - \left\langle \rho - \sigma, \nabla \mathbf{F}(\sigma) \right\rangle \\ &= \rho^2 - \sigma^2 - \left\langle \rho - \sigma, 2\sigma \right\rangle = \rho^2 + \sigma^2 - 2\rho\sigma \\ &= \rho^T \rho + \sigma^T \sigma - 2\rho^T \sigma = \left\| \rho - \sigma \right\|^2. \end{aligned} \tag{5.56}$$

In Fig. 5.9, we have illustrated the squared Euclidean distance function $D(\rho\|\sigma)$, with Euclidean generator function $\mathbf{F}(x) = x^2 = \sum_{i=0}^{d-1} x_i^2$.

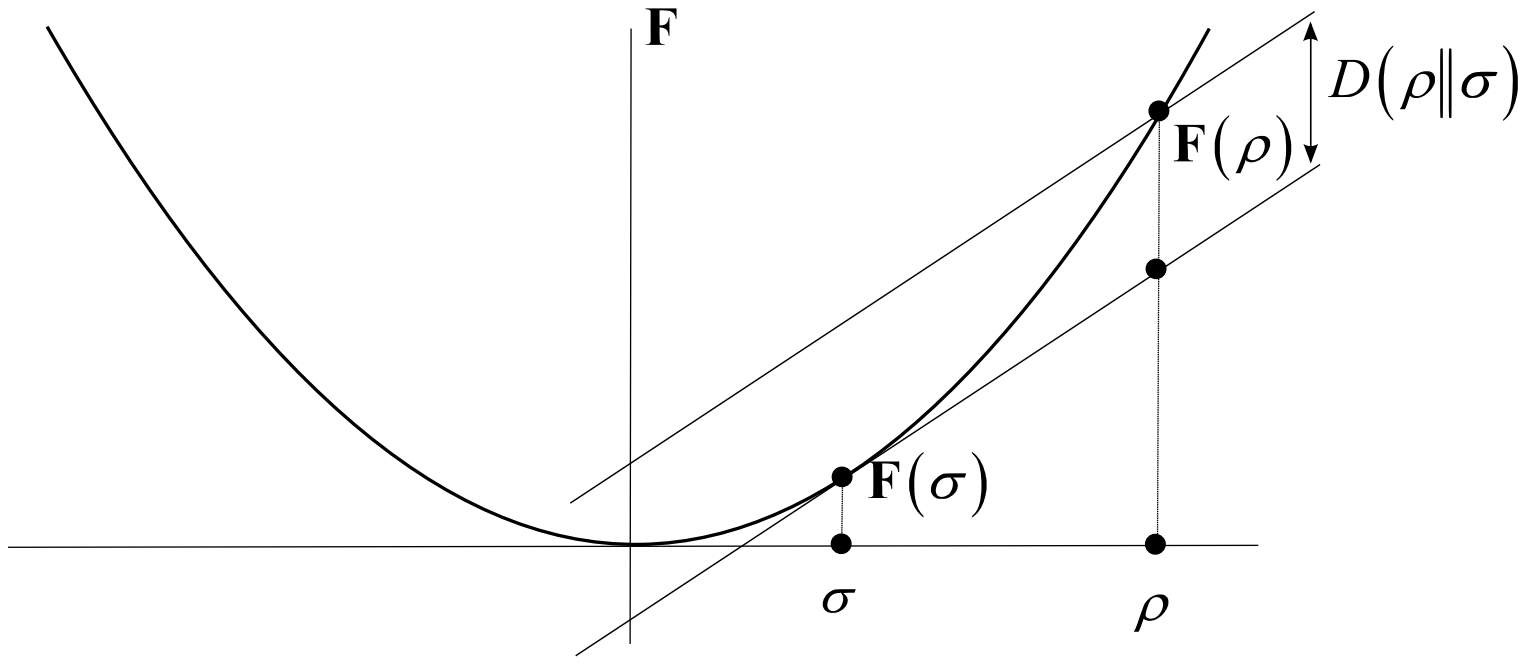

**Fig. 5.9.** The squared Euclidean distance function with Euclidean generator function $\mathbf{F}$.

As we have concluded previously, the density matrices of quantum states can be represented by 3D points in the Bloch sphere. If we compute the distance between two quantum states in the 3D Bloch sphere representation, we compute the distance between two density matrices $\rho$ and $\sigma$. The transformation of the quantum channel $\mathcal{N}$ is modeled by an affine map, that maps quantum states to quantum states.

We have used an Euclidean generator function (5.55) in (5.56). Now, we turn our attention to the quantum informational distance function. In this case the generator function is the negative von Neumann entropy function $-\mathrm{S}$, (see (5.44)), hence the properties of the generator function will differ from (5.55). The quantum informational distance function $D(\rho\|\sigma)$ with generator function $\mathbf{F}(\rho) = -\mathrm{S}(\rho)$ is illustrated in Fig. 5.10.

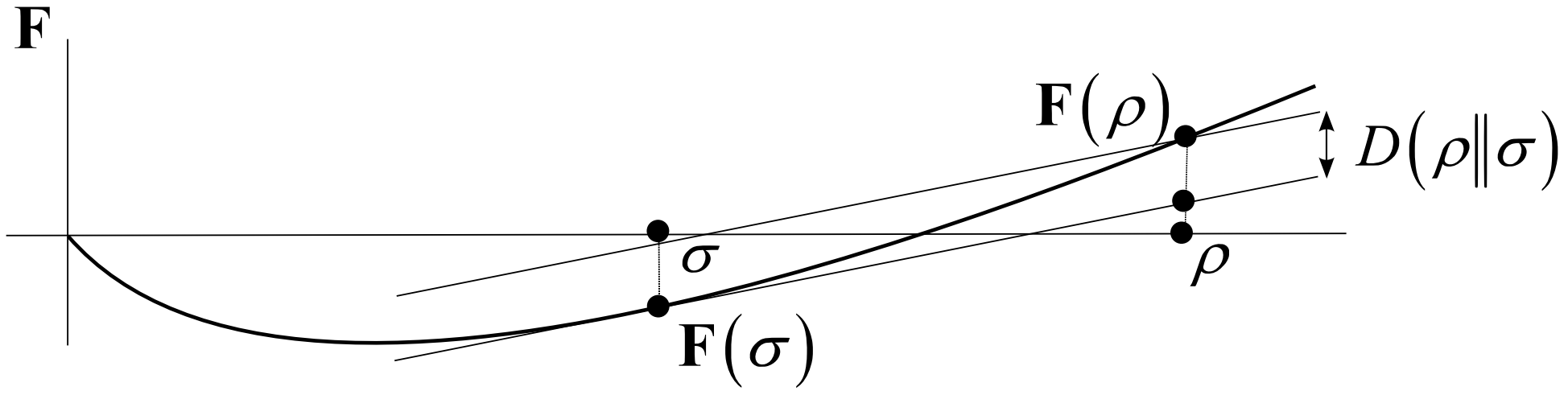

**Fig. 5.10.** Negative von Neumann generator function.

The quantum informational distance function is a linear operator, thus for convex functions $\forall \mathbf{F}_1 \in \mathcal{C}$ and $\forall \mathbf{F}_2 \in \mathcal{C}$, $D_{\mathbf{F}_1+\lambda\mathbf{F}_2}(\rho\|\sigma) = D_{\mathbf{F}_1}(\rho\|\sigma) + \lambda D_{\mathbf{F}_2}(\rho\|\sigma)$, for any $\lambda \geq 0$.

### 5.4.2 Quantum Informational Ball

In this section we discuss the properties of quantum relative entropy based "quantum informational balls". The output states of the quantum channel can be enclosed by a ball - however between the density matrices we cannot use the Euclidean distance. If we would like to determine the capacity of the quantum channel using a geometric interpretation, then we have to seek the smallest enclosing *quantum informational ball*, which is the smallest among all possible balls. Moreover, as we have stated earlier, the computation of quantum channel capacity is numerically very hard, since it is an NP-Complete problem [Beigi07]. Using the smallest quantum informational ball representation the problem can be solved without the extremely high computational costs - we just have to construct an algorithm to fit the quantum ball, and we will have very good approximation of the channel capacity in our hands.

Based on [Nielsen07-07a] the geometric structure of these balls significantly differs from the geometric structure of classical Euclidean balls. The quantum informational ball $B$ with center $c$ can be defined in the Bloch sphere representation for left-sided and right-sided bisectors with respect to quantum informational distances as

$$B\left(c,r\right)=\left\{\rho\in\mathbb{C}^{d}\middle|D\left(\rho\middle\|c\right)\leq r\right\}\text{ and }B'\left(c,r\right)=\left\{\rho\in\mathbb{C}^{d}\middle|D\left(c\middle\|\rho\right)\leq r\right\}. \tag{5.57}$$

The left-sided quantum informational ball $B\left(c,r\right)$ is a convex ball, while the right-sided ball $B'\left(c,r\right)$ is not necessarily convex, see the book of Imre and Gyongyosi [Imre12]. Using inverse transformation and relation

$$D\left(\rho\middle\|\sigma\right)=\mathbf{F}\left(\rho\right)+\mathbf{F}^{*}\left(\sigma'\right)+\left\langle\rho,\sigma'\right\rangle=D^{*}\left(\sigma'\middle\|\rho'\right), \tag{5.58}$$

the connection between of left sided $B\left(c,r\right)$ and right-sided $B'\left(c,r\right)$ quantum informational balls can be expressed as

$$B'\left(c,r\right)=\nabla^{-1}\mathbf{F}\left(B\left(c',r\right)\right), \tag{5.59}$$

where $c'=\nabla\mathbf{F}\left(c\right)$. The two distances are neither *necessarily convex* nor *identical*, however, the right-sided information balls can be transformed into left-sided balls using the *inverse* transformation, we can further define a third-type quantum informational ball, by taking the symmetric distance [Nielsen07]

$$D^{symm}=\frac{D\left(\rho\middle\|\mathbf{c}\right)+D\left(\mathbf{c}\middle\|\rho\right)}{2}. \tag{5.60}$$

As we will see later, to compute the smallest enclosing quantum informational ball, we use quantum relative entropy-based Delaunay tessellation, which is *symmetric* only for pure states and *asymmetric* for mixed states. The quantum information theoretical distance is neither symmetric, nor do they satisfy the triangular inequality of metrics. The spherical Delaunay triangulation between *pure* states and between pure and mixed states with equal radii can be simply obtained as the 3D Euclidean Delaunay tessellation restricted to the Bloch sphere.

Using the results of Petz [Petz96,08], Cortese [Cortese02,03], Hayashi [Hayashi05], Ruskai [Ruskai01] and King [King99-03] we refer to channel capacity as the radius $r^{*}$ of the smallest enclosing ball as follows

$$C\left(\mathcal{N}\right) = r^* = \max_{all\ p_i,\rho_i} \chi = \max_{all\ p_i,\rho_i} \mathrm{S}\left(\mathcal{N}\left(\sum_{i=0}^{n-1} p_i\rho_i\right)\right) - \sum_{i=0}^{n-1} p_i \mathrm{S}\left(\mathcal{N}\left(\rho_i\right)\right). \quad (5.61)$$

A quantum state can be described by its density matrix $\rho \in \mathbb{C}^{l\times l}$, which is an $l \times l$ matrix, where $d$ is the level of the given quantum system, i.e., for example for a qubit $d = 2$. For an $n$ qubit system, the level of the quantum system is $l = d^n = 2^n$. We use the fact that particle state distributions can be analyzed probabilistically by means of density matrices. A two-level quantum system can be defined by its density matrices in the following way:

$$\rho = \frac{1}{2}\begin{pmatrix} 1+z & x-iy \\ x+iy & 1-z \end{pmatrix},\ x^2+y^2+z^2 \le 1,\ x,y,z \in \mathbb{R}., \quad (5.62)$$

which also can be rewritten as

$$\rho = \begin{pmatrix} \frac{1+z}{2} & \frac{x-iy}{2} \\ \frac{x+iy}{2} & \frac{1-z}{2} \end{pmatrix},\ x^2+y^2+z^2 \le 1,\ x,y,z \in \mathbb{R}. \quad (5.63)$$

where $i$ denotes the complex imaginary $i^2 = -1$. The eigenvalues $\lambda_1, \lambda_2$ of $\rho\left(x,y,z\right)$ are given by

$$\lambda_1, \lambda_2 = \frac{1 \pm \sqrt{x^2+y^2+z^2}}{2}, \quad (5.64)$$

the eigenvalue decomposition $\rho$ is

$$\rho = \sum_i \lambda_i E_i, \quad (5.65)$$

where $E_i E_j$ is $E_i$ for $i = j$ and $\mathbf{0}$ for $i \neq j$. For a mixed state $\rho\left(x,y,z\right)$, $\log\left(\rho\right)$ defined by

$$\log\left(\rho\right) = \sum_i \left(\log\left(\lambda_i\right)\right) E_i. \quad (5.66)$$

The Bloch vectors $\mathbf{r}_1$ and $\mathbf{r}_2$ are real 3-dimensional vectors with length $m = 1$ for pure states, and $m < 1$ for mixed states. They can be expressed as

$$\mathbf{r} = \begin{bmatrix} r_x \\ r_y \\ r_z \end{bmatrix}. \tag{5.67}$$

Now, we define an alternate version of (5.61), using the quantum relative entropy function as distance measure between the channel output states. We will use the concept of *optimal channel output state* and *average output state (mixed state).* The optimal channel output states are a subset of output states, i.e., they are the most distance from the origin of the Bloch sphere (The minimal von Neumann entropy channel output state also belongs to this set.). Or with other words a channel output state is called optimal, if it maximizes the Holevo quantity of the quantum channel. The *optimal average* output state is the average of the set of optimal states [Holevo98], [Schumacher97].
The Holevo quantity can be represented geometrically, using the quantum relative entropy function as a distance measure as [Petz96,08], [Schumacher99,2000], [Cortese02]

$$\chi = D\left(\rho_k \middle\| \sigma\right), \tag{5.68}$$

where $\rho_k$ denotes an *optimal* (for which the Holevo quantity will be maximal) *output state* and $\sigma = \sum p_k \rho_k$ is the mixture of the optimal output states [Schumacher99]. For non-optimal output states $\delta$ and optimal $\sigma = \sum p_k \rho_k$ we have

$$\chi = D\left(\delta \middle\| \sigma\right) \leq D\left(\rho_k \middle\| \sigma\right). \tag{5.69}$$

Using the optimal channel output density matrices $\rho_k$ and they average $\sigma$, the geometric interpretation of quantum channel capacity using the quantum relative entropy function as a distance measure can be expressed as follows [Petz96,08], [Schumacher99,2000], [Cortese02]

$$C\left(\mathcal{N}\right) = \max_{all\ p_i,\rho_i} \chi = \chi\left(\mathcal{N}\right) = r^* = \min_{\{\sigma\}} \max_{\{\rho_k\}} D\left(\rho_k \middle\| \sigma\right), \tag{5.70}$$

where the *quantum informational radius*

$$r^* = \left| \mathbf{r}^* \right| \tag{5.71}$$

is the length (with respect to quantum informational distance) of the Bloch vector $\mathbf{r}^*$. Schumacher and Westmoreland have also proven [Schumacher99], that there exists a *unique* optimum output state $\left\{ p_k, \rho_k \right\}$ for every $\sigma$ that satisfies the maximization (i.e., maximizes the HSW capacity), such that $\sigma = \sum p_k \rho_k$. For this the Holevo information $\mathcal{X}$ can be derived in terms of the quantum relative entropy in the following way [Petz96,08], [Schumacher99,2000], [Cortese02]

$$\begin{aligned} \sum_k p_k D\left(\rho_k \| \sigma\right) &= \sum_k \left( p_k Tr\left(\rho_k \log\left(\rho_k\right)\right) - p_k Tr\left(\rho_k \log\left(\sigma\right)\right)\right) \\ &= \sum_k \left( p_k Tr\left(\rho_k \log\left(\rho_k\right)\right)\right) - Tr\left( \sum_k \left( p_k \rho_k \log\left(\sigma\right)\right)\right) \\ &= \sum_k \left( p_k Tr\left(\rho_k \log\left(\rho_k\right)\right)\right) - Tr\left(\sigma \log\left(\sigma\right)\right) \\ &= \mathrm{S}\left(\sigma\right) - \sum_k p_k \mathrm{S}\left(\rho_k\right) = \mathcal{X}. \end{aligned} \tag{5.72}$$

The result of (5.72) will have great importance later since it describes the connection between the numerical and geometric methods. The fact that the Holevo information can be described in terms of quantum relative entropy function, (see (5.68)) will provide the base of the geometric computation of quantum channel capacity.

It can therefore be concluded that the HSW channel capacity $C\left(\mathcal{N}\right)$ in terms of the quantum relative entropy can be expressed as [Petz96,08], [Schumacher99,2000], [Cortese02]

$$C\left(\mathcal{N}\right) = \max_{all\ p_k, \psi_k} \mathcal{X} = \max_{all\ p_k, \psi_k} \sum_k p_k D\left(\mathcal{N}\left(\psi_k\right) \| \mathcal{N}\left(\psi\right)\right) \tag{5.73}$$

where $\psi_k$ denotes the *pure input* quantum states of channel $\mathcal{N}$ and $\psi = \sum_k p_k \psi_k$.

### 5.4.3 Quantum Relative Entropy in the Bloch Sphere Representation

As we have seen in (5.73), the HSW capacity $C\left(\mathcal{N}\right)$ of quantum channel $\mathcal{N}$ can be given in a geometric representation by quantum relative entropy function $D(\cdot\|\cdot)$. The radius of the smallest quantum informational ball uses the result that the Holevo information can be measured in terms of quantum relative entropy function, (see (5.72)), and its maximized value will be equal to the capacity of the analyzed quantum channel. This is the first main element. The second: the quantum relative entropy function has a strict geometric analogy.
This quantum ball contains the channel ellipsoid, and inside the quantum ball the distances between the quantum states are measured by the quantum relative entropy function (see (5.70)), and in the geometric representation it uses the negative von Neumann entropy generator function (see (5.44)), and its extended version, the quantum relative entropy function, see (5.45). From now on, we refer the relative entropy based ball as the *smallest quantum informational ball*. The radius of this quantum ball is already defined in (5.70).
We show an example of a two-dimensional smallest enclosing quantum informational ball in Fig. 5.11. This quantum relative entropy ball is a deformed ball, thus the approximation algorithm has to be tailored for quantum informational distance. The center $\mathbf{c}^*$ of the smallest enclosing quantum informational ball differs from the center of an Euclidean ball.

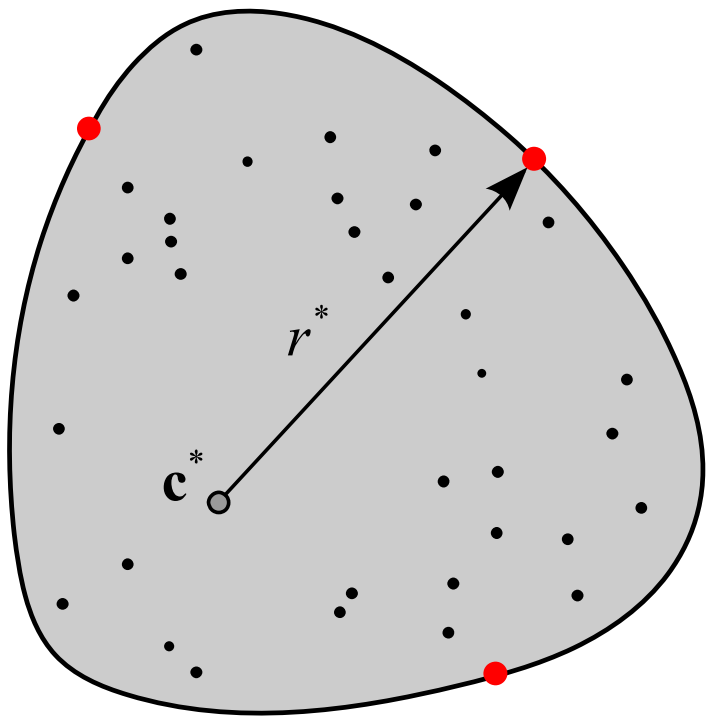

**Fig. 5.11.** The smallest enclosing quantum informational ball.

In the geometric representation of $C(\mathcal{N})$, the *maximum* is taken over the *surface* of the channel ellipsoid, and the *minimum* is taken over the *interior* of the ellipsoid. The distance calculations between the quantum states are based on the quantum relative entropy function $D(\rho\|\sigma)$. As it was shown by Cortese [Cortese02] using the work of Schumacher and Westmoreland [Schumacher99, 2000] and later by Kato et al. [Kato06] and Nielsen et al. [Nielsen07-07a,08], the quantum relative entropy function $D(\cdot\|\cdot)$ for an arbitrary quantum state $\rho = (x, y, z)$ and mixed state $\sigma = (\tilde{x}, \tilde{y}, \tilde{z})$, with radii $r_\rho = \sqrt{x^2 + y^2 + z^2}$ and $r_\sigma = \sqrt{\tilde{x}^2 + \tilde{y}^2 + \tilde{z}^2}$ is given by

$$\begin{aligned} D(\rho \| \sigma) = \frac{1}{2}\log\left(\frac{1}{4}\left(1 - r_\rho{}^2\right)\right) + \frac{1}{2} r_\rho \log\left(\frac{\left(1 + r_\rho\right)}{\left(1 - r_\rho\right)}\right) \\ - \frac{1}{2}\log\left(\frac{1}{4}\left(1 - r_\sigma{}^2\right)\right) - \frac{1}{2r_\sigma}\log\left(\frac{\left(1 + r_\sigma\right)}{\left(1 - r_\sigma\right)}\right)\langle \rho, \sigma \rangle, \end{aligned} \tag{5.74}$$

where $\langle \rho, \sigma \rangle = (x\tilde{x} + y\tilde{y} + z\tilde{z})$. For a maximally mixed state $\sigma = (\tilde{x}, \tilde{y}, \tilde{z}) = (0,0,0)$ and $r_\sigma = 0$, the quantum relative entropy can be expressed as

$$D(\rho \| \sigma) = \frac{1}{2}\log\left(\frac{1}{4}\left(1 - r_\rho{}^2\right)\right) + \frac{1}{2} r_\rho \log\left(\frac{\left(1 + r_\rho\right)}{\left(1 - r_\rho\right)}\right) - \frac{1}{2}\log\left(\frac{1}{4}\right). \tag{5.75}$$

The quantum relative entropy between two mixed quantum states depends on the lengths of their Bloch vectors and the angle $\theta$ between them, as illustrated it in Fig. 5.12.

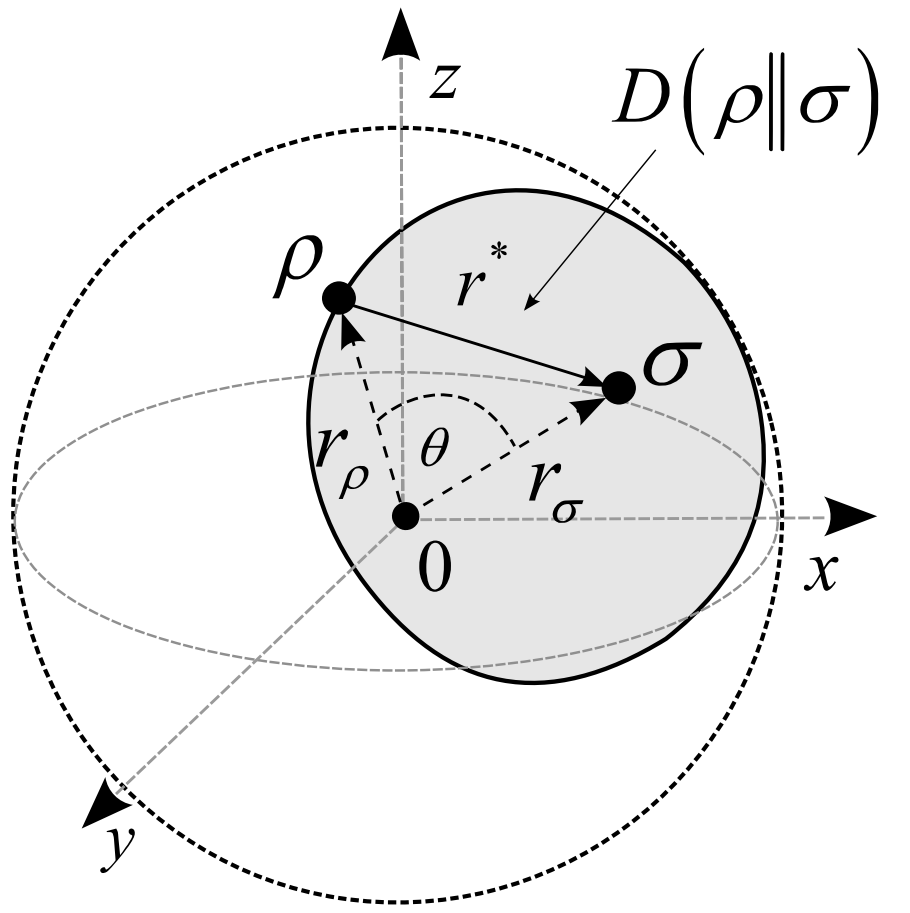

**Fig. 5.12.** The quantum relative entropy between two mixed quantum states depends on the lengths of their Bloch vectors and the angle between them.

Let assume, we have a maximally mixed state $\sigma = \frac{1}{2}I$ with $r_\sigma = 0$, in this case the radius of the quantum ball (i.e., the capacity of the analyzed quantum channel) will be equal to

$$D(\rho\|\sigma) = D\left(\rho\left\|\frac{1}{2}I\right.\right) = \frac{1}{2}\log\left(1 - r_\rho^{\ 2}\right) + \frac{r_\rho}{2}\log\left(\frac{1 + r_\rho}{1 - r_\rho}\right) = 1 - \mathrm{S}(\rho). \quad (5.76)$$

If we have a unital quantum channel, then the average state is equal to $\sigma = \frac{1}{2}I$, hence we can use (5.76) to determine the HSW capacity of the quantum channel [Schumacher99], [Cortese02], [Kato06]. It has to be emphasized that based on (5.74) and (5.75), the relative entropy function between density matrices is equivalent to the relative entropy between the Bloch vectors $\mathbf{r}_\rho$, $\mathbf{r}_\sigma$ for the quantum states $\rho$ and $\sigma$

$$D(\rho\|\sigma) = D(\mathbf{r}_\rho\|\mathbf{r}_\sigma). \quad (5.77)$$

The results of Schumacher and Westmoreland [Schumacher2000] are based on the same fact, hence a geometric approach can be defined to measure distances on the Bloch sphere, using quantum relative entropy as distance measure function.

### 5.4.4 Derivation of the Quantum Relative Entropic Formula

In this section we derive the quantum relative entropy based distance measure formula. From now on, based on (5.77), we will refer to the density matrices $\rho$ and $\sigma$ as follows:

$$\mathbf{r}_\rho = [x, y, z], \quad (5.78)$$

and

$$\mathbf{r}_\sigma = [\tilde{x}, \tilde{y}, \tilde{z}]. \quad (5.79)$$

The Bloch vectors $\mathbf{r}_\rho$ and $\mathbf{r}_\sigma$ are real, 3D vectors, have unit length for pure states and less than one for mixed states. Using (5.78) and (5.79), the density matrices can be expressed as

$$\rho = \frac{1}{2}\left(I + \mathbf{r}_\rho \cdot \sigma\right), \quad (5.80)$$

and

$$\sigma = \frac{1}{2}\left(I + \mathbf{r}_\sigma \cdot \sigma\right). \tag{5.81}$$

The density matrices for quantum states with Bloch vectors $\mathbf{r}_\rho = \left[x, y, z\right]$ and $\mathbf{r}_\sigma = \left[\tilde{x}, \tilde{y}, \tilde{z}\right]$, and radii $r_\rho = \sqrt{x^2 + y^2 + z^2}$ and $r_\sigma = \sqrt{\tilde{x}^2 + \tilde{y}^2 + \tilde{z}^2}$, in terms of Bloch vector representation are

$$\rho = \begin{bmatrix} \frac{1}{2}\left(1 + z\right) & \frac{1}{2}\left(x - iy\right) \\ \frac{1}{2}\left(x + iy\right) & \frac{1}{2}\left(1 - z\right) \end{bmatrix}, \text{ and } \sigma = \begin{bmatrix} \frac{1}{2}\left(1 + \tilde{z}\right) & \frac{1}{2}\left(\tilde{x} - i\tilde{y}\right) \\ \frac{1}{2}\left(\tilde{x} + i\tilde{y}\right) & \frac{1}{2}\left(1 - \tilde{z}\right) \end{bmatrix}. \tag{5.82}$$

The eigenvalues $\left(\lambda_\rho^{(1)}, \lambda_\rho^{(2)}\right)$, $\left(\lambda_\sigma^{(1)}, \lambda_\sigma^{(2)}\right)$ of the density matrices $\rho$ and $\sigma$ can be expressed as

$$\begin{aligned} \lambda_\rho^{(1)} &= \frac{1}{2}\left(1 + \sqrt{x^2 + y^2 + z^2}\right) = \frac{1 + r_\rho}{2}, \\ \lambda_\rho^{(2)} &= \frac{1}{2}\left(1 - \sqrt{x^2 + y^2 + z^2}\right) = \frac{1 - r_\rho}{2}, \end{aligned} \tag{5.83}$$

and for density matrix $\sigma$

$$\begin{aligned} \lambda_\sigma^{(1)} &= \frac{1}{2}\left(1 + \sqrt{\tilde{x}^2 + \tilde{y}^2 + \tilde{z}^2}\right) = \frac{1 + r_\sigma}{2}, \\ \lambda_\sigma^{(2)} &= \frac{1}{2}\left(1 - \sqrt{\tilde{x}^2 + \tilde{y}^2 + \tilde{z}^2}\right) = \frac{1 - r_\sigma}{2}. \end{aligned} \tag{5.84}$$

We would like to derive a formula for quantum relative entropy $D\left(\rho \| \sigma\right)$ in terms of Bloch sphere vectors $\mathbf{r}_\sigma$ and $\mathbf{r}_\rho$, (see (5.78) and (5.79)) with vector lengths $r_\rho = \sqrt{x^2 + y^2 + z^2}$ and $r_\sigma = \sqrt{\tilde{x}^2 + \tilde{y}^2 + \tilde{z}^2}$.

We can express the quantum relative entropy formula in two terms [Cortese03] $D_A$ and $D_B$

$$D\left(\rho \| \sigma\right) = D_A - D_B. \tag{5.85}$$

To expand $D_A$, we use the eigenvalues $\left(\lambda_\rho^{(1)}, \lambda_\rho^{(2)}\right)$ of $\rho$:

$$\begin{aligned} D_A &= Tr\left(\rho \log \rho\right) = \lambda_\rho^{(1)} \log\left(\lambda_\rho^{(1)}\right) + \lambda_\rho^{(2)} \log\left(\lambda_\rho^{(2)}\right) \\ &= \left(\frac{1+r_\sigma}{2}\right)\log\left(\frac{1+r_\sigma}{2}\right) + \left(\frac{1-r_\sigma}{2}\right)\log\left(\frac{1-r_\sigma}{2}\right) \\ &= -1 + \left(\frac{1+r_\sigma}{2}\right)\log\left(1+r_\sigma\right) + \left(\frac{1-r_\sigma}{2}\right)\log\left(1-r_\sigma\right). \end{aligned} \tag{5.86}$$

As it can be concluded from (5.83) and (5.86), $D_A = -\mathrm{S}\left(\rho\right)$, where $\mathrm{S}$ is the von Neumann entropy of the density matrix $\rho$. The second term $D_B$ can be expressed in the *basis* which diagonalizes the density matrix $\sigma$ as

$$D_B = Tr\left(\rho \log \sigma\right) = \log\left(\lambda_\rho^{(1)}\right) Tr\left(\rho \left|e_1\right\rangle\left\langle e_1\right|\right) + \log\left(\lambda_\rho^{(2)}\right) Tr\left(\rho \left|e_2\right\rangle\left\langle e_2\right|\right), \tag{5.87}$$

where $\left|e_1\right\rangle$ and $\left|e_2\right\rangle$ are the eigenvectors of density matrix $\sigma$. After some numerical calculations, the two eigenvectors can be expressed as

$$\left|e_1\right\rangle = N_1 \begin{bmatrix} 1 \\ \left(-1\right)\left[ -2\left(\frac{1}{2}\left(1+\sqrt{x^2+y^2+z^2}\right)\right) r_x^\rho - \frac{2i\left(\frac{1}{2}\left(1+\sqrt{x^2+y^2+z^2}\right)\right) y + x + iy + xz + iyz}{x^2+y^2} \right] \end{bmatrix} \tag{5.88}$$

and

$$\left|e_2\right\rangle = N_2 \begin{bmatrix} 2\left(\frac{1}{2}\left(1+\sqrt{x^2+y^2+z^2}\right)\right) x - \frac{2i\left(\frac{1}{2}\left(1+\sqrt{x^2+y^2+z^2}\right)\right) y - x + iy + xz + iyz}{x^2+y^2} \\ 1 \end{bmatrix}, \tag{5.89}$$

where parameters $N_1$ and $N_2$ are the normalization constants

$$N_1 = N_2 = \sqrt{2\left(\frac{x^2 + \left(\sqrt{x^2 + y^2 + z^2}\right)z + y^2 + z^2}{x^2 + y^2}\right)}. \tag{5.90}$$

Since, from (5.81) we know that the Bloch sphere representation of density matrix $\sigma$ is $\sigma = \frac{1}{2}\left(I + \mathbf{r}_\sigma \cdot \vec{\sigma}\right)$, we can use this result to express $D_B$

$$\begin{aligned} D_B = \frac{1}{2}\log\left(\lambda_\sigma^{(1)}\right)\left(Tr\left(\left|e_1\right\rangle\left\langle e_1\right|\right) + \sum_{i=x,y,z} \rho_{(x,y,z)} Tr\left(\sigma_i \left|e_1\right\rangle\left\langle e_1\right|\right)\right) + \\ \frac{1}{2}\log\left(\lambda_\sigma^{(2)}\right)\left(Tr\left(\left|e_2\right\rangle\left\langle e_2\right|\right) + \sum_{i=x,y,z} \rho_{(x,y,z)} Tr\left(\sigma_i \left|e_2\right\rangle\left\langle e_2\right|\right)\right). \end{aligned} \tag{5.91}$$

The values of $Tr\left(\left|e_1\right\rangle\left\langle e_1\right|\right)$ and $Tr\left(\left|e_2\right\rangle\left\langle e_2\right|\right)$ are 1, because $\left|e_1\right\rangle\left\langle e_1\right|$ and $\left|e_2\right\rangle\left\langle e_2\right|$ define density matrices with unit trace. We can define parameter $\Upsilon_i^{(j)}$ as

$$\Upsilon_i^{(j)} = Tr\left(\sigma_i \left|e_j\right\rangle\left\langle e_j\right|\right) = \left\langle e_j\right|\sigma_i\left|e_j\right\rangle, \tag{5.92}$$

from which

$$D_B = \frac{1}{2}\log\left(\lambda_\sigma^{(1)}\right)\left(1 + \sum_{x,y,z} \rho_{(x,y,z)} \Upsilon_{(x,y,z)}^{(1)}\right) + \frac{1}{2}\log\left(\lambda_\sigma^{(2)}\right)\left(1 + \sum_{x,y,z} \rho_{(x,y,z)} \Upsilon_{(x,y,z)}^{(2)}\right), \tag{5.93}$$

where the parameters $\Upsilon_i^{(1)}$ can be given as follows

$$\Upsilon_x^{(1)} = \frac{\tilde{x}\left(\sqrt{\tilde{x}^2 + \tilde{y}^2 + \tilde{z}^2} + \tilde{z}\right)}{\tilde{x}^2 + \left(\sqrt{\tilde{x}^2 + \tilde{y}^2 + \tilde{z}^2}\right)\tilde{z} + \tilde{y}^2 + \tilde{z}^2} = \frac{\tilde{x}\left(r_\sigma + \tilde{z}\right)}{\left(r_\sigma\right)^2 + r_\sigma \tilde{z}} = \frac{\tilde{x}}{r_\sigma},$$
$$\Upsilon_y^{(1)} = \frac{\tilde{y}\left(\sqrt{\tilde{x}^2 + \tilde{y}^2 + \tilde{z}^2} + \tilde{z}\right)}{\tilde{x}^2 + \left(\sqrt{\tilde{x}^2 + \tilde{y}^2 + \tilde{z}^2}\right)\tilde{z} + \tilde{y}^2 + \tilde{z}^2} = \frac{\tilde{y}\left(r_\sigma + \tilde{z}\right)}{\left(r_\sigma\right)^2 + r_\sigma \tilde{z}} = \frac{\tilde{y}}{r_\sigma}, \quad (5.94)$$
$$\Upsilon_z^{(1)} = \frac{\tilde{z}\left(\sqrt{\tilde{x}^2 + \tilde{y}^2 + \tilde{z}^2} + \tilde{z}\right)}{\tilde{x}^2 + \left(\sqrt{\tilde{x}^2 + \tilde{y}^2 + \tilde{z}^2}\right)\tilde{z} + \tilde{y}^2 + \tilde{z}^2} = \frac{\tilde{z}\left(r_\sigma + \tilde{z}\right)}{\left(r_\sigma\right)^2 + r_\sigma \tilde{z}} = \frac{\tilde{z}}{r_\sigma}.$$

Similarly, parameters of $\Upsilon_i^{(2)}$ can be expressed as

$$\Upsilon_x^{(2)} = \frac{\tilde{x}\left(\sqrt{\tilde{x}^2 + \tilde{y}^2 + \tilde{z}^2} - \tilde{z}\right)}{-\tilde{x}^2 + \left(\sqrt{\tilde{x}^2 + \tilde{y}^2 + \tilde{z}^2}\right)\tilde{z} - \tilde{y}^2 - \tilde{z}^2} = -\frac{\tilde{x}\left(r_\sigma - \tilde{z}\right)}{\left(r_\sigma\right)^2 - r_\sigma\left(\tilde{z}\right)} = -\frac{\tilde{x}}{r_\sigma},$$
$$\Upsilon_y^{(2)} = \frac{\tilde{y}\left(\sqrt{\tilde{x}^2 + \tilde{y}^2 + \tilde{z}^2} - \tilde{z}\right)}{-\tilde{x}^2 + \left(\sqrt{\tilde{x}^2 + \tilde{y}^2 + \tilde{z}^2}\right)\tilde{z} - \tilde{y}^2 - \tilde{z}^2} = -\frac{\tilde{y}\left(r_\sigma - \tilde{z}\right)}{\left(r_\sigma\right)^2 - r_\sigma\left(\tilde{z}\right)} = -\frac{\tilde{y}}{r_\sigma}, \quad (5.95)$$
$$\Upsilon_z^{(2)} = \frac{\tilde{z}\left(\sqrt{\tilde{x}^2 + \tilde{y}^2 + \tilde{z}^2} - \tilde{z}\right)}{-\tilde{x}^2 + \left(\sqrt{\tilde{x}^2 + \tilde{y}^2 + \tilde{z}^2}\right)\tilde{z} - \tilde{y}^2 - \tilde{z}^2} = -\frac{\tilde{z}\left(r_\sigma - \tilde{z}\right)}{\left(r_\sigma\right)^2 - r_\sigma\left(\tilde{z}\right)} = -\frac{\tilde{z}}{r_\sigma}.$$

Using the results for $\Upsilon_i^{(1)}$ and $\Upsilon_i^{(2)}$, $D_B$ can be expressed as

$$\begin{aligned} D_B &= \frac{1}{2}\log\left(\lambda_\sigma^{(1)}\right)\left[1 + \sum_{x,y,z} \rho_{(x,y,z)} \Upsilon_{(x,y,z)}^{(1)}\right] + \frac{1}{2}\log\left(\lambda_\sigma^{(2)}\right)\left[1 + \sum_{x,y,z} \rho_{(x,y,z)} \Upsilon_{(x,y,z)}^{(2)}\right] \\ &= \frac{1}{2}\left[1 + \sum_{x,y,z} \rho_{(x,y,z)} \frac{\sigma_{(\tilde{x},\tilde{y},\tilde{z})}}{r_\sigma}\right]\log\left(\lambda_\sigma^{(1)}\right) + \frac{1}{2}\left[1 + \sum_{x,y,z} \rho_{(x,y,z)} \frac{-\sigma_{(\tilde{x},\tilde{y},\tilde{z})}}{r_\sigma}\right]\log\left(\lambda_\sigma^{(2)}\right) \\ &= \frac{1}{2}\left[1 + \frac{\mathbf{r}_\rho \mathbf{r}_\sigma}{r_\sigma}\right]\log\left(\lambda_\sigma^{(1)}\right) + \frac{1}{2}\left[1 - \frac{\mathbf{r}_\rho \mathbf{r}_\sigma}{r_\sigma}\right]\log\left(\lambda_\sigma^{(2)}\right). \end{aligned} \quad (5.96)$$

Substituting the previously derived results for the eigenvalues $\lambda_\sigma^{(1)}$ and $\lambda_\sigma^{(2)}$, $D_B$ can be given by

$$\begin{aligned} D_B &= \frac{1}{2}\left[1+\frac{\mathbf{r}_\rho \mathbf{r}_\sigma}{r_\sigma}\right]\log\left(\frac{1+r_\sigma}{2}\right)+\frac{1}{2}\left[1-\frac{\mathbf{r}_\rho \mathbf{r}_\sigma}{r_\sigma}\right]\log\left(\frac{1-r_\sigma}{2}\right) \\ &= \frac{1}{2}\log\left(1-\left(r_\sigma\right)^2\right)-1+\frac{\mathbf{r}_\rho \mathbf{r}_\sigma}{2r_\sigma}\log\left(\frac{1+r_\sigma}{1-r_\sigma}\right). \end{aligned} \tag{5.97}$$

Now we can calculate the quantum relative entropy $D\left(\rho\|\sigma\right)$ between density matrices $\rho$ and $\sigma$ in the Bloch-sphere representation

$$\begin{aligned} D\left(\rho\|\sigma\right) &= D_A - D_B \\ &= \frac{1}{2}\log\left(1-\left(r_\rho\right)^2\right)+\frac{r_\rho}{2}\log\left(\frac{1+r_\rho}{1-r_\rho}\right)-\frac{1}{2}\log\left(1-\left(r_\sigma\right)^2\right)-\frac{\mathbf{r}_\rho \mathbf{r}_\sigma}{2r_\sigma}\log\left(\frac{1+r_\sigma}{1-r_\sigma}\right) \\ &= \frac{1}{2}\log\left(1-\left(r_\rho\right)^2\right)+\frac{r_\rho}{2}\log\left(\frac{1+r_\rho}{1-r_\rho}\right)-\frac{1}{2}\log\left(1-\left(r_\sigma\right)^2\right)-\frac{r_\rho\cos\left(\theta\right)}{2}\log\left(\frac{1+r_\sigma}{1-r_\sigma}\right), \end{aligned} \tag{5.98}$$

where $\theta$ is the angle between Bloch vectors $\mathbf{r}_\rho^T = \left[x,y,z\right]$ and $\mathbf{r}_\sigma^T = \left[\tilde{x},\tilde{y},\tilde{z}\right]$, and the length of the radii are $\left|\mathbf{r}_\rho\right| = r_\rho = \sqrt{x^2+y^2+z^2}$ and $\left|\mathbf{r}_\sigma\right| = r_\sigma = \sqrt{\tilde{x}^2+\tilde{y}^2+\tilde{z}^2}$. We note, if $\sigma = \frac{1}{2}I$, then $r_\sigma = 0$, thus the von Neumann entropy of state $\sigma$ will be equal to one, i.e., we use the following relation

$$\mathrm{S}\left(\sigma\right) = H\left(\frac{1}{2}\left(1-r_\sigma\right)\right), \tag{5.99}$$

which will result in

$$\begin{aligned} D\left(\rho\|\sigma\right) &= D\left(\rho\middle\|\frac{1}{2}I\right) \\ &= \frac{1+r_\rho}{2}\log\left(\frac{1+r_\rho}{2}\right)+\frac{1+r_\rho}{2}+\frac{1-r_\rho}{2}\log\left(\frac{1-r_\rho}{2}\right)+\frac{1-r_\rho}{2} \\ &= 1-\mathrm{S}\left(\rho\right). \end{aligned} \tag{5.100}$$

For a $d$-dimensional system, $D\left(\rho\middle\|\frac{1}{2}I\right)$ can be expressed as

$$D\left(\rho \middle\| \frac{1}{2}I\right) = Tr\left(\rho\left(\log\left(\rho\right) - \log\left(\frac{1}{2}I\right)\right)\right)$$
$$= Tr\left(\rho\left(\log\left(\rho\right) - \log\left(d\right)\left(\frac{1}{2}I\right)\right)\right) = \log\left(d\right)Tr\left(\rho\right) - \mathrm{S}\left(\rho\right) \qquad (5.101)$$
$$= \log\left(d\right) - \mathrm{S}\left(\rho\right).$$

The relative entropy between two density matrices $\rho$ and $\sigma$, can be expressed in the Bloch sphere representation. The derived formula is not symmetric in general, hence

$$D\left(\rho \| \sigma\right) \neq D\left(\sigma \| \rho\right), \qquad (5.102)$$

except, if $\mathbf{r}_\rho = \mathbf{r}_\rho$. For the geometric meaning of (5.102) see Fig. 5.13. The contours of $D\left(\rho \| \sigma\right)$ changes in function of $\rho$, the average state $\sigma$ is the fixed maximally mixed state in the center of the Bloch sphere. The average quantum state, $\sigma = \sum_k p_k \rho_k$, is denoted by $\mathbf{c}^*$, the quantum informational radii are denoted by $r_i$, $i = 1,2,3,4$. The quantum informational distances are measured by the length of the vectors indexed from 1 to 4, with the following values $r_1 = 0.01$, $r_2 = 0.2$, $r_3 = 0.4$ and $r_4 = 0.7$.

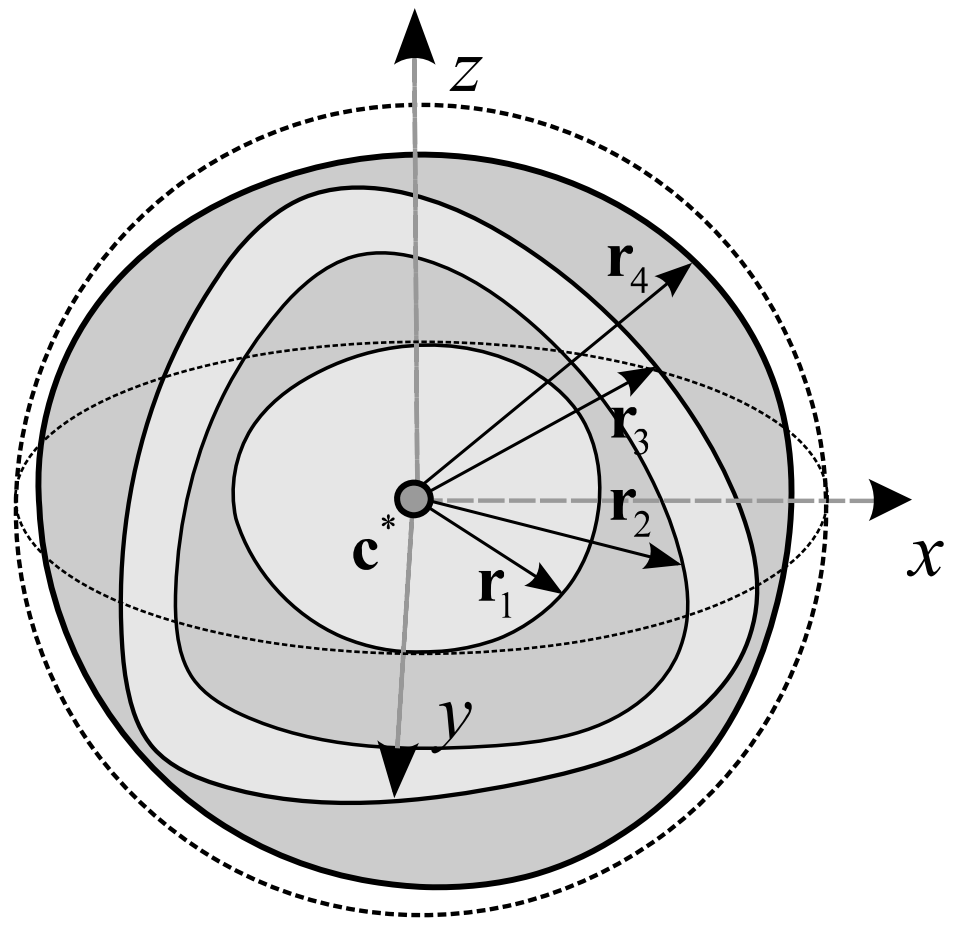

**Fig. 5.13.** The smallest enclosing quantum informational ball for different radii lengths.

To see clearly the connection between the Euclidean length and the quantum informational distance of the radii, we give their comparison in Table 5.1.

| *Quantum Informational length* | *Euclidean length* |
|---|---|
| 0 | 0 |
| 0.01 | 0.1 |
| 0.2 | 0.5 |
| 0.4 | 0.7 |
| 0.7 | 0.9 |
| 1 | 1 |

**Table 5.1.** Comparison of Euclidean and quantum informational distances.

As it can be concluded from Fig. 5.13 and Table 5.1, the quantum informational distance differs from the Euclidean distance function. The length of the quantum informational radius cannot be described by the length of the Bloch vector. Moreover, the quantum ball has a distorted structure which roots in the quantum relative entropy function-based distance calculations.

### 5.4.5 Analytical Derivation of the HSW Channel Capacity of Depolarizing Quantum Channel

The depolarizing quantum channel was introduced in Section 3.6.2. Here, we derive analytically the HSW capacity of this channel. In the next section we show, that the same result can be obtained in a geometric way. The first term of the HSW capacity is the von Neumann entropy of the average quantum system $\sigma$, which in this case can be derived as

$$\begin{aligned}\mathrm{S}\left(\sigma\right) = \mathrm{S}\left(\mathcal{N}\left(\sum_i p_i\rho_i\right)\right) = -\Bigg[p\frac{1}{2} + \left(1-p\right)p_0 \log\left(p\frac{1}{2} + \left(1-p_0\right)p_0\right) \\ +\left(p\frac{1}{2} + \left(1-p\right)\left(1-p_0\right)\right)\log\left(p\frac{1}{2} + \left(1-p\right)\left(1-p_0\right)\right)\Bigg]=1.\end{aligned} \tag{5.103}$$

The second term would be equal to $H\left(0\right)=0$ if the channel parameter would be $p$=0. In general, we have to calculate with the form

$$\begin{aligned}\sum_i p_i \mathrm{S}(\mathcal{N}(\rho_i)) &= p_0 \mathrm{S}(\mathcal{N}(\rho_0)) + (1-p_0)\mathrm{S}(\mathcal{N}(\rho_1)) \\ &= p_0 \mathrm{S}\left(p\frac{1}{2}I\right) + (1-p_0)\mathrm{S}\left(p\frac{1}{2}I\right) \\ &= p_0 H\left(p\frac{1}{2}\right) + (1-p_0)H\left(p\frac{1}{2}\right) = H\left(p\frac{1}{2}\right).\end{aligned} \tag{5.104}$$

From these results, and using the fact, that

$$\sum_i p_i \mathrm{S}(\mathcal{N}(\rho_i)) = \mathrm{S}(\mathcal{N}(\rho_0)) = \mathrm{S}(\mathcal{N}(\rho_1)) = H\left(p\frac{1}{2}\right), \tag{5.105}$$

since in the case of depolarizing channel the von Neumann entropies of channel output states are equal to $H\left(p\frac{1}{2}\right)$. For the depolarizing channel every output state will have the same von Neumann entropy, independently what the original input was. Or in other words, the depolarizing channel defined in (3.70) does the following: it works either as a completely useless channel with a maximally mixed output, or it transmits the input with the same output entropy, based on the depolarizing parameter $p$. We obtain for the Holevo quantity

$$\begin{aligned}\mathcal{X} = -\Bigg\{&\left(p\frac{1}{2} + (1-p)p_0\right)\log\left(p\frac{1}{2} + (1-p)p_0\right) \\ &+ \left(p\frac{1}{2} + (1-p)(1-p_0)\right)\log\left(p\frac{1}{2} + (1-p)(1-p_0)\right)\Bigg\} - H\left(p\frac{1}{2}\right).\end{aligned} \tag{5.106}$$

Let us consider $p = 1$ i.e., a maximally depolarizing channel. In the geometric interpretation it means, that the average state $\sigma$ shown in (5.103) will be equal to the origin of the Bloch sphere. In this case (5.103) reaches its maximum if

$$p_0 = p_1 = \frac{1}{2}, \tag{5.107}$$

which will turn (5.103) to

$$\mathrm{S}\left(\mathcal{N}\left(\sum_i p_i \rho_i\right)\right) = -\left(\frac{1}{2}\log\left(\frac{1}{2}\right) + \frac{1}{2}\log\left(\frac{1}{2}\right)\right) = \mathrm{S}\left(\frac{1}{2}I\right) = H\left(\frac{1}{2}\right) = 1 \tag{5.108}$$

which leads to HSW capacity

$$C(\mathcal{N}) = \mathrm{S}\left(\frac{1}{2}I\right) - \mathrm{S}\left(\frac{1}{2}I\right) = 1 - 1 = 0\,, \tag{5.109}$$

i.e., we have a zero-capacity quantum channel, which transforms Alice's every state into the maximally mixed state $\frac{1}{2}I$. Setting $p = 0$ we produce a completely noiseless channel, hence we get

$$C(\mathcal{N}) = -\left(\frac{1}{2}\log\left(\frac{1}{2}\right) + \frac{1}{2}\log\left(\frac{1}{2}\right)\right) - \overbrace{\mathrm{S}\left(\rho_{pure}\right)}^{0} = \mathrm{S}\left(\frac{1}{2}I\right) = 1. \tag{5.110}$$

Finally, the HSW capacity of the depolarizing quantum channel in the function of parameter $p$ can be expressed as

$$C(\mathcal{N}) = 1 - \mathrm{S}\left(p\frac{1}{2}I\right) = 1 - H\left(\frac{1}{2}p\right). \tag{5.111}$$

### 5.4.6 Geometric Way to Determine the Classical Capacity of Depolarizing Quantum Channels

Here we show the steps of geometrical iteration for unital quantum channels, since the same results can be applied to depolarizing channels. The density matrix $\sigma$ must be expressible as a convex combination of $\rho_k$ as $\sigma = \sum_k p_k\rho_k$, which satisfies the min-max criteria:

$$C(\mathcal{N}) = \min_{\sigma}\max_{\rho} D\left(\rho\|\sigma\right) = \min_{\sigma}\max_{\rho_k} D\left(\rho_k \middle\| \sigma = \sum_k p_k\rho_k\right). \tag{5.112}$$

To find the best solution, the we seeking the optimum $\sigma$ and the quantum informational ball corresponding to the HSW channel capacity $C(\mathcal{N})$. The optimal initial state $\sigma$ can be given as a state vector $\mathbf{r}_\sigma$, in the iteration process, the algorithm modifies vector $\mathbf{r}_\sigma$ to find its optimum value. In Fig. 5.14 we illustrated an unacceptable situation, where the center $\mathbf{c}^* = \sigma = \sum_k p_k\rho_k$ of the smallest quantum informational ball does not lie inside the channel ellipsoid.

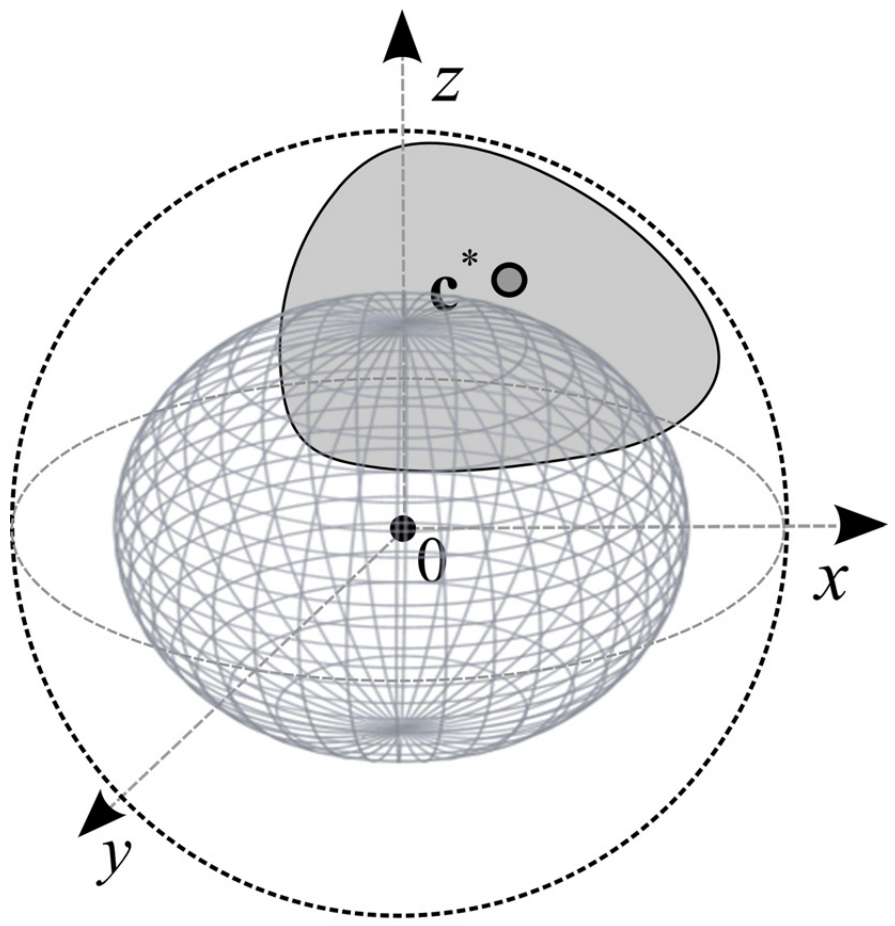

**Fig. 5.14.** An unacceptable situation, where the average state does not lie inside the channel ellipsoid.

In Fig. 5.15, we present another unacceptable configuration, in which there are no permissible $\rho_k$ channel output state, because the smallest enclosing quantum informational ball does not intersect the channel ellipsoid.

In the next example, the computed length of the radius is not acceptable, because state $\sigma$ must be expressible as a convex combination of density matrices $\rho_k$ which satisfy the min-max criteria.

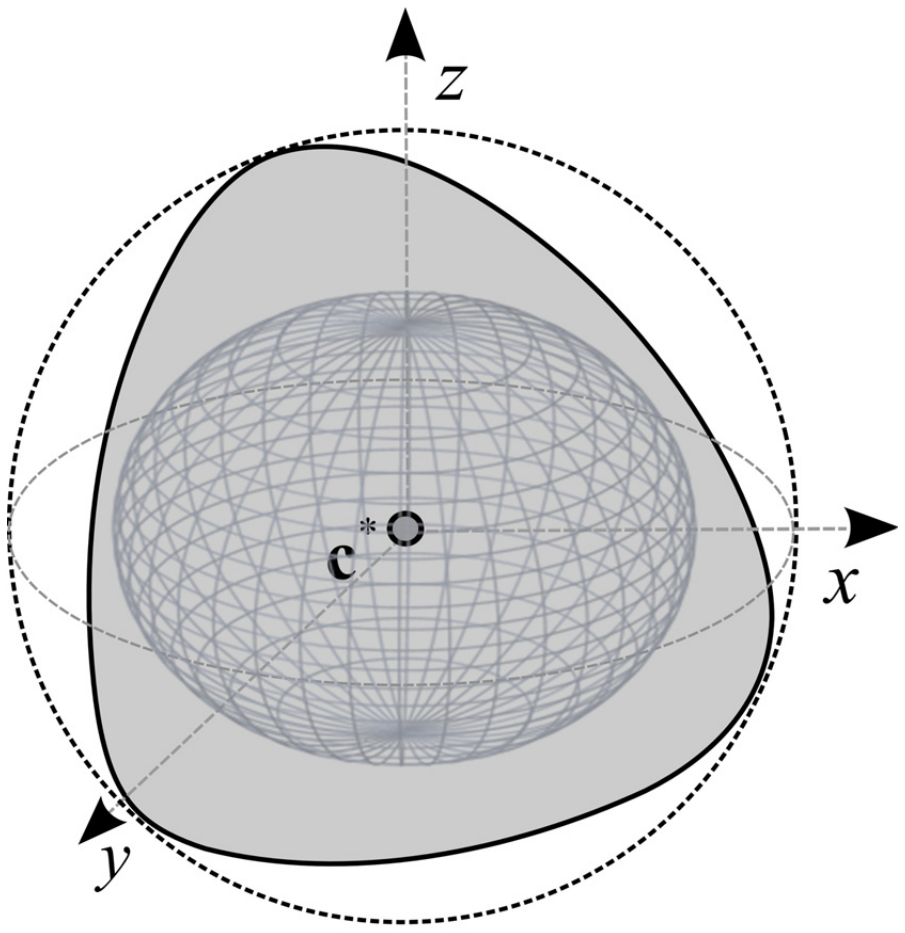

**Fig. 5.15.** Unacceptable situation, in which the smallest enclosing quantum informational ball does not intersect the channel ellipsoid.

As we have illustrated in Fig. 5.16, there is only one permissible $\rho_k$ density matrix denoted by $\rho_2$, and $\sigma \neq \rho_1$. It means, that only one optimal vector $\mathbf{r}_\sigma$ exists. If we find the point, where any movement of $\mathbf{r}_\sigma$ will increase $D_{\max}\left(\sigma\right) = D_{\max}\left(\mathbf{c}^*\right)$, we have found the final state of $\sigma = \mathbf{c}^*$. In this situation, the smallest enclosing quantum informational ball can not be used in the capacity analysis of the unital channel $\mathcal{N}$.

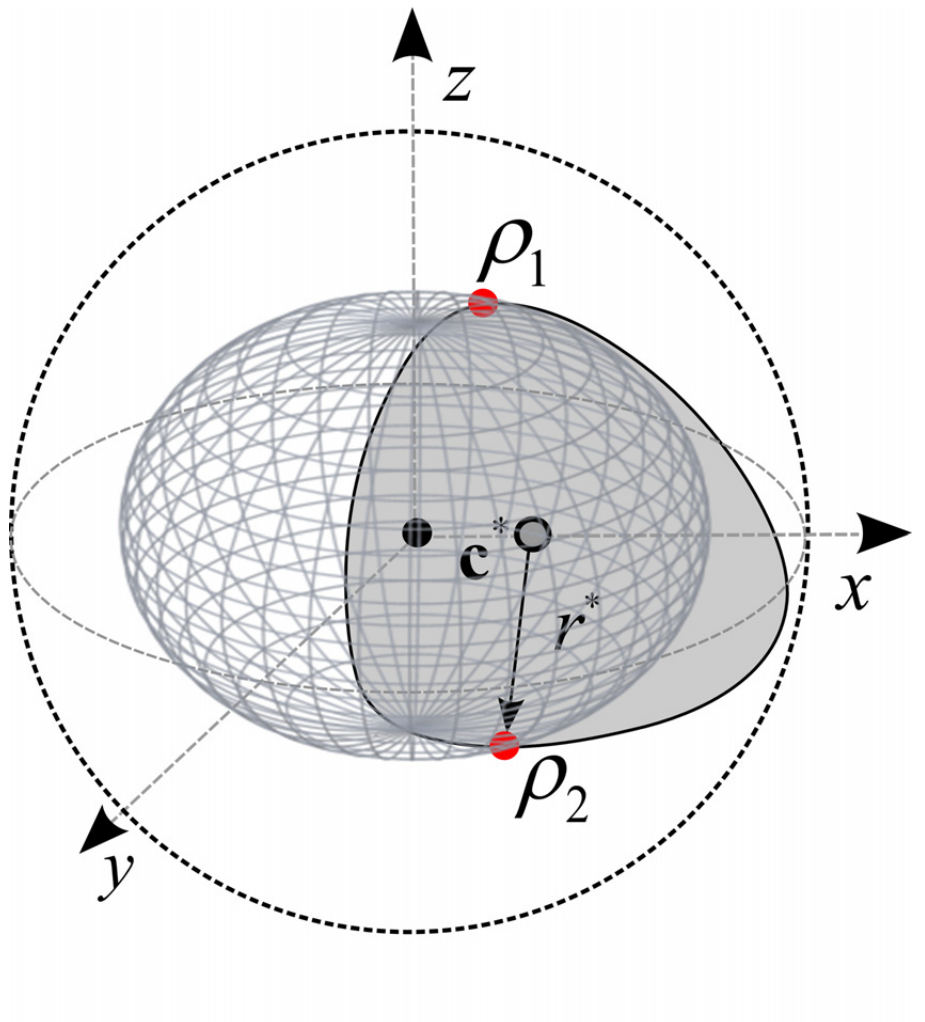

**Fig. 5.16.** The computed length of the radius is not acceptable, the average has to be expressible as a convex combination of density matrices of output states, which satisfies the min-max criteria.

In Fig. 5.17, we depict the situation, where the radius of the smallest enclosing informational ball is also not acceptable, because of the algorithm did not realize the maximization criteria in the iteration. Choosing a smallest enclosing quantum informational ball with larger radius $r^*$ could solve the problem.

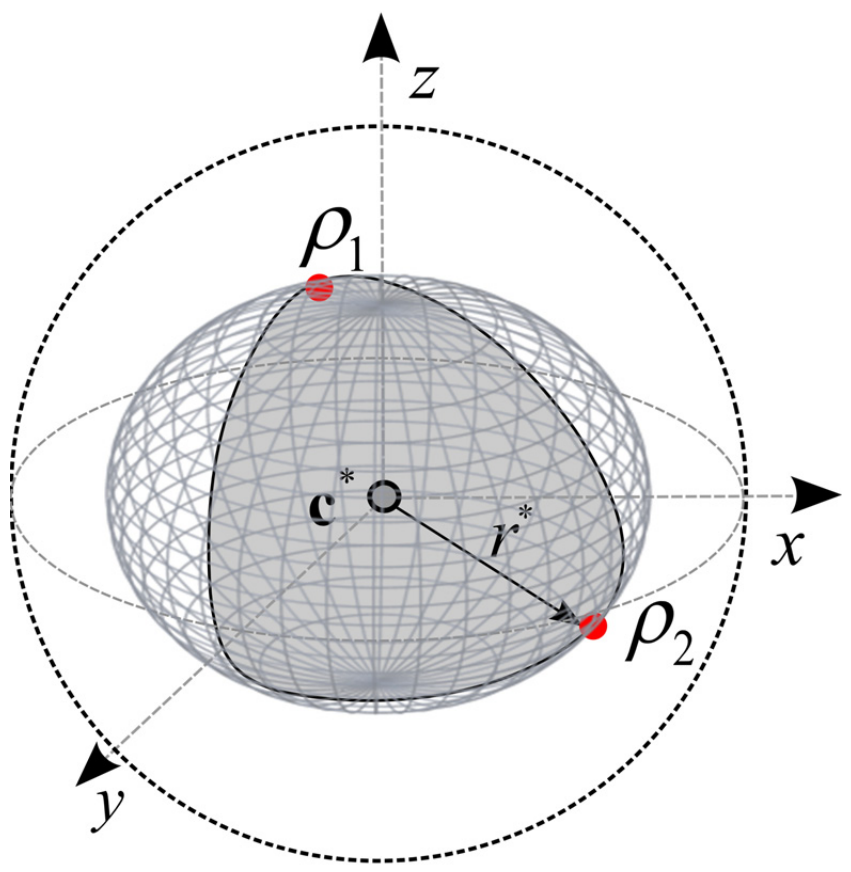

**Fig. 5.17.** The radius of the smallest enclosing informational ball is not acceptable, because of the algorithm did not realize the maximization criteria in the iteration.

In Fig. 5.18, we illustrated an acceptable situation, where smallest enclosing quantum informational ball intersects both of the channel endpoints $\rho_1$, and $\rho_2$. The optimal average point is denoted by $\sigma = \mathbf{c}^*$. The two states denoted by $\rho_1, \rho_2$ lie at the intersection of the smallest quantum informational ball and the channel ellipsoid, and these states could be used to form a convex combination $\rho_k$ that is equivalent to $\sigma = \mathbf{c}^* = \sum_k p_k \rho_k$. For this state, the quantum informational ball satisfies the requirements of Schumacher and Westmoreland [Schumacher99], therefore the smallest enclosing quantum informational ball can be used to analyze the capacity of unital quantum channel $\mathcal{N}$.

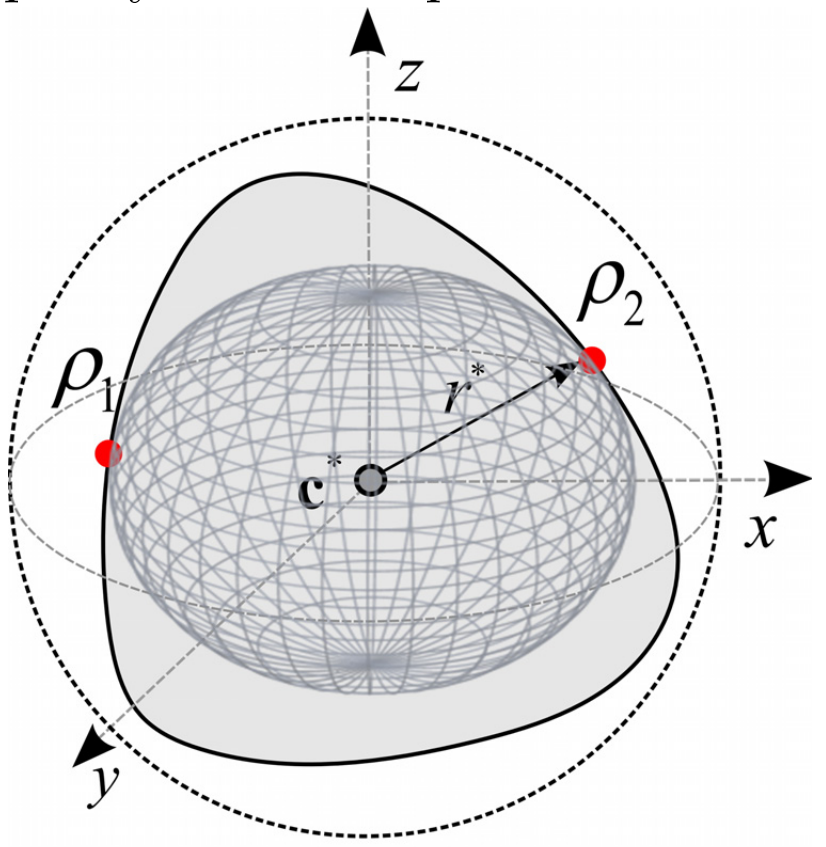

**Fig. 5.18.** An acceptable situation, where smallest enclosing quantum informational ball intersects both of the channel endpoints.

As it can be verified easily, we get the smallest enclosing quantum informational ball with maximum radius $r^*$. The length of radius is equal to HSW channel capacity

$$\begin{aligned} r^* &= C(\mathcal{N}) = \min_{\sigma} \max_{\rho} D(\rho \| \sigma) \\ &= \min_{\sigma} \max_{\rho_k} D\left(\rho_k \middle\| \sigma = \sum_k p_k \rho_k\right). \end{aligned} \tag{5.113}$$

The vector $\mathbf{r}_\sigma$ must lie in the channel ellipsoid between the two endpoints of the channel denoted by $\rho_1$ and $\rho_2$. We note, the optimum states found by the iteration for a unital quantum channel, are equal to the $\min_{\rho_{12}} \mathrm{S}(\rho_{12})$ minimum output von Neumann entropy states. Using the optimal state $\sigma$, we can increase the radius of the length of the smallest enclosing quantum informational ball by the moving of the state. In this case we will get the unacceptable situation, as we have illustrated it in Fig. 5.19. The original position and the optimal ball is denoted by dashed line and state with channel output state $\rho_1$ and center average state $\sigma = \mathbf{c}^*$. The length of the non-optimal radius is denoted by $r$.

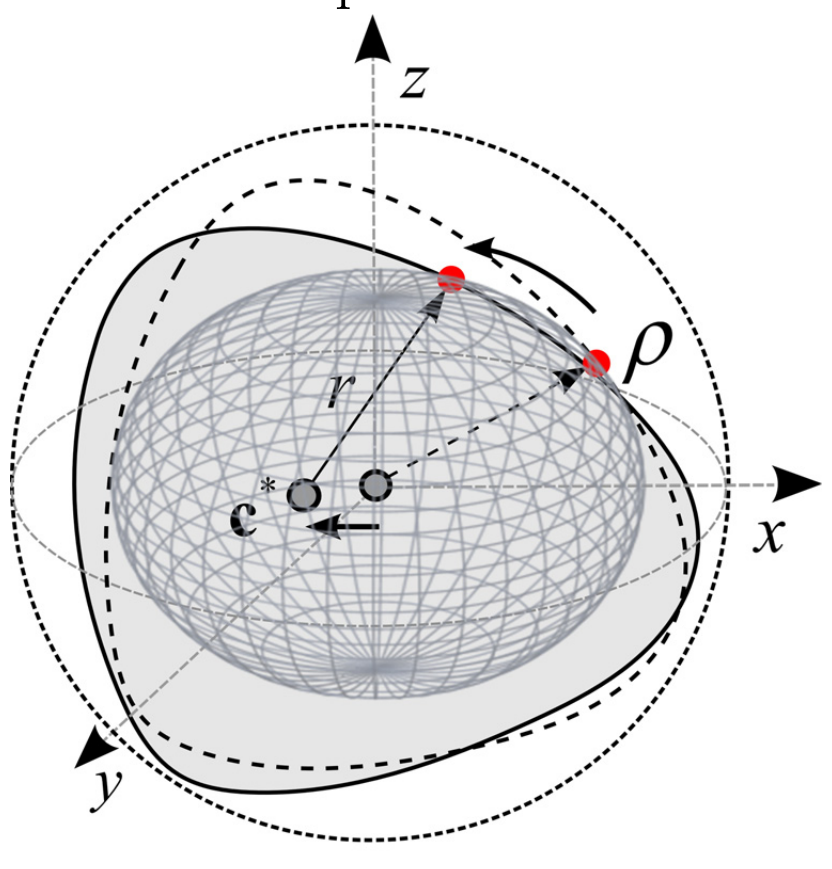

**Fig. 5.19.** By the moving of the optimal average state, the radius of the smallest enclosing quantum informational ball will increase, which will result an unacceptable situation.

As the quantum ball is moved, the relative entropy function $D(\rho_1 \| \sigma)$ between center $\sigma = \mathbf{c}^*$ and channel output state $\rho_1$ will increase. If we apply these iterations to a depolarizing quantum channel, then at the end of the fitting of the quantum informational ball, the radius of the quantum ball will be equal to

$$\begin{aligned} r^* &= C(\mathcal{N}) = \min_{\sigma} \max_{\rho} D(\rho \| \sigma) = \min_{\sigma} \max_{\rho_k} D\left(\rho_k \middle\| \sigma = \sum_k p_k \rho_k\right) \\ &= \sum_{all\ p_i} p_i D\left(\rho_i \middle\| \frac{1}{2} I\right) = 1 - \sum_i p_i \mathrm{S}(\mathcal{N}(\rho_i)) \\ &= \max_{all\ p_i, \rho_i} \chi = \max_{all\ p_i, \rho_i} \left[ \mathrm{S}\left(\mathcal{N}\left(\sum_i p_i \rho_i\right)\right) - \sum_i p_i \mathrm{S}(\mathcal{N}(\rho_i)) \right] \\ &= 1 - \mathrm{S}\left(p \frac{1}{2} I\right) = 1 - H\left(\frac{1}{2} p\right). \end{aligned} \tag{5.114}$$

As can be concluded from (5.114), the capacity (5.111) of the depolarizing channel derived in an analytical way is equivalent to the formula of (5.114), which describes the capacity using the geometric interpretation. As follows, there is a very elegant alternative way to compute the capacity of a unital depolarizing channel and to avoid the hard numerical computations. Of course these results can be applied to all quantum channel models [Cortese02] [Gyongyosi12].

### 5.4.7 Geometric Way to Determine the Capacity of Amplitude Damping Quantum Channel

The geometric interpretation of amplitude damping channel model firstly was shown in Section 3.6.3. Here, we parameterize the ellipsoid of the channel, and we describe the capacity of the channel. In Fig. 5.20(a) the Euclidean distances from the origin of the Bloch sphere to center $\mathbf{c}^*$ and to point $\rho$ are denoted by $m_\sigma$ and $m_\rho$, respectively. To determine the optimal length of vector $\mathbf{r}_\sigma$, the average state $\sigma$ is moved, which is denoted by $\mathbf{c}^*$ in the figures. As it moves vector $\mathbf{r}_\sigma$ from the optimum position, the larger quantum ball corresponding to a larger value of quantum relative entropy will intersect the channel ellipsoid surface, thereby increasing $\max_{\mathbf{r}_\rho} D(\mathbf{r}_\rho \| \mathbf{r}_\sigma)$.

The optimal quantum informational ball is illustrated in light-grey in Fig. 5.20(b).

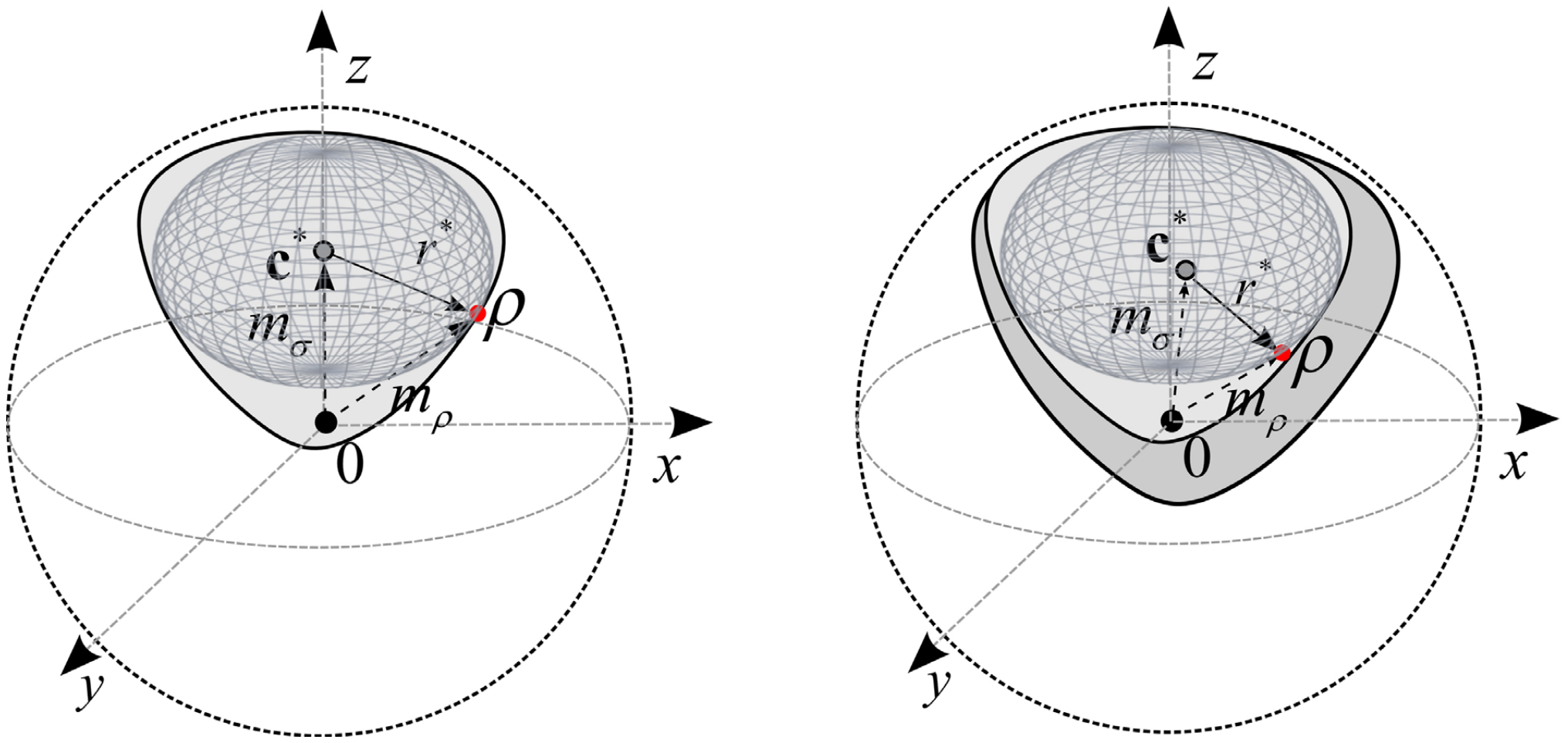

**Fig. 5.20.** a: Intersection of quantum informational ball and channel ellipsoid of amplitude damping channel, b: the movement of the center from the optimal position increases the radius of the quantum informational ball.

Using the results of Fig. 5.20, in case of the amplitude damping channel, $m_\sigma$ measures the Euclidean distance between the average and the center of the Bloch sphere, while the second one, $m_\rho$, represents the Euclidean distance from the center to the optimal channel output state. The average state $\sigma$ (i.e., the center $\mathbf{c}^*$ of the smallest quantum ball) of the amplitude-damping channel will differ from the center of the Bloch sphere. Using the angle $\theta$ between Bloch vectors $\mathbf{r}_\rho$ and $\mathbf{r}_\sigma$, the Euclidean length of the radii $m_\rho = \sqrt{x^2 + y^2 + z^2}$ and $m_\sigma = \sqrt{\tilde{x}^2 + \tilde{y}^2 + \tilde{z}^2}$, the quantum informational radius $r^*$ for an amplitude damping quantum channel is

$$
\begin{aligned}
r^* &= C(\mathcal{N}) = \min_{\sigma} \max_{\rho} D(\rho \| \sigma) \\
&= \min_{\sigma} \max_{\rho_k} D\left(\rho_k \middle\| \sigma = \sum_k p_k \rho_k\right) \\
&= \max_{all\ p_i, \rho_i} \chi = \max_{all\ p_i, \rho_i} \left[ \mathrm{S}\left( \mathcal{N}\left( \sum_i p_i \rho_i \right)\right) - \sum_i p_i \mathrm{S}\left(\mathcal{N}(\rho_i)\right)\right] \\
&= \frac{1}{2}\log\left(1 - (m_\rho)^2\right) + \frac{m_\rho}{2}\log\left(\frac{1+m_\rho}{1-m_\rho}\right) - \frac{1}{2}\log\left(1-(m_\sigma)^2\right) - \frac{\mathbf{r}_\rho \mathbf{r}_\sigma}{2 r_\sigma}\log\left(\frac{1+m_\sigma}{1-m_\sigma}\right) \\
&= \frac{1}{2}\log\left(1 - (m_\rho)^2\right) + \frac{m_\rho}{2}\log\left(\frac{1+m_\rho}{1-m_\rho}\right) - \frac{1}{2}\log\left(1-(m_\sigma)^2\right) - \frac{r_\rho \cos(\theta)}{2}\log\left(\frac{1+m_\sigma}{1-m_\sigma}\right).
\end{aligned}
\tag{5.115}
$$

## 5.5 The Classical Zero-Error Capacities of some Quantum Channels

Having presented some examples related to the classical and quantum capacities of the bit flip and depolarizing quantum channels we show two illustrations for zero error capacities.

The zero-error capacity of the bit flip channel (see Section 3.6.1) can be reached for the following two non-adjacent orthogonal input states (see the channel ellipsoid in Fig. 3.23)

$$
|\psi_1\rangle = \frac{1}{\sqrt{2}}(|0\rangle + |1\rangle) \text{ and } |\psi_2\rangle = \frac{1}{\sqrt{2}}(|0\rangle - |1\rangle) \tag{5.116}
$$

Using these non-adjacent input states for the encoding, the single-use classical zero-error capacity of the bit flip channel is

$$
C_0^{(1)}(\mathcal{N}) = \frac{1}{1}\log(2) = 1. \tag{5.117}
$$

It can be verified that for these inputs the non-adjacent property holds:

$$
\begin{aligned}
&Tr(\mathcal{N}(|\psi_1\rangle\langle\psi_1|)\mathcal{N}(|\psi_2\rangle\langle\psi_2|)) \\
&= Tr((p(\sigma_X|\psi_1\rangle\langle\psi_1|\sigma_X) + (1-p)|\psi_1\rangle\langle\psi_1|)(p(\sigma_X|\psi_2\rangle\langle\psi_2|\sigma_X) + (1-p)|\psi_2\rangle\langle\psi_2|)) \\
&= 0.
\end{aligned}
\tag{5.118}
$$

For any other two inputs the classical zero-error capacity of the bit flip channel is trivially zero. To describe the classical zero-error capacity of the depolarizing channel, we use the channel map already shown in Section 3.6.2, i.e., $\mathcal{N}\left(\rho_i\right) = p\frac{I}{2} + \left(1-p\right)\rho_i$. For a depolarizing channel any two inputs $\left\{\left|\psi_1\right\rangle\left\langle\psi_1\right|,\left|\psi_2\right\rangle\left\langle\psi_2\right|\right\}$ are *adjacent*; that is, there are no inputs for which the channel will have positive classical zero-error capacity, since

$$
\begin{aligned}
&Tr\left(\mathcal{N}\left(\left|\psi_1\right\rangle\left\langle\psi_1\right|\right)\mathcal{N}\left(\left|\psi_2\right\rangle\left\langle\psi_2\right|\right)\right) \\
&= Tr\left(\left(p\left|\psi_1\right\rangle\left\langle\psi_1\right| + \left(1-p\right)\frac{1}{2}I\right)\left(\left(p\left|\psi_1\right\rangle\left\langle\psi_1\right| + \left(1-p\right)\frac{1}{2}I\right)\right)\right) \\
&= Tr\left(p^2 Tr\left(\left|\psi_1\right\rangle\left\langle\psi_1\right|\left|\psi_2\right\rangle\left\langle\psi_2\right|\right) + \frac{p\left(1-p\right)}{2}Tr\left(\left|\psi_1\right\rangle\left\langle\psi_1\right| + \left|\psi_2\right\rangle\left\langle\psi_2\right|\right) + \frac{\left(1-p\right)^2}{2}\right) \\
&> 0,
\end{aligned}
\tag{5.119}
$$

where $0 < p < 1$, which means that the required non-adjacent condition $Tr\left(\mathcal{N}\left(\left|\psi_1\right\rangle\left\langle\psi_1\right|\right)\mathcal{N}\left(\left|\psi_2\right\rangle\left\langle\psi_2\right|\right)\right) = 0$ is not satisfied for the output states; that is, no inputs exist for which the channel can produce maximally distinguishable outputs, and thus for the classical zero-error capacity of the depolarizing channel we have

$$C_0^{(1)}\left(\mathcal{N}\right) = C_0\left(\mathcal{N}\right) = 0. \tag{5.120}$$

## 5.6 Related Work

We summarize the most important works regarding on the calculation of capacity and the geometrical interpretation of the capacities of a quantum channel.

**Fundamentals of Geometry of Quantum Channel Capacities**

In the first part of this section we overviewed the most important quantum channel models, and we showed that their ability for carrying information can be measured geometrically. We analyzed just the most important channel models; further information about the various channel maps can be found in the great book of Hayashi [Hayashi06]. The description of amplitude damping channels (see

Section 3.6.3) with their various capacities can be found in the work of Giovanetti and Fazio [Giovannetti05].

The applications of computational geometry in the quantum space were studied in the works of Gyongyosi and Imre [Gyongyosi11a-d], in Kato's paper [Kato06] and also in the works of Nielsen et al. [Nielsen07], [Nielsen08], [Nielsen08a], [Nielsen08b], [Nielsen09], and Nock and Nock [Nock05]. On Voronoi diagrams by the Kullback-Leibler divergence see the works of [Onishi97] and [Nielsen07a]. The first paper on the application of computational geometry on 1-qubit quantum states was presented by Oto et al. in 2004 [Oto04]. The properties of quantum space with regard to different notions of "distance," and the computational cost of these constructions were studied by Onishi et al. [Onishi97]. The Laguerre diagrams in the quantum space were first mentioned by Nielsen et al. [Nielsen07]. Nielsen et al. have published many interesting results on the geometrical interpretation of the capacity of a quantum channel [Nielsen08a], [Nielsen08b]. An introduction to the measuring processes of continuous observables can be found in Ozawa's work [Ozawa84]. For the mathematical description of the Bloch vector for $n$-level systems see [Kimura03].

On the continuity of quantum channel capacities, a work was published by Leung and Smith in 2009 [Leung09]. On the quantum capacities of noisy quantum channels see the work of DiVincenzo, Shor and Smolin from 1998 [DiVincenzo98] and [Wilde11]. The relation of quantum capacity and superdense coding of entangled states was shown in the work of Abeyesinghe et al. [Abeyesinghe06]. The connection of single-use quantum capacity and hypothesis testing was studied by Wang and Renner [Wang10].

**Geometry of Quantum Channels**

The geometrical interpretation of the Holevo-Schumacher-Westmoreland channel capacity and the proof of the relative entropy based capacity formula can be found in the works of Cortese [Cortese02] who gave some very useful results on the connection between the quantum relative entropy function and the geometric interpretation of the channel output states. The classical channel capacity and its geometric interpretation have also been analyzed by Kato [Kato06] and Nielsen et al. [Nielsen08b].

In the literature, many articles have investigated the question of how many input states are required to achieve the maximal channel capacity of a quantum channel [Hayashi05], [Hayashi06], [King01a], [King01b], [Ruskai01]. Ruskai et al. also demonstrated a very nice solution to the determination of the number of input quantum states required to achieve the optimal channel capacity [Ruskai01]. Hayashi's work is very important from the viewpoint of the geometric interpretation of HSW capacity and the maximally achievable rate for the various quantum channel models. As Hayashi et al. have concluded, in the cases of some quantum channel models, the optimal channel capacity can be achieved by two optimal input states, although for some other channels, the optimal capacity requires more input states [Hayashi03], [Hayashi05]. The geometry of single-qubit maps was studied by Oi [Oi01].

**Quantum Informational Distance**

More information about the *Fundamental Information Inequality*, which states that the relative entropy function is always non-negative, and about its important mathematical corollaries, can be found in the works of Jaynes in 1957 [Jaynes57a], [Jaynes57b], and 2003 [Jaynes03]. Its "quantum version", the Fundamental Quantum Information Inequality states the monotonicity of the quantum relative entropy function. The continuity property of the quantum relative entropy function was showed by Fannes in 1973 [Fannes1973]. Further details about the properties of the quantum relative entropy function can be found in Hayashi's book [Hayashi06], or see the work of Mosonyi and Datta [Mosonyi09]. The works of Petz et al. [Petz96-Petz10a] also clearly presents the mathematical background of these questions. Bounds on the quantum relative entropy function and the trace distance function were determined by Schumacher and Westmoreland in 2002 [Schumacher02]. An introduction to the application of the difference distance measures can be found in Winter's work [Winter99], who showed how these measures can be used in the construction of coding theorems such as the well-known HSW-theorem and in the capacity measure of multiple access channel techniques [Czekaj08], [Yard05b], [Yard08]. The difference distance measures also can be used to quantify different parameters of the quantum communication protocols (similar to a performance measure), hence these metrics have great importance in various fields of Quantum Information Processing. Ogawa and Nagaoka in 2007 published an article [Ogawa07] in which they showed how good codes can be constructed using these various difference measures. In the computation of the classical capacities and the quantum capacity of a quantum channel, convex optimization is a very important problem. Further information about these topics can be found in the book of Boyd and Vandenberghe [Boyd04].

**Early Days**

The original works of G. Voronoi from the very beginning of the twentieth century are [Voronoi1907], [Voronoi1908]. On the convex sets see [Bregman67], and the work of Bures from a different field [Bures69]. A work on the convexity of some divergence measures based on entropy functions was published by Burbea and Rao in 1982 [Burbea82]. The generalization of the complex numbers to three dimensional complex vectors originally was made by W.R. Hamilton [Goodman04]. Hamilton extended the theory of complex numbers to the three dimensional space. The formula of outer product between vectors was discovered by Grassmann. Grassmann's work was inspiration to William Kingdom Clifford, who introduced geometric algebra [Isham99]. In the nineteenth century, Clifford's motivation was to connect Grassmann's and Hamilton's work. Clifford united the inner product and outer product into a single geometric product, and he introduced the Clifford-algebra. Clifford's results have many applications in classical and quantum mechanics, and quantum physics.

**Computational Geometry**

In the literature of computational geometry, many very efficient and robust algorithms exist for computing Delaunay triangulations in two or three dimensions, whose can be applied to quantum space, with respect to quantum informational distance. In three dimensions, the combinatorial and algorithmic complexity can be computed by robust and efficient methods [Amari93], [Amato01], [Arge06], [Clarkson89a], [Kitaev97], [Goodman04], [Pach02], [Rajan94]. The complexity of a Delaunay triangulation is $\mathcal{O}\left(n \log n\right)$, while the worst-case bound on the complexity of a Delaunay triangulation is $\mathcal{O}\left(n^2\right)$.

Computation of convex hulls was one of the first problems in computational geometry [Aluru05], [Brodal02], [Buckley88], [Chan01], [Seidel04]. The amount of literature about convex hull calculations is huge, and there are many computational geometric algorithms to solve this problem. One of the earliest algorithms was constructed by Graham and Andrew [Goodman04], then a divide-and-conquer approach was designed by Preparata and Hong [Goodman04], and later an incremental method was constructed [Aurenhammer92]. A more efficient approach was given by Overmars and van Leeuwen, and similar results were introduced by Hershnerger and Suri, Chan, and Brolat and Jacob [Goodman04]. An algorithm on finding the smallest enclosing ball, and on construction of smallest enclosing disks was published by Welzl in 1991 [Welzl91].

**Complexity of Calculations**

The lower bound on the complexity of a convex hull computation was thought to $\Omega\left(n\log n\right)$, and this was not improved for a long time, until finally Jarvis introduced his “Jarvis’s march” [Goodman04], [Isham99] that computes the convex hull in $\mathcal{O}\left(cn\right)$ time, where $c$ is the complexity of the convex hull [Goodman04]. The same result was obtained by Overmars and van Leeuwen, Nykat, Eddy, and finally by Green and Silverman [Goodman04], [Isham99]. The next relevant result was shown by Kirkpatrick and Seidel who further reduced the complexity of the calculation to $\mathcal{O}\left(\log\left(c\right)n\right)$, and later Chan [Goodman04] showed a simpler algorithm. As has been shown, a convex hull in three-dimensional space can be computed in $\mathcal{O}\left(\log\left(n\right)n\right)$, and for higher dimensions the complexity of the convex hull is no longer linear in the number of points [Chan01], [Chen06], [Chen07], [Clarkson89], [Clarkson89a], [Cormen01], [Cornwell97], [Feldman07].

The computation of the common intersection of half planes is dual to the computation of the convex hull of points in the plane, and it is also a well studied problem. The convex hull computation between quantum states can be derived from the problem of half-plane intersections. The problem of the common intersection of half planes was studied by Preparata and Shamos [Goodman04], and [Isham99] and they gave many solutions to the problem in $\mathcal{O}\left(\log\left(n\right)n\right)$ time. As has been shown, computing the common intersection of half-spaces is a harder problem if the dimension increases, since the common intersection can be as large as $\mathcal{O}\left(n^{\lfloor d/2 \rfloor}\right)$. Many linear programming approaches have been developed [Bregman67], [Goodman04], [Seidel04], [Shewchuck02], [Shirley05], [Wein07], [Worboys04], [Yoshizawa99]. Welzl showed different methods to convex hull calculation problem of a set of points [Welzl85], [Welzl88], [Welzl91]. We also suggest the works of Sharir [[Sharir94], [Sharir04].

**Application of Computational Geometry**

As an introduction to basic theories and methods of computational geometry can be found in the book of Goodman and O’Rourke [Goodman04]. An interesting paper on random quantum channels, and their graphical calculus and the Bell state phenomenon was published by Collins and Nechita in 2009, for details see [Collins09]. About the core-set approaches and the properties of smallest enclosing balls, see the works of Nielsen and Nock [Nielsen07], [Nielsen08], [Nielsen08a], [Nielsen08b], [Nielsen09], [Nock05], and Kato et al. [Kato06]. The approximate clustering via core-sets was studied by Badoiu et al. see [Badoiu02].
On the optimality of Delaunay triangulation see the work of Rajan [Rajan94]. A technique on clustering based on weak coresets was published by Feldman et al. [Feldman07]. On the mathematical background of finding the smallest enclosing ball of balls see [Fischer04]. The problem of smaller coresets for clustering was studied by Har-Peled and Kushal [Har-Peled05]. The role of coresets in dynamic geometric data streams was studied by Frahling and Sohler in [Frahling05].

For clustering for metric and non-metric distance measures, see the work of Ackermann et al. [Ackermann08]. On range searching see the work of Agarwal et al. [Agarwal04] and on the properties of some important geometrical functions see [Agarwal07]. An algorithm for calculating the capacity of an arbitrary discrete memoryless channel was shown in [Arimoto72]. On the applications of Voronoi diagrams, see [Asano06], [Asano07], [Asano07a] or the works of Aurenhammer et al. [Aurenhammer84,87,91,92,98,2000], or [Boissonnat07]. An important work on clustering with Bregman divergences was published by Banerjee et al. [Banerjee05]. About triangulation and mesh generation see [Bern04], [Bern99].

The Delaunay triangulation and the complexity of its construction was proven by Rajan in 1994 [Rajan94]. On Delaunay triangulation and Voronoi diagram on the surface of a sphere see the work of Renka [Renka97]. A conference paper on the computation of Voronoi diagrams by divergences with additive weights was published by Sadakane et al. [Sadakane98]. On Delaunay refinement algorithms for triangular mesh generation see [Shewchuck02]. On streaming computation of Delaunay triangulations a work was published by Isenburg et al. [Isenburg06]. The minimum enclosing polytope in high dimensions was studied by Panigrahy [Panigrahy04]. A work on finding the center of large point sets in statistical space was published by Pelletier [Pelletier05].

**Comprehensive Surveys**

The mathematical background of quantum informational divergences can be found in Petz's works [Petz96], [Petz08]. A great work on modern differential geometry was published by Isham [Isham99]. A handbook on Discrete and Computational Geometry was published by Goodman and O'Rourke [Goodman04]. We also suggest the work of [Janardan04]. A summarization on convex hull computations was published by Seidel [Seidel04]. A work on dynamic planar convex hull was published by Brodal and Jacob [Brodal02]. On matrix analysis see the book of Horn and Johnson [Horn86]. A great work on modern graph theory was published by Bollobás in 1998 [Bollobas98]. The methods of information geometry were also summarized by Amari [Amari2000].

For further supplementary information see the book of Imre and Gyongyosi [Imre12].

# 6. Superadditivity of Classical and Quantum Capacities of Quantum Channels

Additivity of quantum channels in terms of their capacity is one of the very important questions in quantum information theory. The problem statement is whether entangled input states and/or joint measurements could improve the joint capacity of the quantum channels? In the case of classical channels, the correlation between the inputs of the channel does not improve the channel capacity, hence strict additivity holds. Furthermore, additivity is strongly related to the single-use and asymptotic capacities. Obviously these capacities for a certain channel are equal if additivity holds. The additivity conjectures have emerged from an attempt to find a closed form expression for the capacity of a noisy quantum channel. The accessible classical information from continuous quantum degrees of freedom is limited, which is the key issue behind the additivity of quantum channel capacity.

The equality of the various channel capacities (i.e., classical, private and quantum) is known for some special cases, but the generalized rule is still *unknown*. At present, the main challenges connected the quantum channel additivity have not solved yet, some of them are only confirmed for some classes of quantum channels. Recently, the question of the additivity property of a quantum channel has been studied exhaustively, however the most basic question on the classical capacity of a quantum channel - namely, the additivity of classical channel capacity for *different channel maps* - still remain open.

Section 6 is organized as follows. In the first we introduce the problem of additivity of quantum channel capacities. We give the definition of the degradable, the entanglement-breaking and the noiseless ideal quantum channels. We also show an important property of quantum coherent information for some channels. In the second part, we analyze the additivity of the classical and quantum capacities of quantum channels. The description of the most relevant works can be found in the Related Work subsection.

## 6.1 The Additivity Problem of Quantum Channels

In 2009, a counterexample to strict additivity was shown by Matt Hastings [Hastings09]. Hastings has analyzed the additivity of Holevo information. He has constructed a channel pair, using a random construction scheme, which can produce an output which is a counterexample to the additivity of the minimum output entropy—a theorem which was introduced in the middle of the 1990s. The counterexample of Hastings's implies that the entangled states are more resistant to noise, and the output entropy of the channel output states will be lower. Hastings used entangled inputs and joint measurement previously shown in Section 3, see Fig. 3.7.

As was presented by Shor [Shor04a] a few years before Hastings's discovery, the additivity problem of the classical capacity of the quantum channel can be analyzed from the viewpoint of the additivity of the minimum von Neumann entropy output states. Shor has also shown that if there exists a combination for which the additivity of minimum entropy channel output states of the channels is violated, then the additivity of the Holevo information is also violated. Later, Hastings have found a possible combination [Hastings09], for which the additivity of the Holevo information was violated. We note, the violation in the minimum output entropies was very small, and it appears only for *sufficiently large* input dimensions. However, for the asymptotic setting the question remains opens.

To define additivity of the various capacities, we introduce a generalized notation of the various capacities of quantum channels. For this purpose, we will use the generalized "*all-in-one*" notation $C_{ALL}\left(\mathcal{N}\right)$ to describe all capacities of the quantum channel. This generalized notation involves the classical capacities and the quantum capacity of quantum channel.

The non-additivity property consist of two categories: if *subadditivity* holds for quantum channels then the joint capacity of the channels is smaller than the sum of the individual capacities of the quantum channels

$$C_{ALL}\left(\mathcal{N}_1 \otimes \mathcal{N}_2\right) < C_{ALL}\left(\mathcal{N}_1\right) + C_{ALL}\left(\mathcal{N}_2\right). \tag{6.1}$$

On the other hand in case of *superadditivity* the joint capacity is greater than the sum of the individual capacities of the channels

$$C_{ALL}\left(\mathcal{N}_1 \otimes \mathcal{N}_2\right) > C_{ALL}\left(\mathcal{N}_1\right) + C_{ALL}\left(\mathcal{N}_2\right). \tag{6.2}$$

Finally, if strict additivity holds then

$$C_{ALL}\left(\mathcal{N}_1 \otimes \mathcal{N}_2\right) = C_{ALL}\left(\mathcal{N}_1\right) + C_{ALL}\left(\mathcal{N}_2\right). \tag{6.3}$$

*To this day, there are very many conjectures on this subject. Three aspects are clear, namely additivity depends on the encoding scheme, the decoding (measurement) scheme and the properties of the channels maps.*

The additivity holds for the maximized quantum mutual information $\max_{all\ p_i,\rho_i} I\left(A:B\right)$, thus, if we have tensor product input states and single measurement setting, then for the classical capacity $C\left(\mathcal{N}\right)$ we will have

$$\begin{aligned} C\left(\mathcal{N}_{12}\right) &= \lim_{n\to\infty}\frac{1}{2}C\left(\mathcal{N}_{12}\right) = C\left(\mathcal{N}_1\right) + C\left(\mathcal{N}_2\right) \\ &= \max_{all\ p_i,\rho_i} I\left(A_1:B_1\right) + \max_{all\ p_i,\rho_i} I\left(A_2:B_2\right), \end{aligned} \tag{6.4}$$

where $\mathcal{N}_{12} = \mathcal{N}_1 \otimes \mathcal{N}_2$ and $\left\{A_i, B_i\right\}$ denotes the input and output of the $i$-th quantum channel in the tensor product structure.

On the other hand, neither the Holevo quantity $\chi$, nor the quantum coherent information $I_{coh}\left(\rho_A : \mathcal{N}\left(\rho_A\right)\right)$ are additive in general, only in case of some special examples. Moreover, the picture also changes dramatically if we talk about the entanglement assisted capacity or private classical capacity of the quantum channel.

During the past decade, many new techniques were developed and many questions on the additivity of the different capacities have arisen. Unfortunately, most conjectures on the additivity of the capacities failed, and roughly speaking because most capacities are non-additive. There is no general formula to describe the additivity property of every quantum channel model, but one of the main results of the recent researches was the "very simplified" picture, that

$$C_{ALL}\left(\mathcal{N}_{12}\right) \neq C_{ALL}\left(\mathcal{N}_1\right) + C_{ALL}\left(\mathcal{N}_2\right). \tag{6.5}$$

for different capacities of the most quantum channel models. From this viewpoint, the strict additivity in quantum communication can be viewed as a special case, or a counterexample. The non-additive property is deeply woven into the essence of quantum mechanical systems, as can be elucidated by the fact that, for us, and thus for an external observer, a quantum system is rather different from a classical - simply additive – system [Smith10].

Furthermore, the non-additive property has many advantages and disadvantages, too. We start with the most important *disadvantage*: the different capacities *cannot be described* by a simple single-use formula, like we did in the case of classical systems. Instead, we have to use an approximation, – called the asymptotic capacity. This approximation is based on the assumption, that *unlimited number* of channel copies is available. This approximation used to compute the asymptotic capacity of the quantum channel is called the *regularization* of the channel capacities, which leads us to the fact that the computation of the different quantum capacities is much more sophisticated than for classical systems.

Now, we present the *advantageous side* of the regularization. The non-additive property and the regularization of the channel capacities make it possible to achieve higher channel capacities and lower error probabilities than for classical systems. Thus, these phenomena make it possible to enhance the information transmission through quantum channels or to use zero-capacity quantum communication channels by the exploitation of entangled input states, and by the combination of different channel maps.

### 6.1.1 The Four Propositions for Additivity

For a given quantum channel $\mathcal{N}$, the minimal output entropy $\mathrm{S}$ of $\mathcal{N}$ can be defined as

$$\mathrm{S}_{min}\left(\mathcal{N}\right)=\min_{\rho\in\mathcal{S}}\mathrm{S}\left(\mathcal{N}\left(\rho\right)\right). \tag{6.6}$$

For the additivity of the *minimum* entropy output of two quantum channels $\mathcal{N}_1$ and $\mathcal{N}_2$, the following property holds

$$\mathrm{S}_{min}\left(\mathcal{N}_1\otimes\mathcal{N}_2\right)=\mathrm{S}_{min}\left(\mathcal{N}_1\right)+\mathrm{S}_{min}\left(\mathcal{N}_2\right). \tag{6.7}$$

Using quantum systems $\mathbb{S}_1$, $\mathbb{S}_2$, an entangled state is an element of the set $\mathbb{K}$ defined as

$$\mathbb{K}=\mathbb{S}_1\otimes\mathbb{S}_2-\left\{\rho_1\otimes\rho_2\middle|\rho_1\in\mathbb{S}_1,\rho_2\in\mathbb{S}_2\right\}, \tag{6.8}$$

where $\rho_1 \otimes \rho_2$ denotes the decomposable tensor product states. The entanglement of a quantum state can be defined by the entanglement of formation. For a state $\rho$ in a bipartite system $\mathbb{S}_A \otimes \mathbb{S}_B$, the entanglement of formation $E_F$ can be defined as

$$E_F\left(\rho\right) = \min_{\rho} \sum_i p_i \mathrm{S}\left(Tr_B \left|i\right\rangle\left\langle i\right|\right), \tag{6.9}$$

where the minimization is over all possible mixed $\rho$ such that $\rho = \sum_i p_i \left|i\right\rangle\left\langle i\right|$, and $\sum_i p_i = 1$. The additivity property of the entanglement formation can be defined for two states $\rho_1 \in \mathbb{S}_{A1} \otimes \mathbb{S}_{B1}$ and $\rho_2 \in \mathbb{S}_{A2} \otimes \mathbb{S}_{B2}$ in the following way

$$E_F\left(\rho_1 \otimes \rho_2\right) = E_F\left(\rho_1\right) + E_F\left(\rho_2\right), \tag{6.10}$$

where $E_F$ is calculated over bipartite $A$-$B$ partition.

We can define the *strong superadditivity* of the entanglement of formation for $\rho \in \mathbb{S}_{A1} \otimes \mathbb{S}_{A2} \otimes \mathbb{S}_{B1} \otimes \mathbb{S}_{B2}$ as

$$E_F\left(\rho\right) \geq E_F\left(Tr_1(\rho)\right) + E_F\left(Tr_2(\rho)\right), \tag{6.11}$$

where $Tr_i(\cdot)$ means tracing out the space $\mathbb{S}_{Ai} \otimes \mathbb{S}_{Bi}$. The four propositions for additivity presented in this section – for the minimum output entropy, for the classical capacity, for the entanglement formation and for the strong superadditivity – *are equivalent*.

## 6.2 Additivity of Classical Capacity

In this subsection we overview the additivity property of classical capacity of quantum channels. The classical capacities of the joint channel structure $\mathcal{N}_{12} = \mathcal{N}_1 \otimes \mathcal{N}_2$ will be referred as the *joint classical capacity*, and will be denoted by $C\left(\mathcal{N}_{12}\right) = C\left(\mathcal{N}_1 \otimes \mathcal{N}_2\right)$. We illustrated the measurement setting for additivity analysis of tensor-product channel capacity $C\left(\mathcal{N}_1\right) + C\left(\mathcal{N}_2\right)$, with *single measurement* setting in Fig. 6.1. Because of the single measurement setting – which destroys the possible benefits of entanglement – the properties of the input are irrelevant from the additivity point of view. As follows, in this case strict additivity $C(\mathcal{N}_1) + C(\mathcal{N}_2)$ holds for the joint capacity $C(\mathcal{N}_1 \otimes \mathcal{N}_2)$.

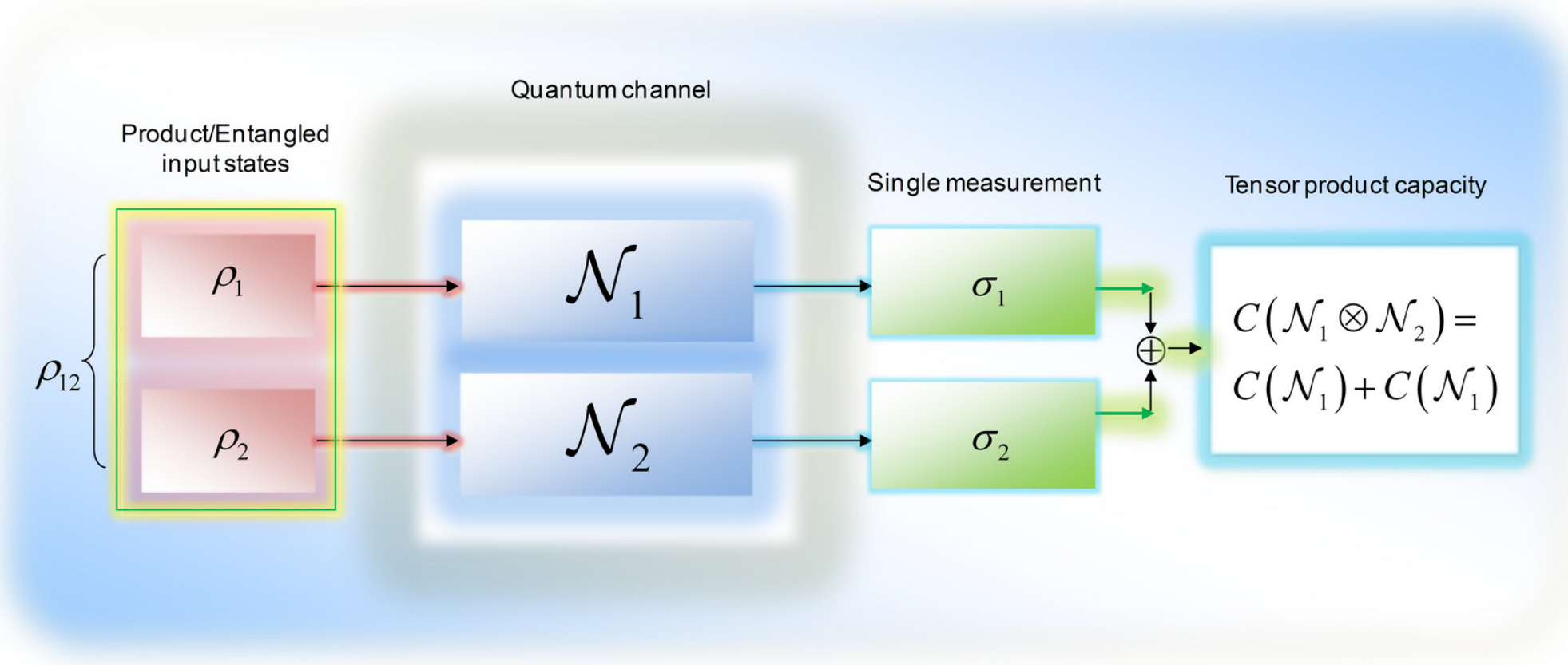

**Fig. 6.1.** Setting for tensor-product channel capacity analysis, using single measurement setting. The single measurement destroys the possible benefits of entangled inputs.

Let us try to enhance the analysis of tensor-product channel capacity by *joint measurement*. $C_{PROD.}(\mathcal{N}_1 \otimes \mathcal{N}_2)$ refers to the capacity when *product input* states and *joint measurement* setting are used, as is illustrated in Fig. 6.2.

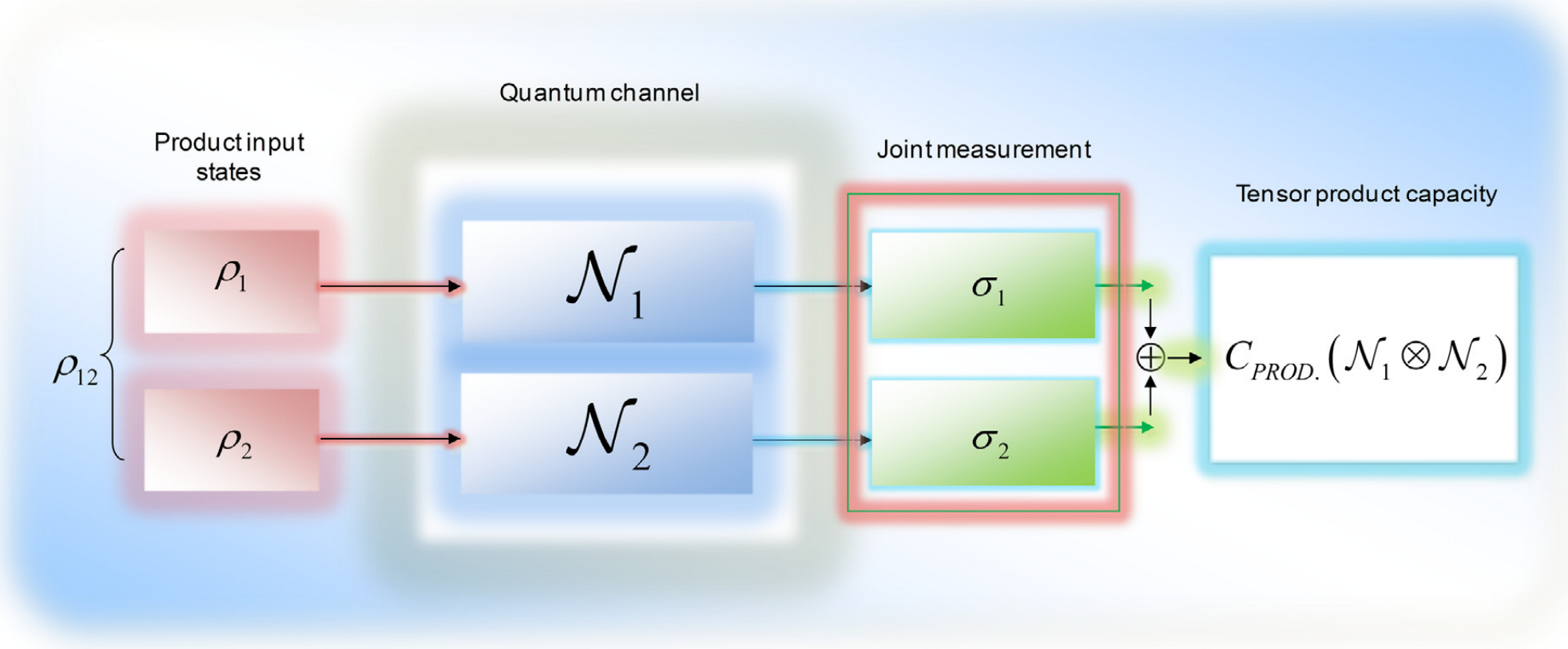

**Fig. 6.2.** Setting for tensor-product channel capacity analysis, using joint measurement setting.

Finally we consider the case of EPR input states and joint measurement setting in Fig. 6.3 for the classical capacity $C_{ENT.}(\mathcal{N}_1 \otimes \mathcal{N}_2)$.

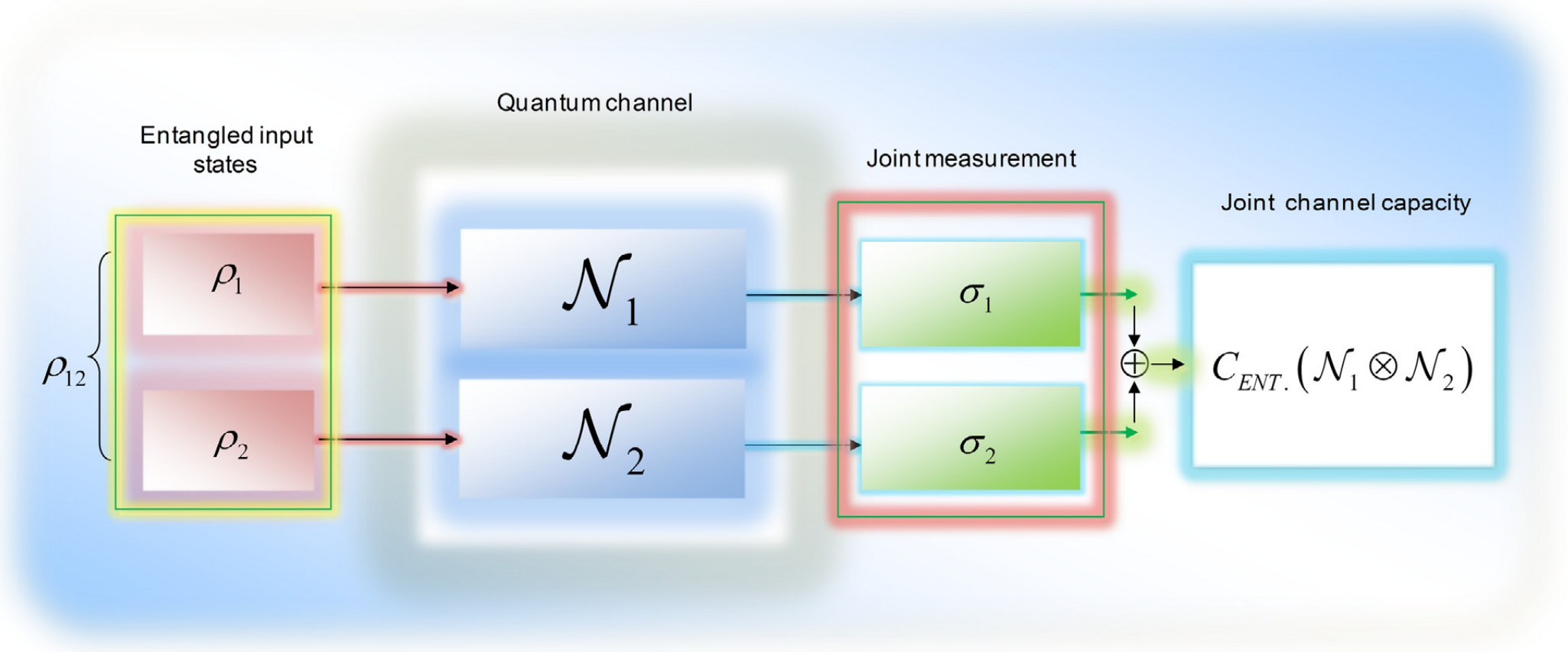

**Fig. 6.3.** Setting for joint-capacity analysis using entangled input states and joint measurement setting. The joint measurement setting is required to exploit the possible benefits of entanglement.

The main question on the different classical capacities can be stated as

$$C_{ENT.}\left(\mathcal{N}_1 \otimes \mathcal{N}_2\right) \overset{?}{\geq} C_{PROD.}\left(\mathcal{N}_1 \otimes \mathcal{N}_2\right). \tag{6.12}$$

## 6.3 Additivity of Quantum Capacity

Some very important channels – such as the erasure channel, the amplitude damping channel, or some Gaussian channels – are degradable quantum channels, and hence for this channels the single-use quantum capacity *will be equal* to the asymptotic quantum capacity. We note, that it might also be the case that for some non-degradable quantum channels the capacity also can be characterized by the single-use formula (i.e., quantum coherent information), however this question is still open.

### 6.3.1 The Degradable Quantum Channel

A quantum channel $\mathcal{N}_{\mathcal{D}}$ is a *degradable* channel if the amount of information leaked to the environment is less than the amount of information which can be transmitted over it.

The concept of a degradable quantum channel is illustrated in Fig. 6.4. Bob simulates the environment $E$ with $\mathcal{N}_1$ and $\mathcal{D}$. The input of the first channel is denoted by $\rho_A$ and the output of channel $\mathcal{N}_1$ is by $\sigma_B$. The simulated environment is represented by $\sigma_E$. If the resultant channel $\mathcal{N}_2 = \mathcal{D}\mathcal{N}_1$ is "noisier" than $\mathcal{N}_1$, then $\mathcal{N}_1$ is a degradable channel, where $\mathcal{D}$ is a degrading quantum channel.

The precise definition is the following: If the channel between Alice and Bob is denoted by $\mathcal{N}_1$, and the channel between Alice and the environment is $\mathcal{N}_2$, and for the noise of the two quantum channels the following relation holds

$$Noise\left(\mathcal{N}_1\right) < Noise\left(\mathcal{N}_2\right), \tag{6.13}$$

where $\mathcal{N}_2$ can be expressed from $\mathcal{N}_1$ as

$$\mathcal{N}_2 = \mathcal{D}\mathcal{N}_1, \tag{6.14}$$

then $\mathcal{N}_1$ is a degradable channel, and $\mathcal{D}$ denotes the so called *degrading channel.* Bob, having $\mathcal{N}_1$ and $\mathcal{D}$ in his hands can realize (i.e., he can "simulate") the noisier $\mathcal{N}_2$. In this case, $\mathcal{N}_1$ is called a *degradable* quantum channel.

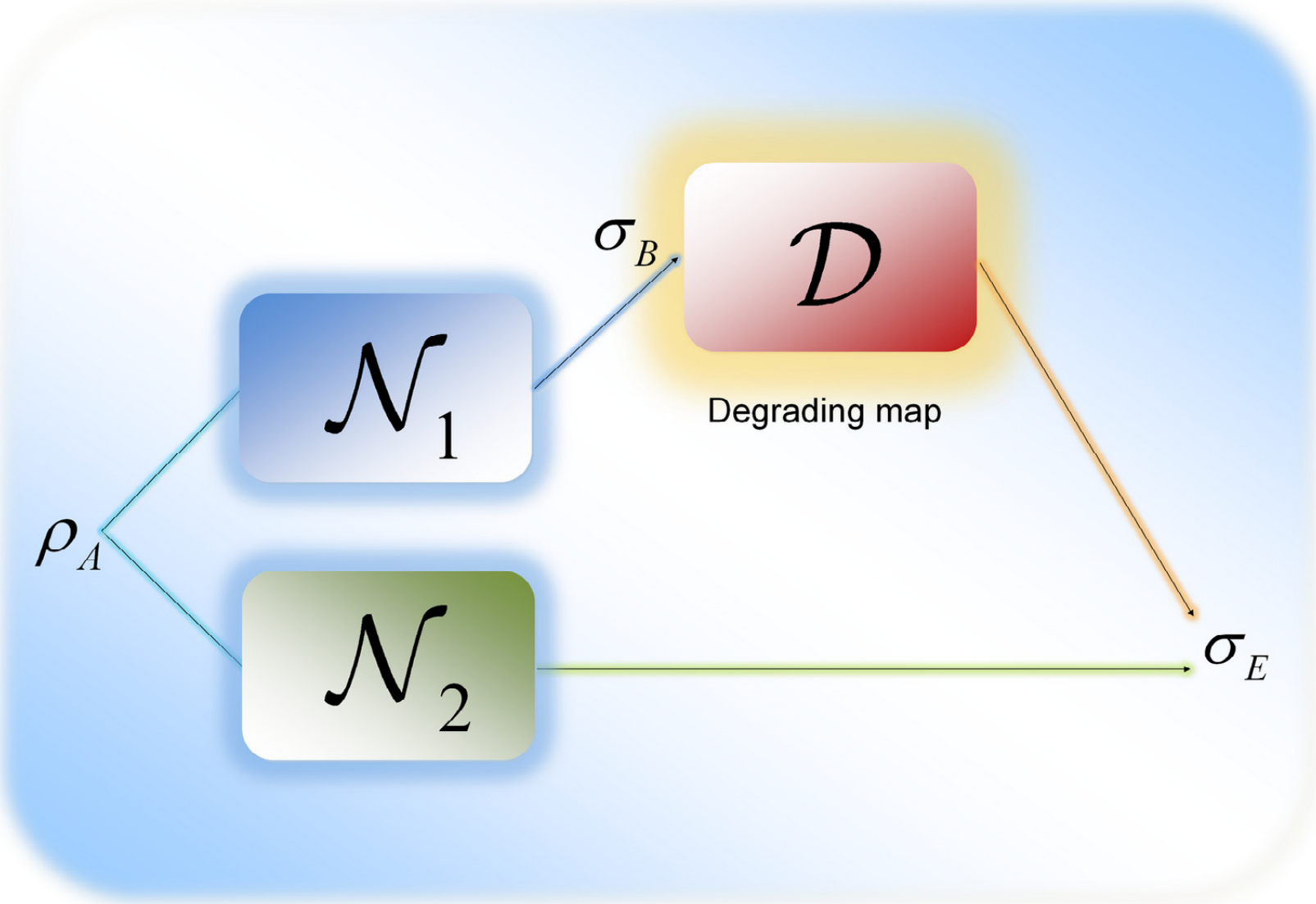

**Fig. 6.4.** The concept of a degradable quantum channel. Bob can simulate the environment by means of his degrading channel $\mathcal{D}$. $\mathcal{N}_1$ is degradable if the simulated $\mathcal{N}_2$ is noisier than $\mathcal{N}_1$. Bob's input is the output of the $\mathcal{N}_1$; the output of $\mathcal{N}_2$ is the environment state. The environment state also can be generated by Bob with his degrading channel.

We note that the degradability of a quantum channel $\mathcal{N}_1$ can be checked easily, since if we take the inverse of $\mathcal{N}_1$ then $\mathcal{N}_1^{-1}$ is not a completely positive trace preserving map. On the other hand, if $\mathcal{N}_1$ is degradable, then there exists degrading quantum channel $\mathcal{D}$, which is equal to

$$\mathcal{D}= \mathcal{N}_2\mathcal{N}_1^{-1}, \tag{6.15}$$

since $\mathcal{N}_2 = \mathcal{D}\mathcal{N}_1$. Or, in other words, if degrading quantum channel $\mathcal{D}$ exists, then $\mathcal{N}_2\mathcal{N}_1^{-1}$ has to be a completely positive trace preserving map. In this case, $\mathcal{N}_1$ is a *degradable* channel.

If $\mathcal{N}_2\mathcal{N}_1^{-1}$ is not a completely positive trace preserving map, then the *degrading* quantum channel $\mathcal{D}$ does not exist, and hence $\mathcal{N}_1$ is not degradable.

#### 6.3.1.1 Description of Degrading Maps

In (6.14) the degrading map $\mathcal{N}_2 = \mathcal{D}\mathcal{N}_1$ means that if the input quantum system $\rho_A$ evolutes a first-type of noise and then, a second-type of noisy transmission, the output state will be equal to the output of the "simulated" channel $\mathcal{N}_2\left(\rho_A\right) = \sigma_E$. We have concatenated two noisy quantum channels.

For the input density matrix $\rho$, the output of a quantum channel $\mathcal{N}$ can be given by its Kraus representation as $\mathcal{N}\left(\rho\right) = \sum_i N_i \rho N_i^{\dagger}$. If we would like to describe the output of the concatenated structure of two quantum channels, then it can be done as follows:

$$\mathcal{D}\mathcal{N}_1\left(\rho\right) = \sum_i D_i \mathcal{N}_1\left(\rho\right) D_i^{\dagger} = \sum_{i,i'} D_i N_{i'} \rho N_{i'}^{\dagger} D_i^{\dagger},\ i = 1,2, \tag{6.16}$$

where $\left\{N_i\right\},\left\{D_i\right\}$ are the Kraus operators of the two channels, $\mathcal{N}_1$ and $\mathcal{D}$.

### 6.3.2 On the Additivity of Coherent Information

For a *degradable* quantum channel $\mathcal{N}_{\mathcal{D}}$, the following relation holds between the single-use quantum capacity $Q^{(1)}\left(\mathcal{N}_{\mathcal{D}}\right)$, the asymptotic quantum capacity $Q\left(\mathcal{N}_{\mathcal{D}}^{\otimes n}\right)$, and the maximized quantum coherent information $\max_{all\ p_i,\rho_i} I_{coh}\left(\mathcal{N}_{\mathcal{D}}\right)$ of $\mathcal{N}_{\mathcal{D}}$:

$$nQ^{(1)}\left(\mathcal{N}_{\mathcal{D}}\right) = Q\left(\mathcal{N}_{\mathcal{D}}^{\otimes n}\right) = n \max_{all\ p_i,\rho_i} I_{coh}\left(\mathcal{N}_{\mathcal{D}}\right). \tag{6.17}$$

However, if we choose a non-degradable quantum channel, then we cannot say anything about the asymptotic quantum capacity from the knowledge of the single-use quantum capacity, that is, whether it is additive or non-additive; but the *quantum coherent information* for a non-degradable quantum channel could be *superadditive*.

For example, the depolarizing quantum channel (see Section 3.6.2) is a *non-degradable* channel, and the following relation holds for its quantum capacity

$$nQ^{(1)}\left(\mathcal{N}_{\mathcal{D}}\right) < Q\left(\mathcal{N}_{\mathcal{D}}^{\otimes n}\right), \tag{6.18}$$

which was proven for $n = 5$ [DiVincenzo98]. In the other cases, we have no superadditivity in the coherent information, or we have no knowledge about the violation of additivity of quantum coherent information. It is an interesting result, since the classical capacity of depolarizing quantum channel $\mathcal{N}$ is proven to be strictly additive, i.e.,

$$nC^{(1)}\left(\mathcal{N}\right) = C\left(\mathcal{N}^{\otimes n}\right). \tag{6.19}$$

This means that independently whether we use unentangled inputs or we have entangled states, the classical capacity of the depolarizing channel will be additive for every possible $n$. The proof strategy of superadditivity of quantum coherent information in the case of a depolarizing quantum channel for $n = 5$ is based on the usage of a special coding strategy, called the "*repetition code concatenation*". The reader does not have to consider a complicated and obscure coding strategy, since it simply means that the quantum information is encoded into an $n = 5$ length quantum code. This encoding strategy uses random codes concatenating with a repetition code. The random code is an EPR-state which will be concatenated with other qubits.

Thus, let us assume that Alice encodes her EPR state using the *repetition code concatenation* strategy, using code length $n = 5$ . The concatenation will result in the following special codeword:

$$\frac{1}{\sqrt{2}}\left(\left|00\right\rangle + \left|11\right\rangle\right) \to \frac{1}{\sqrt{2}}\left(\left|00000\right\rangle + \left|11111\right\rangle\right). \tag{6.20}$$

If Alice uses this strategy, then the quantum coherent information will become *superadditive* and the single-use quantum capacity and the asymptotic quantum capacity will differ.

In Fig. 6.5 we compare the cases of Alice using a "normal" encoding scheme and using the special repetition code concatenation scheme. The normal encoding scheme is denoted by the normal line; the special coding scheme is depicted by the dashed line.

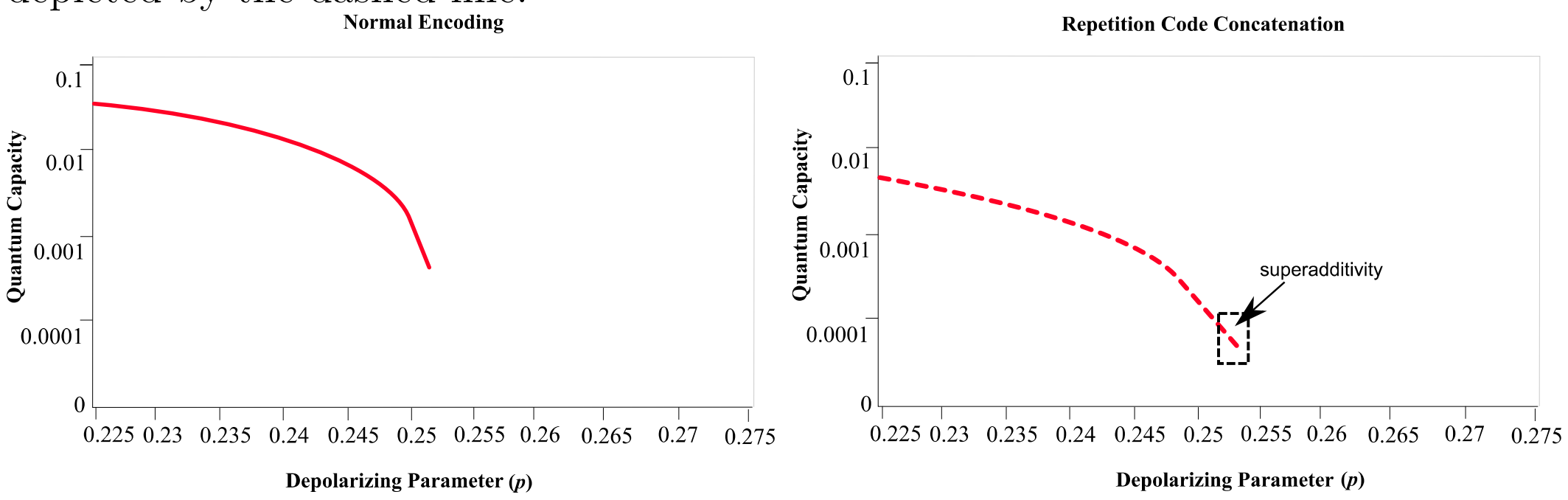

**Fig. 6.5.** The superadditivity of quantum coherent information for a depolarizing quantum channel. The violation of additivity requires special encoding and code length and it occurs only in a very limited domain.

Using the special repetition code concatenation encoding scheme for the depolarizing channel and quantum code of length $n = 5$ , the quantum coherent information will be superadditvite.

The non-additivity will occur only in a very small domain if the depolarizing parameter is in the following range:

$$0.25239 < p < 0.25408\,. \tag{6.21}$$

In this case, the single-use quantum capacity will be zero:

$$5Q^{(1)}\left(\mathcal{N}\right) = 0\,, \tag{6.22}$$

while the asymptotic version will be greater than zero:

$$Q\left(\mathcal{N}^{\otimes 5}\right) > 0\,. \tag{6.23}$$

In both cases, the quantum capacities decrease as the depolarizing parameter of the quantum channel increases; however in some domains the repetition code concatenation encoding scheme still produces positive quantum capacity, where the quantum capacity with the normal coding strategy completely vanishes. This domain can be found at the end of the possible range; it is within a very small and tight range – but it exists. Using a special encoding technique the quantum capacity of the depolarizing quantum channel can be increased, the additivity of quantum coherent information can be violated, and the maximum amount of transmittable quantum information can be increased.

Having the results of Fig. 6.5 in our hands, the reader may ask: what is the background of this phenomenon? Well, we arrive again at the same reasoning that we have seen in many cases – the *usage of entanglement, and a special property of the repetition coding scheme* which is known as the degeneracy of the quantum code. It means that for a degenerate codeword the characteristics of the errors will differ from the error characteristics which occur for a non-special (i.e., non-degenerate) input codeword. Simply, the noise of the same quantum channel has different effect on a codeword which uses this special encoding scheme than on a "normal" codeword. An interesting phenomenon is that this difference arises significantly only if the noise of the channel becomes high, and it can be obtained only for a very limited range of lengths of codeword.

### 6.3.3 The Hadamard and Entanglement-breaking Channels

The Hadamard channel is an important channel model, since the *noiseless ideal* qubit channel $\mathcal{N} = I$ belongs to this class. We note that there are several other channels which are also Hadamard channels, such as the cloning channel, the generalized dephasing channel, etc [Bradler09], [King07].

The $\mathcal{N}_{Had.}$ Hadamard channel is a quantum channel whose complementary channel is an entanglement-breaking channel (see Fig. 6.6), i.e., a channel which destroys the input entanglement on its output. More formally: a noisy quantum channel $\mathcal{N}_{EB}$ is *entanglement-breaking* if for a half of a maximally entangled input, state the output of the channel is a separable state. The $\mathcal{N}_{Had.}$ Hadamard channel got its name from the property that the output of this type of channel can be written in the form of *elementwise multiplication* of the input density matrix $\Upsilon$ and another operator with matrix $\Gamma$ with respect to the basis $\left\{\left|i\right\rangle^{B}\right\}$ as the as follows:

$$\mathcal{N}_{Had.}\left(\rho\right) = \Upsilon * \Gamma^{\dagger}, \tag{6.24}$$

where $*$ is the *Hadamard-product* [King07].
Let us assume that the maximally entangled input system of an $\mathcal{N}_{EB}$ entanglement-breaking channel is $\left|\Psi\right\rangle_{AA'} = \frac{1}{\sqrt{d}}\sum_{i=0}^{d-1}\left|i\right\rangle_{A}\left|i\right\rangle_{A'}$. For $\mathcal{N}_{EB}$ the following property holds

$$\mathcal{N}_{EB}\left(\left|\Psi\right\rangle\left\langle\Psi\right|_{AA'}\right) = \sum_{x} p_X\left(x\right)\rho_x^A \otimes \rho_x^B, \tag{6.25}$$

where $p_X\left(x\right)$ represents an arbitrary probability distribution, while $\rho_x^A$ and $\rho_x^B$ are the separable density matrices of the output system.

In other words, for a Hadamard channel, the map to its environment is realized by an entanglement-breaking map. The Hadamard quantum channels have many important properties but the most important of these is the following: the classical capacity $C\left(\mathcal{N}_{Had.}\right)$ of a Hadamard channel is always can be determined without the computation of the asymptotic formula. Furthermore, the maximized Holevo information $\mathcal{X}\left(\mathcal{N}_{Had.} \otimes \mathcal{N}\right)$ for $\mathcal{N}_{Had.} \otimes \mathcal{N}$ *is always additive*, i.e.,

$$\mathcal{X}\left(\mathcal{N}_{Had.} \otimes \mathcal{N}\right) = \mathcal{X}\left(\mathcal{N}_{Had.}\right) + \mathcal{X}\left(\mathcal{N}\right), \tag{6.26}$$

and therefore the asymptotic HSW capacity is

$$C\left(\mathcal{N}_{Had.} \otimes \mathcal{N}\right) = C^{(1)}\left(\mathcal{N}_{Had.} \otimes \mathcal{N}\right), \tag{6.27}$$

where $\mathcal{N}$ can be any quantum channel. It also means that the classical capacity of a Hadamard quantum channel can be maximized with product input states.

To describe the properties of the quantum Hadamard channel, we need the following ingredients:

- definition of $\mathcal{D}$ degrading map (see Section 6.3.1),

- definition of $\mathcal{N}_{EB}$ entanglement-breaking channel (see in short above, Section 6.3.3).

Since. the quantum Hadamard channels are degradable channels. In other words, Bob can simulate the map of the environment, i.e., the second nosier channel $\mathcal{N}_2$, with his channel $\mathcal{N}_1$ and degrading map $\mathcal{D}$, where $\mathcal{D}$ is an entanglement-breaking map as $\mathcal{N}_2 = \mathcal{N}_1\mathcal{D}$. The degrading map $\mathcal{D}$ is an entanglement-breaking map can be decomposed as two maps:

- *The first map:* Bob performs a measurement on his received qubit $B$, which yields a classical variable $Y$. This degrading map is denoted by $\mathcal{D}^{B\to Y}$.

- *The second map*: Bob, using the measurement result $Y$, prepares a new quantum state and sends it to his environment $E$. This second degrading map is denoted by $\mathcal{D}^{Y\to E}$.

For the $\mathcal{D}^{B\to E}$ degrading map between Bob and the environment, the following relation using the two above degrading maps holds:

$$\mathcal{D}^{B\to E} = \mathcal{D}^{B\to Y}\mathcal{D}^{Y\to E}. \tag{6.28}$$

These statements are summarized in Fig. 6.6. The quantum channel between Alice and Bob is denoted by $\mathcal{N}_1$. The second, noisier channel between and Alice and the environment $E$ is called the *complementary channel* and is depicted by $\mathcal{N}_2$. For the Hadamard channel $\mathcal{N}_1$, the noisier complementary channel $\mathcal{N}_2$ is an entanglement-breaking channel. The Hadamard channel $\mathcal{N}_1$ is degradable, because Bob can simulate $\mathcal{N}_2$ by a degrading map; moreover, the degrading map $\mathcal{D}^{B\to E}$ is an entanglement-breaking map.

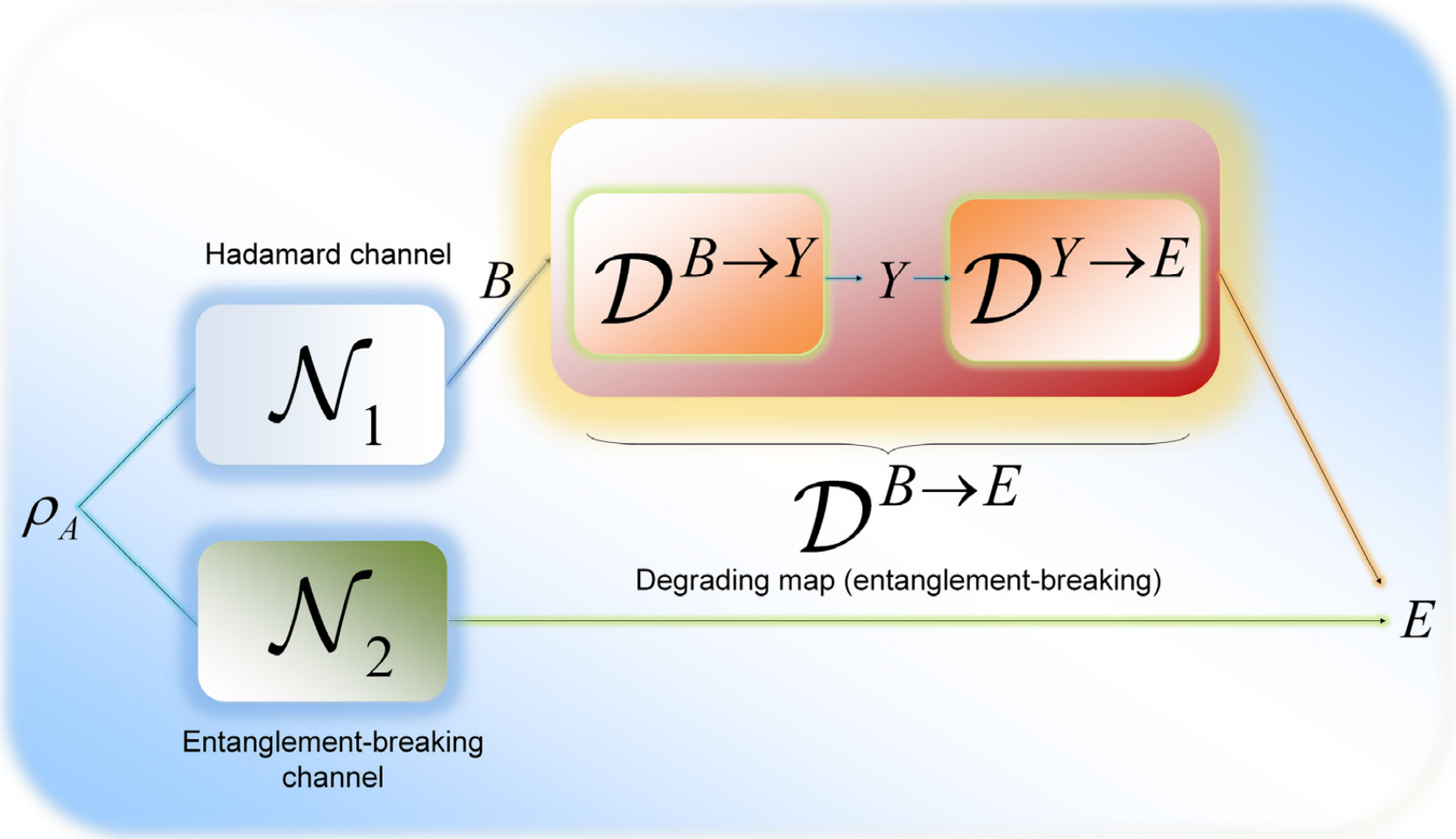

**Fig. 6.6.** The complementary channel of a Hadamard channel is an entanglement-breaking channel. The Hadamard channel is a degradable channel.

Now, we turn our attention to a rather natural but important type of quantum channel— the ideal noiseless quantum channel.

#### 6.3.3.1 The Noiseless Ideal Quantum Channel

After we have derived the most important properties of a Hadamard channel, the reader might ask: "why is the noiseless ideal quantum channel also a Hadamard channel?" Let us assume that Alice sends a qubit to Bob. Bob receives the qubit and makes a measurement on his quantum state.

Next, he can send an arbitrary quantum state to the environment - or in other words, the map to the environment of the ideal qubit channel $I$ is an entanglement-breaking map.

The ideal noiseless qubit channel $\mathcal{N}_1$, i.e., whichever quantum channel realizes the $I$ identity transformation is also a Hadamard channel, since for every Hadamard channel, the map to the environment of the channel is an entanglement-breaking map. Similar to Fig. 6.6, Bob can simulate the map of the environment, i.e., the second channel $\mathcal{N}_2$, with his ideal channel $\mathcal{N}_1 = I$ and his degrading map $\mathcal{D}$ as $\mathcal{N}_2 = \mathcal{N}_1\mathcal{D} = I\mathcal{D}$, where $\mathcal{D}$ is an entanglement-breaking map. Bob measures the received qubit and sends an arbitrary state to its environment.

The second, complementary channel $\mathcal{N}_2$ realizes an entanglement-breaking channel. After we have seen that the noiseless ideal quantum channel $I$ is a Hadamard channel, we can make an important statement regarding on its additivity: the maximized Holevo information of the ideal quantum channel $\mathcal{X}(I)$ is always additive, i.e., the classical capacity of an ideal channel is always can be computed by the single-use formula, and $C(I)$ can be maximized with product state inputs:

$$\mathcal{X}(I \otimes \mathcal{N}) = \mathcal{X}(I) + \mathcal{X}(\mathcal{N}), \tag{6.29}$$

and

$$C(I \otimes \mathcal{N}) = C^{(1)}(I \otimes \mathcal{N}), \tag{6.30}$$

where $\mathcal{N}$ can be any other quantum channel.
We note there are several other channels for which the complementary channel $\mathcal{N}_2$ is also entanglement-breaking channels, such as the generalized dephasing channel, the cloning channel or the Unruh channels [Bradler09], [Bradler09a].

## 6.4 Summary

In order to conclude the additivity related result we compared the additivity violation of classical and quantum channels, see Table 6.1.

| *Capacity* | *Single - use* $\overset{?}{=}$ *Asymptotic* | *Superadditivity (Asymptotic capacity)* |
|---|---|---|
| Classical channel $C(N)$ (i.e., not a quantum) | *Single - use* $=$ *Asymptotic* | *No* |
| Classical Capacity $C(\mathcal{N})$ | *Single - use* $\neq$ *Asymptotic* | Yes, for *same* channel maps (Unknown for *different* channel maps.) |
| Quantum Capacity $Q(\mathcal{N})$ | *Single - use* $\neq$ *Asymptotic* | Yes (For same and different channels) |
| Classical Zero-Error $C_0(\mathcal{N})$ | *Single - use* $\neq$ *Asymptotic* | Yes (For same and different channels) |
| Quantum Zero-Error $Q_0(\mathcal{N})$ | *Single - use* $\neq$ *Asymptotic* | Yes (For same and different channels) |

**Table 6.1.** The current knowledge on the additivity of different capacities of quantum channels.

The strict additivity of the classical capacity of classical communication channels was proven by Shannon [Shannon48].

The classical capacity $C(\mathcal{N})$ of quantum channels (for the same channel usage) is proven to be non-additive, since Hastings has shown in 2009 [Hastings09] that there are *identical* (same channel maps) quantum channels $\mathcal{N}_1$ and $\mathcal{N}_2$, such that $C(\mathcal{N}_1 \otimes \mathcal{N}_2) > C(\mathcal{N}_1) + C(\mathcal{N}_1)$. On the other hand, if we use two different channels $\mathcal{N}$ and $\mathcal{M}$, then the additivity of $C(\mathcal{N})$ is still not clarified.

The private classical capacity $P(\mathcal{N})$ has been proven to be non-additive, for the same and for different channels, too. The question has been analyzed by Smith and Smolin [Smith09b], and by Li et al [Li09], and they have all found that the private classical capacity is *non-additive*. The properties of the asymptotic private classical capacity have been investigated by Cai-Winter-Young [Cai04].

Interestingly, the entanglement assisted capacity $C_E(\mathcal{N})$ has been shown to be additive, for both of the channel constructions. The main results on the additivity of entanglement assisted capacity were found by Bennett, Shor, Smolin and Thapliyal [Bennett02].

Smith and Yard showed [Smith08], that the quantum capacity is also non-additive, for two different channels $\mathcal{N}$ and $\mathcal{M}$.

The superadditivity of the asymptotic zero-error classical capacity was proven by Duan [Duan09] and Cubitt et al. [Cubitt09]. The asymptotic quantum zero-error capacity was also found to be non-additive, as was shown by and Cubitt and Smith [Cubitt09a].

As we can conclude from these results, the classical, the quantum and the private classical capacities of the quantum channels are not additive, - at least, in general sense. This means, that counterexamples (i.e., special examples which are additive) can be found, however, this is still an open question.

## 6.5 Related Work

We summarize the most important works regarding on the additivity problem of the quantum channel capacities.

### Additivity Problem in quantum information theory

The problem of the additivity of quantum channels is rooted in the advanced properties of quantum communication channels. As we have seen in Section 3, in the description of the classical capacity of general quantum channels, the HSW theorem states different capacities of the single-use and the asymptotic capacities. The HSW theorem is a very elegant tool to describe the transmission of classical information over a quantum channel with product input states, however it does not give an answer to the additivity problem. It left open many questions, since in the general case the classical capacity of quantum channel is not additive, and the Holevo information cannot be used to describe the maximal classical capacity—for a general quantum channel, the asymptotic formula will give an explicit answer.

The role of additivity problems in quantum information theory was summarized in 2000 by Amosov et al. [Amosov2000]. An interesting connection between the Holevo information and the minimum output entropy of a quantum channel was shown by Shor [Shor04a]. Shor proven that the additivity of Holevo information implies the additivity of the minimum output entropy, and this connection holds in the reverse direction, too. As Shor found, these various questions on additivity are the same, in particular, if one is false then all are false—for details see Shor's article from 2004 [Shor04a].

**Additivity of Holevo Information**

Shor's result on the additivity of Holevo information and minimum output entropy states made the picture so much simpler, since it made it possible to analyze the question of additivity.

The strict additivity of unital quantum channels and the capacities of the most important quantum channels, such as depolarizing quantum channels (see Section 3.6.2), was first proven, by King, in 2002 for unital quantum channels [King02] and in 2003, specially for depolarizing quantum channels [King03b] (a summary can be found in King's remarks [King09]). King also revealed the fact that the classical capacity of the depolarizing channel can be maximized without a joint measurement setting or entangled input states (hence without any "special" quantum influences, according to the channel's strict additivity). The additivity of depolarizing channel for higher dimension was studied by Amosov [Amosov07].

We also mentioned the depolarizing quantum channel regarding on the superadditivity of quantum coherent information. The depolarizing channel map has been studied exhaustively in the literature [Amosov04], [Bennett98], [Bruss2000], [Cortese02], [Datta04b], [Fujiwara02], [King03b], [Michalakis07]. The properties of erasure quantum channels can be found in the work of Bennett [Bennett97] and of Grassl from the same year [Grassl97]. The various aspects of the additivity problem in quantum information theory was studied by Datta et al., see their works [Datta04-04b] and [Datta05]. The additivity for covariant quantum channels was analyzed by Datta et al. [Datta06].

Beside the properties of encoding and decoding of quantum states, the additivity of Holevo information depends on the map of the quantum channel, and for a general quantum channel the additivity of Holevo information was an open question. King in 2003 found a quantum channel for which the Holevo information is additive, but at this time (in 2003) the general formula was still an open question.

Then *something happened* in 2009.

**Superadditivity of Holevo Information**

Until 2009, it was conjectured that the Holevo information was additive in general, however the complete theoretical background was not clear to the researchers. Then, in 2009, the picture changed, since Hastings gave a proof that the strict additivity of Holevo information does not hold in the general case (using same channel maps), and the Holevo information is superadditive. More precisely, Hastings gave a counterexample for which channel (using two identical channel maps) the classical capacity will be superadditive—hence for entangled inputs the additivity of Holevo information fails. For the details see the proof of Hastings [Hasting2009]. (We note that, besides the existence of this counterexample, it cannot be generalized—hence, the answer for the additivity of classical capacity in the general case is still open.)

We note that the preliminaries of Hastings's proof were laid down by Winter in 2007 [Winter07], and by Hayden and Winter [Hayden08]. Hastings's proof from 2009 also gave an answer to Shor's conjectures: all of the additivity conjectures—as stated in Shor's paper from 2004, see [Shor04a]—*are false*. After Hastings's proof, in 2009, Brandao and Horodecki published a paper on Hastings's counterexamples to the minimum output entropy additivity conjecture. Later, in 2010, another work was published by Aubrun et al., on Hastings's additivity counterexample via Dvoretzky's theorem, see [Aubrun10]. Hastings's proof was also analyzed by Fukuda and King and Moser in 2010, see [Fukuda10], and by Fukuda and King, for details see [Fukuda10a].

While for the general case the Holevo information was found to superadditive, there are some quantum channels, for which the Holevo information remains additive. These channels are, for example, the Hadamard channels, the entanglement-breaking channel, or the identity quantum channel, which latter is also a Hadamard channel. The description of Hadamard channels and their Kraus-representation can be found in detail in the King et al.'s work [King07]. The capacities of these channels were also analyzed by Bradler et al [Bradler10]. The details and the properties of entanglement-breaking quantum channels can be found in the work of Horodecki, Shor and Ruskai [Horodecki03]. The additivity of classical capacity for entanglement-breaking channels is proven by Shor's [Shor02a].

**Additivity of Degradable Channels**

In this section we also introduced the definition of *degradable* quantum channel. The degradable quantum channels have many important properties. *First*, any quantum channel for which the information which can be transmitted from Alice to Bob is greater than the information leaked to the environment satisfies the requirements of a degradable quantum channel [Cubitt08b]. *Second*, the most important quantum channel models belong to this set, such as the erasure channel, the amplitude damping channel, the Hadamard quantum channel, and the bosonic quantum communication channel: these channels all have tremendous importance in practical optical communications [Wolf06]. *Third* (which is the most important for us), Devetak and Shor in 2005 [Devetak05a] have also proved that for degradable quantum channels, *quantum coherent information* (see Section 4) *is additive*. This has deep relevance from the viewpoint of the computation of the capacities of degradable quantum channels, since their capacity can be derived without the computation of the asymptotic formula. (This means that degradable quantum channels are special cases, for which it is enough to know the single-use capacity. The regularization would be necessary if quantum coherent information were non-additive.)

A complementary quantum channel is also a channel, however it is an abstract channel and focuses on the information leaked to the environment, i.e., it describes the "environment's output". The description of this abstract channel model can be found in [Devetak03], [Smith08d], and [Smith09a]. The structure of degradable quantum channels was also studied by Cubitt et al. [Cubitt08b].

**Superadditivity of Quantum Coherent Information**

In 1997, Bennett, DiVincenzo and Smolin derived the quantum capacities of the erasure quantum channels (for details see [Bennett97]). This was an important paper, since it stated that the quantum coherent information is superadditive for the depolarizing quantum channel in a very limited domain, which was first mentioned by Shor and Smolin in 1996, in [Shor96b]. In this initial report, Shor and Smolin explicitly gave the very limited range for which the quantum coherent information will be *superadditive*. Originally, the authors studied special quantum error-correcting codes for which it is not necessary to completely reveal the error syndrome, and they finally arrived at a very important conclusion: the superadditivity of quantum coherent information. This work was a very important milestone in quantum information theory, but the picture was only completed in 1998. In this year, another important step in the history of quantum capacity was made by DiVincenzo, Shor and Smolin, who analyzed the quantum capacities of various quantum channels [DiVincenzo98]. In this paper, the authors proved that the quantum coherent information is superadditive for the depolarizing quantum channel. The authors showed an example in which they used five qubit length quantum codewords and a special encoding scheme called the repetition code concatenation; for a very limited domain, using this encoding scheme, the quantum coherent information will be additive. For details see the work of DiVincenzo, Shor and Smolin [DiVincenzo98]. This was an important result in the characterization of the quantum capacity of the quantum channel. In the same year, Schumacher and Westmoreland derived the connection between the private information and the quantum coherent information (see [Schumacher98a]), whose connection was also used in the exact definition of the private classical capacity of the quantum channel by Devetak and Shor in 2005 [Devetak05a] and in the superactivation of quantum capacity by Smith and Yard in 2008 (see [Smith08]). The connection between the Holevo information and the quantum coherent information was shown by Schumacher and Westmoreland [Schumacher2000].

The quantum capacity of another important quantum channel,—the amplitude damping channel—was proven by Giovannetti and Fazio in 2005. The quantum capacity of this channel has great relevance in practical quantum communications, since this channel describes the energy dissipation due to losing a particle. In their work, the authors studied the information-capacity description of spin-chain correlations [Giovannetti05]. On the classical and quantum capacities of Gaussian quantum channels see the works of Wolf and Eisert [Wolf05] and Wolf et al. [Wolf06]. About the properties of the Gaussian quantum channels see the work of Eisert and Wolf [Eisert05]. The quantum capacities of some bosonic channels were proven by Wolf and Perez-Garcia in [Wolf07]. A great paper on the classical capacity of quantum Gaussian channels was published by Lupo et al. in 2011 [Lupo11]. For about ten years after the superadditivity of quantum coherent information of the depolarizing quantum channel was discovered by DiVincenzo et al. in 1998 [DiVincenzo98], no further quantum channels were found for which the quantum coherent information is superadditive. Finally, the picture has broken in 2007, when Smith and Smolin showed new examples for the superadditivity of quantum coherent information. They introduced a new encoding scheme, called the *degenerate quantum codes*, and they proved that with the help of use of these quantum codes for some Pauli channels the quantum coherent information will be superadditive [Smith07].

# 7. Superactivation of Quantum Channel Capacities

In the first decade of the 21st century, many revolutionary properties of quantum channels were discovered. These phenomena are purely quantum mechanical and completely unimaginable in classical systems. Recently, the most important discovery in quantum information theory was the possibility of transmitting quantum information over zero-capacity quantum channels. The phenomenon called superactivation is rooted in the extreme violation of additivity of the channel capacities of quantum channels. The superactivation of zero-capacity quantum channels makes it possible to use two zero-capacity quantum channels with a positive joint capacity for their output. Currently, we have no theoretical background to describe all possible combinations of superactive zero-capacity channels. In practice, to discover such superactive zero-capacity channel-pairs, we must analyze an extremely large set of possible quantum states, channel models, and channel probabilities. An efficient algorithmical method of finding such combinations was shown by Gyongyosi and Imre in [Gyongyosi11b].

Section 7 is organized as follows. In the first part we overview the superactivation of quantum capacity of zero-capacity quantum channels. Next we show, that there is a huge difference between the superactivation of single-use and asymptotic quantum capacity. Then, we start to discuss the superactivation of classical and quantum zero-error capacities. The important works are summarized in the Related Work subsection.

## 7.1 Superactivation of Quantum Capacity

In 2008, Smith and Yard have found only one possible combination for superactivation of quantum capacity. Duan [Duan09] and Cubitt et al [Cubitt09] found a possible combination for the superactivation of the classical zero-error capacity of quantum channels, which has opened up a debate regarding the existence of other possible channel combinations. Later, these results were extended to the superactivation of quantum zero-error capacity by Cubitt and Smith [Cubitt09a]. Initially, the superactivation property was proven for the transmission of quantum information over zero-capacity quantum channels. In this combination, each quantum channel has zero quantum capacity individually; however, their joint quantum capacity is strictly greater than zero. On the other hand, these results did not allow transmitting classical information over combination of quantum channels in which each channel has zero *classical* capacities. Gyongyosi et al. [Gyongyosi12a] gave the mathematical proof that by the adding of quantum entanglement, two quantum channels each with zero *classical capacity* can be combined together to transmit classical information – which result previously was thought to be impossible.

As shown in Fig. 7.1, the problem of superactivation can be discussed as part of a larger problem – the problem of quantum channel additivity. The superactivation property can be discussed from the viewpoint of the superactivation of quantum capacity or the classical and quantum zero-error capacities of the quantum channel.

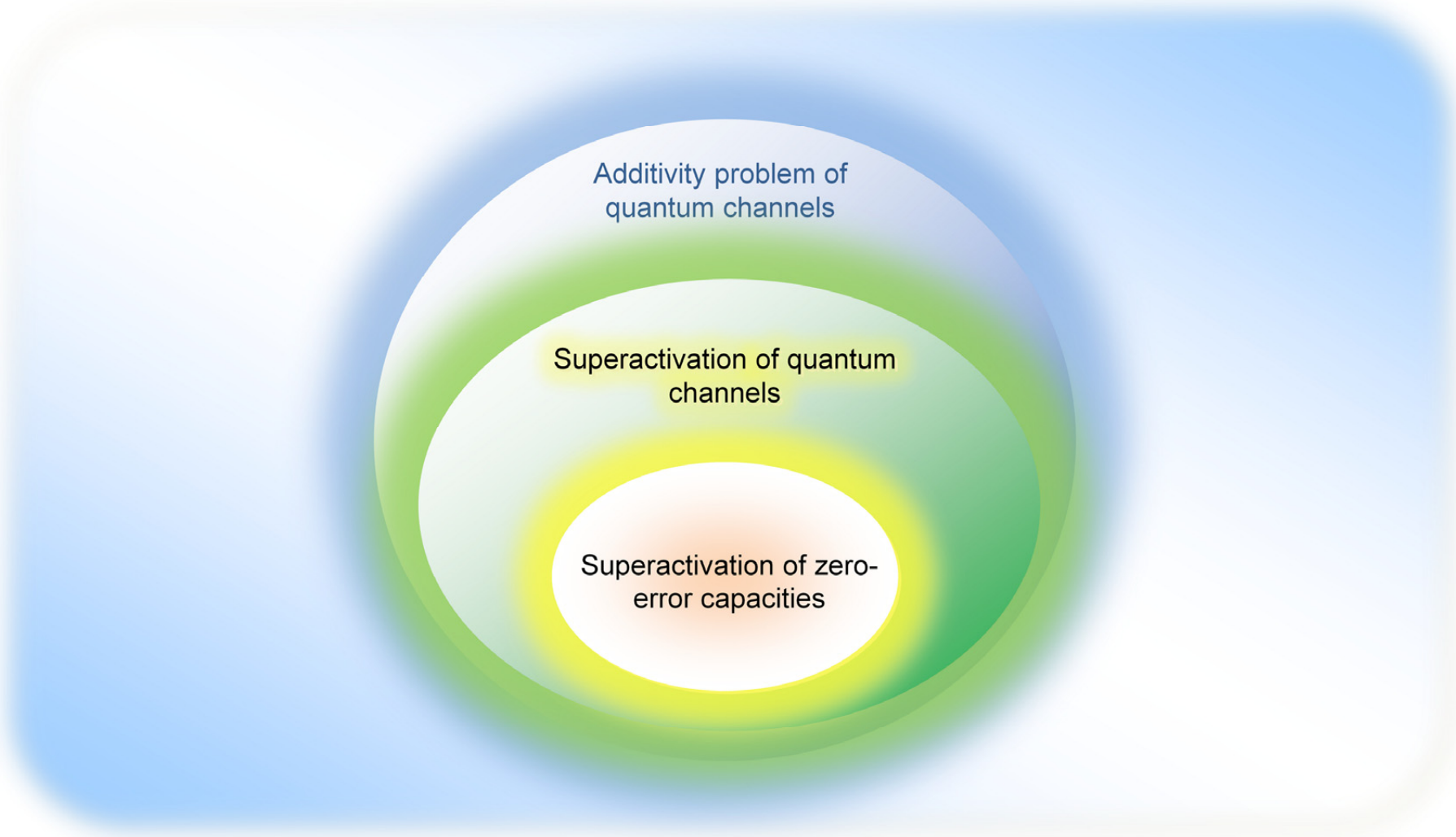

**Fig. 7.1.** The problem of superactivation of zero-capacity quantum channels as a sub domain of a larger problem set.

For the combination of any quantum channel $\mathcal{N}_1$ that has some private classical capacity $P(\mathcal{N}_1) > 0$ and a "fixed" 50% erasure symmetric channel $\mathcal{N}_2$, the following connection holds between the asymptotic quantum capacity of the joint structure $\mathcal{N}_1 \otimes \mathcal{N}_2$, and the private classical capacity $P(\mathcal{N}_1)$:

$$Q\left(\mathcal{N}_1 \otimes \mathcal{N}_2\right) \geq \frac{1}{2} P\left(\mathcal{N}_1\right). \tag{7.1}$$

The channel combination for the superactivation of the asymptotic quantum capacity of zero-capacity quantum channels is shown in Fig. 7.2.

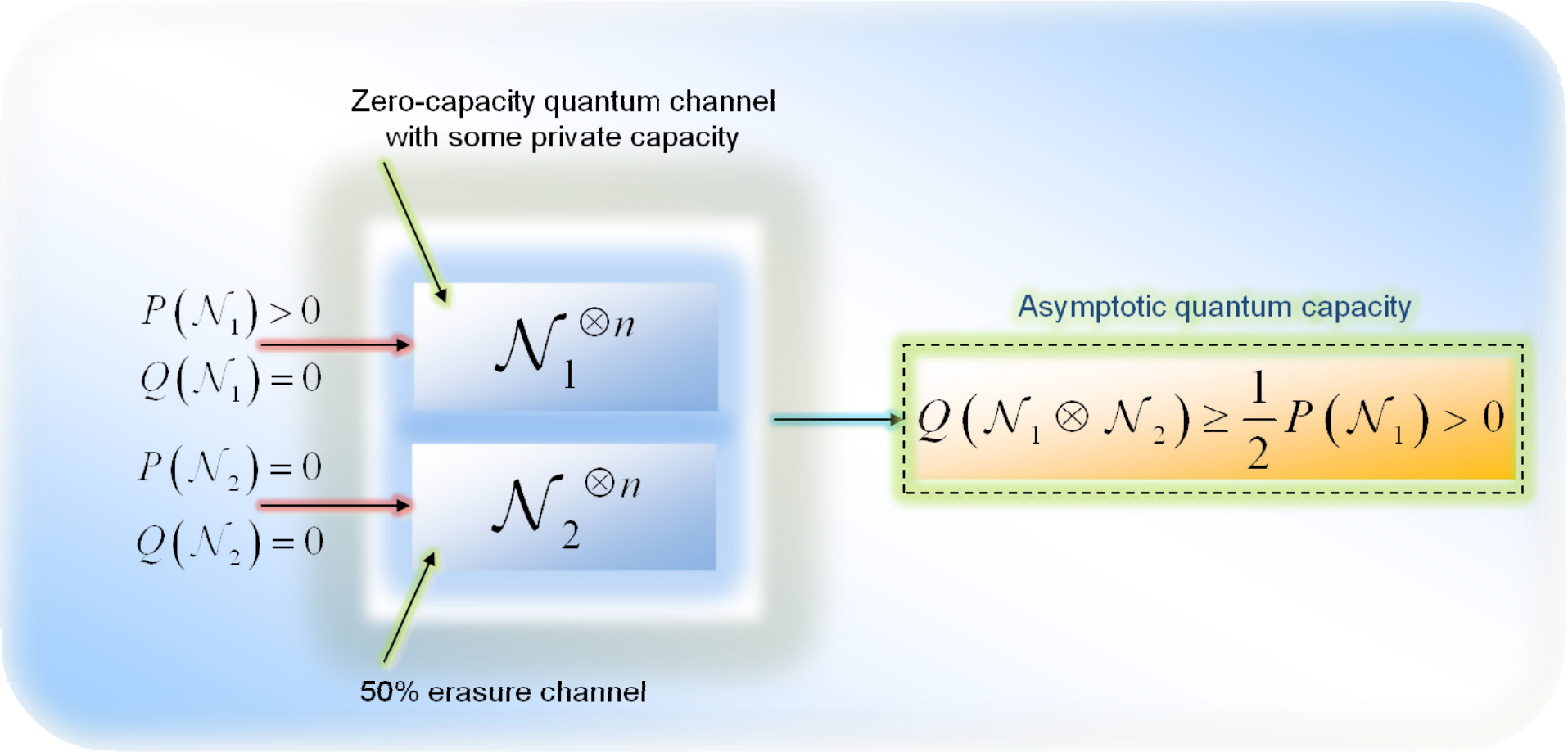

**Fig. 7.2.** The first channel has some positive private classical capacity, and the second quantum channel is a 50% erasure channel with zero quantum capacity.

In what follows, we will see that it is possible to find other combinations of quantum channels $\mathcal{N}_1$ and $\mathcal{N}_2$, which has individually "zero-capacity" in the sense that

$$Q\left(\mathcal{N}_1\right) = Q\left(\mathcal{N}_2\right) = 0, \tag{7.2}$$

and still satisfy

$$Q\left(\mathcal{N}_1 \otimes \mathcal{N}_2\right) > 0. \tag{7.3}$$

This rather strange phenomena is called *superactivation*. For the channel combination $\mathcal{N}_1 \otimes \mathcal{N}_2$ the positive single-use quantum capacity $Q^{(1)}\left(\mathcal{N}_1 \otimes \mathcal{N}_2\right)$ was proven using simple algebra [Smith08]. In the channel construction of Smith and Yard's, the superactivation of the quantum capacity of the two quantum channels requires two EPR states (In their proof, the first channel is the four-dimensional Horodecki channel $\mathcal{N}_H$ with $P\left(\mathcal{N}_H\right) > 0$, the second is the four-dimensional 50% erasure channel $\mathcal{A}_e$.).

Next, we will show the difference between the superactivated single-use quantum capacity $Q^{(1)}\left(\mathcal{N}_1 \otimes \mathcal{N}_2\right)$ and superactivated asymptotic quantum capacity $Q\left(\mathcal{N}_1 \otimes \mathcal{N}_2\right)^{\otimes n}$ can be made arbitrarily high, if we use a different channel combination. The results on the superactivation of quantum capacity also implied the fact that the quantum capacity is not convex, hence for the combination of two quantum channels $\mathcal{N}_1$ and $\mathcal{N}_2$, the following property holds between their joint quantum capacity $Q^{(1)}\left(\mathcal{N}_1 \otimes \mathcal{N}_2\right)$ and their individual capacities $Q^{(1)}\left(\mathcal{N}_1\right)$ and $Q^{(1)}\left(\mathcal{N}_2\right)$

$$Q^{(1)}\left(\left(1-p\right)\mathcal{N}_1 + p\mathcal{N}_2\right) > \left(1-p\right)Q^{(1)}\left(\mathcal{N}_1\right) + pQ^{(1)}\left(\mathcal{N}_2\right), \tag{7.4}$$

with probability $0 \leq p \leq 1$, which means the following: the single-use joint quantum capacity of the channel combination could be greater than the sum of their individual quantum capacities.

Finally, we show a channel combination example for which there is a large difference between the *single-use* and the *asymptotic* quantum capacity. This can be achieved by the combination of a $d$ dimensional random phase coupling channel $\mathcal{R}_d$ (defined by random unitary maps) [Smith09b], and a 50% erasure channel, denoted by $\mathcal{N}_2$. The random phase coupling channel and is defined as follows:

$$\mathcal{N}_1 = \mathcal{R}_d = \mathcal{R}^{U_1,U_2}{}_d \otimes \left|U_1U_2\right\rangle\left\langle U_1U_2\right|, \tag{7.5}$$

where $\mathcal{R}^{U_1,U_2}{}_d$ denotes the *d dimensional* random phase coupling channel. The random phase coupling channel consists of two unitary transformations $U_1$ and $U_2$, where both unitary transformations are unknown to Alice, while Bob knows both $U_1$ and $U_2$.

The asymptotic quantum capacity of this structure will be denoted by

$$Q\left(\mathcal{N}_1 \otimes \mathcal{N}_2\right) = Q\left(\mathcal{R}_d \otimes \mathcal{N}_2\right), \tag{7.6}$$

where $\mathcal{R}_d$ is the random phase coupling channel and $\mathcal{N}_2$ is the 50% erasure channel. In the case of the single-use quantum capacity, this channel realizes the following map as depicted in Fig. 7.3:

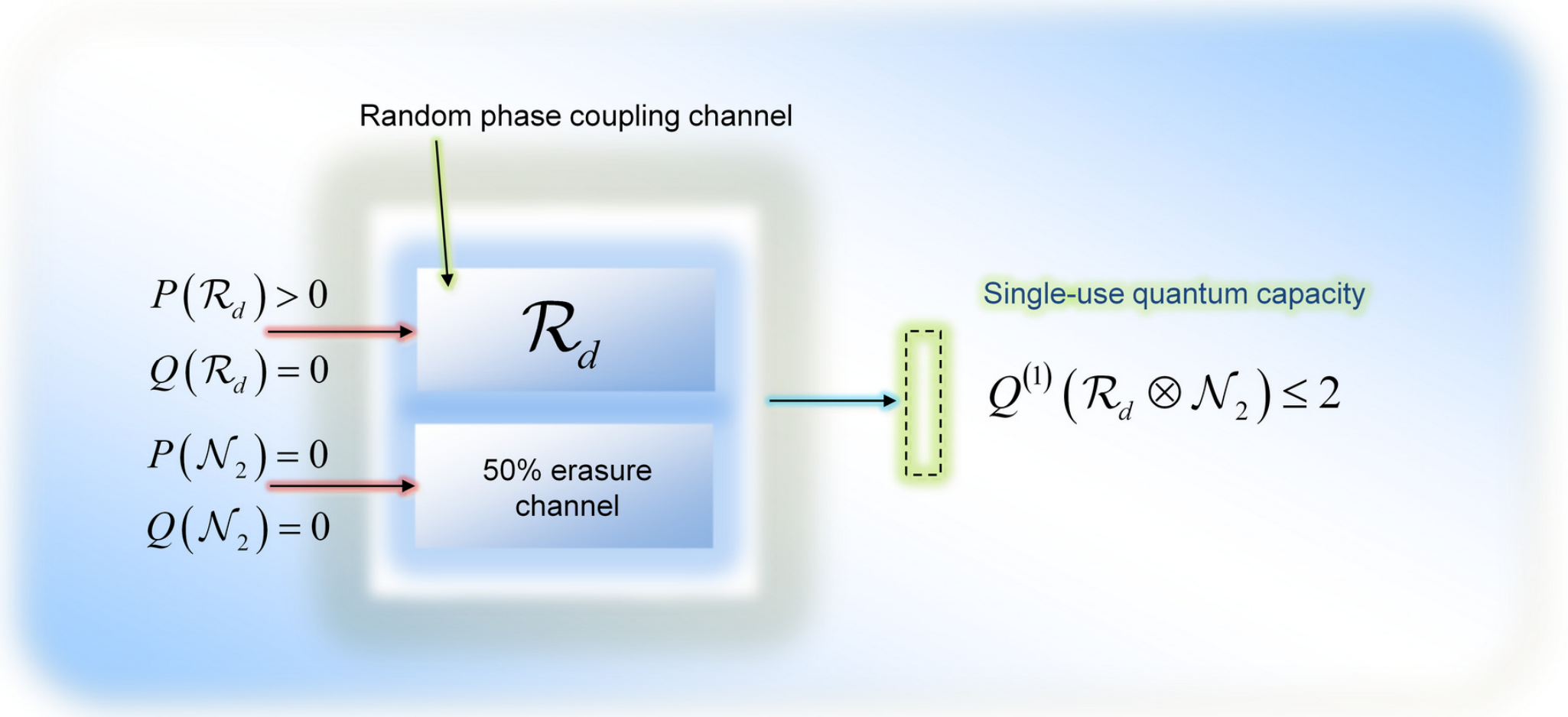

**Fig. 7.3.** The single-use quantum capacity of the channel construction, which consist of the random phase coupling channel and the 50% erasure quantum channel.

In this case, the *single-use* quantum capacity of the joint channel is measured by the maximized quantum coherent information, as

$$Q^{(1)}\left(\mathcal{R}_d \otimes \mathcal{N}_2\right) = \max_{all\ p_i,\rho_i} I_{coh}\left(\rho_A : \mathcal{R}_d \otimes \mathcal{N}_2\right) \leq 2\,, \tag{7.7}$$

since we know that $Q^{(1)}\left(\mathcal{R}_d\right) \leq 2$ or $Q^{(1)}\left(\mathcal{N}_2\right) = 0$.

On the other hand, if we measure the *asymptotic* quantum capacity for the same channel construction $\left(\mathcal{R}_d \otimes \mathcal{N}_2\right)$, then we will find that

$$Q\left(\mathcal{R}_d \otimes \mathcal{N}_2\right) \geq \frac{1}{8}\log\left(d\right) \gg Q^{(1)}\left(\mathcal{R}_d \otimes \mathcal{N}_2\right), \tag{7.8}$$

where $d$ is the input dimension. This asymptotic version of the previously seen construction is shown in Fig. 7.4.

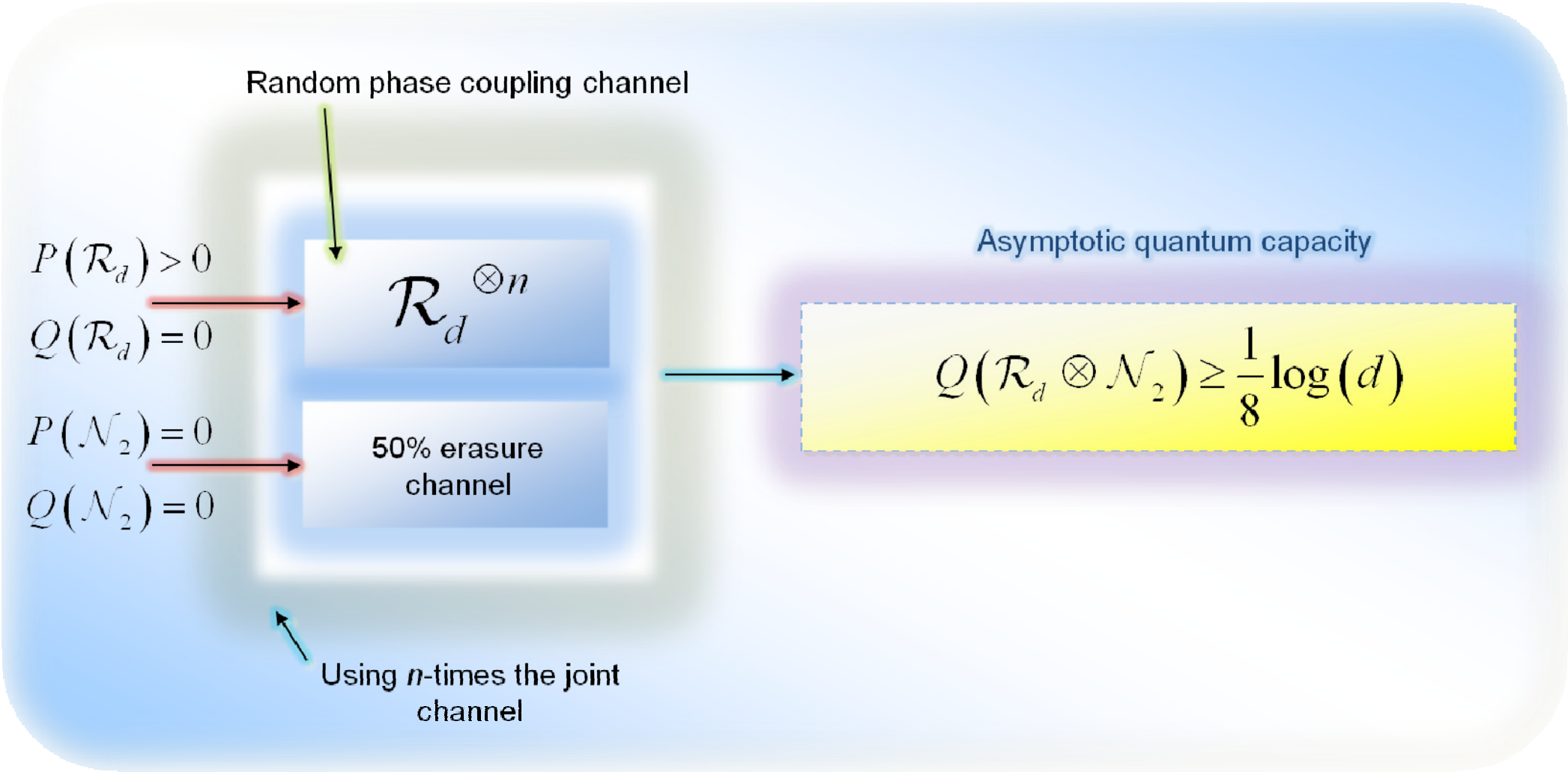

**Fig. 7.4.** The asymptotic capacity of the channel combination. There is a big gap between the quantum maximized coherent information and the asymptotic quantum capacity of the analyzed channel construction.

Summarize, if we have a joint channel combination which contains a random phase coupling channel and a 50% erasure channel, then the convexity of quantum capacity (see (7.4)) will be also satisfied, since for this combination the joint quantum capacity greater than the sum of individual capacities. Moreover, for this channel combination while the single-use quantum capacity of the structure is bounded by 2, the asymptotic quantum capacity can be significantly increased.

## 7.2 Superactivation of Classical and Quantum Zero-Error Capacities

As we have seen in Section 3, the zero-error capacity of the quantum channel describes the amount of information which can be transmitted perfectly through a noisy quantum channel. The superactivation of the zero-error capacity of quantum channels makes it possible to use noisy quantum channels with perfect information transmission [Duan09], [Cubitt09], [Cubitt09a].

The superactivation of quantum channels may be the starting-point of a large-scale revolution in quantum information theory and in the communication of future quantum networks where quantum channels are extremely noisy. With the help of the superactivation of zero-error capacity of quantum channels, the perfect information transmission can be realized in a very noisy network environment. Moreover, it can be used to enhance the security of quantum communication over a very noisy quantum environment; hence the application of superactivation in future quantum communication networks is very diverse. The superactivation of quantum zero-error capacity can be used to build more efficient quantum repeaters in the future, and the entanglement purification process can almost completely be removed from the current approaches [Gyongyosi11e-f]. In Fig. 7.5, we show possible ways of applying superactivation in the quantum repeater structure [Gyongyosi11c]. The superactivation makes it possible to transmit information perfectly between the repeater stations, using very noisy quantum channels.

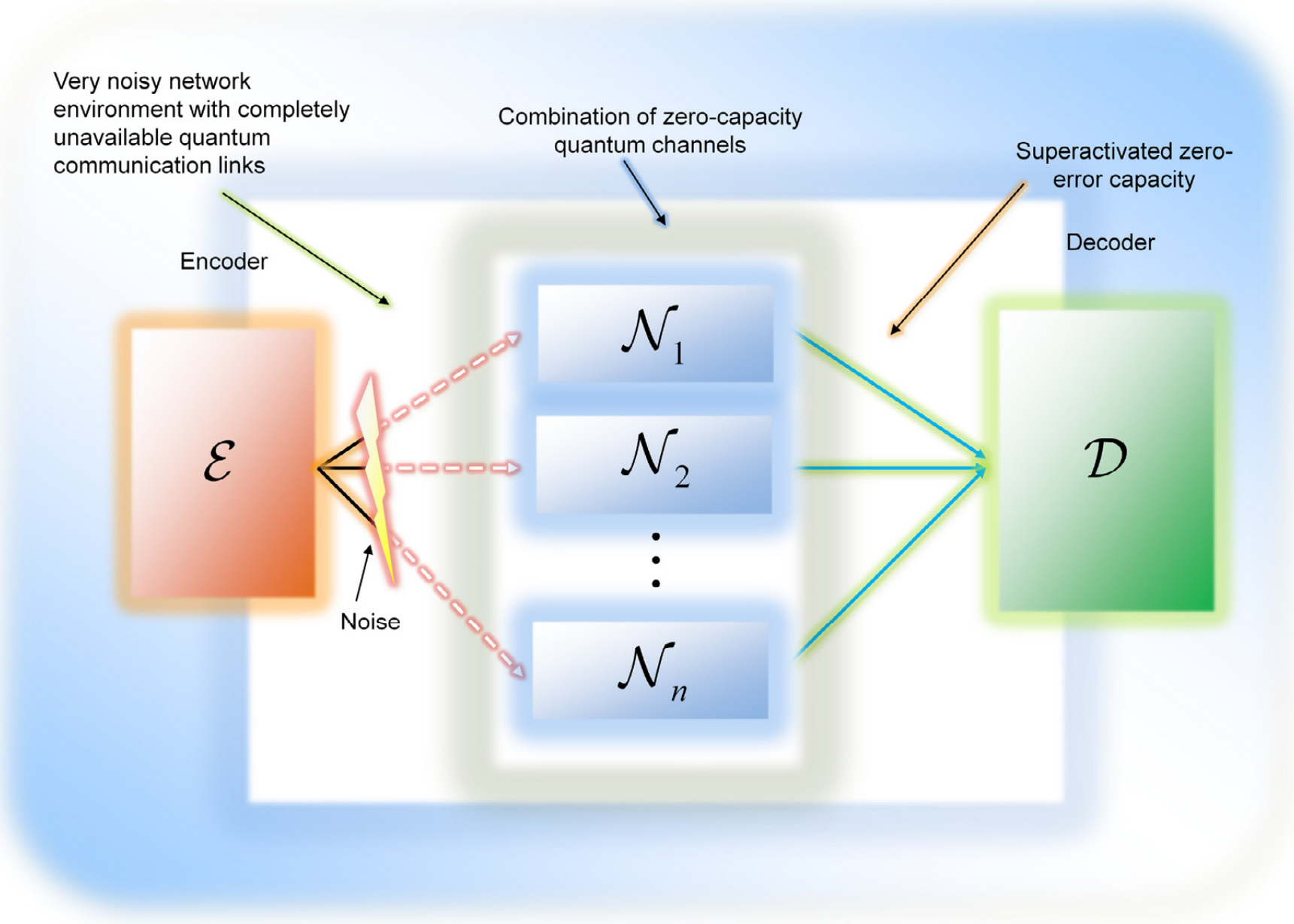

**Fig. 7.5.** The superactivation of zero-error capacity can help to transmit information through a very noisy quantum network, and the information can be transmitted perfectly through a temporarily unavailable quantum channel.

Using the superactivation of quantum channels, the efficiency of the quantum repeater can be increased, since the purification steps can be completely removed, and the long-distance quantum communication techniques can be revolutionized [Gyongyosi11f].

In order to give deeper insight we show an example how superactivation of classical zero-error capacity $C_0\left(\mathcal{N}\right)$ can be realized. The encoder setting requires EPR-states to encode the non-adjacent codewords and joint measurement so that we can distinguish the codewords. Let assume, the information source emits classical binary symbols $x_i \in \left\{0,1\right\}$ which are encoded into EPR pairs $X = \left[x_1, x_2, \ldots, x_n\right] \rightarrow \left[\left|\Psi_1\right\rangle \otimes \left|\Psi_2\right\rangle \otimes \left|\Psi_3\right\rangle \cdots \otimes \left|\Psi_n\right\rangle\right]$, where $n$ is the number of input *EPR photon pairs in the quantum blockcode*, while

$$\left|\Psi_i\right\rangle \in \left\{\left|\beta_{00}\right\rangle, \left|\beta_{01}\right\rangle\right\} = \left\{\frac{1}{\sqrt{2}}\left(\left|00\right\rangle + \left|11\right\rangle\right), \frac{1}{\sqrt{2}}\left(\left|01\right\rangle + \left|10\right\rangle\right)\right\} \tag{7.9}$$

denotes the $i$-th entangled input photon pair. The $n$ entangled states define a $\left|\Psi_{IN}\right\rangle$ input system which contains EPR states:

$$\left|\Psi_{IN}\right\rangle = \left[\left|\Psi_1\right\rangle \otimes \left|\Psi_2\right\rangle \otimes \left|\Psi_3\right\rangle \cdots \otimes \left|\Psi_n\right\rangle\right]. \tag{7.10}$$

The $i$-th sent codeword $\left|\Psi_{X_i}\right\rangle = \left[\left|\Psi_{i,1}\right\rangle \otimes \left|\Psi_{i,2}\right\rangle \otimes \left|\Psi_{i,3}\right\rangle \cdots \otimes \left|\Psi_{i,n}\right\rangle\right]$ encoded by the encoder will be decoded by decoder $\mathcal{D}$, using the POVM operators $\left\{\mathcal{M}_j\right\}$, where $\sum_j \mathcal{M}_j = I$. For each $n$-length input codeword $X \in \left\{X_1, X_2, \ldots, X_K\right\}$, the output words $X_i' \in \left\{1, \ldots, m\right\}^n$ are generated by the measurement operators $\left\{\mathcal{M}_1, \ldots, \mathcal{M}_m\right\}$. The decoder associates each output word with integers 1 to $K$ representing input messages. In the encoding process, Alice chooses a message $X_i$ from the set of $K$ input messages, and prepares an $n$ length quantum blockcode $\left|\Psi_{X_i}\right\rangle = \left[\left|\Psi_1\right\rangle \otimes \left|\Psi_2\right\rangle \otimes \left|\Psi_3\right\rangle \cdots \otimes \left|\Psi_n\right\rangle\right]$. These entangled states are sent through the joint channel construction, in which each quantum channel have zero zero-error capacities individually. Bob uses his decoder $\mathcal{D}$, to obtain an output integer number using the POVM measurement, which will identify the input codeword.

The general view of the required channel setting for the superactivation of two quantum channels each with zero zero-error capacities is shown in Fig. 7.6. The $i$-th entangled photon pair $\left|\Psi_i\right\rangle$ consists of entangled particles $\left|\Psi_i\right\rangle \in \left\{\left|\beta_{00}\right\rangle, \left|\beta_{01}\right\rangle\right\}$, where $\left|\beta_{00}\right\rangle$ and $\left|\beta_{01}\right\rangle$ are the Bell states (see Appendix)

and $\rho_i^{(1)} = \left|\Psi_i^{(1)}\right\rangle\left\langle\Psi_i^{(1)}\right|, \rho_i^{(2)} = \left|\Psi_i^{(2)}\right\rangle\left\langle\Psi_i^{(2)}\right|$ denote the first and the second qubits of the EPR state.

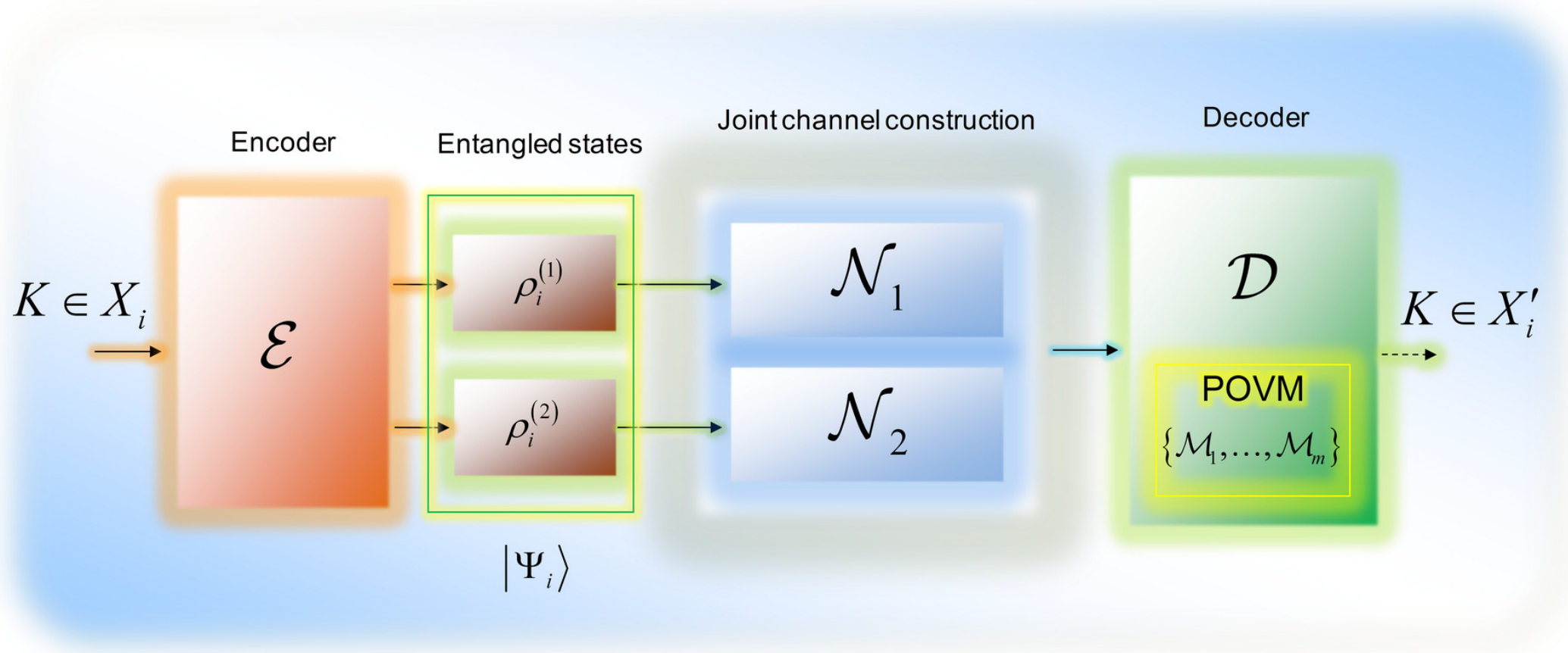

**Fig. 7.6.** Information transmission with zero-error over superactivated quantum channels. Each quantum channels have zero zero-error capacities individually.

Using $n$ EPR states for the transmission the single-use and asymptotic superactivated zero error capacities for the joint structure $(\mathcal{N}_1 \otimes \mathcal{N}_2)$ can be expressed as follows:

$$C_0^{(1)}(\mathcal{N}) = \frac{1}{2}\log\left(K\left(\mathcal{N}_1 \otimes \mathcal{N}_2\right)\right) \tag{7.11}$$

and

$$C_0(\mathcal{N}) = \lim_{n\to\infty}\frac{1}{2n}\max_n \log\left(K\left(\mathcal{N}_1 \otimes \mathcal{N}_2\right)^{\otimes n}\right), \tag{7.12}$$

where $K\left(\left(\mathcal{N}_1 \otimes \mathcal{N}_2\right)^{\otimes n}\right)$ is the maximum number of classical $n$-length messages that the superactivated joint channel construction $(\mathcal{N}_1 \otimes \mathcal{N}_2)$ can transmit with zero error. Since we will encode the $n$-length input codewords with EPR input states, we use $2n$ in the denominator.

According to the current results, the superactivation of classical zero-error [Duan09], [Cubitt09] and quantum zero-error capacities [Cubitt09a] works only for very special channel pairs with strict requirements on the input and output dimensions of the channels.

In 2012 Gyongyosi et al. [Gyongyosi12b] proved that the superactivation of all channel capacities for which the superactivation is possible is limited by the mathematical properties of the quantum relative entropy function.

## 7.3 Related Work

The discovery of superactivation was a very important result in the characterization of the capability of the quantum channel to transmit information. The effect of superactivation roots in the extreme violation of additivity, i.e., in the superadditivity property of channel capacities (see Section 6). Both the superadditivity of classical and quantum channel capacities were already shown before the possibility of the superactivation would had been brought to the surface. While in the case of the superactivation of quantum capacity the superadditivity of quantum coherent information, in the case of the superactivation of classical zero-error capacity, the superadditivity of Holevo information provides the theoretical background for the effect. Up to 2011, the superactivation of general classical capacity of the quantum channels is not possible.

The superactivation is nothing more than an extreme violation of additivity property of quantum channels. As the inventors of the HSW theorem in 1997 have conjectured, entanglement among the input states cannot help to enhance the rate of classical communication. However later, in 2009, Hastings proved that entanglement can help to increase the classical capacity, and showed that the additivity of the Holevo information can fail, i.e., the additivity works only for some very special channels and cannot be extended to the general case [Hastings09].

The first important discovery—which also gave a strong background to these advanced properties—was made by Horodecki et al. in 2005 [Horodecki05]. As they showed, quantum information can be negative (see Section 2), and they have also constructed a protocol which uses this fact. It was an important milestone from the viewpoint of the discovery of the advanced—classically unimaginable—properties of quantum channels. Further details about the meaning of the negativity of quantum information can be found in the proof of Horodecki et al. [Horodecki05] and 2007 [Horodecki07]. Before their results appeared, a paper about the role of negative entropy and information in quantum information theory was published by Cerf in 1997 [Cerf97]. In the paper of Horodecki et al.

[Horodecki05], the authors also defined a protocol which can exploit the negativity of quantum information. The continuity of quantum conditional entropy function was proved by Alicki and Fannes in 2004 [Alicki04]. An attempt for giving a uniform framework for the currently known different quantum protocols was made by Devetak et al. in 2004 [Devetak04a], by Devetak and Shor in 2005 [Devetak05a], and by Devetak et al. in 2008 [Devetak08]. However, as they concluded, there are still many open questions. Later, in 2006, Abeyesinghe et al. published a paper in which they tried to give a more generalized picture of the various quantum communication protocols and their various capacities [Abeyesinghe06].

**Discovery of Superactivation**

The possibility of the superactivation of the quantum channels was discovered by Graeme Smith and Jon Yard in 2008 [Smith08]. They have shown that the quantum capacity of zero-capacity quantum channels can be superactivated, and in 2011 they demonstrated in laboratory environment that the superactivation of the quantum capacity also works in practice [Smith11]. In 2009 and 2010, Duan and Cubitt et al. showed that the classical zero-error capacity [Duan09], [Cubitt09], and the quantum zero-error capacity can also be superactivated [Cubitt09a]. In 2011 the effect of superactivation was extended to more general classes of quantum channels. Brandao, Oppenheim and Strelchuk have demonstrated that the superactivation of depolarizing quantum channels (see Section 3.6.2) is also possible and can be extended for more general classes, for details see [Brandao11]. In 2010, Brandao et al. have also studied the connection between the public quantum communication and the effect of superactivation, and the possible impacts on quantum error correction and entanglement distillation, for details see [Brandao10]. A very good overview on the working mechanism of superactivation was published by Oppenheim [Oppenheim08].

In 2008 Smith and Smolin showed that the non-additivity of the private classical capacity can be extended in a different way [Smith08a], since this property can be used in the superactivation of zero-capacity quantum channels. Li et al. [Li09] constructed a channel combination for which the entanglement-assisted quantum capacity is greater than the classical capacity. The results presented by Li and Winter et al., demonstrated that the private classical capacity is also non-additive [Li09]. The very strong non-additivity of private classical capacity has been shown by Smith [Smith09b], and by Li et al [Li09]. Many of these discoveries were made in 2008 and 2009, and the some results were discovered just in 2010 and 2011. In both of the superadditivity property of private classical information and the superactivation of quantum capacity the erasure channel has great importance. The fact that any symmetric channel (i.e., for example a 50% erasure channel) has zero quantum capacity was shown in [Bennett97]. The proof of that any positive capacity would violate the no-cloning theorem was shown in [Bennett96a].

**Algorithmical Solution for Superactivation**

Smith et al. have also shown in 2011 that there exist a channel combination—using optical fiber quantum channels—for which superactivation of quantum capacity can be realized in practice [Smith11]. In 2010, a geometrical method for the discovery of further combinations of these superactive quantum channels was developed by Gyongyosi and Imre [Gyongyosi11b]. Gyongyosi and Imre in 2011 showed that the phenomenon of superactivation can be used to develop more efficient quantum repeaters with the elimination of the inefficient entanglement purification process [Gyongyosi11c], [Gyongyosi11f], [Gyongyosi12].

# 8. Quantum Channels in Practice

Section 8 discusses quantum networks from long-distance quantum communications point of view.

This section is organized as follows. The first part of the section gives a brief introduction to the working mechanism of the quantum repeater. Next, we introduce the reader to the performance of the practical long-distance quantum communications. In the end, Further Reading summarizes the historical background with the description of the most important works.

## 8.1 Long-distance Quantum Communications

Because of the no cloning theorem which makes it hard to amplify signals carrying quantum information, the success of future long-distance quantum communications and global quantum key distribution systems depends on the development of efficient quantum repeaters. It is based on the transmission of entangled quantum states between the repeater nodes. In the quantum communication networks of the future, besides long-distance communication, other networks structures could be implemented, such as self-organizing, truly probabilistic quantum networks.

Obviously, idealistic quantum channels do not require any repeater function. The noise introduced by the real quantum channels is the reason why repeaters are required at all. A long-distance communication chain between Stations A and B with a repeater station is depicted in Fig. 8.1. Considering the terminology from now on, network entities are called *nodes*; *stations* refer to information source and destination nodes while those relaying nodes which only receive and send the payload information without processing it are the *repeaters*.

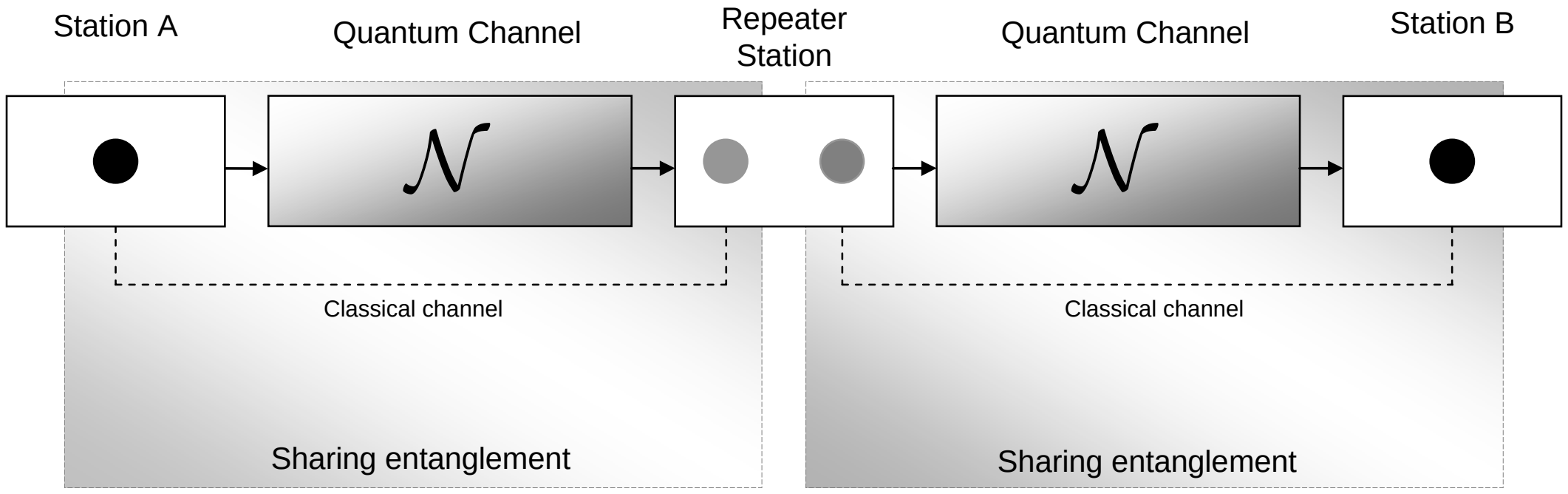

**Fig. 8.1.** Long-distance communication chain with one repeater.

The quantum repeater itself can be regarded as a quantum computer, which realizes entanglement purification and quantum teleportation [Loock08]. The purification of the entangled quantum states requires the following steps: single qubit rotations, measurements and a controlled phase gate, which controls the quantum operations. The quantum channels between the quantum repeater are noisy, the noise of the channels can be compensated with the help of quantum teleportation protocol.

The role of teleportation [Imre05] is obviously to deliver unknown quantum states carrying payload information between neighboring network entities. During teleportation see Fig. 8.2, the sender's input quantum state is destroyed and recovered at the receiver's side, using shared entanglement between the parties. The receiver needs two classical bits to recover the unknown quantum state. This is the reason why auxiliary classical channels are included in Fig. 8.1. Both the sender and the receiver shall use only *local* quantum operations, which are based on this classical information.
The steps of the quantum teleportation protocol are summarized in Fig. 8.2. Repeater stations A and B share an EPR pair, whose qubits are denoted by density matrices $\Psi_A$ and $\Psi_B$. The unknown qubit interacts with the first half of the EPR state and then both qubits are measured which yields two classical bits. The classical bits are sent through a classical communication channel. In Repeater station B the second half of the EPR state is locally transformed according to the two classical bits, and finally the unknown quantum state is obtained.

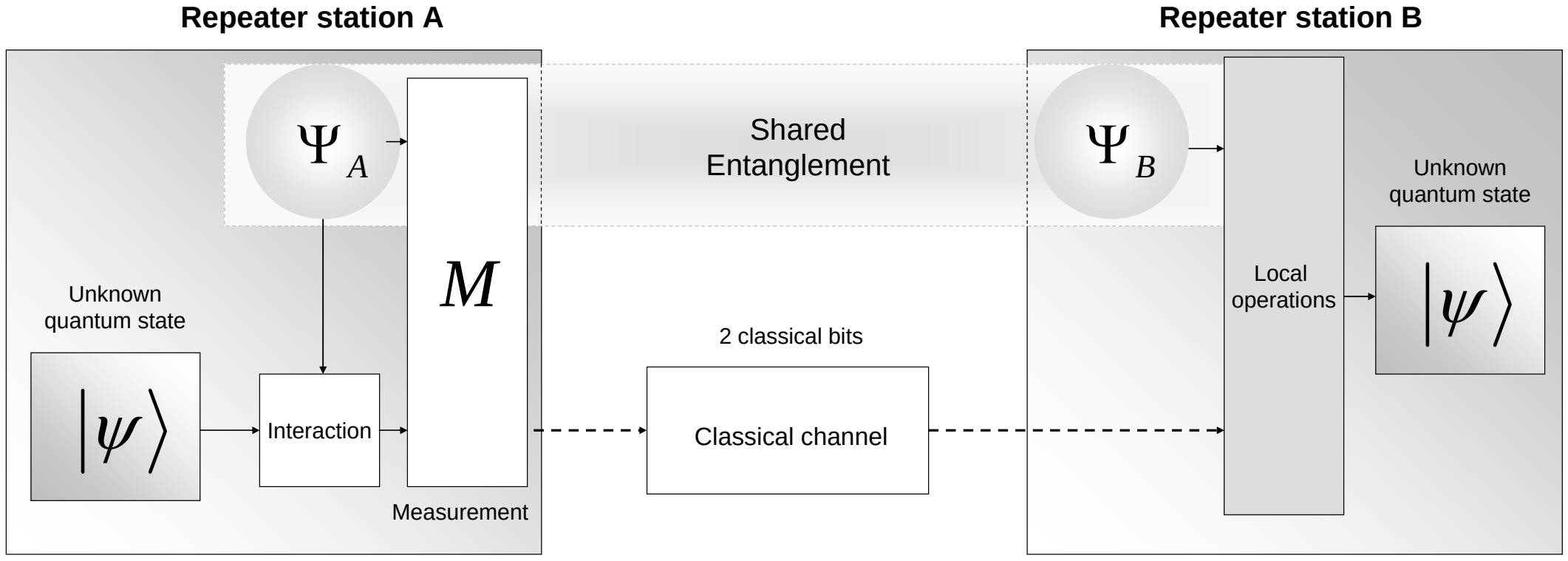

**Fig. 8.2.** The process of quantum teleportation in the quantum repeater.

Prior to the teleportation an entanglement sharing process has to be performed over the quantum channel since entanglement cannot be established between distant locations by means of only classical channels and local transforms. The sender station entangles a given quantum state with another separate physical qubit, and then it is sent over the quantum channel. If the entangling operation and the transmission of the half EPR pair prove to be successful, then the sender and the receiver have shared an entangled pair and they are ready to perform teleportation.

Unfortunately the entanglement sharing uses a noisy quantum communication channel; hence the transmitted half EPR pair will be distorted. Therefore, the shared and disturbed entanglement has to be purified. The purification is an error-correcting scheme, and it uses local quantum operations only – hence these operations can be realized in the separated stations locally. Roughly speaking the purification takes several disturbed EPR pairs and by means of local quantum transformation and classical communication, it combines these pairs into one, higher-fidelity EPR state.

Quantum repeaters use the purification protocol to increase the fidelity of transmission. The rate of entanglement purification depends on the fidelity of the shared quantum states, since the purification step is a probabilistic process. Another important bottleneck of the purification algorithm is that it requires a lot of classical information exchange between the quantum nodes. In the work of Van Meter et al. [VanMeter08], a new algorithm has been introduced for this purpose, they called it *banded purification*. The banded purification scheme could improve the utilization of the resources.

Sharing of quantum entanglement plays critical role in quantum repeaters. The fidelity of the entanglement decreases during the transmission through the noisy quantum channel. Therefore, in practical implementations, the quantum entanglement cannot be distributed over very long-distances; instead, the EPR states are generated and distributed between smaller segments. While teleportation links neighboring stations, *entanglement swapping* is a consecutive set of quantum teleportation steps – i.e., it is an “extended teleportation protocol”, which allows bridging the gap between the physically separated stations in long-distances. The entanglement swapping connects the EPR states, which states are generated independently between the neighboring repeater stations into an EPR state, i.e., an entangled pair has been produced between two distance stations connected over several repeater stations. The entanglement swapping consists of local measurement at each repeater station. The measurement at the repeater frees up the two local qubits, which results in two classical bits. After the measurement Alice and Bob can determine the local operations to adjust the distant EPR state into the corresponding Bell state, i.e., the remote EPR state between Alice and Bob is generated successfully. The entanglement swapping can be applied simultaneously for all repeater stations, i.e., the entanglement swapping between to distant stations can be realized fast. The physical apparatus behind the purification and swapping of the quantum states is rather simple: it requires only controlled phase gates and a simple Hadamard gate, with a measurement apparatus.

### 8.1.1 General Model of Quantum Repeater

As we will present in this section, there are several differences between classical and a quantum repeaters. For example a quantum repeater is not simply a signal amplifier, in contrast to classical repeaters. The design of a quantum repeater has been studied by Van Meter et al. [VanMeter08-09], and in 2008, Munro et al. constructed a system design for a practical quantum repeater [Munro08].

As we mentioned before the working mechanism of the quantum repeater is based on two protocols:

- *purification of quantum states to restore the effect of noisy quantum channels on EPR pairs.*
- *entanglement swapping: set of quantum teleportation steps using purified EPR pairs.*

These protocols are involved when long-distance quantum communication is implemented by means of repeaters:

1. Sharing of Bell states among the quantum nodes i.e., *Entanglement sharing* = *Creation* of EPR pairs at a certain node and the *delivery of* one half pair to another neighboring node.
2. *Purification* of the shared EPR states which consist of *entanglement distillation* and *scheduling* of distillation steps.
3. *Teleportation* between neighboring nodes and *entanglement swapping* between the source and destination stations.

The first two steps – entanglement sharing and purification – together are called *entanglement distribution.*

*Step 1. - Sharing of EPR states among the quantum nodes*

The creation of EPR-states uses the combination of atomic qubits systems and coherent light pulses. The quantum states initially are unentangled among the two stations. In the first phase of entanglement creation, the single qubit system interacts with the coherent pulse, which pulse being transmitted over the quantum channel between the two stations. In the next phase, the receiver station interacts the incoming coherent pulse with a second single qubit system. Finally, in the receiver node the coherent pulse is measured. After the measurement the entanglement is generated between the two single qubit systems of the sender and the receiver stations.

The EPR-pairs are shared between the neighboring repeaters independently and simultaneously. These stations could be a few tens of kilometers from each other. Using optical devices for quantum communication, nowadays the entangled quantum states can be created with a fidelity of 0.63 for 20 kilometers [VanMeter08], which can even be increased by the purification step. In the sharing process, the sender station entangles a quantum state with another separate physical qubit, then it is multiplexed into the optical-fiber - or the quantum channel. At the receiver side, the multiplexed pulses are demultiplexed, and the receiver entangles each pulse with a free quantum state. If the entangling operation succeeds, then the sender and the receiver share an EPR state.

The entangled quantum states can be sent through the quantum channel as single quantum states or as multiple photons. In the first case the fidelity of the shared entanglement could be higher, however, it has lower probability of success in practice, since these quantum states can be lost easily on the *noisy* quantum channel.

*Step 2. - Purification of the shared EPR states*

The *purification* step takes two EPR pairs and by the use of local quantum transformation and classical communication, it combines the two EPR states into one EPR pair, which has greater fidelity. The theoretical working mechanism of purification between Station A and the Repeater is illustrated in Fig. 8.3.

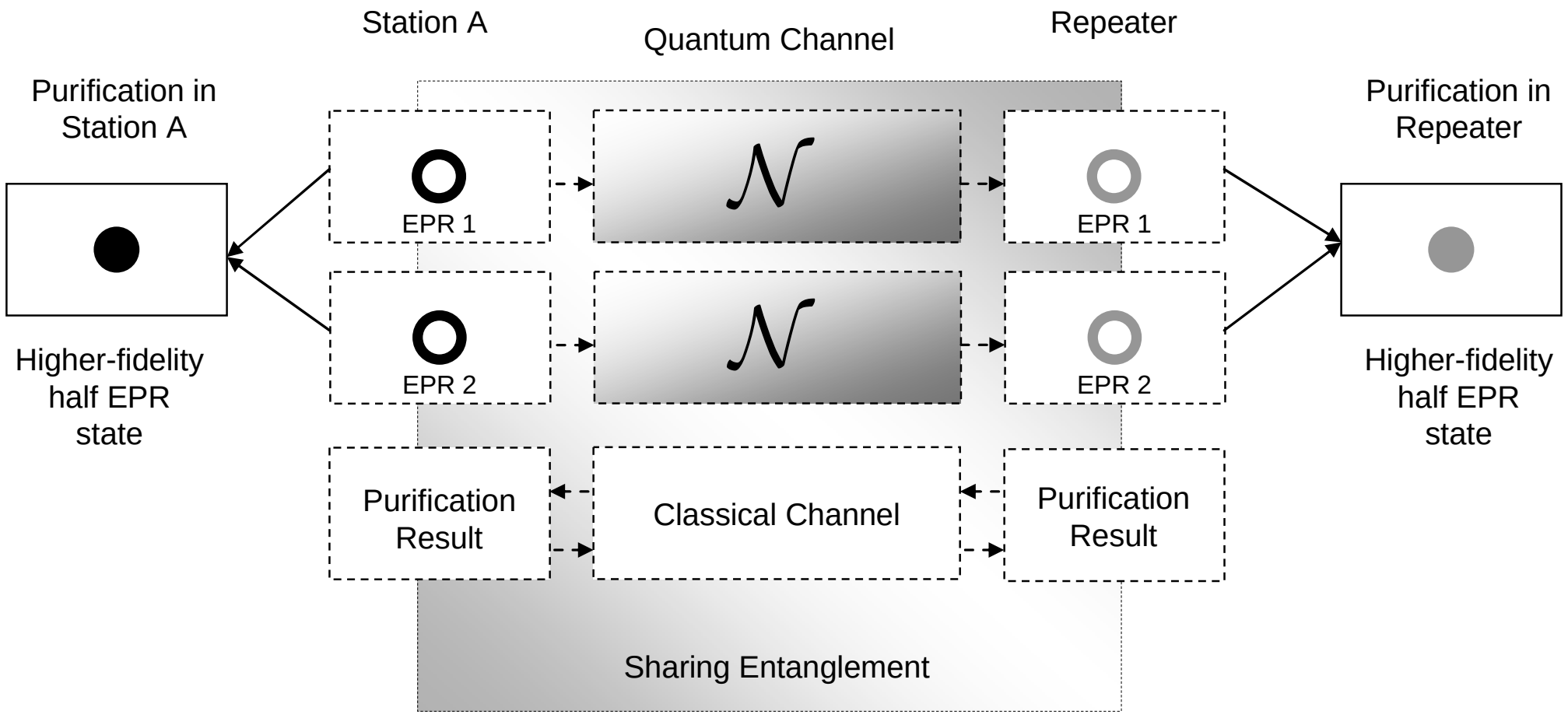

**Fig. 8.3.** The noisy entangled quantum states have to be purified. This step requires a lot of resources and classical communication.

The probabilistic purification could result in two possible outcomes:

1. The purification operation fails, and both EPR pairs are then freed as uncorrelated qubits.

2. The purification operation succeeds then the result is one EPR pair, with higher fidelity and an uncorrelated free qubit pair. If the fidelity is high enough, this state can be applied for teleportation and entanglement swapping else it has to be purified again.

*Step 3. - Entanglement swapping*

The entanglement swapping connects the EPR-states shared between the neighboring stations into a distant EPR-state. The aim of entanglement swapping is to create shared entanglement between the source and the destination nodes. Entanglement swapping is equal to a set of quantum teleportation steps: hence entanglement swapping can be viewed as an “extended” teleportation protocol, which is able to bridge the large distances between physically separated stations. The details for two stations A and B and a repeater station between them are illustrated in Fig. 8.4.

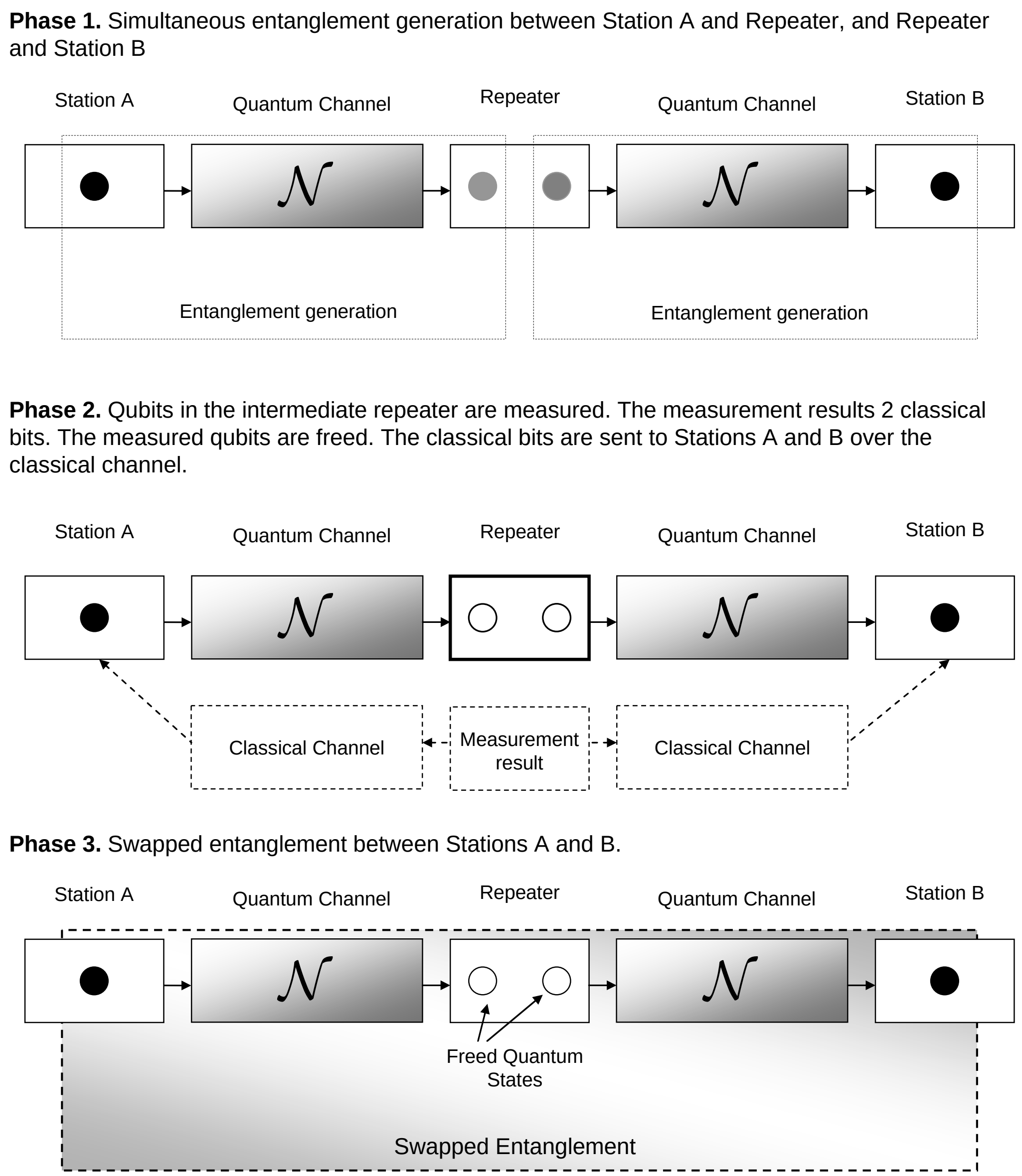

**Fig. 8.4.** Entanglement swapping - The transmission of entanglement from Station $A$ to Station $B$, through a repeater. The transmission is based on shared and purified EPR states and classical communication, with local transformations.

Station A is the information source while Station B plays the role of the destination. They are connected via a repeater. The operations of *entanglement swapping* are:

- entanglement sharing between Station A and Repeater, and Repeater and Station B, purification and quantum teleportation,

- measurement in Repeater, classical communication from Repeater to Stations A and B,

- local operations at Station B. Entanglement between the final stations (A and B) has been created.

This process destroys an EPR pair at the repeater, hence two quantum states have been freed, and they can be reused in the next teleportation. As the result of the swapped entanglement, the quantum entanglement has been transferred from the sender to the receiver, through an arbitrary number of repeater stations. Using the swapped entanglement between Alice and Bob, the unknown data qubit can be transferred from Alice to Bob with the help of quantum teleportation.

### 8.1.2 Brief Summary

Up to this point, we have used so many definitions in this section; hereby we give a short summarization:

- *entanglement sharing*: generation and sharing entanglement between the adjacent nodes, including quantum teleportation,

- *entanglement purification*: entanglement distillation-based protocol, purifies the noisy EPR states,

- *purification scheduling*: schedules the entanglement purification process,

- *entanglement distribution*: consists of the entanglement sharing and the entanglement purification between intermediate repeater nodes.

- *entanglement swapping*: extended entanglement between two distant nodes, Alice and Bob, spanned over many intermediate nodes.

## 8.2 Levels of Entanglement Swapping

Up to this point we assumed that there is one repeater station between the source and the destination. Now, we extend the picture and we assume that there are arbitrarily repeater stations in the system. This chain of repeaters is constructed for the purpose of extending the shorter distance between those stations that share an EPR state into longer distances.

In this section, we give a short account of the working mechanism of an idealized quantum repeater. If the distance between the sender Station $A$ and the receiver

Station $B$ is denoted by $L$, then the communication is realized through $l$ segments, with segment length

$$L_{segm.} = \frac{L}{l}. \tag{8.1}$$

The entanglement is generated between the neighboring repeater stations, then the shared EPR states are purified, finally they are "connected" with each other by the entanglement swapping operation.

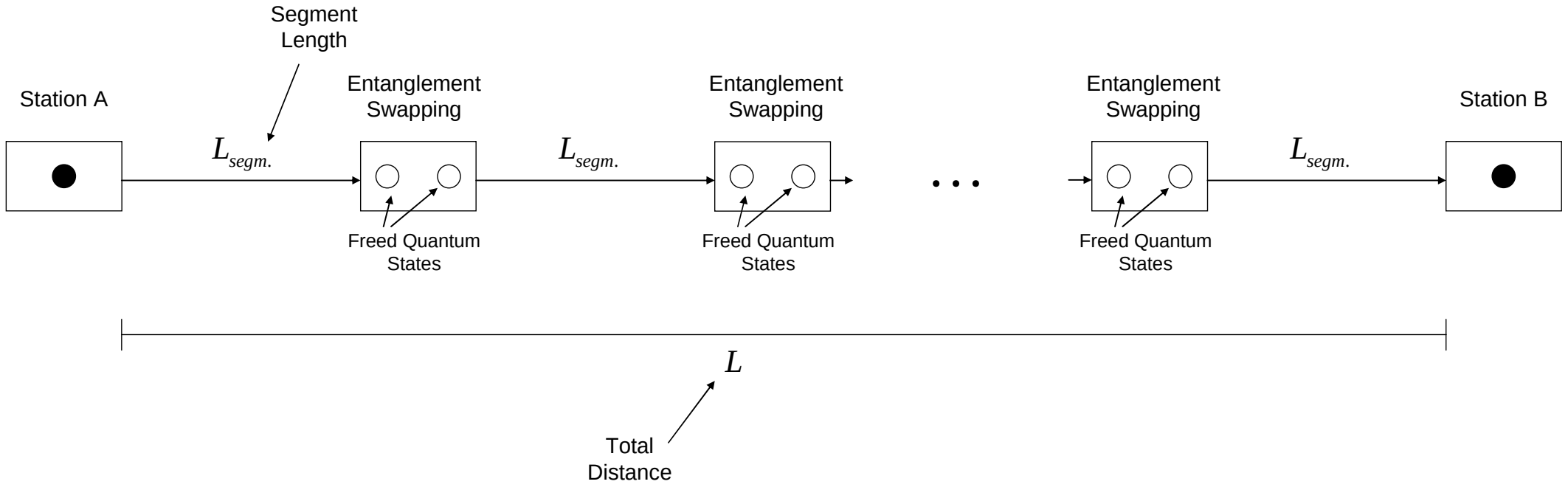

**Fig. 8.5.** Entanglement distribution over the entire distance is realized by the multiple segments between the repeaters.

Having established entanglement between the source and the destination, i.e., Stations A and B, the unknown payload quantum states are transmitted by the quantum teleportation protocol.

In the proposed construction this means that in the swapping process, two $n$-hop Bell pairs are combined into one *2n-hop* EPR state (see Table 8.1). This architecture is called the "doubling architecture" and, as has been shown by Briegel et al. [Briegel98], the performance of the system declines polynomially rather than exponentially as the distance increases.

Using the "doubling architecture," the level $i$ of swapping of a Bell pair spans $2^i$ hops, hence if 3 swaps has been made between the stations, then the connection between the number of swaps and the stations spanned (using $n = 5$ stations) can be summarized according to Figs. 8.6-8.8.

In the *zero level* of entanglement swapping, zero swaps has been made, in the first level one-, while in the second level, two swap transformations have been realized. After the transformations have been finished, the initial EPR pair is stretched to reach all the hops, the EPR states of the repeater stations—which mean three EPR

states—have been destroyed. In this phase, each pair of adjacent stations share an EPR pair with each other.

The zero level of swapping with $n = 5$ (Stations A and B and three repeaters) stations is illustrated in Fig. 8.6.

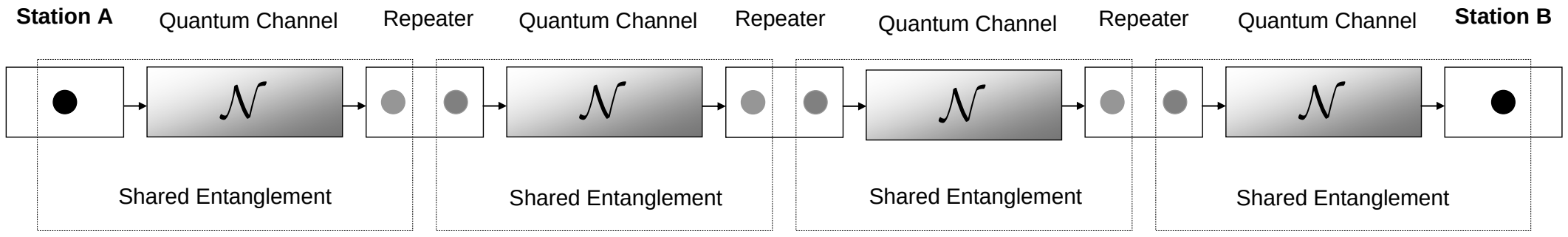

**Fig. 8.6.** Zero level entanglement swapping. The adjacent nodes share entanglement with each other.

In the *first level*, the adjacent stations start to communicate with each other using quantum teleportation and classical communication. As a result of this step, the EPR state will span two stations. The entanglement swapping will free up four quantum states.

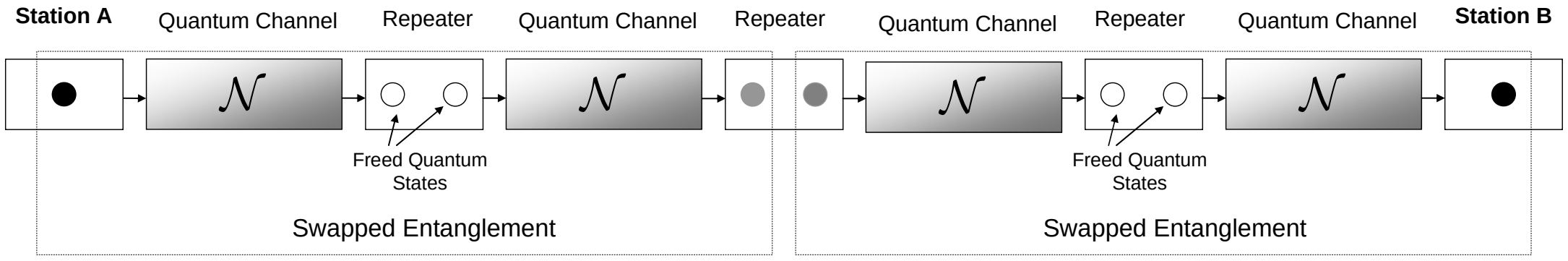

**Fig. 8.7.** First-level entanglement swapping. After the EPR states have been shared, local transformations are made. These transformations free up quantum states in the nodes.

In the *second level*, the entanglement swapping is realized between three stations, which will result in four spanned stations, and in six freed quantum states. The second level of entanglement swapping is illustrated in Fig. 8.8.

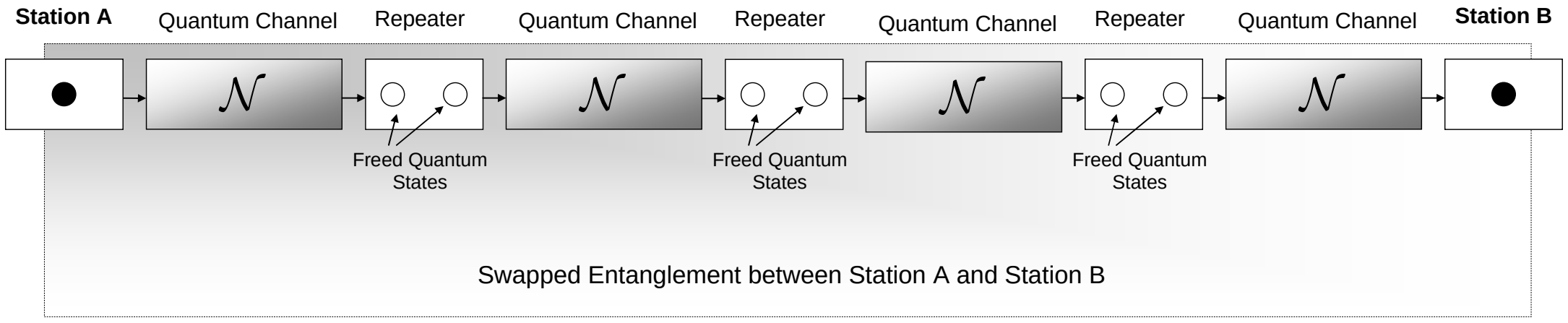

**Fig. 8.8.** Second-level entanglement swapping. After the local transformations have been realized in the repeater stations, the result is an entanglement between node $A$ and node $B$.

The proposed method can be extended to $n$ levels of entanglement swapping and the number of stations spanned to $2^n$. The $i$-th level of swapping connects two $i$-1 level systems. The results of Figs. 8.6-8.8 are summarized in Table 8.1:

| *Level of Swapping (i)* | *Number of Spanned Nodes from Station A (see the size of dashed frame)* | *Freed Quantum States in Repeater Nodes* | *Number of shared EPR pairs between A and B on the i-th swapping level* $\left(2^{n-i}\right)$ |
|---|---|---|---|
| 0 | 1 | 0 | $2^n$ |
| 1 | 2 | 4 | $2^{n-1}$ |
| 2 | 4 | 6 | $2^{n-2}$ |
| ⋮ | ⋮ | ⋮ | ⋮ |
| $n$ | $2^n$ | $\left(2^n - 1\right) \times 2$ | **1** |

**Table 8.1.** The levels of entanglement swapping with the stations spanned and the freed quantum states.

The architecture can be used for long-distance communication and in the quantum networks of the future. Entanglement swapping is realized by the purification step and the quantum teleportation protocol combined with classical communications. The purification scheme is able to correct the errors of the transmission which occurs in the node-switching and the fiber-based communication devices. Entanglement swapping can be extended to long-distances, and with the help of node-to-node quantum communications, a global scale quantum network can be constructed in the future.

## 8.3 Scheduling Techniques of Purification

The sharing of a Bell pairs between two nodes is based on the purification scheme. Due to entanglement distillation, two noisy input Bell pairs are used to produce by the minimization of the noise one output EPR pair. Unfortunately, purification requires a lot of physical resources, since to generate a purer EPR pair, two EPR pairs needed as input.

Another important complexity aspect is the selection of the most appropriate entangled pairs from the set of previously shared pairs, called *scheduling*. Therefore, purification can be regarded as entanglement distillation steps governed by scheduling.

In this subsection, we outline the most important purification algorithms in the realization of quantum repeaters, and their working mechanism is illustrated.

### 8.3.1 Symmetric Scheduling Algorithm

The first purification approaches were developed by Dür et al. [Dür99]. Their method is called "symmetric" *scheduling* which uses a set of EPR pairs, having the same fidelity—in this manner, these pairs are symmetric. The working mechanism of the symmetric purification method, with four EPR noisy pairs initially taken from the set, is shown in Fig. 8.9. The fidelities of the purified EPR-states are denoted by the thickness of the circle in the boxes.

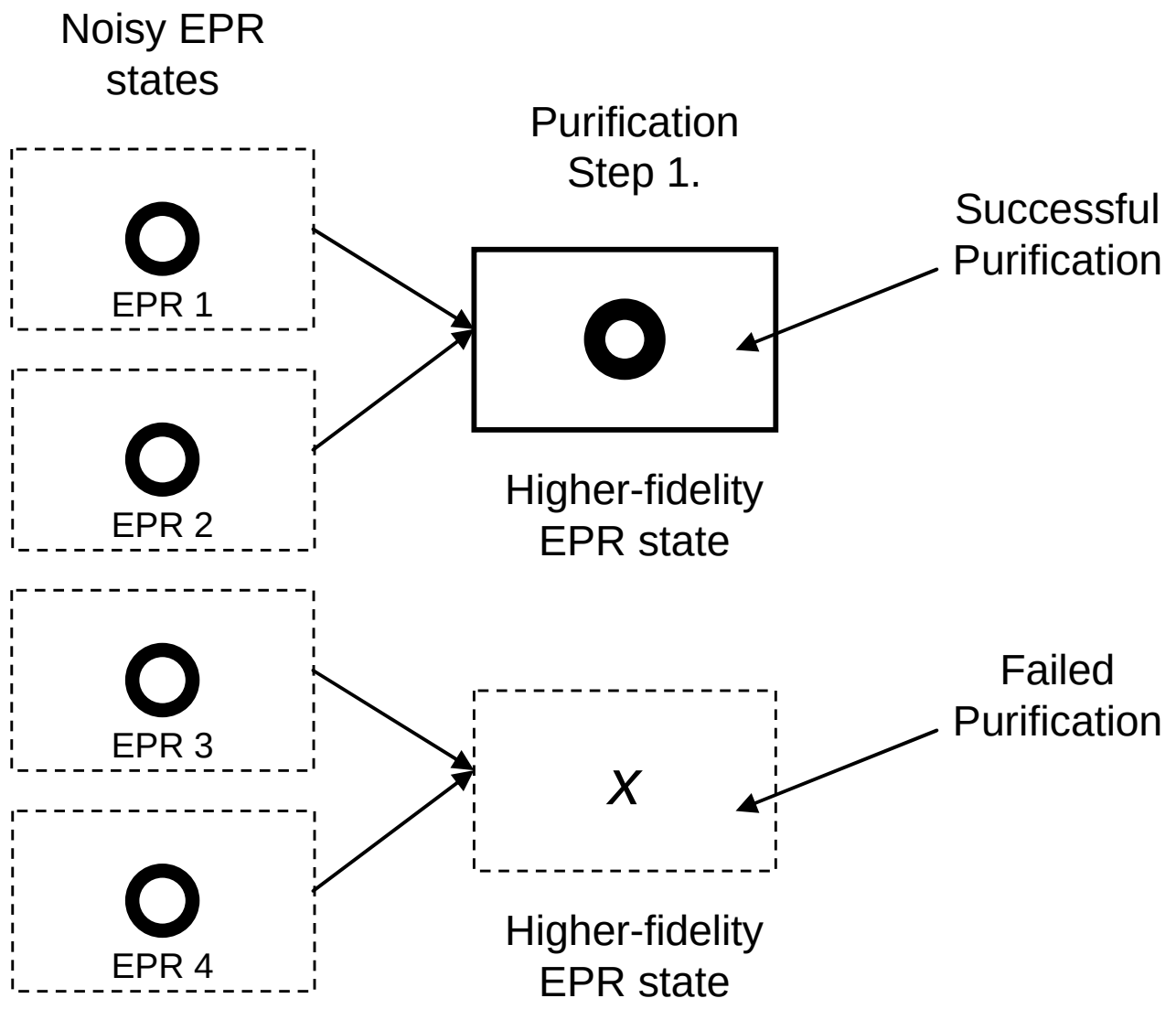

**Fig. 8.9.** The symmetric purification algorithm. The purification is a probabilistic process.

At the first step the first two pair results in a successful purification generating an EPR pair with higher fidelity. According to the figure and the probabilistic nature of purification, unfortunately, the purification of the second two pairs fails.

Since we would like to continue the purification, therefore the algorithm takes two new EPR pair from the set. Supposing the purification of the new two EPR pairs succeeded, then this output can be used again with the previously purified EPR state. Hence, in the next step, the algorithm takes as input the two higher fidelity EPR pairs, and generates a new EPR pair, with even higher fidelity, see Fig. 8.10.

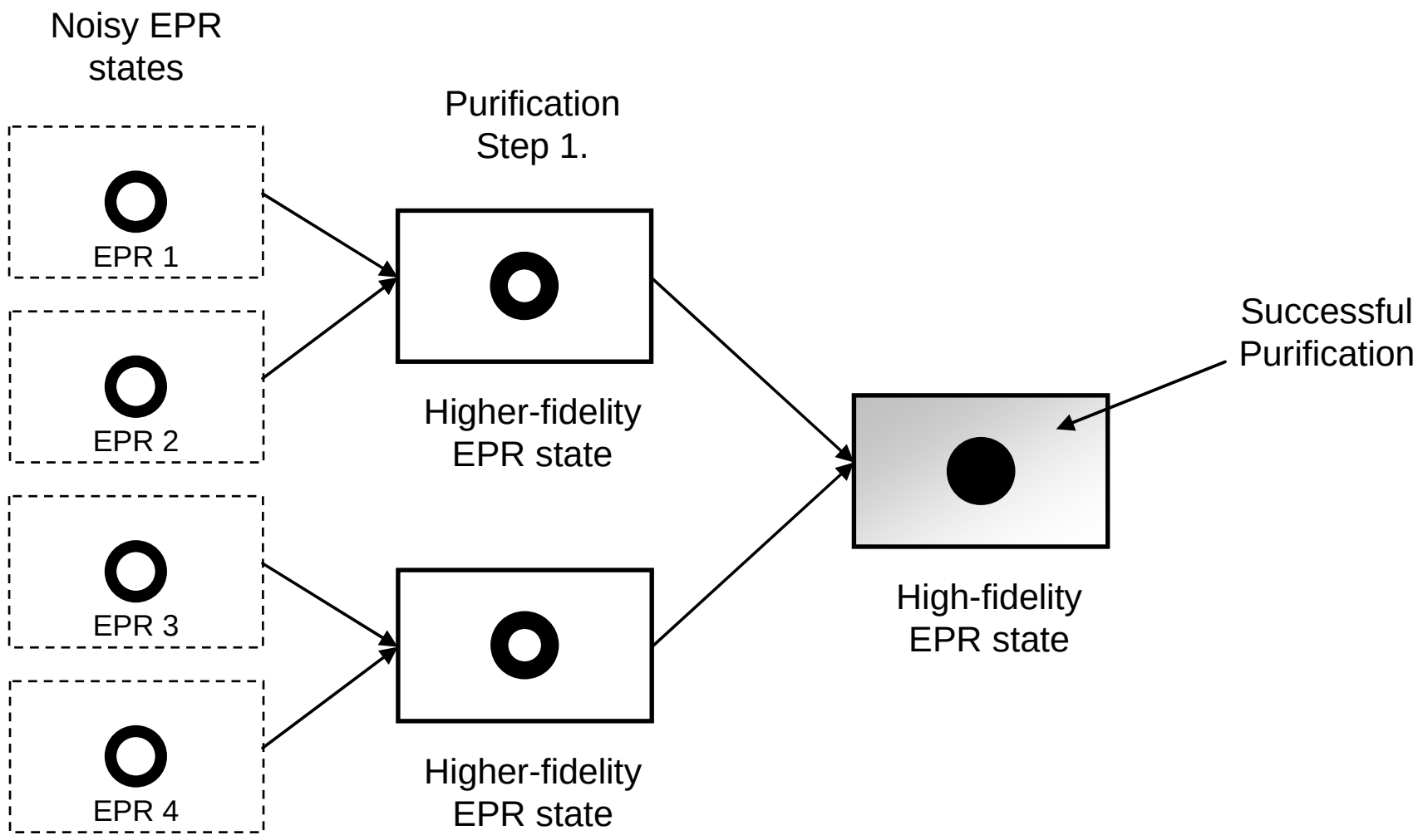

**Fig. 8.10.** From two purified states a higher fidelity EPR state can be generated.

Now, let assume the starting fidelity of the EPR states is for example 0.638, then the fidelity of the final EPR state can be increased to 0.98, however it requires five rounds of the protocol and each steps has to succeeded. This requires $2^5 = 32$ input EPR pairs, which means that in the initialization step, the node $A$ and $B$ have to share 32 Bell states [VanMeter08]. After the nearly perfect EPR state has been constructed, the EPR state can be used to teleport the unknown quantum state.

## 8.3.2 Pumping Scheduling Algorithm

The method of entanglement pumping was introduced by Dür et al. in 2007 [Dür07]. The scheme "pumps up" the fidelities of the states, however this enhancing process requires to involve further noisy EPR-pairs to the process, in each step. The pumping method can be used in case the fidelities of the noisy input EPR states are not equal, and it can be realized by fewer steps, as illustrated in Fig. 8.11.

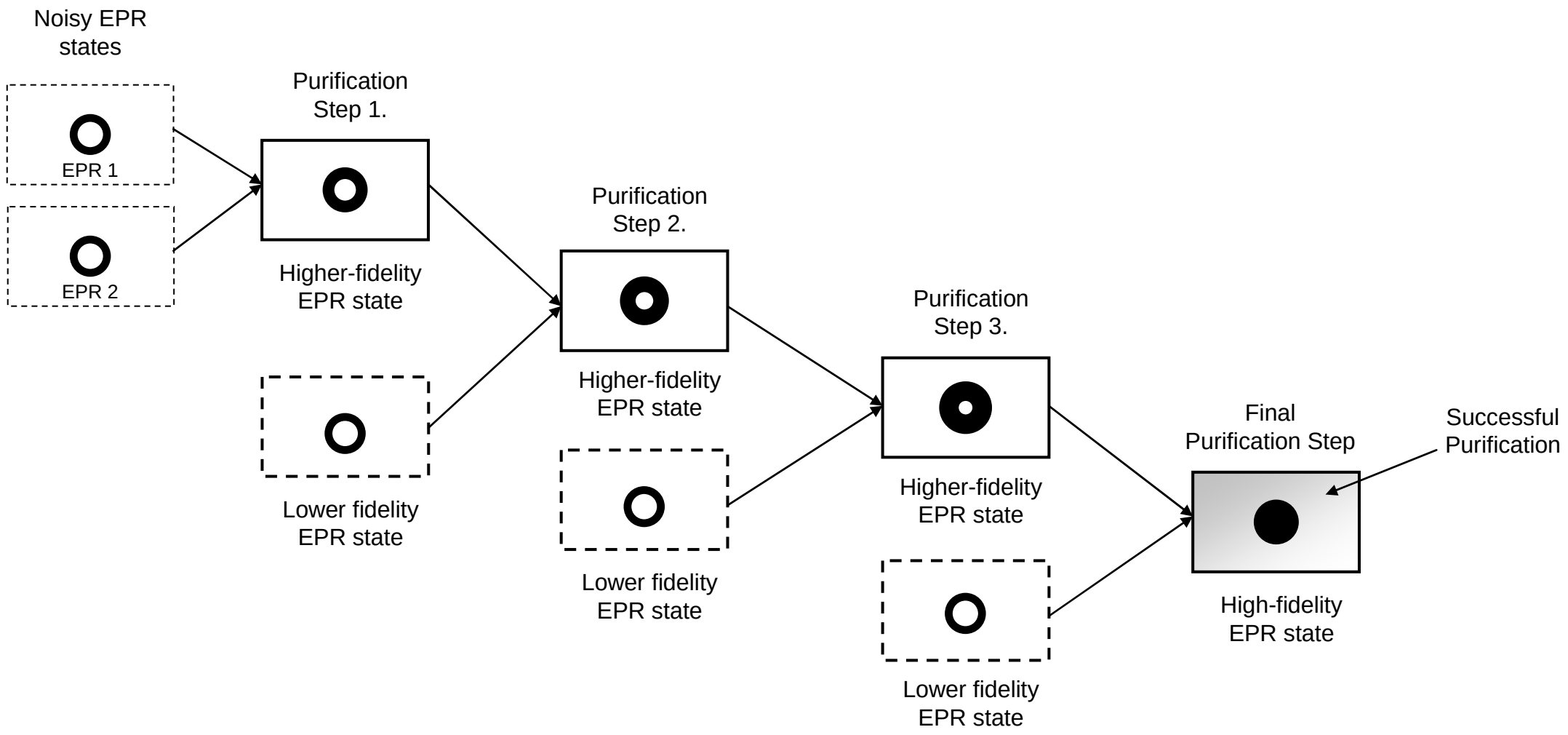

**Fig. 8.11.** The pumping purification method.

This method uses the noisy input EPR pairs in a more efficient way, however the process is slower in practice than the symmetric method.

### 8.3.3 Greedy Scheduling Algorithm

The greedy scheduling method was introduced by Ladd et al. in 2006 [Ladd06]. It was designed for practical reasons, since it always tries to use what it gets as the input. The method purifies all possible resources and it always tries to realize the best output. The greedy purification algorithm is illustrated in Fig. 8.12.

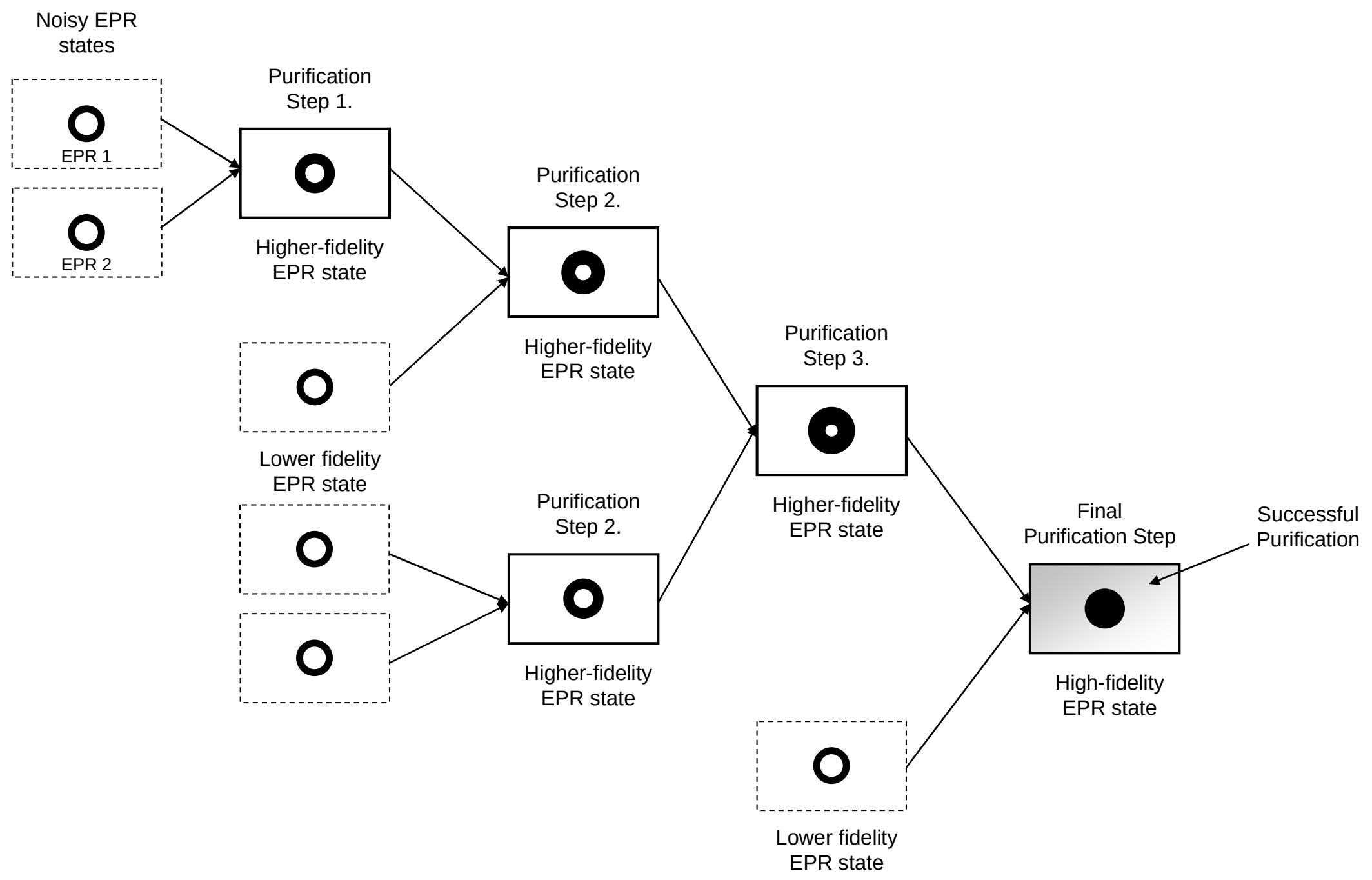

**Fig. 8.12.** The greedy purification method.

In short, the greedy algorithm can be applied in practice with high efficiency if the fidelity of the input pairs is high. However, if the fidelity of the input EPR pairs is low, the algorithm results in lower success probabilities and a lower output fidelity if the algorithm makes the purification.

### 8.3.4 Banded Scheduling Algorithm

A very efficient purification algorithm was developed by Van Meter et al. [VanMeter08]. The purpose of the banded purification is the efficient and flexible purification of noisy input EPR pairs. The scheme uses a rather different approach to purify the states in comparison with the before discussed solutions. They divided the EPR states into groups (i.e., bands) according to their fidelities and the algorithm uses those EPR states that belong to same group. A group (or band) contains a set of quantum states with the same fidelities.

The scheme of the working mechanism of the banded purification scheme is depicted in Fig. 8.13.

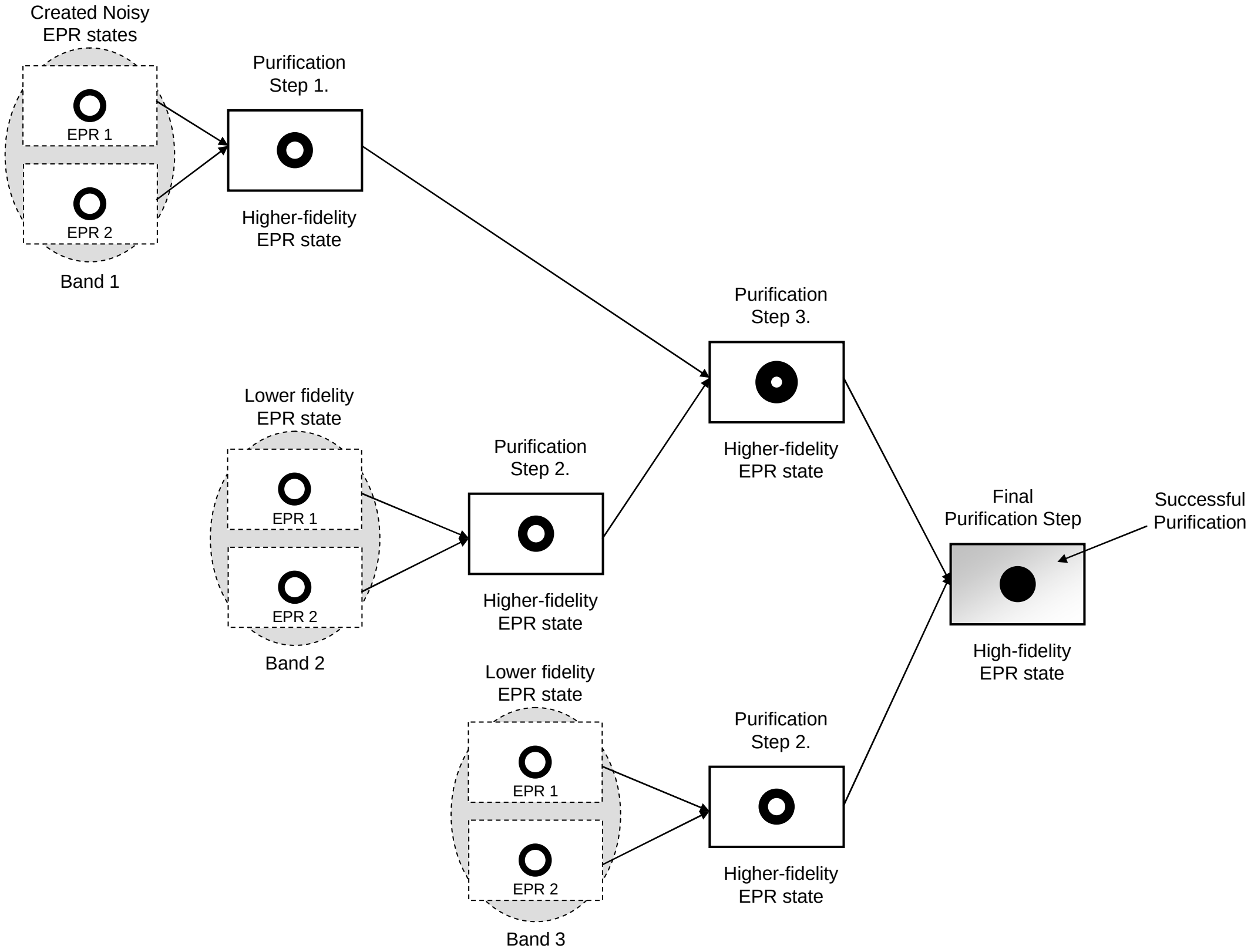

**Fig. 8.13.** The banded purification method.

The banded purification scheme purifies those EPR pairs that have the same fidelity. If in the given step there is no second EPR pair with the required fidelity, then the algorithm waits until a suitable EPR pair is created. Both the banded and the symmetric purification scheduling algorithms can run into deadlock, however in case of the banded purification scheduling, some solutions were developed to avoid this problem. These techniques put an upper bound on the number of possible groups or with other words, the bands. On the other hand, this solution cannot be applied in case of the symmetric purification scheme.

The banded purification algorithm enables the purification of EPR states with different fidelities—hence, it can be implemented in a more flexible and efficient way than the symmetric method. While the symmetric algorithm would be blocked in this situation, the banded purification scheme allows us constructing a purified EPR state from two EPR states which have different fidelities.

The banded purification algorithm prefers the higher fidelity quantum states, hence it does not use high fidelity states with low fidelity quantum states. It follows from this that the purification can be made with a given EPR pair only if the fidelity of the second EPR pair is at least as great as the fidelity of the first EPR pair. Without this condition, the fidelity enhancing of the output EPR would not be possible.

## 8.4 Hybrid Quantum Repeater

One practical approach of the quantum repeater is called the "hybrid quantum repeater." It uses atomic-qubit entanglement and optical coherent state communication. (It is called "hybrid" because the architecture of these repeaters mixes ordinary optical light with atomic cavities.)

The nodes are connected by optical fibers as quantum channels. This channel between two nodes is called the "qubus," which interacts with the quantum state. After the sender node's pulse has interacted with the atomic qubit superposed state, the coherent state is sent through the noisy optical quantum channel. If the channel would be noiseless then the result of the transformation would be a maximally entangled quantum state between the nodes, however in practice this is not possible.

The simplified picture of hybrid quantum repeater is illustrated in Fig. 8.14. In the first repeater node the coherent-state pulse $A_2$ interacts with the superposed atomic qubit system $A_1$. After the interaction between the two systems, the coherent state is transmitted through the noisy quantum channel to Repeater station B. The incoming coherent state interacts with the a second superposed atomic qubit system $B_1$. The interaction results in entanglement between qubits $A_1$ and $B_1$. The entanglement between $A_1$ and $B_1$ is realized after the measurement of the coherent state has been made in Repeater station B. The atomic qubits are stored in cavities. As follows, the hybrid architecture combines atomic qubit systems with simple optical elements.

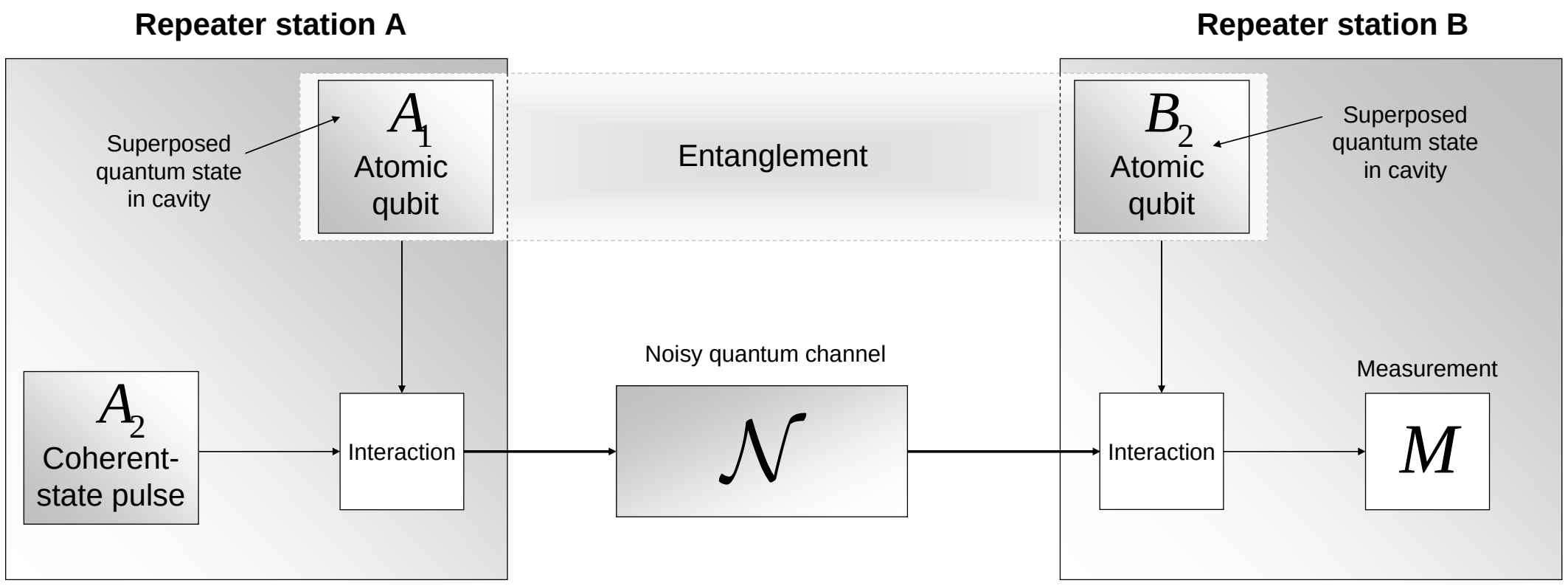

**Fig. 8.14.** The hybrid quantum repeater uses optical elements with atomic cavities.

The main task is the improving of the final fidelities of the shared entangled states. The measurement scheme in the hybrid quantum repeaters can be implemented in practice using linear optics and standard photonics devices, with a relatively low error.

The connection between the photon loss of the optical channel $\eta$ and the optimal probability of success of the entanglement generation can be expressed as

$$p_{succ.} = 1 - \left[ \left( 2F - 1 \right)^{\eta/(1-\eta)} \right], \tag{8.2}$$

where $F$ is the fidelity of the entangled EPR pair. This probability of success can be taken as an upper bound for the probability of success of generating an entangled pair, since (8.2) constitutes the optimal case for the failure probability.
The entanglement purification is the most important part of the quantum repeater, on the other hand it requires the most resources, also. It is made by local, two-qubit unitary gates in the individual nodes.
In the hybrid quantum repeater, the probability of success of the purification process can be expressed as

$$p_{purif.} = F^2 + \left( 1 - F \right)^2, \tag{8.3}$$

and the fidelity after the process becomes

$$F_{purif.} = \frac{F^2}{F^2 + \left(1 - F\right)^2}. \tag{8.4}$$

In practical implementations, where the communication is based on optical quantum channels and coherent states and atomic cavities (see Fig. 8.14), the swapping can be realized in a deterministic way, hence there is no uncertainty in the swap operation. It follows from this fact that the final fidelity after the swapping operation will be

$$F_{swap} = F^2 + \left(1 - F\right)^2. \tag{8.5}$$

As in the case of purification, these fidelities do not take into account the local imperfections of the quantum gates. On the other hand, the main influence of photon losses occurs in the communication channel, hence the main effects can be taken to be due to the distance between the quantum nodes.

From an engineering point of view, one of the most important questions in the development of a practical quantum repeater is the measurement strategy in its stations. With the help of measurement entanglement between the stations can be prepared, however the fidelities are depend on the measurement approach. In the first approaches towards designing a quantum repeater, the measurement process was based on the homodyne projection measurement. (For details see [Duan01] or [Ladd06]). But later it was shown that a quantum repeater can be implemented in a more efficient way using the optimal USD (*Unambiguous State Discrimination*) method [Bernardes10], [Loock08]. While using the homodyne projection measurement strategy the final fidelities are about $F < 0.8$ for relatively small distances, i.e., $L \leq 10\,\mathrm{km}$, in case of the USD approach, the fidelities can be made arbitrarily high, however the price of this improvement is the lower number of shared entangled pairs [Azuma09].

### 8.4.1 Experimental Demonstration of Entanglement Sharing

The success of global quantum communication depends on the development of quantum repeaters. A practical implementation was presented by a group of researchers from the University of Vienna and from China [Yuan08].

The quantum channel is split up into segments and between the segments, the quantum states are entangled with each other. The segments were entangled by ultra cold rubidium atoms which were in entanglement with photons. These entangled photons were sent through the optical fiber, the photons later crossed at a beam splitter. In the end, the photons are measured, which measurement makes the cold atoms entangled. The theoretical background behind their quantum repeater can be summarized as follows: the segments of the quantum channels are entangled individually, however the segments can be entangled together.

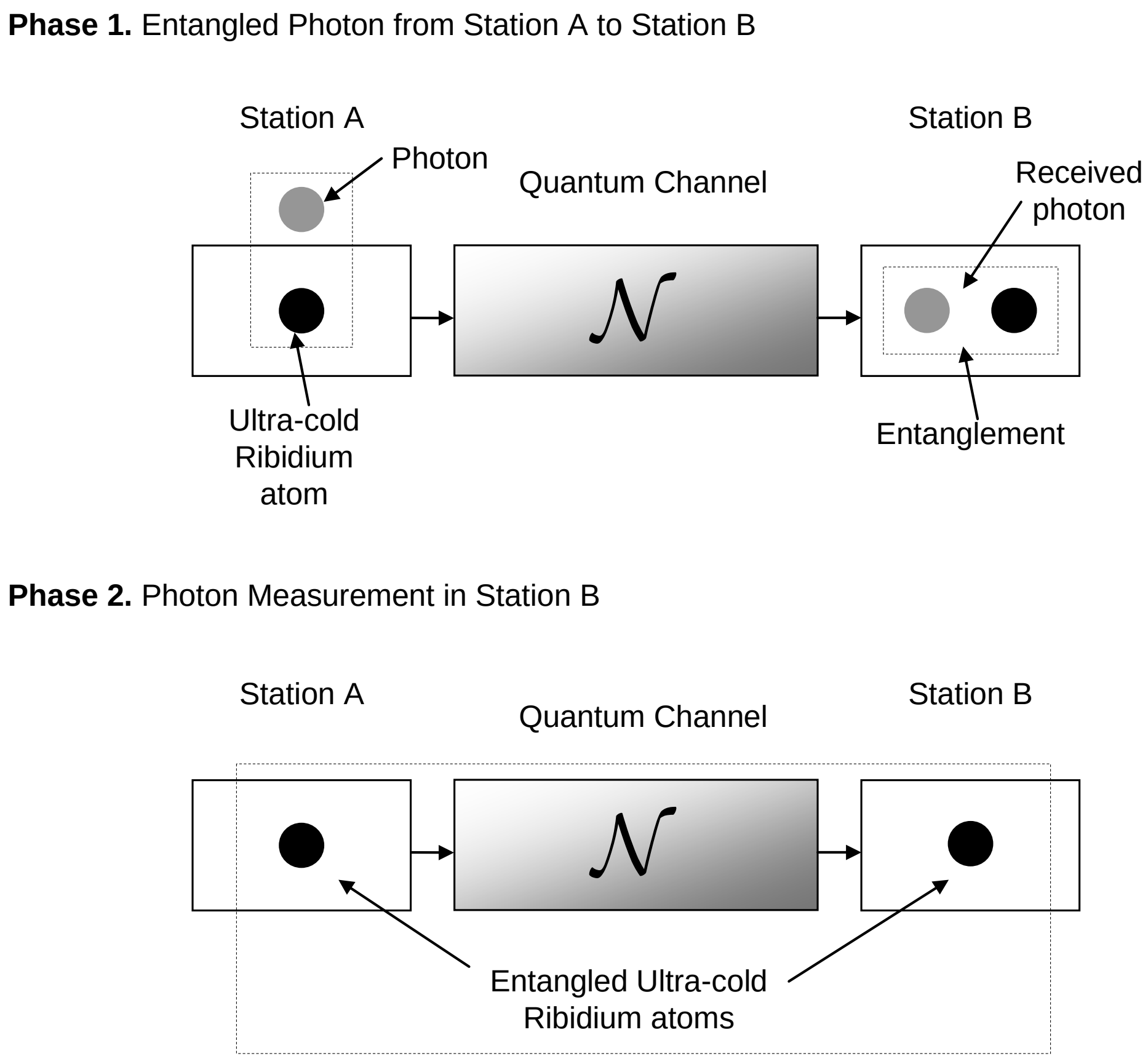

**Fig. 8.15.** In the experimental realization of the quantum repeater, the quantum channel is split up into segments and between the segments, the quantum states are entangled with each other. The segments were entangled by ultra cold rubidium atoms which were in entanglement with photons.

As these researchers found, if they could increase the lifetime of the ultra cold atoms, then the length of the segments of the quantum channel can be extended to a few hundreds kilometers.

### 8.4.2 Performance Analysis of Hybrid Quantum Repeater

The currently available practical ways to optimize the entanglement purification basically were designed for the optimization of the performance of local devices, and a general scalable solution for this problem is still unknown. In 2010, an idea of a possible practical approach of hybrid quantum repeater was presented with nearly maximally entangled EPR pairs over 1280 kilometers, and an analytical approach has been presented by a research group in Europe [Bernardes10].

Now we discuss the performance of thy hybrid quantum repeater introduced in Section 8.4.

The rate of entanglement generation in a hybrid quantum repeater system which contains $2^n$ EPR-pairs (where $n$ is the level of swapping see Table 8.1) over the distance $L$ can be expressed as

$$R_n = \frac{1}{T_0 Z_n\left(P_0\right)}, \tag{8.6}$$

where $T_0$ is the minimum time to generate an EPR state over distance $L_0$, and it can be expressed as

$$T_0 = \frac{2L_0}{c}, \tag{8.7}$$

where $c$ is the speed of light in an optical fiber, i.e., $c = 2 \cdot 10^8\, m/s$, while $Z_n$ is the average number of steps required for generate entanglement in all $2^n$ station-pairs with is success probability $P_0$, can be expressed as

$$Z_n\left(P_0\right) = \sum_{i=1}^{2^n} \binom{2^n}{i} \frac{\left(-1\right)^{i+1}}{1-\left(1-P_0\right)^i}. \tag{8.8}$$

We note, if the success probability $P_0$ is low, then (8.6) can be approximated as

$$R_n = \frac{P_0}{T_0}\left(\frac{2}{3}\right)^n. \tag{8.9}$$

Using these results we can analyze the rate of entanglement generation in a hybrid quantum repeater, which measures well the performance of the whole construction.

Due to a deep analysis on the performance of an practical quantum repeater, with segment length $L_{segm.} = 20$ km and $2^n$ spanned nodes (see Table 8.1), the performance of a quantum repeater was tested over a channel of length $L = 1280$ km. According to the measurement results published in [Bernardes10], the rate of entanglement generation as a function of the fidelity of the entangled pairs *without purification of the entangled pairs and one round of purification*, over the tested distance $L = 1280$ km is depicted in Fig. 8.16.

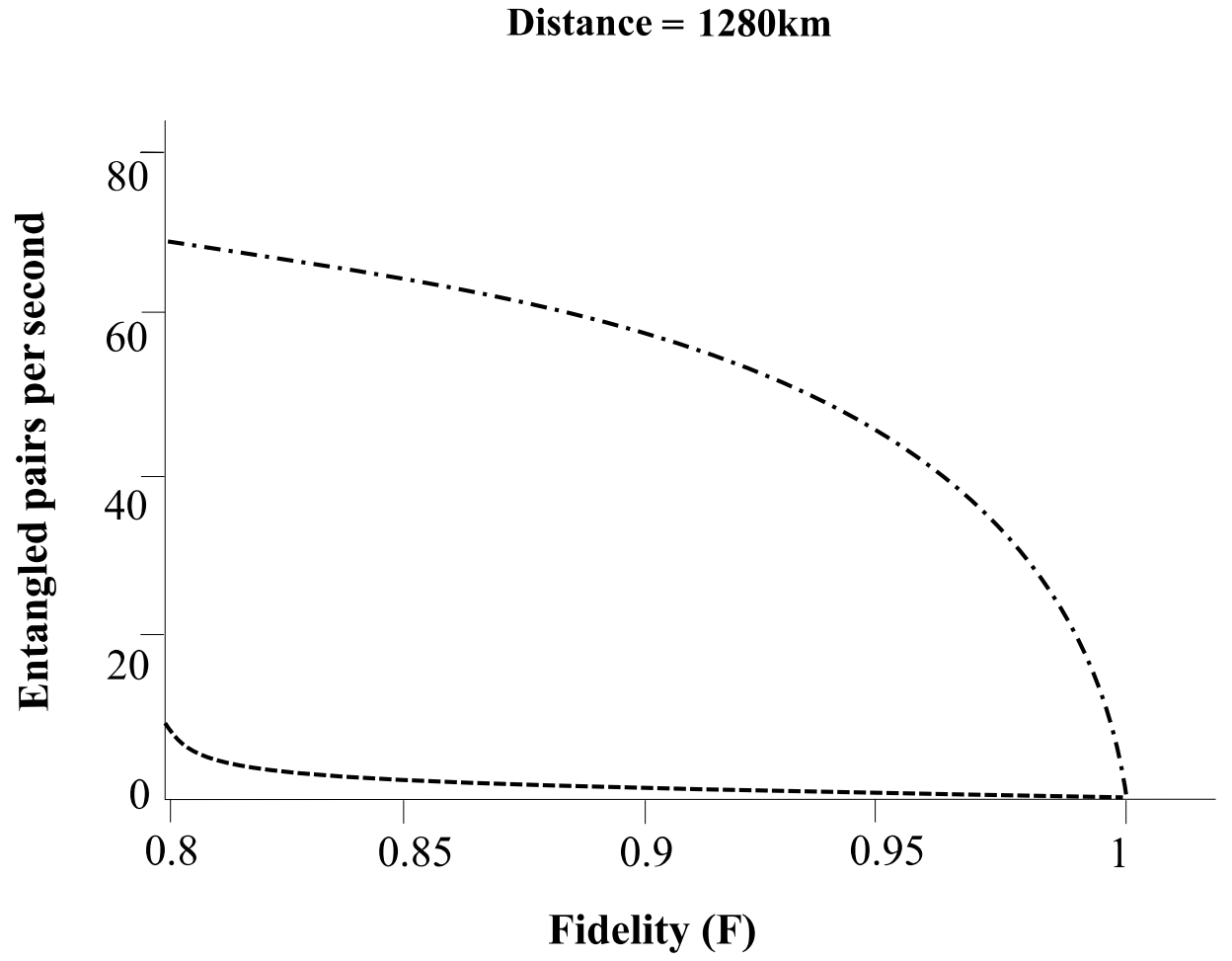

**Fig. 8.16.** The comparison of entanglement generation without purification (dashed line) and with purification (dashed-dotted line).

As can be concluded, the high–fidelity entanglement distribution rate without purification is very low. On the other hand, if the quantum states are *purified*, then the rate of entanglement generation for the hybrid quantum repeater becomes about ten times higher compared to the unpurified case.

Deeper analyses have concluded that the entanglement generation rates between quantum repeaters cannot be increased by the application of more purification steps on shared noisy entangled pairs. Hence, the noise of the quantum channel cannot be remedied by the repetition of the purification. The new rounds of purification steps increase (8.8), i.e., the final rate of entanglement sharing (8.6) decreases. On the other hand, the repetition of the purification step dramatically decreases the efficiency of the entanglement distribution process, hence it cannot be implemented with high efficiency in practice.

In practice a maximum of *two or three repetitions can be tolerated*, however it cannot be used to increase the rate of the quantum repeater, as is illustrated in Fig. 8.17. We note, more purification rounds can help to increase the rate of entanglement generation if the final fidelities of the previous purification steps were very high. The number of rounds of purification also determines the number of initial resources, i.e., if we fix the initial number of available resources the same then the number of purification steps decreases the entanglement sharing rate.

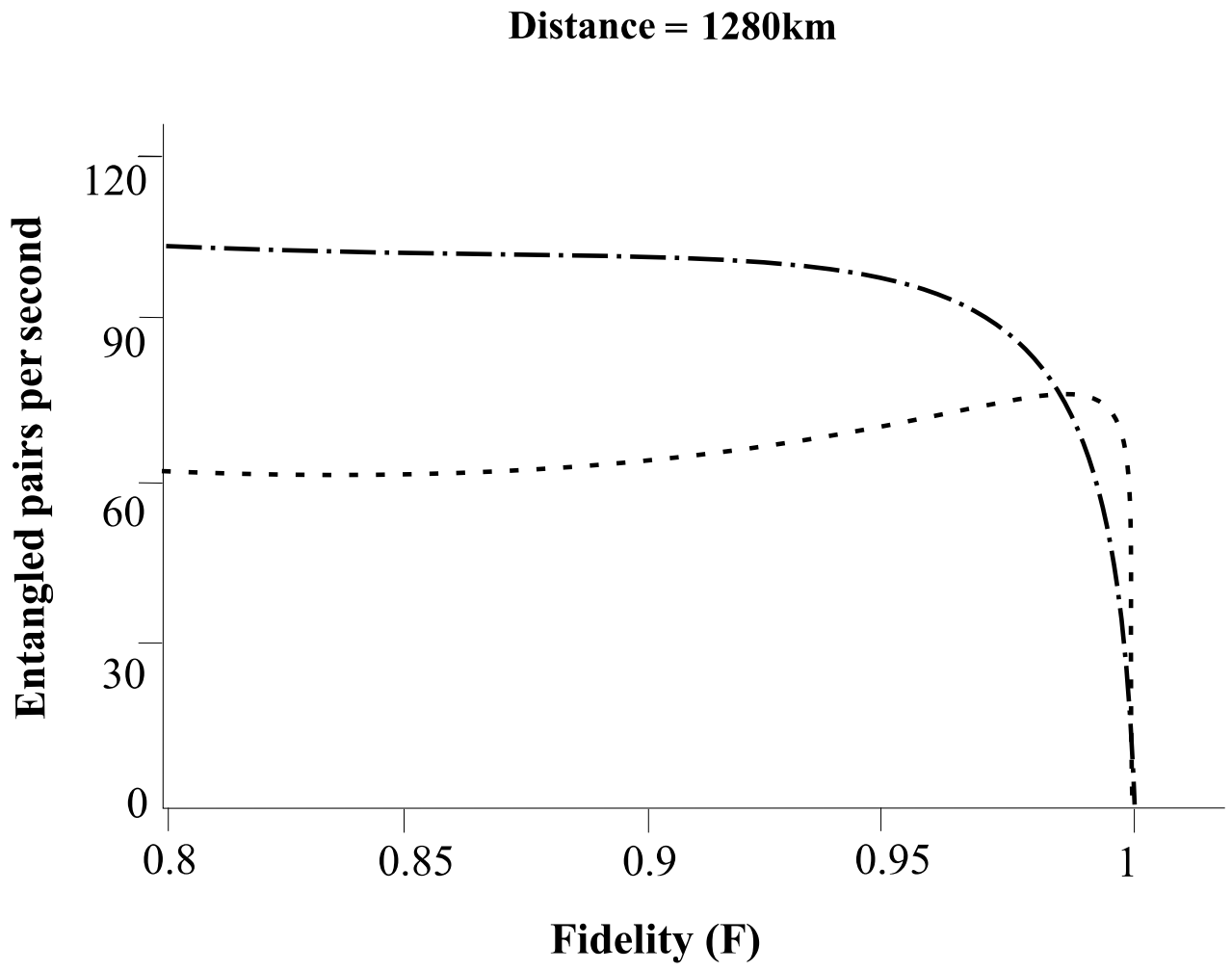

**Fig. 8.17.** The rate of entanglement distribution as a function of the fidelity of the entangled pairs with two (dashed-dotted line) and three repetitions of the purification step (dashed line).

Last but not least we mention that the purification can be made at the beginning of the entanglement distribution or at the end of the entanglement swapping. In Fig. 8.18. we summarize the results related to the hybrid quantum repeater on the rate of entanglement generation between the nodes of the quantum repeater, using the purification method at the first phase (i.e., entanglement sharing between the neighboring stations) of the whole entanglement distribution process and at the last phase (i.e., entanglement sharing between the source and the destination stations) of the process.

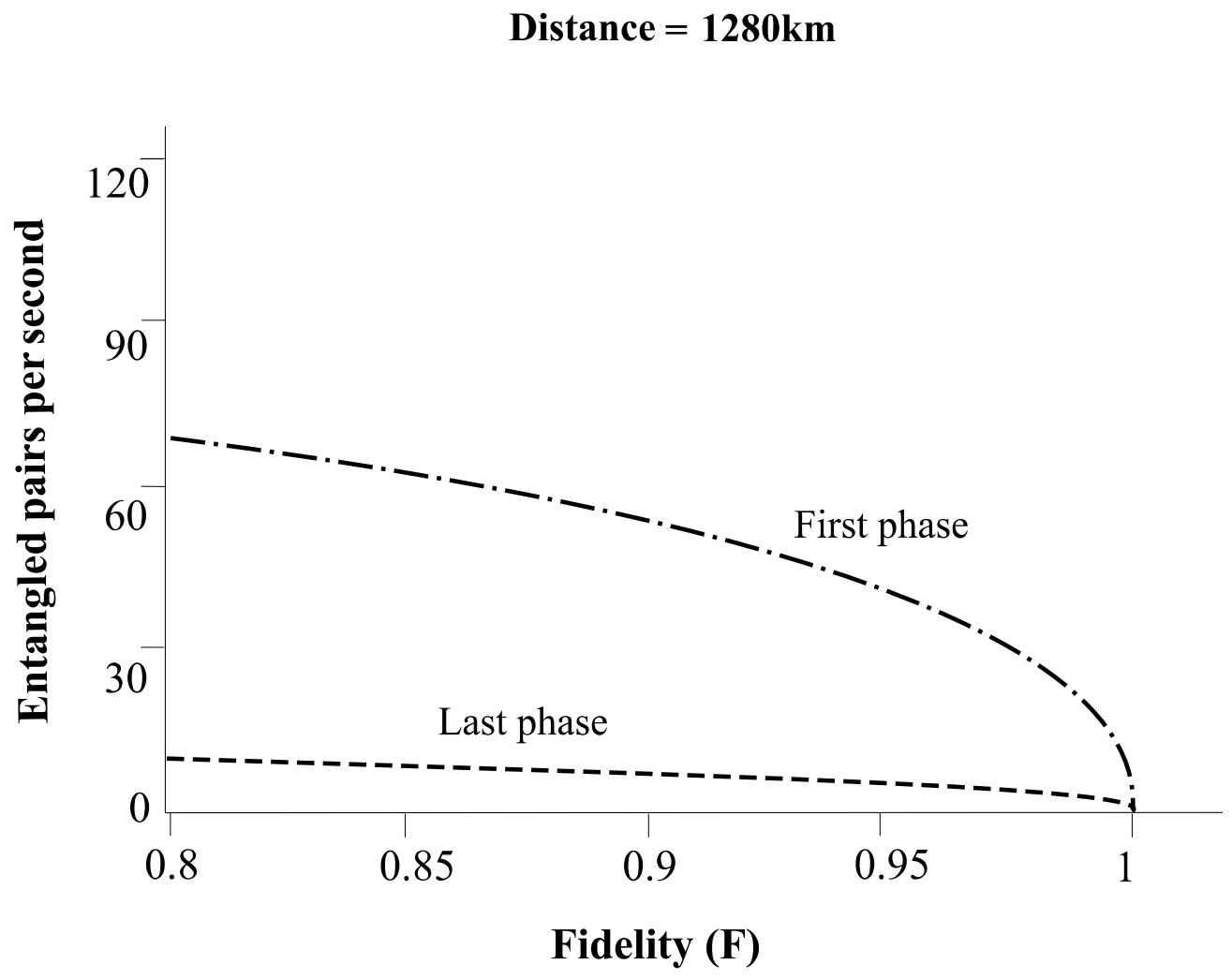

**Fig. 8.18.** The rate of the quantum repeater using the purification method at the beginning (dashed-dotted line) and at the end (dashed line) of the entanglement swapping process.

For practical purposes, it is better to implement the purification *as early as possible*, since the time of purification will then be shorter. If the purification is made only in the last phase of entanglement distribution, the number of the generated pairs per second will be lower, since the distribution of entanglement between the stations is an error-prone process.

### 8.4.3 Experimental Results

A quantum repeater has to amplify the unknown quantum state, and theoretically it has to store these quantum states, hence there are many challenges in this field. In 2008, a group in Germany with researchers from Austria, China, and Germany introduced a very interesting approach, [Eetimes10d] here, we will refer it as ACG repeater.

The properties of quantum states make it impossible to amplify an unknown quantum state without the destruction of the quantum state [Wootters82]. This research group has shown that the problem can be solved if the properties of quantum information processing are combined with the elements of classical signal processing. The group has designed an array which can be used as a quantum repeater. The proposed storage method is based on the BDCZ protocol, after the inventors Briegel, Dür, Cirac, and Zoller [Briegel98].

This group used ultra-cold atoms and a magneto-optical trap, with about –274 C°. This prototype hybrid quantum repeater was based on entanglement interchange, and was realized in practice by combinations of atoms and photons. The storage of the quantum information is based on the generated atom–photon entanglement, however, they found that the quality of the storage cannot be used in a quantum network for communication. Hence many challenges remain. The first result was released only in 2008, but due to the speed of the increase of technical developments, a truly applicable quantum repeater with high quality atom–photon entanglement could be realized within a few more years.

Another interesting photon-based approach to controlling quantum states was reported in 2006 [Chen06a]. The most innovative tools, such as quantum dots, single atoms and ions are very complicated in practice. A controllable single-photon source was developed by a research group from China. They focused on the storing method of the photons, using atomic quantum memories, with the help of laser light and cold rubidium atoms. The photons are generated when the laser light hits the cold atoms. As the inventors have stated, the created photons can generate single spin excitations in a cold atom, and this fact makes it possible to store the quantum states in the form of a quantum memory [Physorg15].
The single photon atomic memory is illustrated in Fig. 8.19.

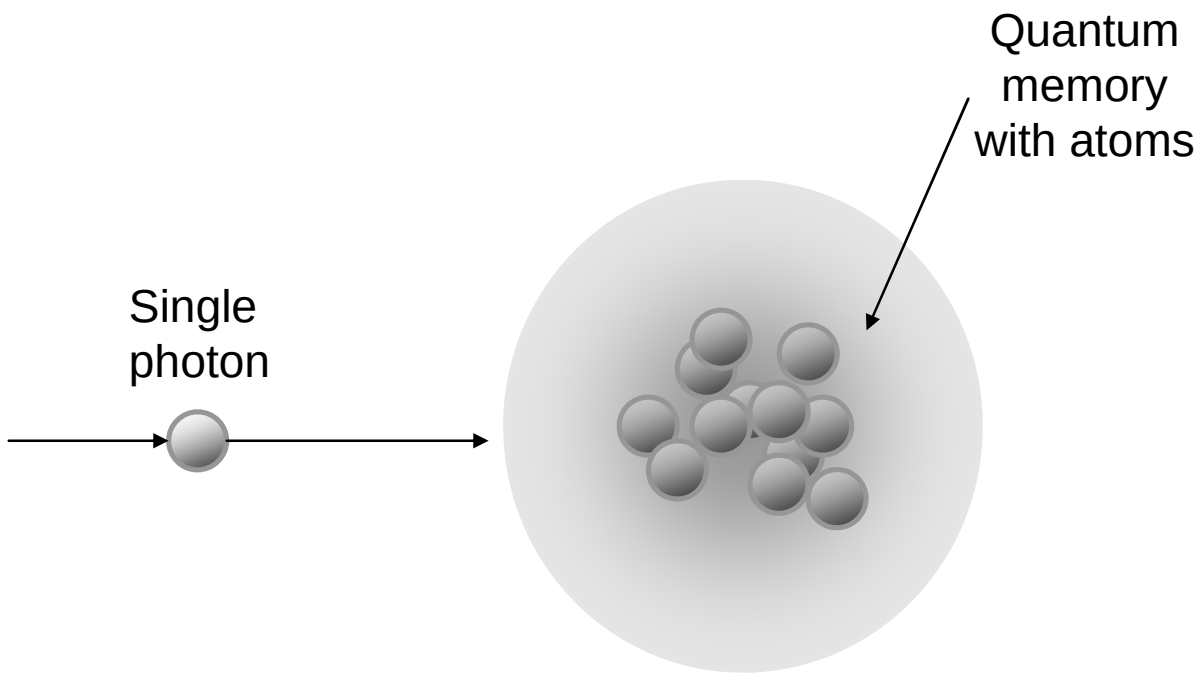

**Fig. 8.19.** The laser beam hits the cold atoms, which transfer results in a single spin excitation. The output can be retrieved by another laser beam.

The group has produced six hundred single spin excitations, which resulted in fifteen single-photon detections per second. As they have shown, the number of available single photon sources can be increased. The method can be used as an efficient quantum-memory based single photon source in the quantum communication devices of the future [WWS11]. A very important practical application of their result could be achieved in the field of the development of quantum repeaters, and long-distance quantum communications can be realized in practice.

In 2010, another very important development was presented by researchers of Georgia Institute of Technology [Physorg16]. They developed a technique which is able to convert quantum states at infrared wavelengths to a longer wavelength domain applied in classical optical networks. At the receiver side, the wavelengths can be converted back to the infrared wavelength. This has really great importance from practical approaches point of view, since it allows transmitting quantum states through very long-distances, with the help of existing optical telecommunication network.

The researchers implemented the conversion using rubidium atoms, which were very close to each other to realize the interaction with the incoming photon. The output of the device was a wavelength which encodes the input single photon. The single incoming photon is encoded in polarization and travels in the infrared light while the outcoming light is a converted telecom wavelength.

The input photon came in as infrared light, the output is a telecom domain light, as we have illustrated in Fig. 8.20.

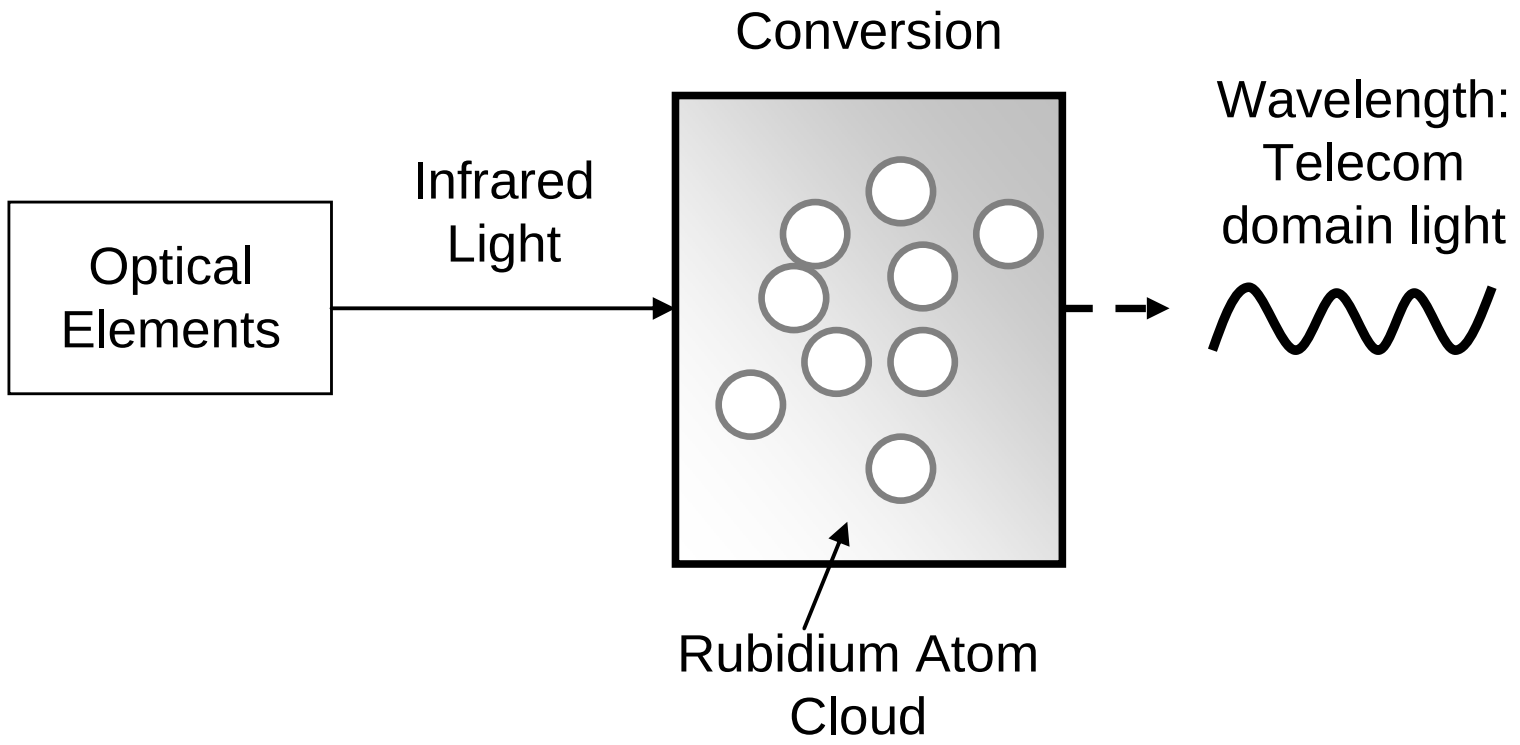

**Fig. 8.20.** As the photons are converted from the infrared domain to the telecom domain, they can be sent through the conventional optical-fiber network, and then, with the help of a magneto-optical trap, the original quantum bit can be restored at the receiver side.

The researchers controlled the efficiency of the transformation by adjusting the size and the density of the rubidium cloud. Unfortunately, the ideal wavelength for the transmission of the quantum states is not equal to the wavelength which can be transmitted at the best rate and with the lowest absorption. Current standard optical networks uses the 1.3 $\mu$m wavelength, however the best results can be obtained at about 800 nanometers. The researchers have also demonstrated that the infrared domain can be used to implement a quantum memory, too. It follows that this technique can be extended to build a working quantum repeater.

A research group at Harvard University used a diamond and a laser beam [Lukin10]. The entanglement generation was realized as follows. The diamond is illuminated with a very powerful leaser beam, in the next step the diamond emits a single photon, which photon is entangled with the spin of the electrons which are contained inside the diamond. Their approach similar to the trapped electron quantum chip. This construction can be used to realize entanglement-based long-distance communication. This entanglement is more reliable than the photon–photon entanglement, since in this case the entanglement is realized between the photons and a solid state material. The method constructed for entanglement generation is shown in Fig. 8.21.

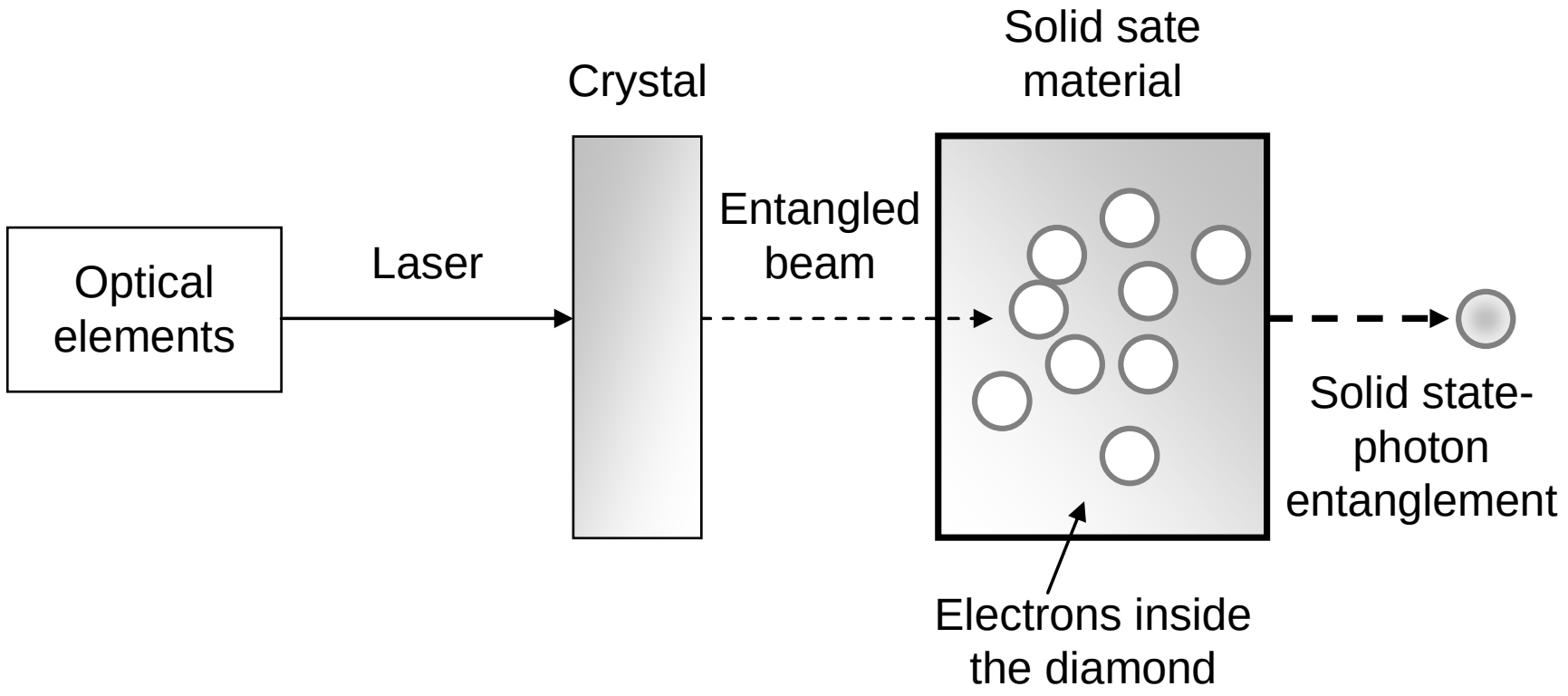

**Fig. 8.21.** Solid state–photon entanglement generation with single optical elements.

This method was the first practical approach in which entanglement was generated between a photon and a solid-state material. The researchers have combined the benefits of the photons and the solid state atoms, since while the photons are not so reliable, they are fast—on the other hand, the solid state atoms are much more reliable than the photons.

## 8.5 Probabilistic Quantum Networks

Beside overcoming long-distances, quantum communication can be extended to other aspects of communication networks. As has been shown, complex quantum networking processes can be realized in the quantum level, between atomic ensembles [Physorg18]. As a researcher group found in 2010, these special quantum networks can be used to generate arbitrary connections inside the network, and truly random behavior can be realized between them. With the help of quantum random networks, complex real-world problems can be analyzed and modeled by this fundamentally new approach. Random quantum networks could help to model real-life communication, in physical or biological systems, and it can be extended to model the behavior of the global communication networks. Using a quantum-based truly random approach, the still unrevealed connections in real-life communication can be also discovered more effectively [Perseguers10]. The random structure can be realized by a measurement performed on a sub-system of the weakly connected network, and the influence of the local measurement determines the global state of the network. The random quantum network firstly introduced at Max Planck Institute of Quantum Optics, Germany, is called the “randum” (i.e., the quantum random) network. The quantum random (i.e., the “randum” network) is illustrated in Fig. 8.22. In the network each node has a given probability to connect with an other node. The nodes which have high probability of being connected form a cluster. If two nodes are connected by a link then they have a common EPR state. The nodes in a given cluster are all entangled with each other.

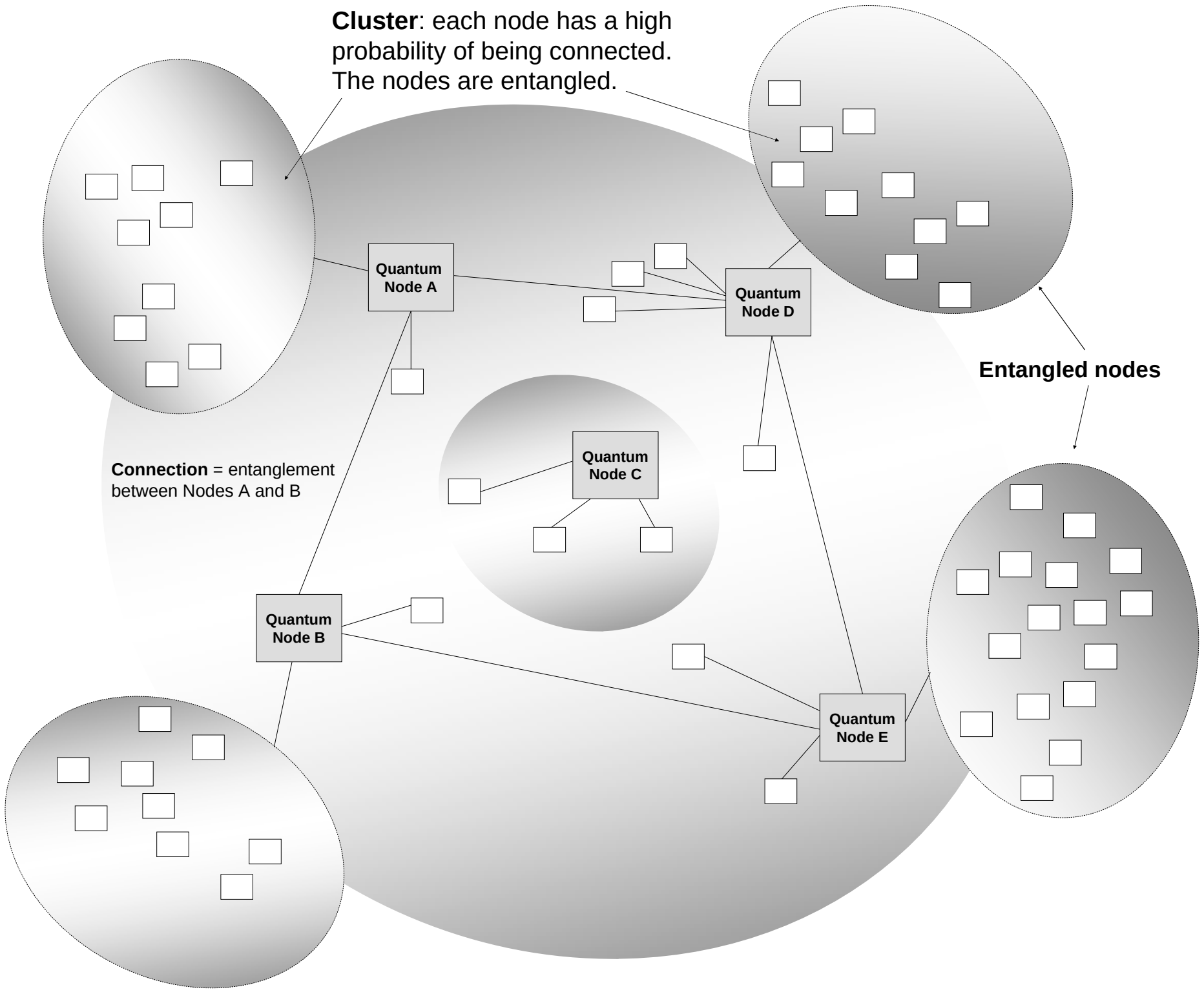

**Fig. 8.22.** The quantum random network. The initial network is weakly connected, but arbitrary connections can be realized by quantum measurements and operations. The shading represents entanglement among the nodes of the network.

The organization of the network is realized on the quantum layer, using true randomness and classically indescribable phenomena. The topology of the quantum network depends on the probabilities of the connections between the nodes, and depending on these probabilities, the network can represent cluster behavior.
There is a very important difference between the random quantum and classical network. In the former case, superposition and the entanglement can be used to generate a fundamentally different network structure. The entanglement between the particles makes it possible to generate more complex connections between the nodes than is possible in the case of a classical network.
In a classical system, the connection probability between the nodes depends on the size of the network. This means that if the probability of the connections between the nodes is small, then the structure of the network will be simple, but if the probability of connection is high, the structure of the network will be more complex. As they have concluded, in the case of a quantum network, the complexity of the network is independent of the connection probabilities between the nodes, and any complexity can be generated between the sub graphs of the quantum network.

## 8.6 Conclusions

Quantum computing offers fundamentally new solutions in the field of computer science. The biggest problem in future quantum communications is the long-distance delivery of quantum information. Since the quantum states cannot be copied, the amplification of quantum bits is a more complex compared to classical communications. There are several differences between a classical and a quantum repeater. The quantum repeater is not a simply signal amplifier, in contrast the classical repeaters. The quantum repeater is based on the transmission of entangled quantum states between the repeater nodes. The main task is the improvement of the fidelities of the shared entangled states. The greatest challenge of long-distance quantum communication is the development of a well-scalable quantum repeater. The entanglement purification is a cardinal question from success point of view during the entanglement sharing process between the repeater stations. The fidelity of the predicated quantum states mostly depends on the noise of the quantum channel. The problem could be solved only if the fidelity can be maximized without the very expensive purification process.
The classical biologically inspired self-organizing systems have increasing complexity and these constructions do not seem to be suitable for handling the service demands of the near future. On the other hand, the cell-organized, quantum mechanics based cellular automata models have many advantages over classical models and circuits, for details see the work of Gyongyosi and Imre from 2010 [Gyongyosi11h]. A truly random quantum network could be a very useful tool for many scientific applications. In the next few years the quantum probabilistic networks could be used to describe the varied phenomena of society, physics, biology and economics.

## 8.7 Related Works

In this section we summarize the most important works regarding on the quantum communication networks.

### The Quantum Repeater

As we have summarized in this section, the biggest problem in the quantum communications of the future is long-distance quantum communication. Since quantum states cannot be copied, the amplification of quantum bits is a more complex question than in the case of classical communication. The success of global quantum communication very strongly depends on the development of a quantum repeater.

The design of quantum repeater mainly based on optical elements. About long-distance quantum communication with optical elements, see the work of Duan et al. [Duan01]. The role of entanglement and the various encoding schemes can be applied in quantum repeaters can be found in the works of Ladd et al. [Ladd06] and Loock et al. [Loock08]. An analysis about the properties of quantum repeaters based on atomic ensembles and linear optics can be found in the article of Sangouard et al. [Sangouard09]. The role of the various photonics modules in the design of quantum repeater was studied by Stephens et al. [Stephens08].

The first entanglement purification approaches were developed by Dür et al. in 1999, [Dür99]. Later, many new purification algorithms were published, such as the greedy scheduling method, introduced by Ladd et al. in 2006 [Ladd06], or the method of entanglement pumping, which also was introduced by Dür et al. in 2007 [Dür07]. A more efficient purification algorithm was developed by Van Meter et al. [VanMeter08] in 2008, called the banded purification scheme. From the viewpoint of practical implementation of quantum repeater, Van Meter et al. have published some very important works in 2008 and 2009, for details see [VanMeter08] and [VanMeter09]. Van Meter et al. have also studied the network integration of quantum repeaters, and the various algorithms for entanglement purification.

Moreover, in 2009 they have defined a new, more powerful algorithm for entanglement purification, for details see [VanMeter09]. Before their result, on entanglement purification and quantum error correction an article was published in 2007, by Dür and Briegel, for details see [Dür07]. For further information about the purification protocol to increase the fidelity of entanglement transmission, see [Sangouard09] and [Stephens08]. A great work on high-fidelity transmission of entanglement over a high-loss free space channel was published by Fedrizzi et al. [Fedrizzi09].

**Quantum Repeaters in Practice**

A practical approach of the quantum repeater is called the “hybrid quantum repeater”, further information about the practical detail can be found in [VanMeter09], [Munro08], [Jiang08]. As we have stated in this section, the *base stations* of the quantum repeaters could be connected by optical fibers, the entangled quantum states are sent through these fibers. The encoding schemes of quantum repeaters were studied by Jiang et al. in 2008, for details see [Jiang08]. Further information about the process of entanglement transmission between the repeater nodes can be found in [Devitt08], [Louis08], [Sangouard09]. About the connection of high-performance networks and the role of quantum repeaters, see the work of Munro et al. [Munro09]. The practical issues of entanglement transmission over optical networks and their relevance in the communication between the repeater nodes are summarized in [Duan01], [Munro10]. A very important work on the performance of practical quantum repeaters was published by Bernardes et al. in 2010, for details see [Bernardes10]. A practical implementation of optical-based quantum repeater was shown in the work of Munro et al. [Munro10]. A practical implementation of quantum repeater was presented by a group with researchers from the University of Vienna and from China [Yuan08]. In 2010, an other important analytical approach was presented with nearly maximally entangled EPR pairs over 1280 kilometers [Bernardes10].

We have also mentioned in this section, that from an engineering point of view, one of the most important questions in the development of a practical quantum repeater is the measurement strategy in its stations. In the first approaches towards designing a quantum repeater, the measurement process was based on the homodyne projection measurement [Dür99]. But later it was shown [Bernardes10] that a quantum repeater can be implemented in a more efficient way using the Unambiguous State Discrimination (USD) method [Duan01]. As was also shown in [Bernardes10], this kind of measurement scheme reduces the number of qubits per half node and increases the rate of the quantum repeater. Another interesting photon-based approach to controlling quantum states was reported in 2006 [Chen06a]. On the phase control of a path-entangled photon state see the great work of Bonato et al. [Bonato10]

**Probabilistic Quantum Networks**

The definition of the Quantum Cellular Automata (QCA) model was formalized by Watrous [Watrous95]. The first implementation of a physical quantum dot QCA machine was introduced by Toth and Lent [Toth01]. In their method, the state of the system converges to the lowest energy state, which state can be regarded as a uniquely defined state in the system. Later, the construction of a physical QCA model was extended to tunnel junctions and other phase changer nano-technological devices, using correlated magnetic and electrical states, or other magnetic and non-magnetic particles such as the metal tunnel junction QCA or the molecular and magnetic QCA [Arrighi07], [Richter96].

About the practical issues of probabilistic quantum networks, see the work Perseguers et al. [Perseguers10]. As has been shown by a research group from Germany, quantum networking can be realized in the microscopic scale—in the quantum regime. As they found in 2010, these quantum networks in the microscopic scale can be used to generate arbitrary connections inside the network, and truly random behavior can be realized between them.

**Error Correction**

The construction of quantum error correcting codes was the first step to a practical quantum computer. The importance of the reduction of the decoherence in practical quantum computing was emphasized by Shor, who was the first to show a scheme for reducing it (but still not yet a quantum error correcting code), in 1995 [Shor95]. In the same year, an important paper was published by Unruh [Unruh95] about maintaining coherence in quantum computers. The first quantum error correcting codes were developed by Shor in the middle of the 1990s [Shor96], and also in the same year, an important step in this field was made by Schumacher and Nielsen [Schumacher96c]. Later, advanced quantum error correcting schemes were developed—known as fault-tolerant quantum computation by Gottesman in 1996 [Gottesman96], and in 1997 [Gottesman97]. A review article about the role of error correcting codes in quantum theory was published by Steane in 1996 [Steane96].

The areas of quantum error correcting and fault tolerant error correcting are the most important fields from the viewpoint of practical quantum computation. A very interesting paper about the possibility of the construction of a perfect quantum error correcting code was published by Laflamme et al. in 1996 [Laflamme96]. About the properties of resilient practical quantum computing an interesting article was published by Knill et al. [Knill98]. Further information about the working mechanisms of quantum error correction and fault-tolerant computation can be found in [Aharonov97], [Shor96], and [Gottesman04a], about the mathematical background of the quantum error-correction codes see the works of Calderbank and Shor [Calderbank96], and of Calderbank et al. [Calderbank97], [Calderbank98].

Our article does not focus on quantum error correcting codes and fault tolerant quantum commutation in detail, however many very good articles and books are available, as presented in the summary list of references, and [Barrett10], [Bennett96a], [Calderbank96], [Dür07], [Kerckhoff10], and [Schumacher96c].

**Handling Quantum Information**

Handling of mixed quantum states has great importance in practical approaches. In 1995, Lo gave a method for the encoding of mixed quantum states [Lo95]. The limits for compression of quantum information carried by ensembles of mixed states were studied in the work of Horodecki [Horodecki98]. The conditions to build reliable quantum computers are summarized in the work of Preskill from 1998 [Preskill98a]. An interesting paper about the quantum coding of mixed quantum states was presented by Barnum et al. in 2001 [Barnum01]. About the engineering correlation and entanglement dynamics in physical systems, see the work of Cubitt and Cirac [Cubitt08c].

An important work on the connection of quantum computing with realistically noisy devices was presented by Knill [Knill05]. About the mechanism of the spin-light coherence for single-spin measurement and control in diamond see the work of Buckley et al. [Buckley10]. A great work on the techniques of designing quantum memories with embedded control, and on the role of photonic circuits for autonomous quantum error correction see the work of Kerckhoff et al. [Kerckhoff10].

On the initialization and manipulation of quantum information stored in silicon by bismuth dopants a great summarization was published by Morley et al. [Morley10]. An experimental realization of the optical readout of single electron spins at nitrogen-vacancy centers in diamond was demonstrated in [Steiner10]. On the physical apparatuses and the experimental background of quantum computing see the work of Hagar et al. [Hagar11].

**Summarize and Outlook**

The properties of long-distance quantum communication with atomic ensembles and linear optics were studied by Duan et al. [Duan01]. On quantum repeaters based on atomic ensembles and linear optics see [Sangouard09]. About deterministic optical quantum computer using photonic module see [Stephens08]. A complete system design for a long-line quantum repeater and the properties of entanglement purification and purification scheduling see the works of Van Meter et al. [VanMeter08], [VanMeter09].

On the experimental demonstration of the quantum repeater we suggest the work of [Yuan08] or see the [WWS11]. Some important results on the rate analysis for a hybrid quantum repeater can be found in the paper of Bernardes et al. [Bernardes10]. On the role of quantum repeaters in high performance quantum communications, see the work of Devitt et al. [Devitt08]. A work on quantum repeater encoding technique was shown by Jiang et al. [Jiang08]. On the properties of the hybrid quantum repeaters see the work of Ladd et al. [Ladd06]. A paper on quantum repeaters using coherent-state communication was published by Loock et al. [Loock08]. About the loss in hybrid qubit-bus couplings and gates see [Louis08]. On high-bandwidth hybrid quantum repeater see the works of Munro et al. [Munro08], [Munro09], [Munro10].

The theory of self-reproducing automata can be found in [Neumann66]. The “game of life” was introduced by Gardner in [Gardner70]. On the mathematical background of the quantum cellular automata see the work of Dam [Dam96], and [Grössing88]. An important work on the microscopic quantum mechanical hamiltonian model of computers see [Benioff80]. On the Invertible cellular automata see [Toffoli90]. On parallel quantum computations with a quantum cellular automata, see the work of Margolus [Margolus91]. The properties of the one-dimensional quantum cellular automata were greatly summarized in [Watrous95]. The connection of quantum cellular automata and quantum lattice gases was shown in [Meyer96]. About quantum computing with quantum-dot cellular automata, see [Toth01]. The various models of quantum cellular automata were summarized by Perez-Delgado and Cheung [Perez-Delgado05]. About the quantum logical functions see [Miller06]. On the $n$-dimensional quantum cellular automata see the work of Arrighi et al. [Arrighi07]. A novel approach for quantum mechanical based autonomic communication was shown by Gyongyosi et al. [Gyongyosi09]. A work on the quantum cellular automata controlled self-organizing networks was published by Gyongyosi and Imre [Gyongyosi11h].

# 9. Conclusions

We reviewed in this paper the colorful palette of communications over quantum channels. In the case of a classical communication channel, we can send only classical information. Quantum channels extend the possibilities, and we can transmit classical information, entanglement assisted classical information, private information and of course, *quantum information*. Contrary to classical channels, quantum channels can be used to construct more advanced communication primitives. Entanglement or the superposed states carry quantum information, which cannot be described classically. Moreover, in the quantum world there exist quantum transformations which can create entanglement or can control the properties of entanglement.

Quantum channels can be implemented in practice very easily e.g. via optical fiber networks or by wireless optical channels, and make it possible to send various types of information. The errors are a natural interference from the noisy environment, and the can be much diverse due to the extended set of quantum channel models.

In the first decade of the 21st century, many revolutionary properties of quantum channels have been discovered. These phenomena were previously completely unimaginable. However, the picture has been dramatically changed forever. In the near future, advanced quantum communication and networking technologies driven by quantum information processing will revolutionize the traditional methods. Quantum information will help to resolve still open scientific and technical problems, as well as expand the boundaries of classical computation and communication systems.

# Acknowledgement

The results discussed above are supported by the grant TAMOP-4.2.2.B-10/1--2010-0009 and COST Action MP1006.

# Appendix

## A.1 Partial Trace

If we have a density matrix which describes only a subset of a larger quantum space, then we talk about the reduced density matrix. The larger quantum system can be expressed as the tensor product of the reduced density matrices of the subsystems, if there is no correlation (entanglement) between the subsystems. On the other hand, if we have two subsystems with reduced density matrices $\rho_A$ and $\rho_B$, then from the overall density matrix denoted by $\rho_{AB}$ the subsystems can be expressed as

$$\rho_A = Tr_B\left(\rho_{AB}\right) \text{ and } \rho_B = Tr_A\left(\rho_{AB}\right), \tag{A.1}$$

where $Tr_B$ and $Tr_A$ refers to the partial trace operators. So, this partial trace operator can be used to generate one of the subsystems from the joint state $\rho_{AB} = \left|\psi_A\right\rangle\left\langle\psi_A\right| \otimes \left|\psi_B\right\rangle\left\langle\psi_B\right|$, then

$$\begin{aligned} \rho_A &= Tr_B\left(\rho_{AB}\right) = Tr_B\left(\left|\psi_A\right\rangle\left\langle\psi_A\right| \otimes \left|\psi_B\right\rangle\left\langle\psi_B\right|\right) \\ &= \left|\psi_A\right\rangle\left\langle\psi_A\right| Tr\left(\left|\psi_B\right\rangle\left\langle\psi_B\right|\right) = \left|\psi_A\right\rangle\left\langle\psi_A\right|\left\langle\psi_B\middle|\psi_B\right\rangle. \end{aligned} \tag{A.2}$$

Since the inner product is trivially $\left\langle\psi_B\middle|\psi_B\right\rangle = 1$, therefore

$$Tr_B\left(\rho_{AB}\right) = \left\langle\psi_B\middle|\psi_B\right\rangle\left|\psi_A\right\rangle\left\langle\psi_A\right| = \left|\psi_A\right\rangle\left\langle\psi_A\right| = \rho_A. \tag{A.3}$$

In the calculation, we used the fact that $Tr\left(\left|\psi_1\right\rangle\left\langle\psi_2\right|\right) = \left\langle\psi_2\middle|\psi_1\right\rangle$. In general, if we have to systems $A = \left|i\right\rangle\left\langle k\right|$ and $B = \left|j\right\rangle\left\langle l\right|$, then the partial trace can be calculated as

$$Tr_B\left(A \otimes B\right) = A\,Tr\left(B\right), \tag{A.4}$$

since

$$\begin{aligned} Tr_2\left(\left|i\right\rangle\left\langle k\right| \otimes \left|j\right\rangle\left\langle l\right|\right) &= \left|i\right\rangle\left\langle k\right| \otimes Tr\left(\left|j\right\rangle\left\langle l\right|\right) \\ &= \left|i\right\rangle\left\langle k\right| \otimes \left\langle l\middle|j\right\rangle \\ &= \left\langle l\middle|j\right\rangle\left|i\right\rangle\left\langle k\right|, \end{aligned} \tag{A.5}$$

where $\left|i\right\rangle\left\langle k\right| \otimes \left|j\right\rangle\left\langle l\right| = \left|i\right\rangle\left|j\right\rangle\left(\left|k\right\rangle\left|l\right\rangle\right)^T$. In this expression we have used the fact that $\left(AB^T\right) \otimes \left(CD^T\right) = \left(A \otimes C\right)\left(B^T \otimes D^T\right) = \left(A \otimes C\right)\left(B \otimes D\right)^T$.

## A.2 Quantum Entanglement

A quantum system $\rho_{AB}$ is separable if it can be written as a tensor product of the two subsystems $\rho_{AB} = \rho_A \otimes \rho_B$. Beside product states $\rho_A \otimes \rho_B$ which represent a composite system consisting of several independent states merged by means of tensor product $\otimes$ similarly to classical composite systems, quantum mechanics offers a unique new phenomenon called *entanglement*. For example the so called *Bell states* (or EPR states, named after Einstein, Podolsky and Rosen) are entangled ones:

$$\begin{aligned} \left|\beta_{00}\right\rangle &= \frac{1}{\sqrt{2}}\left(\left|00\right\rangle + \left|11\right\rangle\right), \\ \left|\beta_{01}\right\rangle &= \frac{1}{\sqrt{2}}\left(\left|01\right\rangle + \left|10\right\rangle\right), \\ \left|\beta_{10}\right\rangle &= \frac{1}{\sqrt{2}}\left(\left|00\right\rangle - \left|11\right\rangle\right), \\ \left|\beta_{11}\right\rangle &= \frac{1}{\sqrt{2}}\left(\left|01\right\rangle - \left|10\right\rangle\right). \end{aligned} \tag{A.6}$$

The characterization of quantum entanglement has deep relevance in quantum information theory. Quantum entanglement is the major phenomenon which distinguishes the classical from the quantum world. By means of entanglement, many classically totally unimaginable results can be achieved in quantum information theory.

## A.3 Fidelity

Theoretically quantum states have to preserve their original superposition during the whole transmission, without the disturbance of their actual properties. Practically, quantum channels are entangled with the environment which results in mixed states at the output. Mixed states are classical probability weighted sum of pure states where these probabilities appear due to the interaction with the environment (i.e., noise). Therefore, we introduce a new quantity, which is able to describe the quality of the transmission of the superposed states through the quantum channel. The quantity which measures this distance is called the *fidelity.*
The fidelity for two pure quantum states is defined as

$$F\left(\left|\varphi\right\rangle, \left|\psi\right\rangle\right) = \left|\left\langle\varphi\middle|\psi\right\rangle\right|^2. \tag{A.7}$$

The fidelity of quantum states can describe the relation of Alice pure channel input state $\left|\psi\right\rangle$ and the received mixed quantum system $\sigma = \sum_{i=0}^{n-1} p_i \rho_i = \sum_{i=0}^{n-1} p_i \left|\psi_i\right\rangle\left\langle\psi_i\right|$ at the channel output as

$$F\left(\left|\psi\right\rangle, \sigma\right) = \left\langle\psi\right|\sigma\left|\psi\right\rangle = \sum_{i=0}^{n-1} p_i \left|\left\langle\psi\middle|\psi_i\right\rangle\right|^2 . \tag{A.8}$$

Fidelity can also be defined for *mixed* states $\sigma$ and $\rho$

$$F\left(\rho,\sigma\right) = \left[Tr\left(\sqrt{\sqrt{\sigma}\rho\sqrt{\sigma}}\right)\right]^2 = \sum_i p_i \left[Tr\left(\sqrt{\sqrt{\sigma_i}\rho_i\sqrt{\sigma_i}}\right)\right]^2 . \tag{A.9}$$

Next we list the major properties of fidelity

$$0 \le F\left(\sigma,\rho\right) \le 1 , \tag{A.10}$$

$$F\left(\sigma,\rho\right) = F\left(\rho,\sigma\right), \tag{A.11}$$

$$F\left(\rho_1 \otimes \rho_2, \sigma_1 \otimes \sigma_2\right) = F\left(\rho_1,\sigma_1\right) F\left(\rho_2,\sigma_2\right), \tag{A.12}$$

$$F\left(U\rho U^\dagger, U\sigma U^\dagger\right) = F\left(\rho,\sigma\right), \tag{A.13}$$

$$F\left(\rho, a\sigma_1 + (1-a)\sigma_2\right) \ge aF\left(\rho,\sigma_1\right) + (1-a)F\left(\rho,\sigma_2\right),\ \ a \in \left[0,1\right]. \tag{A.14}$$

# List of Notations

| | |
|---|---|
| $\otimes$ | Tensor product. |
| $*$ | Hadamard-product. |
| $\lvert\psi\rangle, \langle\psi\rvert$ | Dirac's ket and bra vectors, bra is dual to vector $\lvert\psi\rangle = (\lvert\psi\rangle)^{\dagger} = \left((\lvert\psi\rangle)^{T}\right)^{*}$. |
| $\alpha^{*}$ | Complex conjugate of probability amplitude $\alpha$ . |
| $\lvert\psi\rangle^{\perp}$ | Vector orthogonal to vector $\lvert\psi\rangle$ . |
| $\alpha$ , $\beta$ | Probability amplitudes of generic state $\lvert\psi\rangle$ . |
| $\rho = \sum_i p_i \lvert\psi_i\rangle\langle\psi_i\rvert$ | Density matrix. |
| $Tr(\cdot)$ | Trace operation. |
| $i$ | The complex imaginary, $i^2 = -1$. |
| $\rho = \frac{1}{2}\begin{pmatrix} 1+z & x-iy \\ x+iy & 1-z \end{pmatrix}$ | Density matrix of two-level quantum system. |
| $P_m$ | Projector, if a quantum system is measured in an orthonormal basis $\lvert m\rangle$, then we make a projective measurement with projector $P_m = \lvert m\rangle\langle m\rvert$ . |
| $\mathcal{Z}$ | The von Neumann operator, $\mathcal{Z}= \sum_m \lambda_m P_m$ , where $P_m$ is a projector to the eigenspace of $\mathcal{Z}$ with eigenvalue $\lambda_m$ . For the projectors $\sum_m P_m = I$ , and they are pairwise orthogonal. |
| $\mathcal{P}$ | Set of POVM operators $\{\mathcal{M}_i\}_{i=1}^{m+1}$ , where $\mathcal{M}_i = \mathcal{Q}_i^{\dagger}\mathcal{Q}_i$ , and $\mathcal{Q}_i$ is the measurement operator. For POVM operators $\mathcal{M}_i$ the completeness relation holds, $\sum_i \mathcal{M}_i = I$ . |
| $H(\cdot)$ | Classical Shannon-entropy. |
| $H(A\vert B)$ | Conditional Shannon entropy. |
| $\mathrm{S}(\rho) = -Tr(\rho\log\rho)$ | The von-Neumann entropy of $\rho$ . |
| $D(\rho\Vert\sigma)$ | Quantum relative entropy between quantum systems $\rho$ and $\sigma$ , $Tr\left(\rho(\log\rho - \log\sigma)\right)$. |
| $F = \langle \psi \vert \rho \vert \psi \rangle$ | Fidelity. |
| $U$ | Unitary transformation, $UU^{-1} = I$ . |
| $U^{\dagger}$ | Operator dual (adjugate) to operator $U$ . |
| $U^{-1}$ | Inverse of operator $U$. |
| $U^{n}$ | Operator $U$ for $n$-times. |
| $I$ | Identity operator, identity matrix. |
| $H$ | Hadamard transformation. |
| $X,\ Y,\ Z$ | Pauli $X$, $Y$ and $Z$ transformations. |
| $\vert \beta_{00} \rangle, \vert \beta_{01} \rangle, \vert \beta_{10} \rangle, \vert \beta_{11} \rangle$ | Bell states (EPR states). |
| $\sigma_x, \sigma_y, \sigma_z$ | Pauli operations $X$, $Y$ and $Z$. |
| $\{ p_k, \rho_k \}$ | Ensemble. |
| $Tr_A(\rho_{AB})$ | Partial trace of the operator $\rho_{AB}$ , tracing out system $A$ from the composite system $AB$. |
| $\{x_1, \ldots, x_n\} = \{x_i\}_{i=1}^{n}$ | The set containing $x_1, \ldots, x_n$ . |
| $M_m$ | Measurement operator. |
| $N$ | Classical channel. |
| $\mathcal{N}$ | Quantum channel. |
| $\mathcal{N}_{EB}$ | Entanglement-breaking channel. |
| $\mathcal{N}_{H}$ | Horodecki channel. |
| $\mathcal{N}_{Had.}$ | Hadamard channel. |
| $\mathcal{A}_e$ | Erasure quantum channel. |
| $\mathcal{D}$ | Degrading quantum channel. |
| $\mathcal{N}_2 = \mathcal{D}\mathcal{N}_1$ | Degradable quantum channel, where $\mathcal{D}$ is the degrading quantum channel, and $\mathcal{N}_1$ has lower noise than $\mathcal{N}_2$ . |
| $\mathcal{E}$ | Encoder operator. |
| $\mathcal{D}$ | Decoder operator. |
| $\mathcal{N}_1 \otimes \mathcal{N}_2$ | Joint structure of quantum channels $\mathcal{N}_1$ and $\mathcal{N}_2$ . |
| $\mathcal{N}^{\otimes n}$ | Multiple uses ($n$) of quantum channel $\mathcal{N}$ (joint structure of $n$ parallel channels). |
| $C(\mathcal{N}_1) + C(\mathcal{N}_2)$ | Joint channel capacity of tensor product quantum channels $\mathcal{N}_1$ and $\mathcal{N}_2$ with single measurement. |
| $C(\mathcal{N}_1 \otimes \mathcal{N}_2)$ | Joint cannel capacity of channel structure $\mathcal{N}_1 \otimes \mathcal{N}_2$ . |
| $C_{\mathrm{PROD.}}(\mathcal{N}_1 \otimes \mathcal{N}_2)$ | Joint channel capacity of quantum channels $\mathcal{N}_1$ and $\mathcal{N}_2$ , for product state inputs and joint measurement. |
| $C_{ENT.}\left(\mathcal{N}_1 \otimes \mathcal{N}_2\right)$ | Joint channel capacity of quantum channels $\mathcal{N}_1$ and $\mathcal{N}_2$, for entangled input states and joint measurement. |
| $C\left(\mathcal{N}\right)$ | Asymptotic classical capacity. |
| $C^{(1)}\left(\mathcal{N}\right)$ | Single-use classical capacity. |
| $C_0^{(1)}\left(\mathcal{N}\right)$, $C_0\left(\mathcal{N}\right)$ | Classical zero-error capacity (single-use, asymptotic). |
| $Q\left(\mathcal{N}\right)$ | Asymptotic quantum capacity. |
| $Q^{(1)}\left(\mathcal{N}\right)$ | Single-use quantum capacity. |
| $Q_0^{(1)}\left(\mathcal{N}\right)$, $Q_0\left(\mathcal{N}\right)$ | Quantum zero-error capacity (single-use, asymptotic). |
| $P\left(\mathcal{N}\right)$ | Asymptotic private classical capacity. |
| $P^{(1)}\left(\mathcal{N}\right)$ | Single-use private classical capacity. |
| $C_E\left(\mathcal{N}\right)$ | Entanglement assisted classical capacity. |
| $Q_{\mathcal{A}}\left(\mathcal{N}\right)$ | Assisted quantum capacity, where channel $\mathcal{A}$ assists with $\mathcal{N}$. |
| $C_{ALL}\left(\mathcal{N}\right)$ | Generalized capacity of quantum channels, involves the classical capacities and the quantum capacity. |
| $H\left(A,B\right)$ | Joint entropy. |
| $I\left(A:B\right)$ | Mutual information, the amount of information between two random variables *A* and *B*. |
| $\mathrm{S}\left(A\middle\|B\right)=\mathrm{S}\left(\rho_A\middle\|\rho_B\right)$ | Quantum conditional entropy of quantum systems $\rho_A$ and $\rho_B$. |
| $\mathrm{S}\left(A\right)=\mathrm{S}\left(\rho_A\right)$ | The von Neumann entropy of density operator $\rho_A$. |
| $\mathrm{S}\left(AB\right)=\mathrm{S}\left(\rho_{AB}\right)$ | Quantum joint entropy of quantum systems $\rho_A$ and $\rho_B$. |
| $I\left(A:B\right)=I\left(\rho_A:\rho_B\right)$ | Quantum mutual information, the amount of classical correlation between quantum systems $\rho_A$ and $\rho_B$. |
| $\chi$ | Holevo quantity. |
| $\chi_{\mathcal{N}}$ | Holevo quantity of quantum channel $\mathcal{N}$. |
| $\chi\left(\mathcal{N}\right)=\max\limits_{all\ p_i,\rho_i}\chi_{\mathcal{N}}$ | Maximized Holevo quantity of quantum channel $\mathcal{N}$. |
| $R$ | Channel rate of the communication channel. |
| $I_{coh}\left(\rho_A:\mathcal{N}\left(\rho_A\right)\right)$ | Quantum coherent information between input and output quantum systems $\rho_A$ and $\mathcal{N}\left(\rho_A\right)$. |
| $\mathcal{L}$ | Eavesdropper’s activity on the quantum channel. |
| $P$ | Purification state. |